\documentclass[prd,a4paper,twocolumn,showpacs,preprintnumbers,amsmath,nofootinbib,amssymb,
               floatfix,superscriptaddress]{revtex4}
\usepackage{graphicx}
\usepackage[sort&compress]{natbib}
\usepackage{amssymb}
\usepackage{ulem}
\usepackage{slashed}
\usepackage[usenames]{color}
\usepackage{amsmath,amssymb,mathrsfs}
\usepackage{dsfont}
\usepackage{bigstrut}
\usepackage{morefloats}
\usepackage{multirow}
\usepackage{longtable}
\usepackage{dcolumn}
\begin{document}

\newcommand{\cor}[2]{{\color{Blue}\sout{#1}} {\bf \color{Red} #2}}
\newcommand{\com}[1]{{\bf \color{Green}Comment: #1}}
\newcommand{\be}{\begin{eqnarray}}
\newcommand{\ee}{\end{eqnarray}}
\newcommand{\non}{\nonumber \\}
\newcommand{\po}{{\rm P}}
\newcommand{\npo}{{\rm NP}}
\newcommand{\resi}{a_{-1}}
\newcommand{\oh}{\frac{1}{2}}
\newcommand{\Nst}{N^{*}}
\newcommand{\Dst}{\Delta^{*}}
\newcommand{\mpi}{m_{\pi}}
\newcommand{\gam}{\gamma^{\mu}}
\newcommand{\gan}{\gamma^{\nu}}
\newcommand{\vtau}{\vec{\tau}}
\newcommand{\delpi}{\partial_{\mu}\vec{\pi}}

\newcommand{\gaf}{\gamma^{5}}
\newcommand{\delmu}{\partial_{\mu}}
\newcommand{\delnu}{\partial_{\nu}}
\newcommand{\kst}{K^{*}}
\newcommand{\Sigst}{\Sigma^{*}}

\title{Coupled-channel dynamics in the reactions $\boldsymbol{\pi N\to\pi N,\,\eta N, K\Lambda, K\Sigma}$}

\author{D.~R\"onchen}
\email{d.roenchen@fz-juelich.de}
\affiliation{Institut f\"ur Kernphysik and J\"ulich Center for Hadron Physics, Forschungszentrum J\"ulich, 
D-52425 J\"ulich, Germany}

\author{M.~D\"oring}
\email{doering@hiskp.uni-bonn.de}
\affiliation{Helmholtz-Institut f\"ur Strahlen- und Kernphysik (Theorie) and Bethe Center for Theoretical
Physics,  Universit\"at Bonn, Nu\ss allee 14-16, D-53115 Bonn, Germany}

\author{F.~Huang}
\affiliation{Department of Physics and Astronomy, University of Georgia, Athens, Georgia 30602, USA}

\author{H.~Haberzettl}
\affiliation{Institute for Nuclear Studies and Department of Physics, The George Washington University,
Washington, DC 20052, USA}

\author{J.~Haidenbauer}
\affiliation{Institut f\"ur Kernphysik and J\"ulich Center for Hadron Physics, Forschungszentrum J\"ulich, 
D-52425 J\"ulich, Germany}
\affiliation{Institute for Advanced Simulation, Forschungszentrum J\"ulich, D-52425 J\"ulich, Germany}

\author{C.~Hanhart}
\affiliation{Institut f\"ur Kernphysik and J\"ulich Center for Hadron Physics, Forschungszentrum J\"ulich, 
D-52425 J\"ulich, Germany}
\affiliation{Institute for Advanced Simulation, Forschungszentrum J\"ulich, D-52425 J\"ulich, Germany}

\author{S.~Krewald} 
\affiliation{Institut f\"ur Kernphysik and J\"ulich Center for Hadron Physics, Forschungszentrum J\"ulich, 
D-52425 J\"ulich, Germany}
\affiliation{Institute for Advanced Simulation, Forschungszentrum J\"ulich, D-52425 J\"ulich, Germany}

\author{U.-G.~Mei\ss ner}
\affiliation{Institut f\"ur Kernphysik and J\"ulich Center for Hadron Physics, Forschungszentrum J\"ulich, 
D-52425 J\"ulich, Germany}
\affiliation{Helmholtz-Institut f\"ur Strahlen- und Kernphysik (Theorie) and Bethe Center for Theoretical
Physics,  Universit\"at Bonn, Nu\ss allee 14-16, D-53115 Bonn, Germany}
\affiliation{Institute for Advanced Simulation, Forschungszentrum J\"ulich, D-52425 J\"ulich, Germany}

\author{K. Nakayama}
\affiliation{Institut f\"ur Kernphysik and J\"ulich Center for Hadron Physics, Forschungszentrum J\"ulich, 
D-52425 J\"ulich, Germany}
\affiliation{Department of Physics and Astronomy, University of Georgia, Athens, Georgia 30602, USA}

\begin{abstract} 
Elastic $\pi N$ scattering and the world data of the family of reactions $\pi^- p \to \eta n,$ $K^0\Lambda$, $K^0\Sigma^0$,  $K^+\Sigma^-$, and
$\pi^+p\to K^+\Sigma^+$ are described simultaneously in an analytic, unitary, coupled-channel approach. SU(3) flavor symmetry is used to relate
both the $t$- and the $u$- channel exchanges that drive the meson-baryon interaction in the different channels. Angular distributions,
polarizations, and  spin-rotation parameters are compared with available experimental data. Partial-wave amplitudes are determined and the
resonance content is extracted from the analytic continuation, including resonance positions and branching ratios, and possible sources of
uncertainties are discussed. The results provide the final-state interactions for the ongoing analysis of photo- and electroproduction data. 
\end{abstract}


\maketitle

%%%%%%%%%%%%%%%%%%%%%%%%%%%%%%%%%%%%%%%%%%%%%%%%%%%%%%%%%%%%%%%%%%%%%%%%%%%%%%%%%%%%%%%%%%%%%%%%%%%%%%%%%%%%%%%%%%%%%%%%%%%%%%%%%%%%%%%%%%%%%%%%

\section{Introduction}
To gain insight into the non-perturbative sector of Quantum Chromodynamics (QCD) the knowledge of the excited hadron spectrum is essential,
providing the connection between experiment and QCD. An ab-initio approach to resonance physics is provided by lattice QCD gauge simulations.
Spectra of excited baryons observed on the lattice~\cite{Durr:2008zz, Alexandrou:2008tn, Burch:2006cc, Gattringer:2008vj,Engel:2010my,
Bulava:2010yg, Menadue:2011pd, Melnitchouk:2002eg, Mathur:2003zf, Mahbub:2010rm, Edwards:2011jj} hold the promise to explain the rich
phenomenology in the first, second, and third resonance regions. Once quark masses drop towards the physical limit and finite volume effects,
increasingly important for lighter quark masses, are under full control, comparison to experimental data will be possible. Dynamical
Dyson-Schwinger approaches \cite{Roberts:1994dr, Roberts:2007ji, Roberts:2011cf, Wilson:2011aa} and quark models~\cite{Isgur:1978xj,
Capstick:1986bm, Loring:2001kx, Ronniger:2011td, Gross:2012si, Ramalho:2012ng, Melde:2008yr, Melde:2008dg, Golli:2011jk, Aznauryan:2012ec} also
rely on a comparison with the extracted resonance spectrum. A complementary picture for some resonances is provided by the unitarization of
chiral interactions~\cite{Kaiser:1995cy, Nieves:2001wt, Inoue:2001ip, Meissner:1999vr, Lutz:2001mi, GarciaRecio:2003ks, Baru:2003qq,
Doring:2005bx, Doring:2006pt, Doring:2007rz, Doring:2009uc, Oset:2009vf, Khemchandani:2011et, Khemchandani:2011mf, Doring:2010fw,
Borasoy:2006sr, Borasoy:2007ku, Bruns:2010sv, Ruic:2011wf,Mai:2012wy},  using directly the hadronic degrees of freedom and, as such approaches
are formulated for at most a few partial waves, one again needs a data analysis in terms of partial waves to compare to.

The reliable extraction of amplitudes and of the resonance spectrum from data, therefore, is a prerequisite for comparing theoretical approaches
to experiment. Most resonances have been identified through elastic $\pi N$ scattering in the past~\cite{Cutkosky, Hoehler1, Hoehler2} and to
this day by the GWU/SAID analysis group~\cite{Arndt:1985vj, Arndt:1995bj, Arndt:1998nm, Arndt:2003if, Arndt:2005dg, Arndt:2006bf,
Workman:2008iv, Arndt:2008zz}. A new experimental window has opened through the recent high-precision photon-beam facilities, e.g., at  ELSA,
GRAAL, JLab, SPring-8 and MAMI, allowing for resonance identification and the determination of their electromagnetic
properties~\cite{Klempt:2009pi, Aznauryan:2011qj}. 

Combining different reactions for resonance extraction enables one to determine those states which couple only weakly to $\pi N$. The
simultaneous analysis of different final states of pion- and photon-induced reactions in coupled-channel approaches is, thus, the method of
choice to study reaction dynamics. Dynamical coupled-channel (DCC) models \cite{Schutz:1994ue, Schutz:1994wp, Schutz:1998jx, Krehl:1999km,
Gasparyan:2003fp, Doring:2009bi, Doring:2009yv, Doring:2010ap, Haberzettl:1997jg, Haberzettl:2006bn, Haberzettl:2011zr, Huang:2011as,
Matsuyama:2006rp, JuliaDiaz:2007kz, Durand:2008es, Paris:2008ig, Suzuki:2008rp, Kamano:2008gr, Kamano:2010ud, Suzuki:2009nj, Chen:2007cy,
Tiator:2010rp, Surya:1995ur, Kondratyuk:2003zm, Wagenaar:2009gr, Pascalutsa:2000bs, Pascalutsa:2004pk,Fuda:2003pd} provide a sophisticated tool
for analyzing excited baryons and extracting resonance parameters as they obey theoretical constraints of the $S$-matrix such as
analyticity.  

The DCC model developed and employed in this study (J\"ulich2012 model) is based on an approach pursued over many years \cite{Schutz:1994ue,
Schutz:1994wp, Schutz:1998jx, Krehl:1999km, Gasparyan:2003fp, Doring:2009bi, Doring:2009yv, Doring:2010ap}.  The scattering amplitude is
obtained as the solution of a Lippmann-Schwinger-type scattering equation defined by time-ordered perturbation theory (TOPT) which automatically
implements unitarity constraints. Also, the important $\pi\pi N$ channels are included dynamically in the sense that the $\pi\pi$ and $\pi N$
subsystems match the respective phase shifts. Based on effective Lagrangians, the non-resonant interactions are given by $t$- and $u$-channel
exchanges of known mesons and baryons, while $s$-channel processes can be considered as bare resonances. The explicit treatment of the 
background in terms of $t$- and $u$-channel diagrams introduces strong correlations between the different partial waves and generates a
non-trivial energy and angular dependence of the observables. Analyticity is respected by including dispersive contributions of intermediate
states, as well as the correct structure of branch points~\cite{Frazer:1964zz, Cutkosky:1990zh, Ceci:2011ae}. Thus, a reliable determination of
resonance properties given in terms of pole positions and residues of the scattering amplitude in the complex energy plane is possible. 

The extension of the J\"ulich model to photoproduction within a gauge-invariant approach \cite{Haberzettl:1997jg, Haberzettl:2006bn,
Haberzettl:2011zr} has been accomplished recently \cite{Huang:2011as}. There are a number of studies of photon-induced reactions within
phenomenological dynamical models   \cite{Pascalutsa:2000bs, Pascalutsa:2004pk, Fuda:2003pd, Caia:2005hz, Chen:2007cy, Matsuyama:2006rp,
JuliaDiaz:2007fa, Kamano:2009im, Tanabe:1985rz, Yang:1985yr, Nozawa:1989pu, Pearce:1990uj, Gross:1992tj, Surya:1995ur, Hung:2001pz}. The vast
majority of them, however, do not satisfy the gauge invariance condition as dictated by the generalized Ward-Takahashi identity
\cite{Kazes59,Haberzettl:1997jg}. Exceptions to this are, e.g., the works of Refs.~\cite{Pascalutsa:2000bs, Pascalutsa:2004pk, Caia:2005hz}. 

The adaptation of DCC models to the finite volume, to allow for the prediction of lattice levels and the calculation of finite volume
corrections, was pioneered in Ref.~\cite{Doring:2011ip}. In principle, such extensions of hadronic approaches allow for the analysis of
experimental and lattice data on the same footing~\cite{Doring:2011vk,Doring:2011nd,Doring:2012eu}.

In Ref.~\cite{Doring:2010ap}, the J\"ulich model is extended to the kaon-hyperon sector.  SU(3) flavor symmetry is exploited to relate different
coupling constants; SU(3) symmetry is broken e.g. by physical masses but also by different cut-offs in the form factors of the vertices. The
latter absorb other sources of SU(3) breaking that are not systematically addressed as possible in chiral perturbation theory. A resonance
analysis of the isospin $I=3/2$ sector was performed by fitting simultaneously the elastic $\pi N$ partial-wave amplitudes of the GWU/SAID
analysis~\cite{Arndt:2006bf}  and $\pi^+ p\to K^+\Sigma^+$ differential cross sections, polarizations, and spin-rotation data
\cite{Candlin:1982yv,Candlin:1983cw,Candlin:1988pn}. To achieve a good description of the $K^+\Sigma^+$ data, the inclusion of resonances with
total spin up to $J=7/2$ turned out to be necessary. 

In the present study, we extend this resonance analysis to the isospin $I=1/2$ sector, considering the world data of the set of reactions $\pi^-
p \to \eta n,$ $K^0\Lambda$, $K^0\Sigma^0$,  $K^+\Sigma^-$, and $\pi^+p\to K^+\Sigma^+$, together with $\pi N\to\pi N$ scattering. Besides the
$\pi N$, $\eta N$, $K\Lambda$ and $K\Sigma$ channels, the approach includes three effective $\pi\pi N$ channels, namely $\pi\Delta$, $\sigma N$
and $\rho N$. The considered energy range has been extended beyond 2~GeV and resonances up to $J=9/2$ are included now. For older
coupled-channel analyses of pion-induced inelastic reactions see, e.g., Refs.~\cite{Abaev:1996vn, Sotona:1988fm, Saxon:1979xu}.

The present effort is a first step towards a global analysis of pion- and photon-induced production of $\pi N$, $\eta N$, $K\Lambda$ and
$K\Sigma$. For example, photoproduction data with unprecedented accuracy  were measured recently at ELSA, MAMI, JLab and other facilities for
the $KY$ final states~\cite{McNabb:2003nf, Bradford:2005pt, McCracken:2009ra, Tran:1998qw, Glander:2003jw, Bockhorst:1994jf, Ewald:2011gw,
Castelijns:2007qt, Sumihama:2005er, Lleres:2007tx, Hicks:2007zz, Zegers:2003ux, Lleres:2008em, Bradford:2006ba, Dey:2010hh, Kohri:2006yx,
thesis, Lawall:2005np}, the latter being also accessible through nucleon-nucleon collisions at COSY~\cite{Valdau:2010kw, AbdelBary:2012vw} or
HADES~\cite{Agakishiev:2012ja, Agakishiev:2012qx, Agakishiev:2012xk}. New data for pion-induced reactions are expected by experiments such as
EPECUR at ITEP for the $K\Lambda$ final state~\cite{Alekseev:2012zu} and E19 at J-PARC for the charged $K\Sigma$ final
states~\cite{Shirotori:2012ka}.

This paper is organized as follows: An overview of the formalism is given in Sec.~\ref{sec:formal}, including a new renormalization scheme in
Sec.~\ref{sec:renor}. Details on the data base and the fitting techniques are documented in Sec.~\ref{sec: results}, together with the
comparison of the present results to data. In particular, results of two fit scenarios (A and B) are presented which differ predominantly in
the strength of the coupling of the $\pi N$ system to the effective $\pi\pi N$ channels,  notably to $\sigma N$.  The description of
observables achieved in the various reaction channels is discussed in detail. Extracted resonance pole properties are presented and discussed in
Sec.~\ref{sec:polpos}. The technical details and values of fit parameters shown in the Appendices ensure the reproducibility of the results.

%%%%%%%%%%%%%%%%%%%%%%%%%%%%%%%%%%%%%%%%%%%%%%%%%%%%%%%%%%%%%%%%%%%%%%%%%%%%%%%%%%%%%%%%%%%%%%%%%%%%%%%%%%%%%%%%%%%%%%%%%%%%%%%%%%%%%%%%%%%%%%%%

\section{Formalism}
\label{sec:formal}
Theoretical requirements on the scattering amplitude include two-body unitarity which is realized in most of the current approaches to data
analysis, including those formulated with the $K$-matrix~\cite{Manley:1992yb, Penner:2002ma, Penner:2002md, Shklyar:2004dy, Shklyar:2009cx,
Shklyar:2012js, Scholten:1996mw, Lee:1999kd, Anisovich:2004zz, Sarantsev:2005tg, Anisovich:2010an, Anisovich:2011fc, Anisovich:2012ct,
Shrestha:2012va,Shrestha:2012ep}. Dispersive contributions of the two-body intermediate states are neglected in $K$-matrix approaches. However,
there are calculations in which dispersive parts are included in a factorized form, e.g. CMB-type approaches \cite{Batinic:1995kr, Vrana:1999nt,
Ceci:2006ra, Ceci:2008zz, Batinic:2010zz} or the Chew-Mandelstam formalism used in the GWU/SAID analysis~\cite{Arndt:1985vj, Arndt:1995bj,
Arndt:1998nm, Arndt:2003if, Arndt:2005dg, Arndt:2006bf, Workman:2008iv, Arndt:2008zz}.

In addition to two-body unitarity, three-body unitarity is an important issue when analyzing the baryon spectrum due to the known large
inelasticities caused by the $\pi\pi N$ channel. In the current approach, the $\pi\pi N$ intermediate states are parameterized as $\pi\Delta$,
$\sigma N$, and $\rho N$ channels. The properties of the unstable particles in these channels are matched to the corresponding phase shifts,
i.e. the $\Delta$ fits the $P_{33}$ partial wave in $\pi N$ scattering, and the $\sigma$ and $\rho$ mesons fit the $\pi\pi$ phase shifts for the
corresponding quantum numbers~\cite{Krehl:1999km,Doring:2009yv}.  It should be stressed that the current approach provides the correct analytic
structure, including branch points in the complex plane from three-particle intermediate states. In Ref.~\cite{Ceci:2011ae} it was shown that
the latter  are indeed of practical relevance, to avoid erroneous resonance signals.

\begin{figure}
\begin{center}
\includegraphics[width=0.3\textwidth]{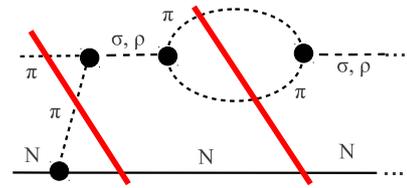}
\end{center}
\caption{(Off-shell) pion exchange as a consequence of three-body unitarity~\cite{Aaron:1969my}.}
\label{fig:threecut}
\end{figure}

For three-body unitarity, the consistent inclusion of three-body cuts is essential~\cite{Aaron:1969my} as illustrated in
Fig.~\ref{fig:threecut}.
In particular, three-body unitarity dictates the inclusion of the discontinuities induced by $t$-channel exchanges simultaneously with the
discontinuities  arising from three-body states propagating in the $s$-channel. Using dispersion relations, the form of the $t$-channel and
$s$-channel propagators can be calculated from these discontinuities~\cite{Aaron:1969my}. It should be noted that this implies the full
off-shell treatment of the exchange potential. In addition to $t$-channel exchanges, it is always possible to add more interactions as long as
they do not induce cuts, such as contact terms.

Thus, meson exchange arises naturally from requirements of the $S$-matrix. While pion exchange is included in the transitions potentials (cf.
Figs.~\ref{fig:dia1} and \ref{fig:dia2}.) it should be noted that full three-body unitarity requires additional diagrams that are beyond the
scope of this work. Those comprise, e.g., the dressing of the nucleon in the $\pi N$ $s$-channel propagator, as induced by the nucleon exchange
term in $\pi N\to\pi N$. This contribution is known to be small, though~\cite{Aaron:1969my}. Baryon-exchange diagrams are also mandatory for the
approximation of the left-hand cut on the amplitude level. The extension to channels involving particles with strangeness then requires
additional diagrams.

All resulting $t$- and $u$-channel processes of the current approach are shown in Figs.~\ref{fig:dia1} and \ref{fig:dia2}.  For the transitions
$MB\to MB$, with the $J^P=0^-$ octet mesons $M$ and the $J^P=1/2^+$ octet baryons $B$, we have included the full set of diagrams required by
SU(3) symmetry. This applies to the classes of diagrams of the exchange of $J^P=1^-$ vector mesons, $J^P=1/2^+$ baryons, and $J^P=3/2^+$
baryons. In particular, non-diagonal transitions such as $\eta N\to KY$ are considered. 
\begin{figure*}
\begin{center}
\includegraphics[width=0.8\textwidth]{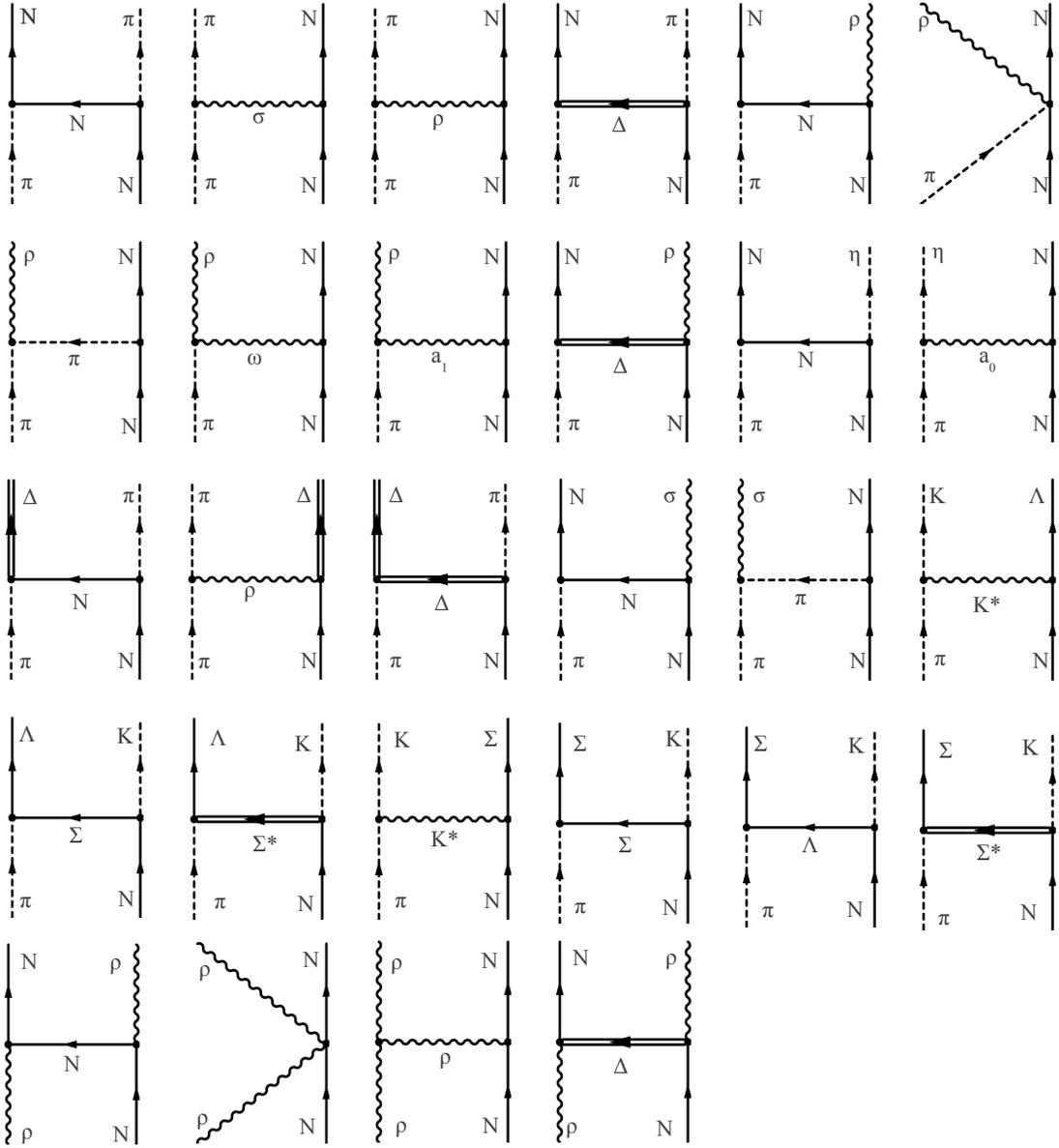}
\end{center}
\caption{Hadron exchanges in the $t$- and $u$-channel employed in the current approach. Not shown are $s$-channel processes (resonances).}
\label{fig:dia1}     
\end{figure*}

\begin{figure*}
\begin{center}
\includegraphics[width=0.8\textwidth]{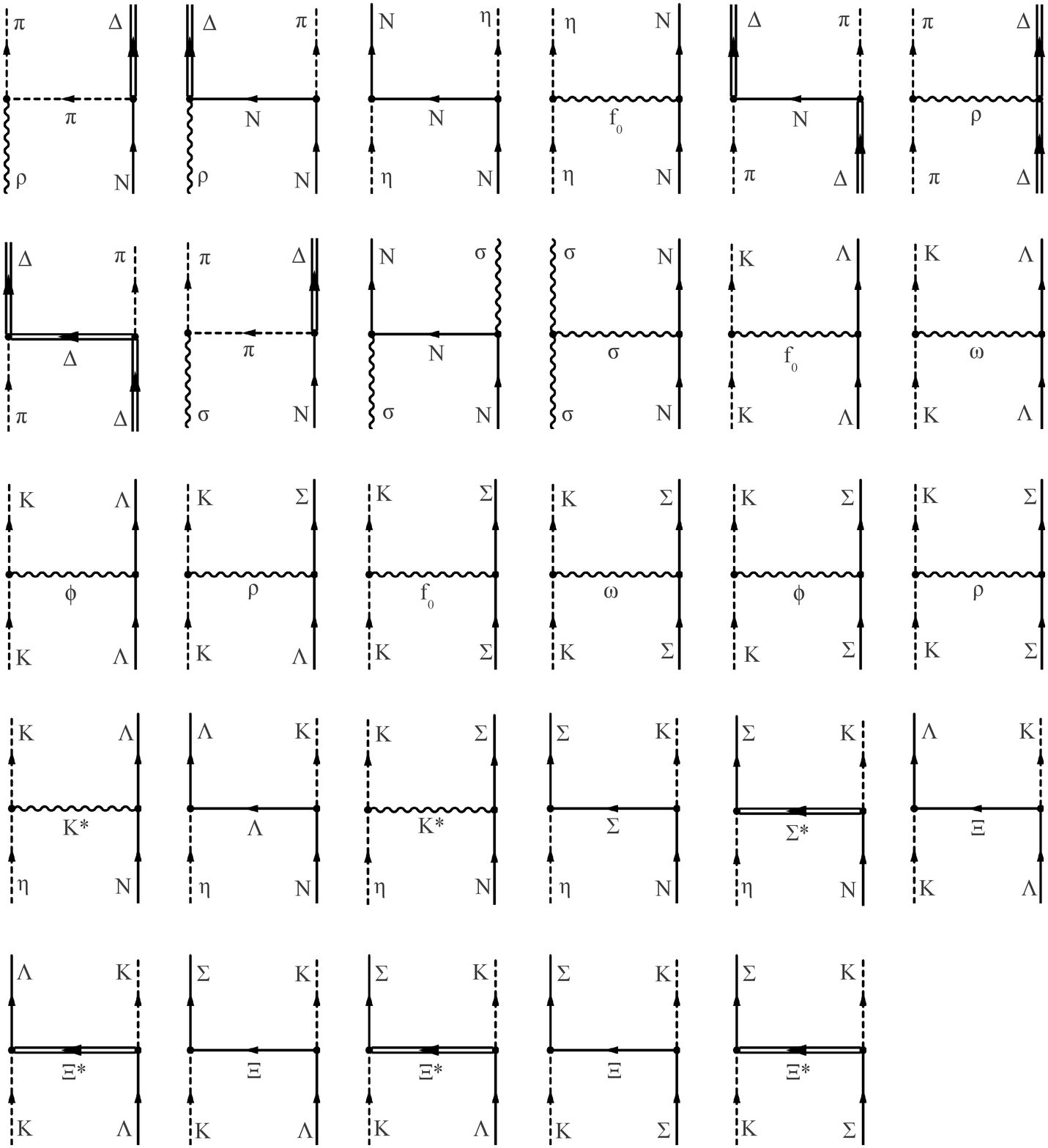}
\end{center}
\caption{Continuation of Fig.~\ref{fig:dia1}.}
\label{fig:dia2}     
\end{figure*}

\begin{figure}
\begin{center}
\includegraphics[width=0.38\textwidth]{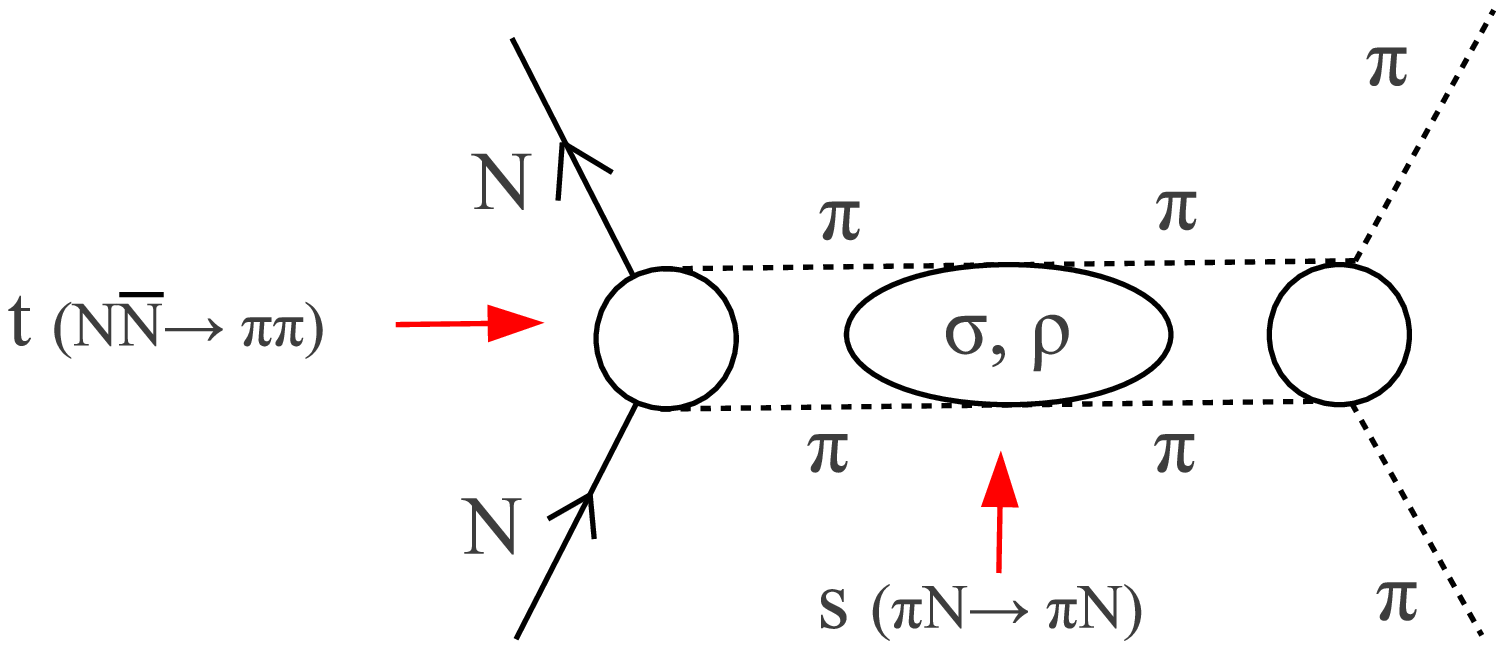}
\end{center}
\caption{Exchanges with $\sigma(600)$ and $\rho(770)$ quantum
numbers in the $t$-channel, determined by using dispersion techniques and crossing
symmetry~\cite{Schutz:1994ue}.
}
\label{fig:dispersion}
\end{figure}

At this point, we would like to stress that in the current approach the $\pi N$ interactions with the $\sigma(600)$ and $\rho(770)$ quantum
numbers in the $t$-channel, as shown in Fig.~\ref{fig:dispersion}, are determined from analytically continued $N\bar N\to \pi\pi$ data, using
dispersive techniques and crossing symmetry (see Ref.~\cite{Schutz:1994ue} and references therein).
Note that for low energies, where the $\pi\pi N$ channel plays no significant role, more systematic schemes using dispersive techniques for the
amplitude extraction have been developed recently~\cite{Ditsche:2012fv,Hoferichter:2012wf}.  Chiral perturbation theory also allows for a
model-independent understanding of the near-threshold behavior~\cite{Fettes:1998ud, Fettes:2001cr, Gasparyan:2010xz, Gasparyan:2011yw,
Griesshammer:2012we}.

The interaction potentials $V$, given by $t$- and $u$-channel exchanges of hadrons as shown in  Figs.~\ref{fig:dia1} and \ref{fig:dia2}, are
derived from the Wess-Zumino interaction~\cite{Wess:1967jq,Meissner:1987ge} plus additional terms. These Lagrangians, in combination with the
form factors, determine also the off-shell behavior of the amplitude. For analytic expressions of the exchange diagrams, see
Appendix~\ref{sec:ExApp}. The Lagrangians and vertices for the $s$-channel exchanges are listed in Appendix B of Ref.~\cite{Doring:2010ap}; see
also Appendix~\ref{sec:respar} of this study.

In the present approach, an integral equation is solved to obtain the $T$-matrix based on the exchange processes. From the resulting $T$-matrix,
the observables in pion-induced reactions can be calculated~\cite{Doring:2010ap} (for photon-induced reactions, see
Refs.~\cite{Haberzettl:2006bn, Haberzettl:2011zr, Huang:2011as}). The scattering equation in the center-of-mass (c.m.) frame, for a given
partial wave, reads \cite{Krehl:1999km, Gasparyan:2003fp, Doring:2009bi, Doring:2009yv, Doring:2010ap} 
\begin{multline}
T_{\mu\nu}(p'',p',z)=V_{\mu\nu}(p'',p',z)\\
+\sum_{\kappa}\int\limits_0^\infty dp\,
 p^2\,\frac{V_{\mu\kappa}(p'',p,z)\,T_{\kappa\nu}(p,p',z)}{z-E_a(p)-E_b(p)+i\epsilon} \ , 
\label{scattering}
\end{multline}
where $p''\equiv|\vec p\,''|$ ($p'\equiv |\vec p\,'|$) is the modulus of the outgoing (incoming)  three-momentum that may be on- or off-shell, 
$z\equiv\sqrt{s}\equiv E$ is the scattering energy, and $\mu,\,\nu,\,\kappa$ are channel indices. Note also that the different possibilities  of
the $\pi\Delta$ and $\rho N$ states to couple to a given $J^P$ are organized in additional channels as listed in Table~\ref{tab:couplscheme}. In
Eq.~(\ref{scattering}), $E_a=\sqrt{m_a^2+p^2}$ and  $E_b=\sqrt{m_b^2+p^2}$ are the energies of the intermediate particles $a$ and $b$ in channel
$\kappa$. Eq.~(\ref{scattering}) is formulated in the partial-wave basis, i.e. the amplitude only depends on the modulus of the incoming,
outgoing, and intermediate  particle momenta. This implies a partial-wave decomposition of the exchange
potentials~\cite{Krehl:1999km,Gasparyan:2003fp} (see also Table~\ref{tab:couplscheme} in Appendix~\ref{sec:ExApp}). The angular dependence of
the full $T$-matrix is provided by the Wigner  $d_{\lambda\lambda'}^J(\vartheta)$-functions in the partial-wave
decomposition~\cite{Krehl:1999km,Gasparyan:2003fp},  where $\vartheta$ is the scattering angle and $\lambda\,(\lambda')$ is the helicity of the
incoming (outgoing) baryon. The denominator in Eq.~(\ref{scattering}) corresponds to the channels with stable particles, $\pi N$, $\eta N$,
$K\Lambda$, and $K\Sigma$; for the effective $\pi\pi N$ channels ($\pi\Delta$, $\sigma N$, $\rho N$), the propagator is  more
complex~\cite{Krehl:1999km,Doring:2009yv}.  While in the present approach the resonances acquire their widths from the solution of the 
scattering equation~(\ref{scattering}), there are also other approaches to unstable particles (see, e.g., the complex-mass scheme of
Ref.~\cite{Djukanovic:2010zz}; see also Ref.~\cite{Cordon:2012xz}).

It should be stressed that throughout this study we use isospin-averaged masses. For the different $K\Sigma$ final states considered, in
principle, the threshold energies are different which could have quantitative effects close to threshold; however, all $K\Sigma$ data available
lie considerably above the $K\Sigma$ thresholds and effects from differences in those thresholds are negligible. Similar considerations hold for
electromagnetic effects. For elastic $\pi N$ scattering, we do not fit to data anyway but to the GWU/SAID amplitudes~\cite{Arndt:2006bf} (cf.
Sec.~\ref{sec:pin}) which include corrections for Coulomb effects.

Having outlined the approach, it is obvious that hadron exchange connects all partial waves of the scattering amplitude, putting strong
constraints on the form of the amplitudes and providing at the same time a background from $t$- and $u$-chan\-nel processes.  Together with a
minimal set of $s$-channel processes (``genuine'' resonances), this allows for a reliable extraction of poles and residues from the analytic
continuation of the amplitude as derived in Ref.~\cite{Doring:2009yv}. 

%%%%%%%%%%%%%%%%%%%%%%%%%%%%%%%%%%%%%%%%%%%%%%%%%%%%%%%%%%%%%%%%%%%%%%%%%%%%%%%%%%%%%%%%%%%%%%%%%%%%%%%%%%%%%%%%%%%%%%%%%%%%%%%%%%%%%%%%%%%%%%%%

\subsection{Decomposition of the amplitude}
\label{sec:deco}
The scattering amplitude $T_{\mu\nu}$ can be decomposed into a non-pole and a pole part as derived in this section. In the present study, this
decomposition merely serves to speed up to the numerical evaluation; no physical meaning is attached to the components of the decomposition.

The sum of the $u$- and $t$-channel diagrams is labeled as $V^\npo$ in the following. Together with the (bare) $s$-channel exchanges $V^\po$,
they constitute the interaction $V$ in Eq.~(\ref{scattering}),
\be
V_{\mu\nu}=V^\npo_{\mu\nu}+V^\po_{\mu\nu}\equiv V^\npo_{\mu\nu}+\sum_{i=0}^{n} \frac{\gamma^a_{\mu;i}\,\gamma^c_{\nu;i}}{z-m_i^b} \ 
\label{blubb}
\ee
with $n$ being the number of bare $s$-channel states in a given partial wave. To simplify the notation, the explicit dependence on momenta and
energy are omitted here and in the following. In Eq.~(\ref{blubb}), $\gamma^c$ [$\gamma^a$] are the bare resonance creation [annihilation]
vertices, indicated by the subscript $c$ [$a$], of the $s$-channel states of bare mass $m_i^b$. The bare vertices for $J\le 3/2$ are derived
from Lagrangians and given in Appendix B of Ref.~\cite{Doring:2010ap}. The vertex functions for $J\geq 5/2$ are given in Eq.~(B.3) of
Ref.~\cite{Doring:2010ap} and Eq.~(\ref{higher1}) in this study.

The $u$-channel diagrams from nucleon, $\Delta(1232)$, $\Lambda$, $\Sigma$, $\Sigma^*(1385)$, $\Xi$ and $\Xi^*(1530)$ exchanges are included
with physically known coupling strengths,  while $u$-channel diagrams from other baryonic resonances are neglected. Those would introduce
additional parameters which are difficult to adjust, for the diagrams do not introduce strong energy dependencies (for a discussion of
$u$-channel contributions, see Ref.~\cite{Lutz:2001yb}). 

The unitarization of $V^\npo$ leads to the definition  of $T^\npo$,
\be
T_{\mu\nu}^\npo&=& V_{\mu\nu}^{\npo}+ \sum_\kappa   V_{\mu\kappa}^{\npo}G_\kappa T_{\kappa\nu}^\npo \ ,
\label{ttnpo}
\ee
the so-called {\it non-pole} part of the full $T$ matrix (projected to a partial wave).  The $G_\kappa$ are the propagators of two- or
three-particle intermediate states [i.e., for the two-particle case, $G_\kappa=(z-E_a-E_b+i\epsilon)^{-1}$ according to Eq.~(\ref{scattering})].
Note that in Eq. (\ref{ttnpo}), and also Eq.~(\ref{dressed}) below, there is an integration over the momentum of the intermediate state, cf.
Eq.~(\ref{scattering}), not written explicitly for simplicity.

Contributions from the $s$-channel exchanges, the {\it pole part} $T^\po$, can be evaluated from the non-pole part $T^\npo$ given in
Eq.~(\ref{ttnpo}).  For this, we define the quantities
\begin{eqnarray}
\Gamma_{\mu;i}^c	&=&\gamma^c_{\mu;i}+\sum_\nu  \gamma^c_{\nu;i}\,G_\nu\,T_{\nu\mu}^\npo \ , \non
\Gamma_{\mu;i}^a	&=&\gamma^a_{\mu;i}+\sum_\nu  T_{\mu\nu}^\npo\,G_\nu\,\gamma^a_{\nu;i} \ , \non
\Sigma_{ij}		&=&\sum_\mu  \gamma^c_{\mu;i}\,G_\mu\,\Gamma^a_{j;\mu} \;,
\label{dressed}
\end{eqnarray}
where $\Gamma^c$ [$\Gamma^a$] are the dressed resonance creation [annihilation] vertices and $\Sigma$ is the self-energy. The indices $i,j$
label the $s$-channel state in the case of multiple resonances. For the two-resonance case, the pole part reads explicitly~\cite{Doring:2009uc}
\be
T^{\po}_{\mu\nu}&=&\Gamma^a_{\mu}\, D^{-1} \, \Gamma^c_\nu \ , \text{where} \non
\Gamma^a_\mu&=&(\Gamma^a_{\mu;1},\Gamma^a_{\mu;2}), \quad
\Gamma^c_\mu=\left(
\begin{matrix}
\Gamma^c_{\mu;1}\\
\Gamma^c_{\mu;2}
\end{matrix}
\right), \non
  D&=&\left(\begin{matrix}
z-m^b_1-\Sigma_{11}&&-\Sigma_{12}\\
-\Sigma_{21}     &&z-m^b_2-\Sigma_{22}
\end{matrix}
\right) \ ,
\label{2res}
\ee
from which the one-resonance case follows immediately.  

It is easy to show that the full scattering $T$ matrix of Eq.~(\ref{scattering}) is given by the sum of pole and non-pole part,
\begin{equation}
T_{\mu\nu}=T_{\mu\nu}^\po+T_{\mu\nu}^\npo \, .
\label{deco1}
\end{equation} 
This decomposition is widely used in the literature,  see e.g. Refs.~\cite{Matsuyama:2006rp,Afnan:1980hp, Lahiff:1999ur}. The non-pole part
$T^\npo$ is sometimes referred to as {\it background}, although the unitarization of Eq.~(\ref{ttnpo}) may lead to dynamically generated poles
in $T^\npo$. In turn, $T^\po$ contains constant and other terms beyond the pole contribution of the Laurent expansion~\cite{Doring:2009bi}. 
Consequently, the conclusion was drawn that the cleanest separation into a background and a resonance part is given by the separation into a
singularity-free part and the part $a_{-1}/(z-z_0)$ that is the leading term in the Laurent expansion of the scattering amplitude in the complex
energy plane about the pole position $z=z_0$ with residue $a_{-1}$.

However, in the present study, we use the decomposition of Eq.~(\ref{deco1}), because the calculation of $T^\po$ is numerically much faster than
that of $T^\npo$. In a fit of only $s$-channel parameters, it is thus convenient to calculate $T^\npo$ once and then fit the resonance
parameters, which only requires the multiple re-evaluation of $T^\po$ (cf. Sec.~\ref{sec:numerics}). 

%%%%%%%%%%%%%%%%%%%%%%%%%%%%%%%%%%%%%%%%%%%%%%%%%%%%%%%%%%%%%%%%%%%%%%%%%%%%%%%%%%%%%%%%%%%%%%%%%%%%%%%%%%%%%%%%%%%%%%%%%%%%%%%%%%%%%%%%%%%%%%%%

\subsection{Renormalization of the nucleon mass and coupling}
\label{sec:renor}
The nucleon is included in our formalism as an explicit $s$-channel state which is dressed via Eq.~(\ref{dressed}) to acquire the correct
physical nucleon mass and $\pi NN$ coupling. This means, to ensure a nucleon pole at the position of the physical nucleon mass $m_N=939$~MeV and
with the physical coupling of $f_{\pi NN}= 0.989$ \cite{Janssen:1996kx}, one has to implement a renormalization procedure in the $P_{11}$
partial wave. 

The renormalization in case of one bare $s$-channel state, coupling to one channel ($\pi N$), is well understood and has been applied, e.g. in
Refs.~\cite{Koch:1985bp,Lahiff:1999ur}. For illustration, we repeat the essential steps in the current notation. 

Following Eq.~(\ref{2res}), the pole part $T^\po$ in the one-resonance case, with an $s$-channel nucleon exchange of bare mass $m_N^b$, can be
written as 
\begin{eqnarray}
 T^\po=\frac{\Gamma^a_{1}\,\Gamma_1^c}{z-m^{b}_N-\Sigma_{11}} 
\label{TP1chan}
\end{eqnarray}
where here and in the following the omission of the channel index means that we refer to the $\pi N$ channel.
We require that $T^\po$ has a pole at the physical nucleon mass $m_N$: 
\begin{eqnarray}
z_0= m^{b}_N+\Sigma_{11}(z_0) \;\;\;\; \text{with} \;\;\; z_0=m_N\,.
\label{renorma1}
\end{eqnarray}
To derive an expression for $m^b_N$ that respects the renormalization condition of Eq.~(\ref{renorma1}) we expand the nucleon self-energy about
$z=z_0$
\begin{eqnarray}
 \Sigma_{11}(z)=\Sigma_{11}(z_0)+(z-z_0)\frac{\partial \Sigma_{11}(z)}{\partial z}\bigg|_{z=z_0}+{\cal O}(z-z_0)^2\non
\end{eqnarray}
and define the reduced quantities $\tilde{\Sigma}_{11}$ and $\tilde{\Gamma}^{a,c}_1$ taking advantage of the fact that the bare coupling
$f_{N}^b$ can be factorized off $\Sigma$ and $\Gamma$,
\begin{eqnarray}
 \Sigma_{11}&=& (f_N^b)^{2}\Sigma_{11}^{\text{red}}=x^{2}f_{\pi NN}^{2}\,\Sigma_{11}^{\text{red}}=x^2\,\tilde{\Sigma}_{11} \non
\Gamma_{1}^{a,c}&=&f_{N}^b \Gamma_{1}^{\text{red}; a,c}=x\,f_{\pi NN}\, \Gamma_{1}^{\text{red}; a,c}=x\,\tilde{\Gamma}_1^{a,c}\, ,
\label{redquan1}
\end{eqnarray}
with $x \in \mathbb{R}$ and the bare $\pi NN$ coupling constant $f_N^b$,
\begin{eqnarray}
 f_N^b&=& x\,f_{\pi NN}\; .
\label{fNb1}
\end{eqnarray}
In Eq.~(\ref{redquan1}), $\tilde{\Sigma}_{11}$ is the nucleon self-energy calculated with the physical nucleon coupling $f_{\pi NN}$, instead of
the bare coupling. The same applies to $\tilde{\Gamma}_1^{a,c}$. $T^\po$ from Eq.~(\ref{TP1chan}) reads now
\begin{eqnarray}
 T^\po=\frac{1}{z-z_0}
 \left(\frac{x^2\tilde{\Gamma}^a_{1}\tilde{\Gamma}_{1}^c}{1-x^{2}\partial_{z}\tilde{\Sigma}_{11}}\right)_{z=z_0}+{\cal O}(z-z_0)^0\, 
\label{tpopo}
\end{eqnarray}
with $\partial_{z}:=\frac{\partial}{\partial z}$.  To determine the bare coupling $f_N^b$, $x$ is calculated: the physical residue of the
nucleon pole in the variable $z$ is given by
\be
(a_{-1})_{\pi N\to \pi N}=\tilde\gamma_{1}^a\,\tilde\gamma_{1}^c
\label{resin}
\ee
where $\tilde\gamma_{1}^{a,c}$ are the bare nucleon vertices calculated at $z=z_0$ with the physical nucleon coupling $f_{\pi NN}$ instead of
the bare coupling (cf. Appendix B.1. of Ref.~\cite{Doring:2010ap}),
\be
\tilde\gamma_{1}^{a}=i\,\sqrt{\frac{3}{8}}\,\frac{f_{\pi NN}\,k}{\pi\, m_\pi}\frac{E_N+\omega_\pi+m_N}{\sqrt{E_N\omega_\pi(E_N+m_N)}} \ ,
\ee
where $k$ is the particle momentum in the center-of-mass. The physical residue of Eq.~(\ref{resin}) has to agree with the residue of $T^\po$
from Eq.~(\ref{tpopo}) at the pole position:
\begin{eqnarray}
\left(\frac{x^2\tilde{\Gamma}^a_{1}\,\tilde{\Gamma}^c_{1}}{1-x^{2}\partial_{z}\tilde{\Sigma}_{11}}\right)_{z=z_0}= 
(a_{-1})_{\pi N\to \pi N}\, ,
\end{eqnarray}
leading to
\begin{eqnarray}
\frac{1}{ x^{2}}=
\left(\partial_{z}\tilde{\Sigma}_{11}+\frac{\tilde{\Gamma}^a_{1}\,\tilde{\Gamma}_{1}^c}{\tilde\gamma^a_{1}\,\tilde\gamma^c_{1}}
\right)_{z=z_0} \ .
\label{x1}
\end{eqnarray}
The bare mass $m_N^b$ and coupling $f_N^b$ of the nucleon are then calculated by inserting Eq.~(\ref{x1}) into (\ref{renorma1}) and
(\ref{fNb1}), which leads for $m_N^b$ to
\begin{eqnarray}
m^b_{N}&=&m_N-x^2\tilde{\Sigma}_{11}(z_0). \non
\label{mb1}
\end{eqnarray}

This matching procedure is valid for the nucleon in case it couples only to the $\pi N$ channel. In Ref.~\cite{Krehl:1999km}, the
renormalization for one bare $s$-channel state coupling to more than one channel is developed. In contrast, in the present study a second bare
$s$-channel state is included that is allowed to couple to multiple channels, while the bare $s$-channel nucleon state is allowed to couple only
to the $\pi N$ channel. The corresponding formalism is developed below. We have also developed the formalism for two bare $s$-channel states
with both of them coupling to multiple channels. The resulting set of equations needs to be solved numerically for $m_N^b$ and $f_N^b$, but we
see no need here to couple the nucleon pole at $z_0=939$~MeV to channels with a significantly higher threshold energy than $\pi N$; at the
nucleon pole, the effect of omitting the  coupling to higher-lying channels is fully absorbed in the value of $x$.

The renormalization for the case of two bare $s$-channel states is derived in the following. The nucleon is allowed to couple to the $\pi N$
channel only, while the second state couples to all channels. That second state leads to a resonance at $z\sim 1650$~MeV in the $P_{11}$ partial
wave as discussed in Sec.~\ref{sec:p11}.

The nucleon pole position is given by a zero of the determinant of $D$ in Eq.~(\ref{2res}),
\begin{eqnarray}
(z_0-m^b_{1}-\Sigma_{11})(z_0-m^b_{2}-\Sigma_{22})-\Sigma_{12}\Sigma_{21}=0 \;,
\label{renorma}
\end{eqnarray}
with $z_0=m_N $ and $\Sigma_{ij}=\Sigma_{ij}|_{z=m_N}$. In Eq.~(\ref{renorma}), $m^b_1\equiv m_N^b$ and $m^b_2$ is the bare mass of
the second genuine resonance; $m^b_2$ is a free parameter of our model which is determined in the fitting process. In addition to
Eqs.~(\ref{redquan1}) and (\ref{fNb1}), we define the following self-energies:
\begin{eqnarray}
\Sigma_{12}&=& f_N^b\,\Sigma_{12}^{\text{red}}= x \,f_{\pi NN}\,\Sigma_{12}^{\text{red}}=x\,\tilde{\Sigma}_{12}  \non
\Sigma_{21}&=& f_2^b\,\Sigma_{21}^{\text{red}}= x \,f_{\pi NN}\,\Sigma_{21}^{\text{red}}=x\,\tilde{\Sigma}_{21}  \ .
\label{redself}
\end{eqnarray}  
Note that $\Sigma_{12}$ ($=\Sigma_{21}$) contains all bare couplings of the second bare state, that are, like $m_2^b$, free fit parameters. 
From Eq.~(\ref{renorma}) one obtains
\begin{eqnarray}
 m_N^b=m_N-x^{2}\frac{\tilde{\Sigma}_{12}\tilde{\Sigma}_{21}}{m_N-m^b_{2}-\Sigma_{22}}-x^{2}\,\tilde{\Sigma}_{11} 
\label{mNb}
\end{eqnarray}
with the $\Sigma$ evaluated at $z=m_N$. Note that for $\tilde\Sigma_{12}=\Sigma_{22}=0$ one recovers Eq.~(\ref{mb1}).

Again, as in the one-resonance case, we calculate $x$ in order to determine the bare nucleon coupling $f_N^b$.  The physical residue, given once
more by Eq.~(\ref{resin}), has to agree with the residue of $T^\po$ from Eq.~(\ref{2res}), 
\begin{multline}
(a_{-1})_{\pi N\to \pi N}=\tilde\gamma_{1}^a\,\tilde\gamma_{1}^c\\
=\frac{ \begin{pmatrix}\Gamma_{1}^a,&\Gamma_{2}^a\end{pmatrix} \begin{pmatrix} z-m^b_{2}-\Sigma_{22} &
\Sigma_{21}\\ \Sigma_{12} &
z-m^b_{1}-\Sigma_{11}\end{pmatrix} \begin{pmatrix} \Gamma^c_{1}\\  \Gamma^c_{2} \end{pmatrix}}
{ \partial_{z}\left[(z-m^b_{1}-\Sigma_{11})(z-m^b_{2}-\Sigma_{22})-\Sigma_{12}\Sigma_{21}\right]}
\end{multline}
with both sides of the equation evaluated at $z=m_N$. Inserting Eqs.~(\ref{redquan1}) and (\ref{redself}), we arrive at an equation for
$x^{2}$ that only depends on known or fitted quantities:
\begin{eqnarray}
\frac{1}{ x^{2}} &=& \partial_{z}\tilde{\Sigma}_{11} +  G_{2}\tilde{\Sigma}_{12}
\left[G_{2}\tilde{\Sigma}_{12}\,(\partial_{z}\Sigma_{22}-1)+2\,\partial_{z}\tilde{\Sigma}_{12} \right] \non & + &
\frac{\left( G_{2}^{-1} \tilde{\Gamma}^a_{1}+\Gamma_{2}^a\tilde{\Sigma}_{12}\right)\left( G_{2}^{-1}
\tilde{\Gamma}^c_{1}+\Gamma_{2}^c\tilde{\Sigma}_{12}\right)}{G_{2}^{-2}  \tilde\gamma_{1}^{a}\,\tilde\gamma_{1}^{c}}
\label{xsq} 
\end{eqnarray}
with $G_{2}^{-1}\equiv m_N-m_{2}-\Sigma_{22}$. Note that for $\tilde\Sigma_{12}=0$, one recovers the one-resonance case of Eq.~(\ref{x1}).

The bare mass and coupling of the nucleon are calculated by inserting Eq.~(\ref{xsq}) in Eq.~(\ref{mNb}) and Eq.~(\ref{fNb1}). The entire
calculation is performed for each step in the fitting process. 

In the scheme derived here we ensure the correct properties of the nucleon in the presence of a second genuine state. In principle, one could
introduce further boundary conditions, such that the residues of the nucleon pole into the channels $\eta N$ and $KY$ satisfy the same SU(3)
constraints as the meson-meson-baryon vertices in the $t$- and $u$-channel exchange diagrams (cf. Appendix~\ref{sec:su3_couplings}). Those
effects are, however, very small, as these channels are separated from the nucleon pole by several hundreds of MeV. Similarly, for internal
consistency one could replace the effective meson-meson-baryon vertices in the $t$- and $u$-channel diagrams by bare ones and then renormalize
them consistently with the nucleon renormalization procedure. As the $t$- and $u$-channel diagrams enter $T^\npo$ which is subsequently used for
the renormalization of the nucleon, such an extension of the scheme would obviously require a self-consistent treatment beyond the scope of the
present work.

At this point, we would like to emphasize that bare nucleon and resonance masses and coupling constants have no physical meaning, because they
depend on the scheme (in our case, the cut-off values of the form factors and their parameterization).

In addition, some bare resonance parameters tend to be strongly correlated: e.g., for the high-lying $\rho N$ channel, the $\rho N$ self energy
$\Sigma$ [cf. Eq.~(\ref{dressed})] has a small imaginary part at the considered energies and thus the bare $\rho N$ coupling to a resonance is
almost fully correlated with the bare (real) mass.  We  quantitatively contrast bare and physical parameters, the latter being pole positions
and residues, in Sec.~\ref{sec:s11_and_bare} and \ref{sec:delta1620}. 

For the sake of reproducibility of the results, the bare resonance parameters are quoted in  Tables~\ref{tab:bare_cou_fit5} and
\ref{tab:bare_cou_20Sep_2000} in Appendix~\ref{sec:respar}. Because of the reasons stated above, they should not be compared to lattice or
Dyson-Schwinger results.

%%%%%%%%%%%%%%%%%%%%%%%%%%%%%%%%%%%%%%%%%%%%%%%%%%%%%%%%%%%%%%%%%%%%%%%%%%%%%%%%%%%%%%%%%%%%%%%%%%%%%%%%%%%%%%%%%%%%%%%%%%%%%%%%%%%%%%%%%%%%%%%%

\section{Results}
\label{sec: results}

%%%%%%%%%%%%%%%%%%%%%%%%%%%%%%%%%%%%%%%%%%%%%%%%%%%%%%%%%%%%%%%%%%%%%%%%%%%%%%%%%%%%%%%%%%%%%%%%%%%%%%%%%%%%%%%%%%%%%%%%%%%%%%%%%%%%%%%%%%%%%%%%

\subsection{Data base}
\label{database}

%%%%%%%%%%%%%%%%%%%%%%%%%%%%%%%%%%%%%%%%%%%%%%%%%%%%%%%%%%%%%%%%%%%%%%%%%%%%%%%%%%%%%%%%%%%%%%%%%%%%%%%%%%%%%%%%%%%%%%%%%%%%%%%%%%%%%%%%%%%%%%%%

The data that enter our calculation are displayed in Figs.~\ref{fig:pin1}-\ref{fig:kpspbeta} together with our results of fits A and B
of the analysis as explained in Sec.~\ref{sec:numerics}. Data sets which differ only by a few MeV in scattering energy are depicted in
the same plot, whereas in the fitting procedure, of course, the exact energy value for every data set is taken. 

While for the reaction $\pi N\to\pi N$ the partial waves from the GWU/SAID analysis~\cite{Arndt:2006bf} are used, for the inelastic channels
$\pi N \to \eta N$ and $\pi N \to KY$ we fit directly to total and differential cross sections and to polarization observables. The bulk of the
existing data for the inelastic channels was obtained in the 1960's and 70's. Though many experiments have been carried out at different
facilities, unfortunately, there are still energy regions where the data situation is not ideal, cf. for example the discussion in
Sec.~\ref{sec:etan} on the reaction $\pi^-p \to \eta n$. For the $\eta N$, $K\Lambda$ and $K^0\Sigma^0$ channels, there is a qualitative
difference between data on differential cross sections $d\sigma/d\Omega$ and polarizations $P$: in general, the polarization measurements
exhibit much larger error bars and sometimes take on unphysical values greater than 1. For certain energies, there exist also conflicting
measurements. The spin-rotation parameter $\beta$ has been measured only for the $K^0\Lambda$ and the $K^+\Sigma^+$ final states, and only for a
few energies.  In the following sections, the data situation in each reaction is critically reviewed. The weights we apply to specific data sets
in the fit are discussed in detail.

Note that a measurement of $d\sigma/d\Omega$, $P$, and $\beta$ represents one possible full set of observables to disentangle the  amplitudes
model-independently for scattering of pseudo-scalar mesons off spin-$1/2$ baryons. In practice, one needs to make model assumptions, as done
here, to obtain sensible results, because the experimental precision is far from being sufficient for a model-independent analysis. To
disentangle the $\pi\pi N$ dynamic model-independently, one would need many more observables anyway.

The combined study of the reactions $\pi^+p\to K^+\Sigma^+$, $\pi^-p\to K^0\Sigma^0$, and in particular $\pi^-p\to\, K^+\Sigma^-$ is  essential
to disentangle the isospin content of the $K\Sigma$ final state. For the $K^+\Sigma^+$ and $K^0\Sigma^0$ final states, the pronounced forward
peak shows the onset of $t$-channel dominance which at energies $>$ 3 GeV is most economically parameterized in terms of Regge
exchanges~\cite{Huang:2008nr, Sibirtsev:2007wk}. In this respect,  the $K^+\Sigma^-$ final state, although omitted in many analyses,  is of
particular interest, because it is sensitive to $u$-channel exchanges,  as $t$-channel processes would require the exchange of a charge 2
particle at the potential level.   Due to its $u$-channel dominance and the small total cross section, the $\pi^- p\to K^+\Sigma^-$ process
imposes strong constraints, despite the somewhat poor data situation. In other words, we found parameter sets that fit $\pi^+p\to K^+\Sigma^+$
and $\pi^-p\to K^0\Sigma^0$ somewhat better than in the present result, but once the data of the $K^+\Sigma^-$ final state were included into
the minimization procedure, the partial-wave content changed. See also Sec.~\ref{sec:kpsp} and Fig.~\ref{fig:nokpsm} where we test the
consequences if the data of the reaction $\pi^-p\to K^+\Sigma^-$ are ignored in the fit.

%%%%%%%%%%%%%%%%%%%%%%%%%%%%%%%%%%%%%%%%%%%%%%%%%%%%%%%%%%%%%%%%%%%%%%%%%%%%%%%%%%%%%%%%%%%%%%%%%%%%%%%%%%%%%%%%%%%%%%%%%%%%%%%%%%%%%%%%%%%%%%%%

\subsection{Numerical details and fit parameters}
\label{sec:numerics}

In the fitting procedure, we consider two different scenarios. In case of fit A, we start from  the parameter set of a preceding J\"ulich $\pi
N$ model~\cite{Gasparyan:2003fp}. The addition  of new channels in the present model and the requirement that many more data have to be
described  ensure that the parameter space is sufficiently explored in the course of the fitting procedure before an acceptable result is
reached.  Still, in their overall properties, the new fit A and the old interaction potential remain fairly similar.  In particular, the sizable
renormalization of the (bare) nucleon mass in the old model  \cite{Gasparyan:2003fp} occurs in the present fit A too; cf. Sec.~\ref{sec:p11}. 
For fit B, we start out from a radically different scenario so that the parameter space has to be searched again completely anew. The prime
reason for such a strategy is that we would like to find out how far our results and, specifically, the extracted resonance parameters are
sensitive to the starting conditions of the fitting procedure.  Certainly, for a solid assessment of the uncertainty of the extracted amplitudes
and  pole properties, one would need to carry out more fits under different conditions. The high numerical effort, however, puts limits on the
number of fits that can be performed.  To make sure that the starting point is indeed quite different in the  fitting procedure for fit B, we
imposed from the beginning that here the renormalization  of the nucleon mass should be as small as possible. Note that the amount of
renormalization  is not an observable and thus, in principle, should not influence the results.  It turned out that also for this scenario a
comparable description of the data can be achieved. Interestingly, in this scenario the coupling of the $\pi N$ system to the effective  $\pi\pi
N$ channels, notably the $\sigma N$ channel, had to be increased significantly, and the strength of the interaction within the $\sigma N$
channel increased, too.  The differences in the scenarios for fit A and B mentioned above  imply consequences for the resonance generation of
the Roper resonance  which will be discussed in detail in Sec.~\ref{sec:p11}.

The present results were obtained in a fit procedure using MINUIT  on the JUROPA supercomputer at the Forschungszentrum J\"ulich. To be able to
handle large amounts of data in combination with time-consuming fitting techniques, an efficient parallelization of the code using MPI (Message
Passing Interface) was mandatory. To find a good local minimum, usually 50-100 parallelized energies were used, while for the final production
runs 300 and more processes ran in parallel. As discussed before, the decomposition of the amplitude into a pole- and a non-pole part according
to Eq.~(\ref{deco1}) provides the possibility to vary resonance parameters without the need to recalculate $T^\npo$ every time. The architecture
of the parallelization takes advantage of this possibility to save CPU time: for every step in the determination of parameters tied to $T^\npo$,
all $T^\po$ parameters are fully optimized, i.e. we perform a nested fit. The computation time can be reduced by two orders of magnitude through
this method.

Apart from the parallelization in energy, some processes perform special tasks such as the renormalization of the nucleon and the control of the
nested fit; combining iterative, nested fitting with the parallel architecture is not trivial.

Compared to Ref.~\cite{Doring:2010ap}, a large amount of additional data for total and differential cross sections and polarization observables
for the $\eta N$ and $KY$ channels was considered as input to the fit. The current set amounts to about 6000 data points. The program is
organized in a way that much larger amounts of data could be handled without slowing the calculation down severely. This will be especially
important once photo- and electroproduction data, that comprise more than $10^5$ points, are included in the analysis. Also, in principle an
additional parallelization in partial waves is possible but was not required so far.

In the previous analysis~\cite{Doring:2010ap}, the reaction $\pi^+ p\to K^+\Sigma^+$ and $\pi N$ scattering were considered and only resonance
parameters, i.e. bare masses and couplings of the resonances to the different channels, were fitted. In the present study, in addition the
important $T^\npo$ parameters are varied. Those are the cut-offs of the form factors in $t$- and $u$-channel exchange diagrams, shown in
Figs.~\ref{fig:dia1} and \ref{fig:dia2} and quoted in Appendix~\ref{sec:ExApp}. 

Bare resonances with a total spin through $J=9/2$ are included, with corresponding new parameters. One bare $s$-channel state is included in
each of the isospin $I=1/2$ partial waves $D_{13}$, $D_{15}$, $F_{15}$, $P_{13}$, $F_{17}$, $H_{19}$ and $G_{19}$, while we have two in $S_{11}$
and $P_{11}$. In the $I=3/2$ sector, one bare $s$-channel state is included in the $S_{31}$, $D_{33}$, $F_{35}$, $P_{31}$, $D_{35}$, $F_{37}$,
$G_{37}$ and $G_{39}$ partial waves and two are included in $P_{33}$. These states couple to all channels $\pi N$, $\rho N$, $\eta N$,
$\pi\Delta$, $K\Lambda$ and $K\Sigma$ if allowed by isospin. Bare $s$-channel states are included as demanded by data. The significance of the
resulting resonances is discussed for each partial wave in Sec.~\ref{casebycase}.

In total, we have 196 free parameters, of which 128 are resonance parameters and 68 belong to the $T^\npo$ part (64 cut-offs in the form factors
of the $t$- and $u$-channel exchanges plus 4 coupling constants). Note that varying a parameter that enters $V^\npo$ changes the amplitude in
all partial waves ---and also for all reactions. While most of the $T^\npo$ parameters are responsible for the part of the amplitude that slowly
varies with energy, the Roper resonance is dynamically generated in the present approach. This means that here the $T^\npo$ parameters have to
be chosen to reproduce the resonance and simultaneously provide the ``background'' in all other partial waves. The values of the resonance
parameters are quoted in Tables~\ref{tab:bare_cou_fit5} and \ref{tab:bare_cou_20Sep_2000} and the $V^\npo$ parameters can be found in
Tables~\ref{tab:coupl} and \ref{tab:cutoff_bg}. 

In the course of fitting, we assigned special weights to some data sets to ensure that the fit respects certain features in the data. For
example, there is a pronounced but narrow forward peak in the reaction $\pi^-p\to K^0\Lambda$ (cf. Fig.~\ref{fig:kldiff1}). At higher energies,
$z>2$~GeV, the peak is consistently present in different experiments, but always with low statistical weight. Without giving those data special
weight in the minimization procedure, some fit results would not reproduce the forward peak. Also, varying the data weights in the course of
fitting helps prevent the fit from getting stuck in shallow minima of the parameter space.

We do not claim a sharp upper limit in energy for the validity of the current approach; for example, the higher partial waves in $\pi N$
scattering have clear  resonances beyond 2 GeV~\cite{Arndt:2006bf} and in such cases it is important to fit even beyond the resonance position.
In any case, in the present Lagrangian-based framework, the amplitude allows, in general, for a well-behaved extrapolation to higher energies;
by contrast, in analyses in which the potential is parameterized purely phenomenologically in terms of polynomials, there may be little control
on the amplitude outside the fitted energy region. As expected from the physics-driven parameterization of our amplitude, the overall agreement
of the fits with the data beyond 2~GeV is, for most observables, quite good. 

In the following, we discuss the data situation and our description of the data for the different reactions. For the reaction $\pi N\to\pi N$,
we show the partial waves of our fit result. For the inelastic reactions $\pi N\to \eta N, \, KY$, there is an overall, energy-dependent phase
that cannot be determined directly from experiments, and therefore we refrain from giving the corresponding amplitudes. Instead, we show the
partial cross sections, i.e. cross sections            for each partial wave (amplitudes can be provided upon request). Due to this unknown
energy-dependent phase, other results should be compared to the present ones at the level of partial cross sections. We stress that for any
meaningful comparison on the level of amplitudes, it is mandatory to take the mentioned phase ambiguity into account.

%%%%%%%%%%%%%%%%%%%%%%%%%%%%%%%%%%%%%%%%%%%%%%%%%%%%%%%%%%%%%%%%%%%%%%%%%%%%%%%%%%%%%%%%%%%%%%%%%%%%%%%%%%%%%%%%%%%%%%%%%%%%%%%%%%%%%%%%%%%%%%%%

\subsection{$\boldsymbol{\pi N\to \pi N}$}
\label{sec:pin}
The energy-dependent partial-wave solution of the GWU/SAID analysis~\cite{Arndt:2006bf} up to $H$ waves is used as input for the elastic $\pi N$
channel, both for isospin $1/2$ and $3/2$.  The points displayed in Figs.~\ref{fig:pin1}-\ref{fig:pin4} and in Appendix~\ref{sec:piNdel},
however, show the energy-independent (single-energy) solution of Ref.~\cite{Arndt:2006bf}. The corresponding error bars represent uncertainties
of the theoretical analysis and cannot be used for calculating the $\chi^2$.  Therefore, to this point no true statistical uncertainties on
extracted quantities (poles and residues) can be determined in our analysis. The inclusion of the actual $\pi N\to\pi N$ experimental data in
the future will make this possible. Note that only the GWU/SAID group fits to $\pi N\to\pi N$ data , while in other analyses (present approach,
EBAC~\cite{Matsuyama:2006rp, JuliaDiaz:2007kz, Durand:2008es, Paris:2008ig, Suzuki:2008rp, Kamano:2008gr, JuliaDiaz:2007kz},
Bonn-Gatchina~\cite{Anisovich:2004zz, Anisovich:2011fc}, Gie{\ss}en~\cite{Penner:2002ma, Penner:2002md, Shklyar:2004dy, Shklyar:2009cx}, Kent
State~\cite{Shrestha:2012va,Shrestha:2012ep}, ...) their result is used as input. For the 20 partial waves of $\pi N\to \pi N$ considered, we
fit to synthetic data generated from the energy-dependent solution~\cite{Arndt:2006bf} in 5~MeV steps in scattering energy $z$ and assign an
error of 0.01 to the real and to the imaginary part. Although we refer to the 2006 paper~\cite{Arndt:2006bf} throughout this study,  we fit to
the unpublished WI08 solution of the SAID group accessible through a web interface~\cite{Arndt:2006bf}. In addition, we fit to the $\pi N$
scattering lengths of Ref.~\cite{Baru:2010xn} and scattering volumes of Ref.~\cite{Fettes:2000xg}.

\begin{figure}
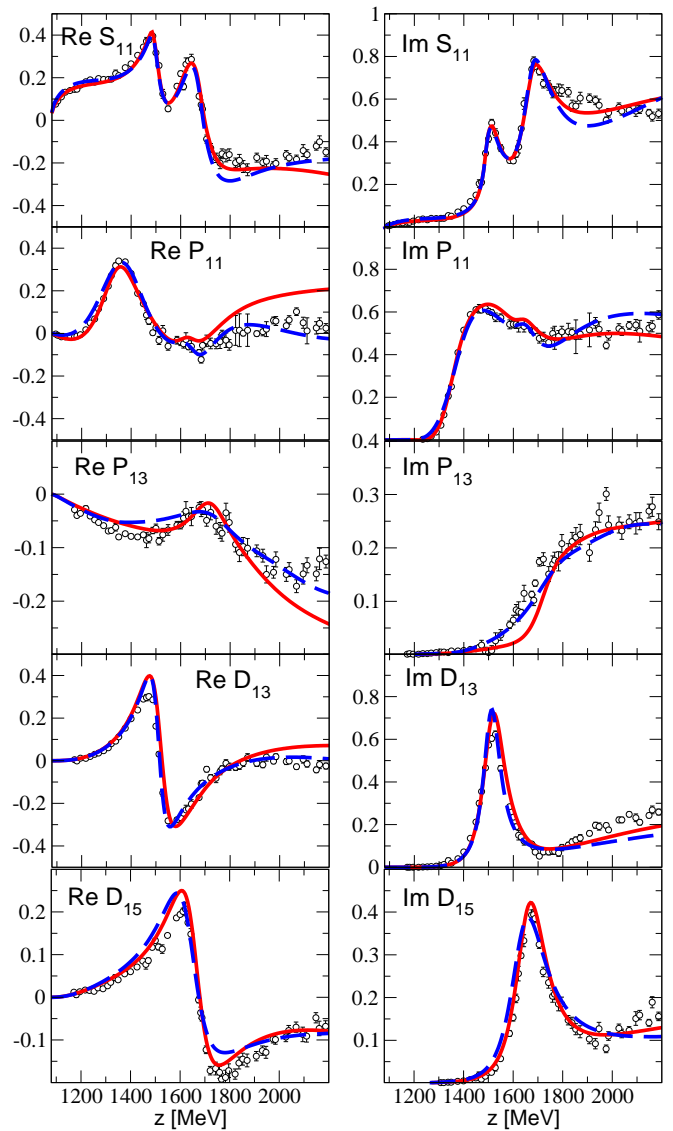

\begin{center}
\includegraphics[width=0.48\textwidth]{s11_p11.eps}\\
\vspace*{-0.2cm}
\includegraphics[width=0.48\textwidth]{p13_d13.eps}\\
\vspace*{-0.15cm}
\hspace*{0.02cm}\includegraphics[width=0.479\textwidth]{d15.eps}
\end{center}
\caption{Reaction $\pi N\to \pi N$, isospin $I=1/2$, $S$- to $D$-waves. Solid (red) lines: fit A; dashed (blue) lines: fit B;
points: GWU/SAID partial-wave analysis (single-energy solution) from Ref.~\cite{Arndt:2006bf}.}
\label{fig:pin1}     
\end{figure}

\begin{figure}
\begin{center}
\hspace*{0.11cm}\includegraphics[width=0.473\textwidth]{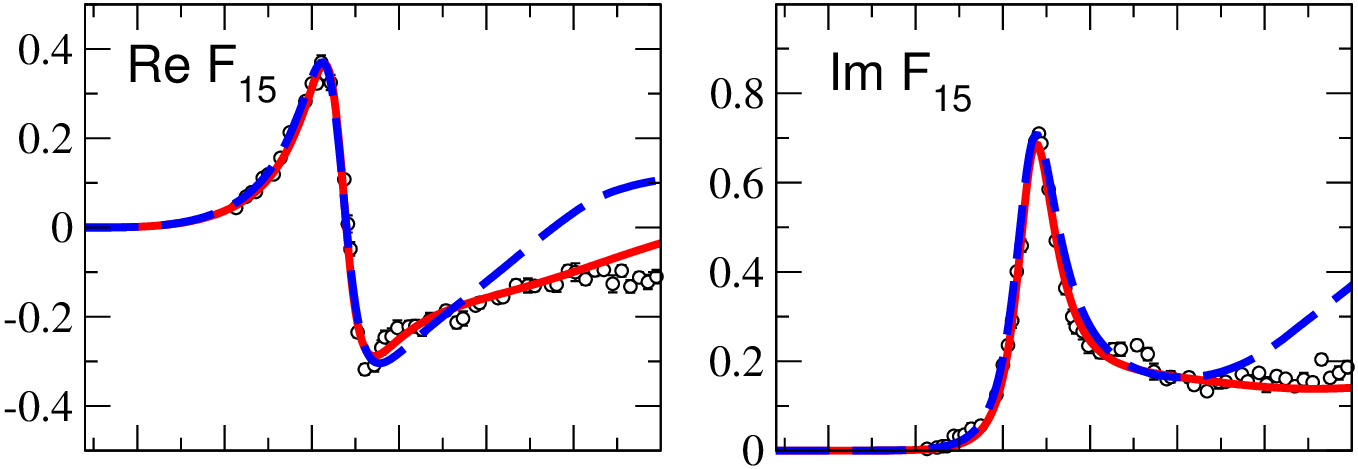}\\
\vspace*{-0.157cm}
\includegraphics[width=0.48\textwidth]{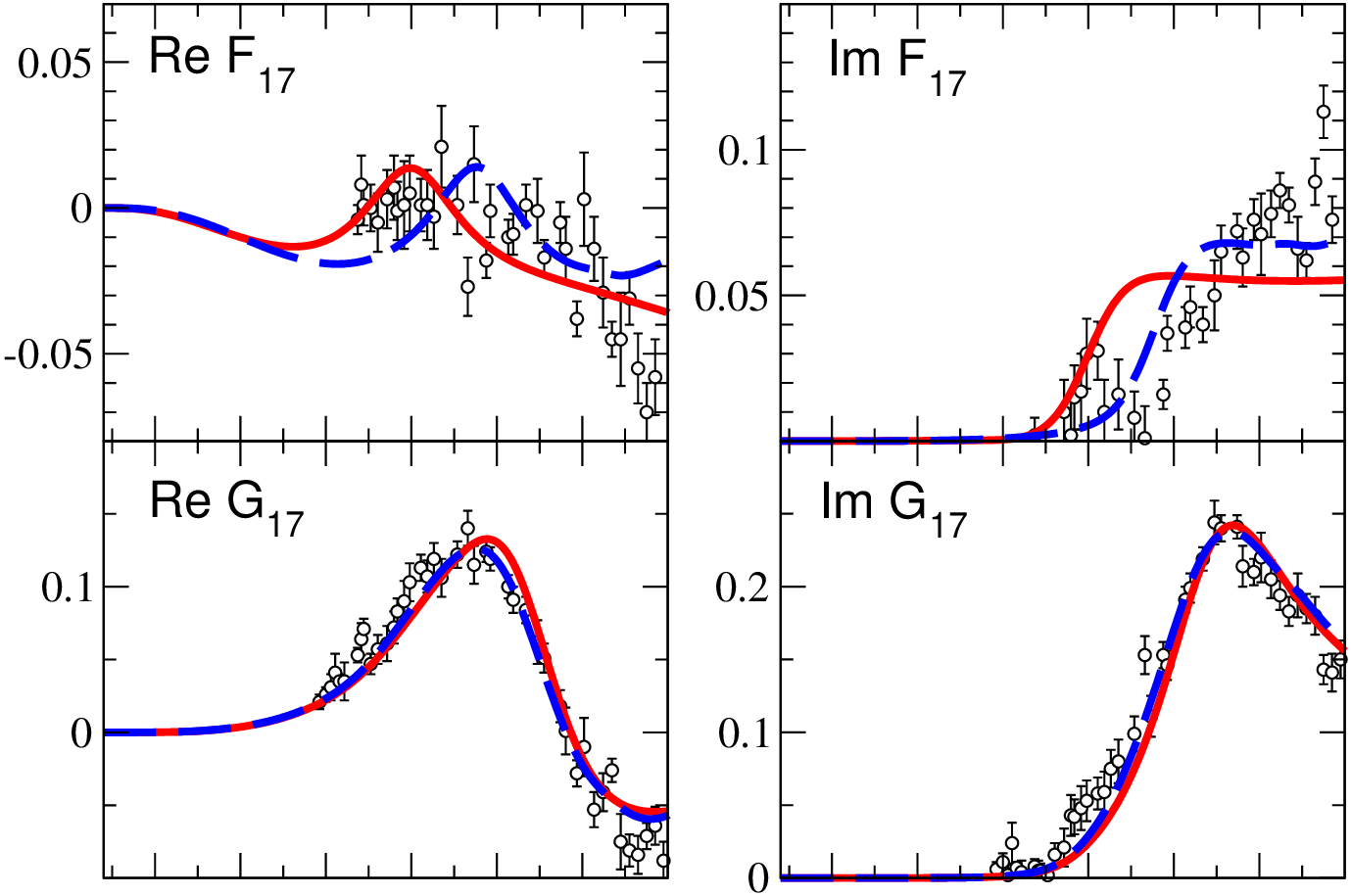}\\
\vspace*{-0.15cm}
\includegraphics[width=0.48\textwidth]{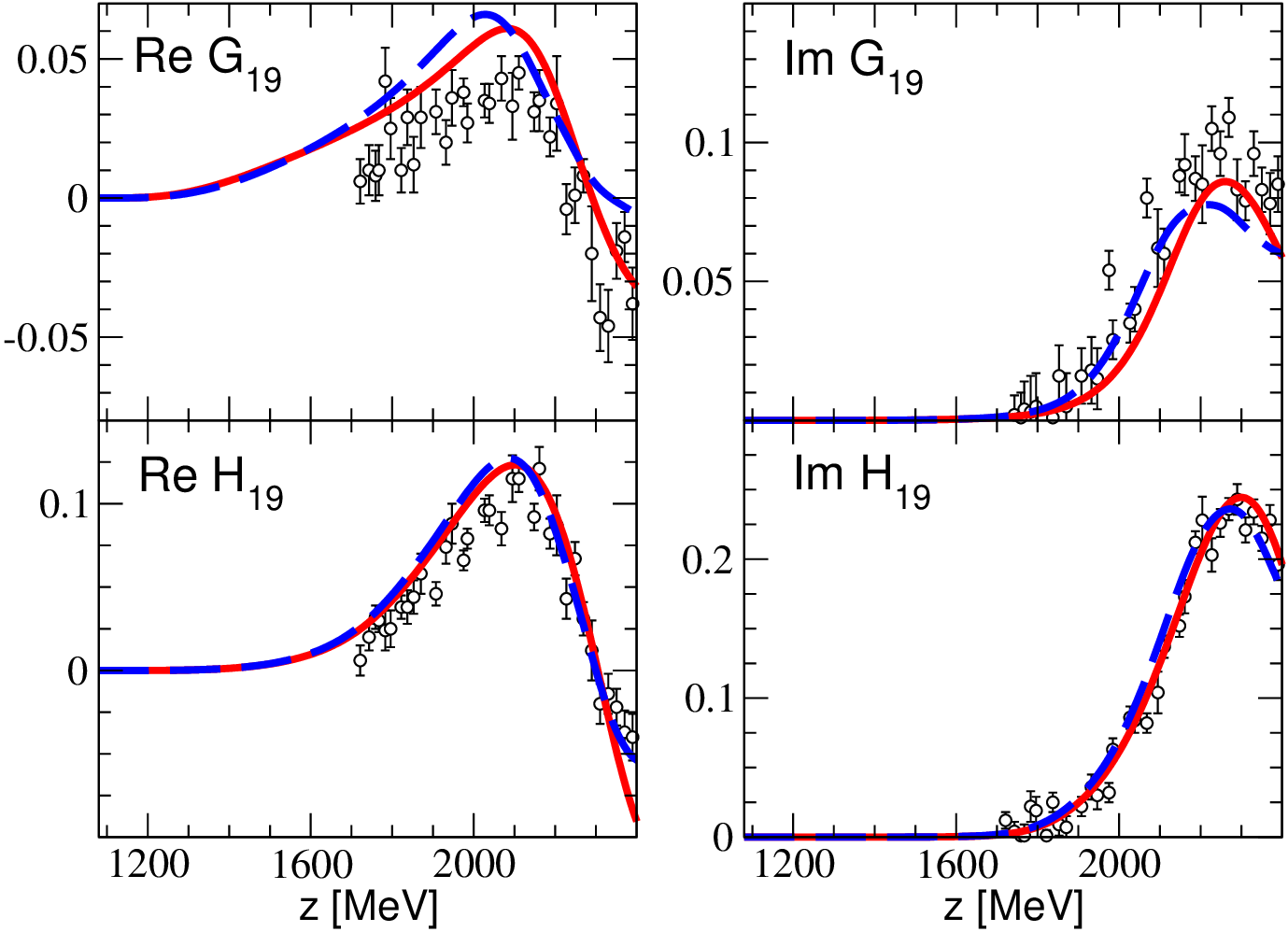}
\end{center}
\caption{Reaction $\pi N\to \pi N$, isospin $I=1/2$, $F$- to $H$-waves. Solid (red) lines: fit A; dashed (blue) lines: fit B;
points: GWU/SAID partial-wave analysis (single-energy solution) from Ref.~\cite{Arndt:2006bf}.}
\label{fig:pin2}     
\end{figure}

\begin{figure}
\begin{center}
\hspace*{0.11cm}\includegraphics[width=0.473\textwidth]{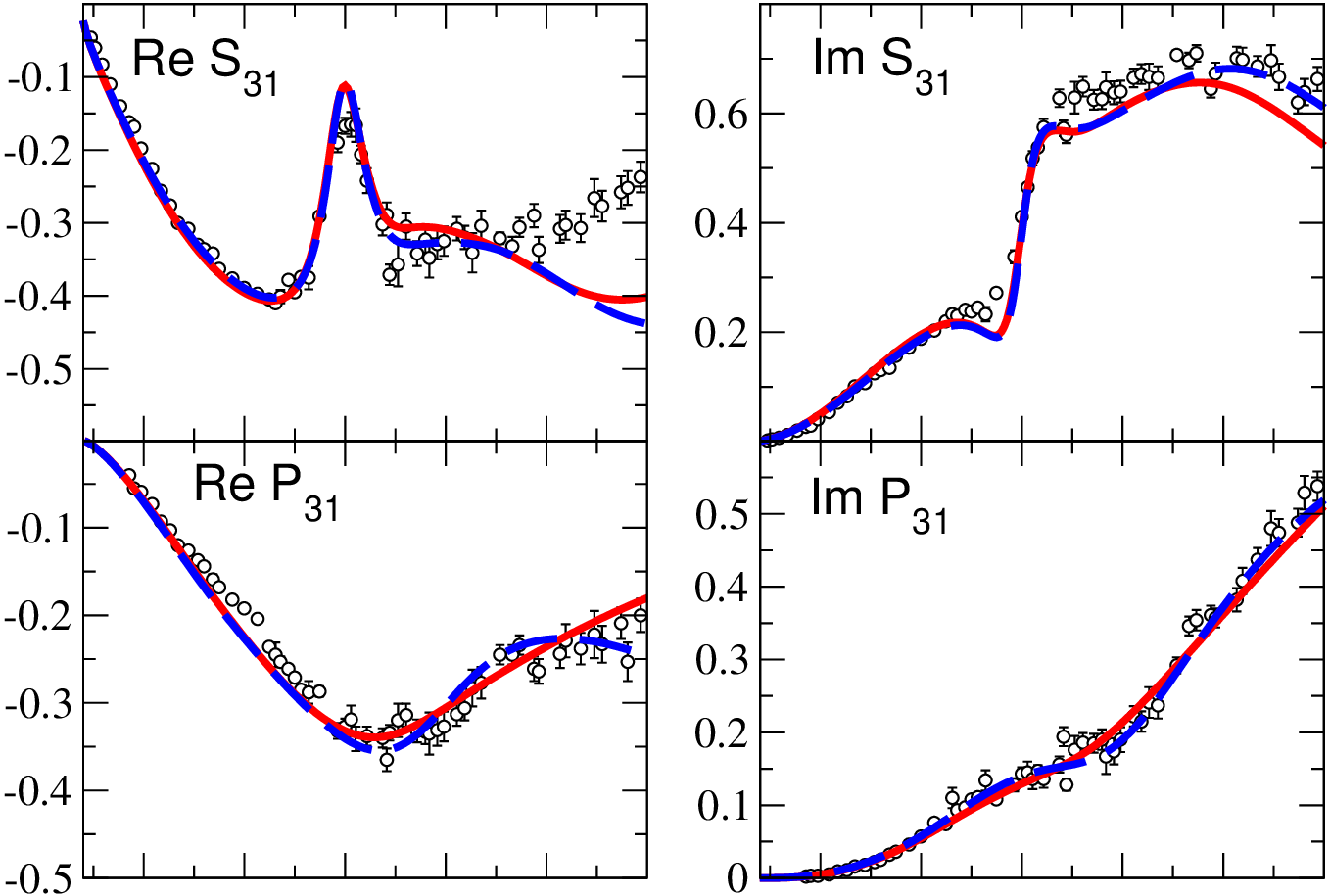}\\
\vspace*{-0.15cm}
\hspace*{0.11cm}\includegraphics[width=0.473\textwidth]{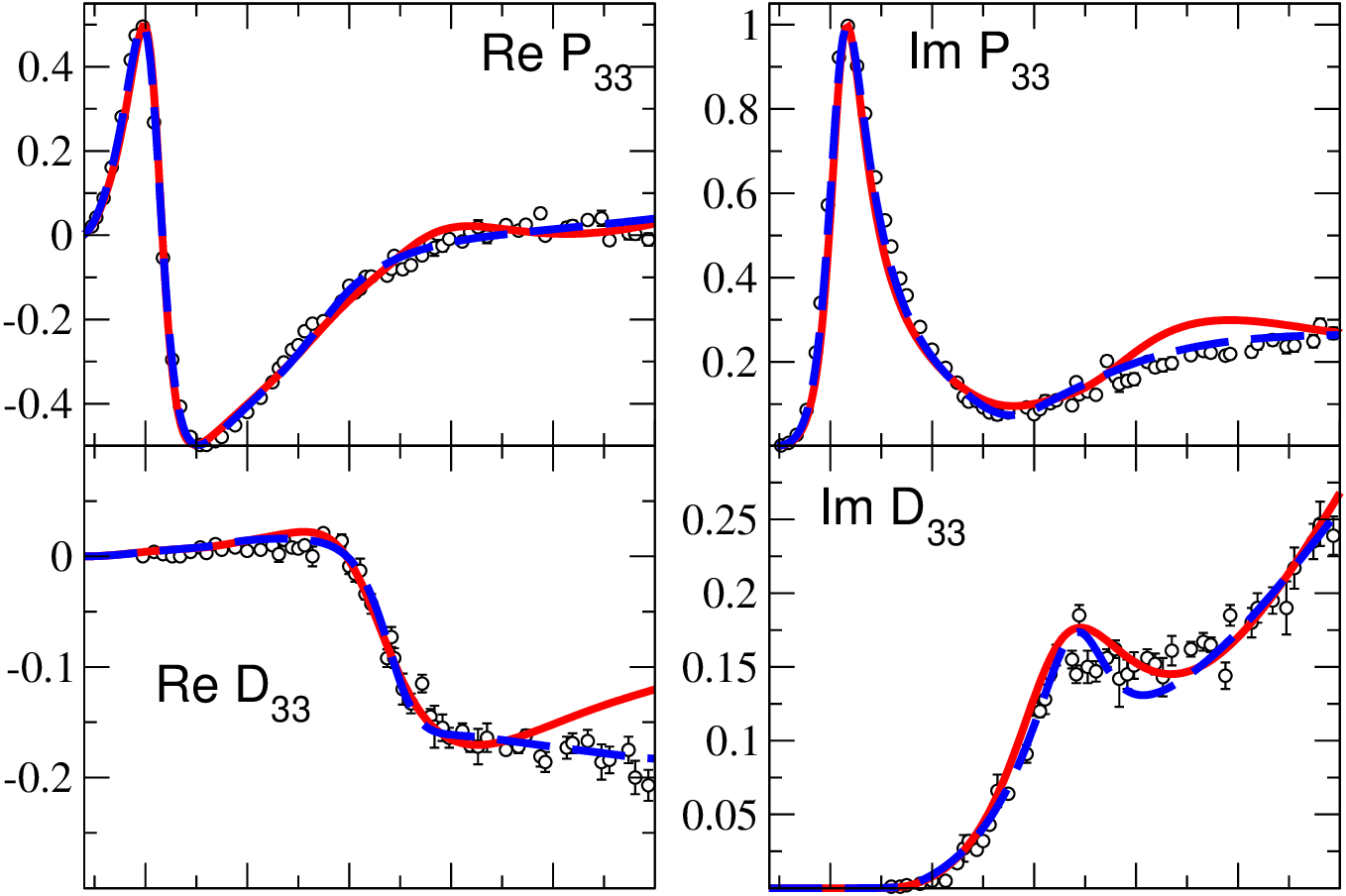}\\
\vspace*{-0.145cm}
\hspace*{0.015cm}\includegraphics[width=0.48\textwidth]{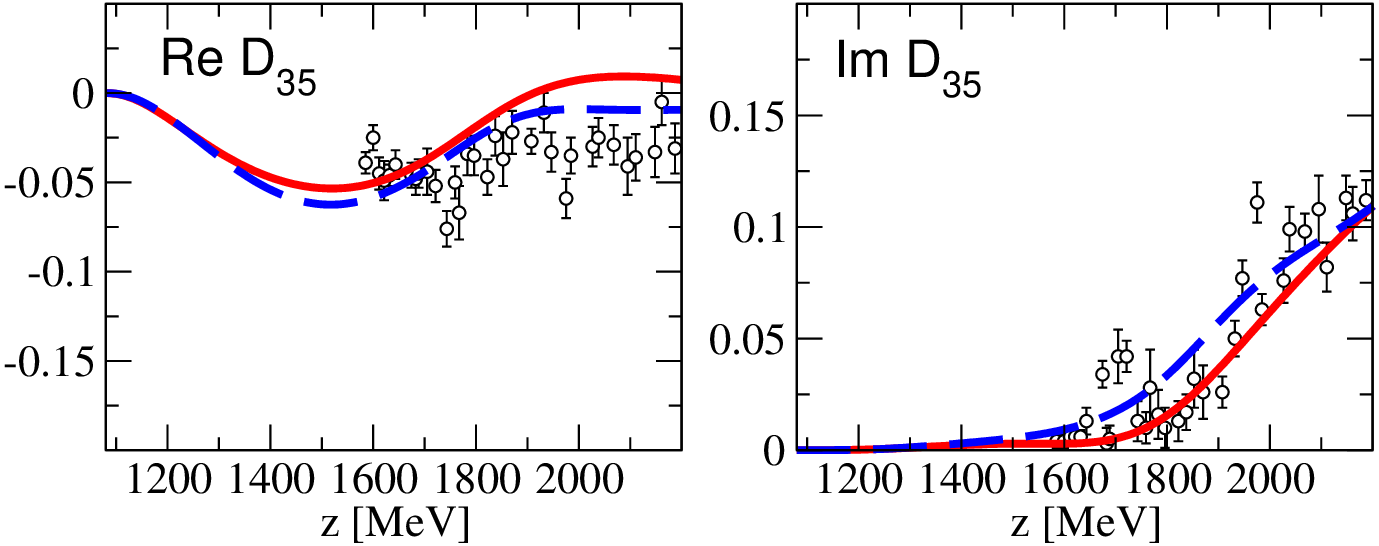}
\end{center}
\caption{Reaction $\pi N\to \pi N$, isospin $I=3/2$, $S$- to $D$-waves. Solid (red) lines: fit A; dashed (blue) lines: fit B; 
points: GWU/SAID partial-wave analysis (single-energy solution) from Ref.~\cite{Arndt:2006bf}.}
\label{fig:pin3}     
\end{figure}

\begin{figure}
\begin{center}
\hspace*{0.09cm}\includegraphics[width=0.473\textwidth]{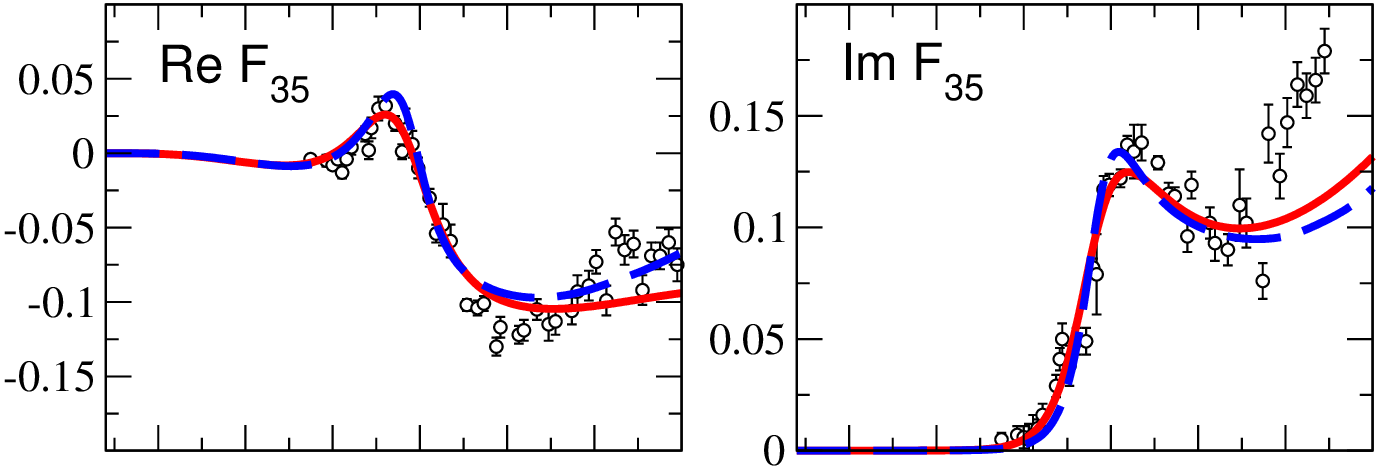}\\
\vspace*{-0.15cm}
\hspace*{0.09cm}\includegraphics[width=0.473\textwidth]{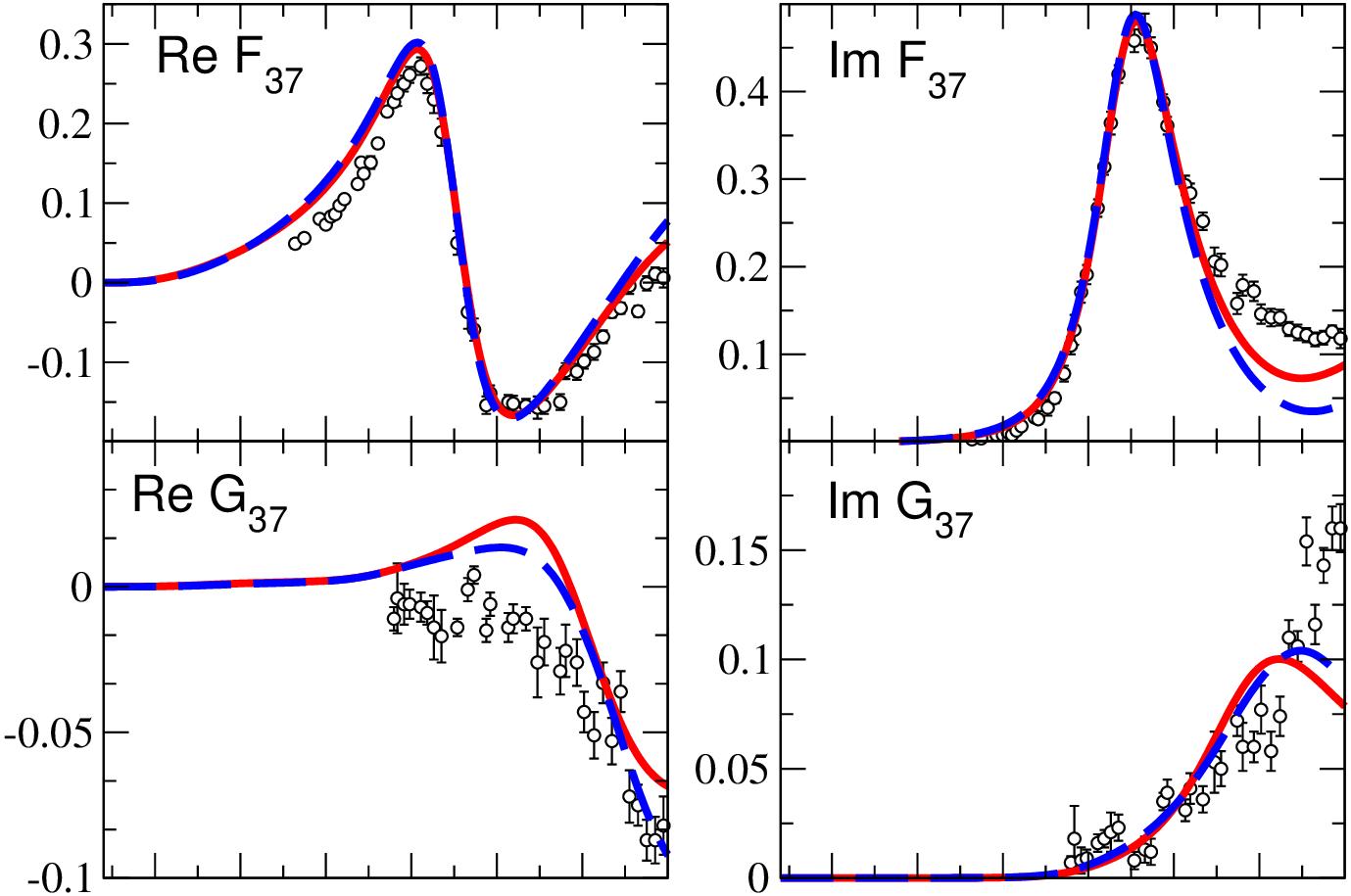}\\
\vspace*{-0.155cm}
\hspace*{0.07cm}\includegraphics[width=0.473\textwidth]{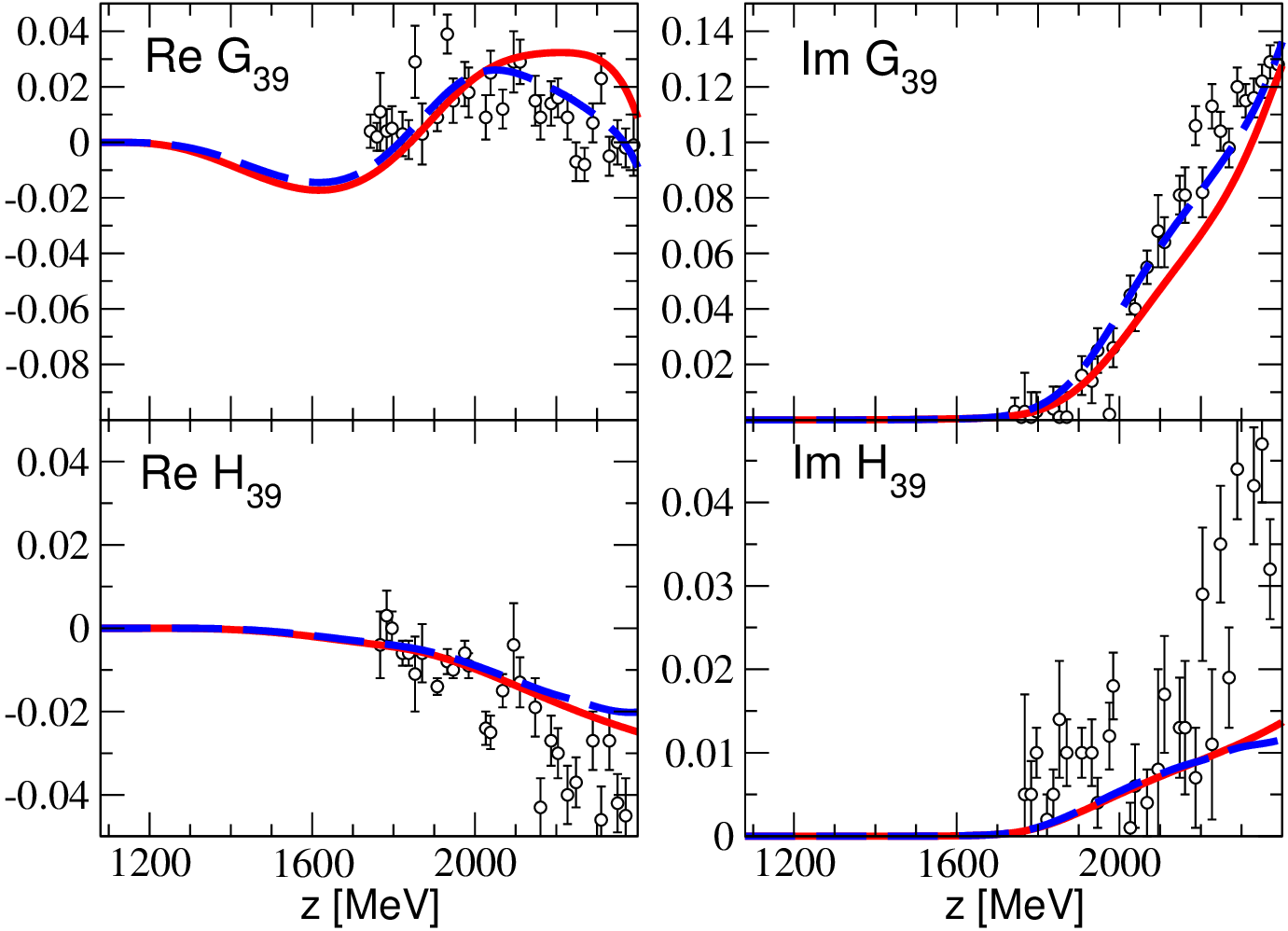}
\end{center}
\caption{Reaction $\pi N\to \pi N$, isospin $I=3/2$, $F$- to $H$-waves. Solid (red) lines: fit A; dashed (blue) lines: fit B; 
points: GWU/SAID partial-wave analysis (single-energy solution) from Ref.\cite{Arndt:2006bf}.}
\label{fig:pin4}     
\end{figure}

Figures~\ref{fig:pin1} to \ref{fig:pin4} show the $I=1/2$ and $I=3/2$ elastic $\pi N\to\pi N$ partial-wave amplitudes. The alternative
representation in terms of phase shifts and inelasticities can be found in Appendix \ref{sec:piNdel}. The result of fit A of this study is
represented by the red solid lines, while the blue dashed lines denote the result of fit B.

The quality of the result has to be seen in perspective: As Fig.~\ref{fig:dia1} shows, there are only four tunable parameters for the $\pi
N\to\pi N$ non-pole transitions, namely the form factor of the $\pi NN$ vertex with $N$ in the $u$-channel, one form factor for the $\sigma$-
and one for the $\rho$-exchange, and one for the $\pi N\Delta$ vertex with $\Delta$ in the $u$-channel. Of these four parameters, two appear
also in different transitions, e.g. $\pi N\to\pi\Delta$ with $N$ or $\Delta$ in the $u$-channel. Thus, these two parameters are also responsible
for the inelasticities.  An additional constraint comes from the very small isoscalar scattering length. There is, therefore, only very little
freedom in the fit of the non-resonant $\pi N\to\pi N$ transition potentials. Some limited additional freedom comes from coupled-channel
effects, through which also other parameters have influence on the non-pole $\pi N\to\pi N$ amplitudes. In any case, the fact that the
non-resonant part of the amplitude in 20 partial waves can be described with very few (2-4) directly relevant parameters, should be considered a
remarkable success.

In some partial waves, there are shortcomings. E.g., for $S_{31}$ (Fig.~\ref{fig:pin3}) we achieve a good description of the imaginary part in
both fits A and B. Still, the description of the real part is not satisfactory at higher energies. As Fig.~\ref{fig:pindel3} shows, the phase
shift in $S_{31}$ is well described, and shortcomings originate from the inelasticities. Similarly, the description of the $S_{11}$ partial wave
at higher energies could be improved. This might be due to channels not yet included in the approach as, e.g., $\omega N$. In this context, one
should also mention pion-induced $\phi$ production on the nucleon, to be considered in the future, which is closely tied to the $KY$ channels as
shown in Refs.~\cite{Doring:2008sv, Meissner:1997qt}. 

For many of the $\pi N$ partial waves, fit A and B provide similar descriptions with only small differences, predominantly at higher energies.
The most apparent deviations appear in the $F_{17}$ partial wave in Fig.~\ref{fig:pin2}. Here, because of the small amplitude, the
resonance position is not easy to be fixed. Also in the $F_{15}$ partial wave, discrepancies between fit A and B appear at higher energies. 
The $P_{11}$ partial wave is of special interest and will be further discussed in Sec.~\ref{sec:p11}.

%%%%%%%%%%%%%%%%%%%%%%%%%%%%%%%%%%%%%%%%%%%%%%%%%%%%%%%%%%%%%%%%%%%%%%%%%%%%%%%%%%%%%%%%%%%%%%%%%%%%%%%%%%%%%%%%%%%%%%%%%%%%%%%%%%%%%%%%%%%%%%%%

\subsection{Reaction $\boldsymbol{\pi N\to\eta N}$}
\label{sec:etan}

Most data of the reaction $\pi^-p \to \eta n$ date back to the 1960's to 1980's and exhibit inconsistencies among each other as scrutinized in
Ref.~\cite{Clajus:1992} where a detailed overview and analysis of the different experiments and the quality of the corresponding data is given.
See also the selection of $\eta N$ data provided by the GWU/SAID group~\cite{Arndt:2005dg} and in Ref.~\cite{Sibirtsev:2001hz}.  Due to a
frequent underestimation of systematic errors and issues concerning beam calibration and normalization, a careful selection and rating of the
data sample is necessary. We mainly follow the estimations made thereto in Ref.~\cite{Clajus:1992}, see also Refs.~\cite{Durand:2008es,
Batinic:1995kr}. 

\begin{figure}[h!]
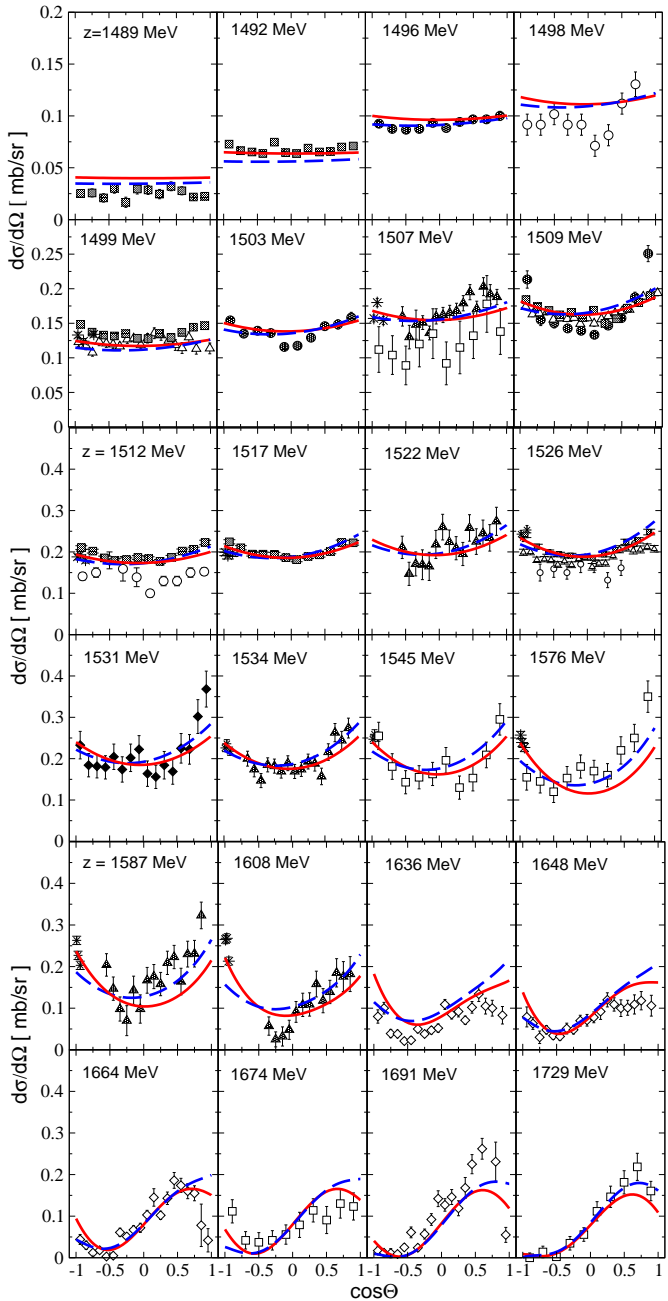

\begin{center}
\includegraphics[width=0.482\textwidth]{dsdo_etaN_1.eps}\\
\vspace*{-0.15cm}
\hspace*{0.02cm}\includegraphics[width=0.48\textwidth]{dsdo_etaN_2.eps}\\
\vspace*{-0.15cm}
\hspace*{0.04cm}\includegraphics[width=0.482\textwidth]{dsdo_etaN_3.eps}
\end{center}
\caption{Differential cross section [1/2] of the reaction $\pi^- p\to \eta n$. Solid (red) lines: fit A; dashed (blue) lines: fit B; data:
filled squares from Ref.~\cite{Prakhov:2005qb}; filled circles from Ref.~\cite{Bayadilov:2008zz}; empty circles from Ref.~\cite{MorrisonPhD};
empty triangles up from Ref.~\cite{Kozlenko:2003hu}; stars from Ref.~\cite{Debenham:1975bj}; filled triangles up from Ref.~\cite{Deinet:1969cd};
empty squares from Ref.~\cite{Richards:1970cy}; filled diamonds from Ref.~\cite{Feltesse:1975nz}; empty diamonds from Ref.~\cite{Brown:1979ii}.}
\label{fig:endiff1}     
\end{figure}

In the present analysis, differential cross section data from Refs.~\cite{Prakhov:2005qb} to \cite{Brown:1979ii} for 38 different energies from
$z=1489$ MeV up to $z=2235\, \text{MeV}$ are included in the fit. For the more recent experiments by Prakhov {\it et al.}~\cite{Prakhov:2005qb}
and Bayadilov {\it et al.}~\cite{Bayadilov:2008zz}, we employ the systematic uncertainties given by the authors. Following
Ref.~\cite{pennerdiss}, a systematic error of $\delta_{sys}=10\%$ of the experimental value is added to the statistical error of the data from
Morrison {\it et al.}~\cite{MorrisonPhD}. As Fig.~\ref{fig:endiff1} shows, the Morrison data (empty circles) are in conflict with the Prakhov
data (filled squares). We have decided to rely on the Prakhov data in this analysis and drastically reduced the weight of other data in the
fitting procedure, in case they are in conflict. E.g., for the data sets at $z=1498$~MeV and $z=1499$~MeV, the data by Prakhov were weighted
with a factor 25 compared to Morrison in both fits.

\begin{figure}
\begin{center}
\includegraphics[width=0.48\textwidth]{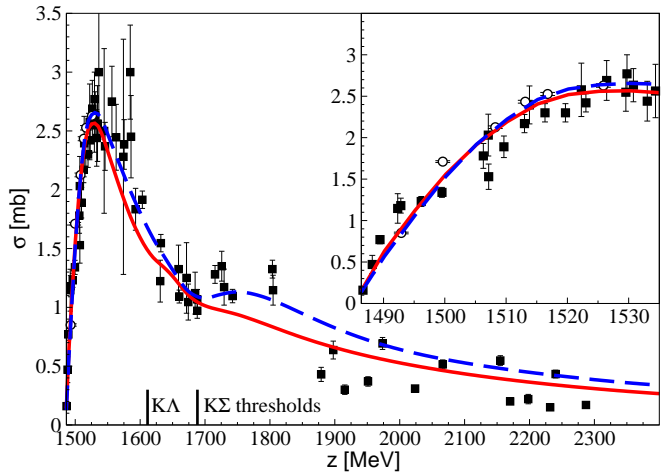}\\
\end{center}
\caption{Total cross section of the reaction $\pi^- p\to \eta n$. Solid (red) lines: fit A; dashed (blue) lines: fit B; data: the filled squares
indicate experiments accepted by the  GWU/SAID group~\cite{Arndt:2003if}; open circles from Prakhov {\it et al.}~\cite{Prakhov:2005qb}.}
\label{fig:entot}     
\end{figure}

The uncertainties stated in Kozlenko {\it et al.}~\cite{Kozlenko:2003hu} are predominantly of statistical nature. Therefore, a systematic error
of $\delta_{sys}=10\%$ was assigned. For Debenham {\it et al.} \cite{Debenham:1975bj} we use the same uncertainties as 
Refs.~\cite{Batinic:1995kr, Durand:2008es} and employ a systematic error of $10\%$ and $0.02\, \text{mb}$.  A systematic error of
$\delta_{sys}=11.4\%$ of the experimental value was added to the statistical error of the data from Deinet {\it et al.}~\cite{Deinet:1969cd} as
advised by the authors themselves. The large energy interval from $z=1636\, \text{MeV}$ to $z=2235\, \text{MeV}$ is almost solely covered by
Brown {\it et al.}~\cite{Brown:1979ii}, supplemented by data from Richards {\it et al.}~\cite{Richards:1970cy} at medium energies. For Richards
{\it et al.}, we assume an additional error of $\delta_{sys}=9\%$ to $14\%$ as given by the authors in the original paper. As argued in
Ref.~\cite{Clajus:1992}, the data from Brown {\it et al.}~\cite{Brown:1979ii} are highly questionable at low energies due to a miscalibration of
the beam momentum and, therefore, are not used as input in the present calculation. At higher energies, on the other hand, the data might be
used if the beam momenta are lowered by $4\%$, together with an additional systematic error of $10\%$. In any case, the Brown data enter the fit
with a much lower weight and we make no attempt to fit those data quantitatively (in the recent analysis of Ref.~\cite{Shrestha:2012va} the data
are quantitatively fitted). For the total cross section, we include the data that are rated as reliable in the GWU/SAID
analysis~\cite{Arndt:2005dg,Arndt:2006bf}. These are shown with filled squares in Fig.~\ref{fig:entot}. Note that in the GWU/SAID approach, the
total cross-section data from Prakhov~\cite{Prakhov:2005qb} are not included to avoid double counting with the respective differential
cross-section data that are included~\cite{strakowsky_private}. 

\begin{figure}
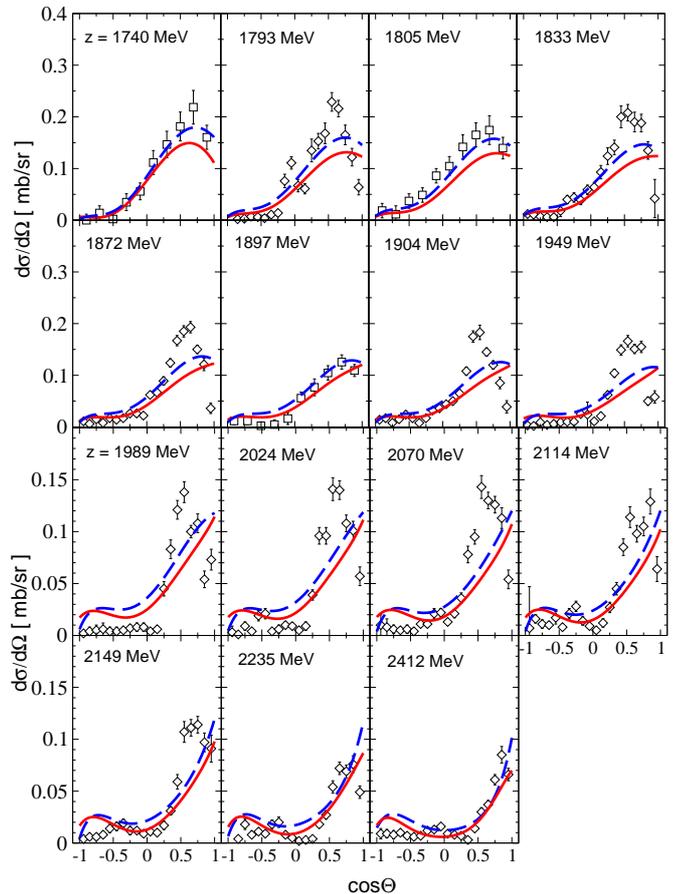

\begin{center}
\hspace*{0.04cm}\includegraphics[width=0.48\textwidth]{dsdo_etaN_4.eps}\\
\vspace*{-0.16cm}
\includegraphics[width=0.484\textwidth]{dsdo_etaN_5.eps}
\end{center}
\caption{Differential cross section [2/2] of the reaction $\pi^- p\to \eta n$. Solid (red) lines: fit A; dashed (blue) lines: fit B; data: empty
squares from Ref.~\cite{Richards:1970cy}; empty diamonds from Ref.~\cite{Brown:1979ii}. Note that the energy for data from
Ref.~\cite{Brown:1979ii} has been lowered (see text). The latter data are unreliable and enter the fit with very low weight.}
\label{fig:endiff2}     
\end{figure}

The results for the differential cross section are shown in Figs.~\ref{fig:endiff1} and \ref{fig:endiff2}, where the red solid line denotes fit
A and the blue dashed line represents fit B. The overall agreement is good for both fits and only slight differences in the data description can
be observed. For the lowest energies, it should be noted that due to the non-monochromatic pion beam and uncertainties in
calibration~\cite{Prakhov:2005qb}, in combination with the very steep rise of the cross section, the errors on the data are larger than those
shown in Fig.~\ref{fig:endiff1}; see the discussion in Ref.~\cite{Mai:2012wy, Prakhov:2005qb}.

For scattering energies of $z\sim 1575$~MeV and higher, the data situation becomes problematic.  In Fig.~\ref{fig:entot} the $\eta N$ total
cross-section data reflect these problems.  The data quality for the differential cross sections is quite low and conflicting at these energies,
and it is difficult to reliably fix the amplitude. Thus, fit A (red solid line) and fit B (blue dashed line) give different descriptions of the
total cross section from $z\sim 1700$~MeV to $z\sim 1850$~MeV, where fit A seems to be too low. This aspect also occurs in the differential
cross sections at the corresponding energies, which are better matched by fit B, although the weight on the different data sets was comparable
to the one in fit A.   

\begin{figure}
\begin{center}
\includegraphics[width=0.48\textwidth]{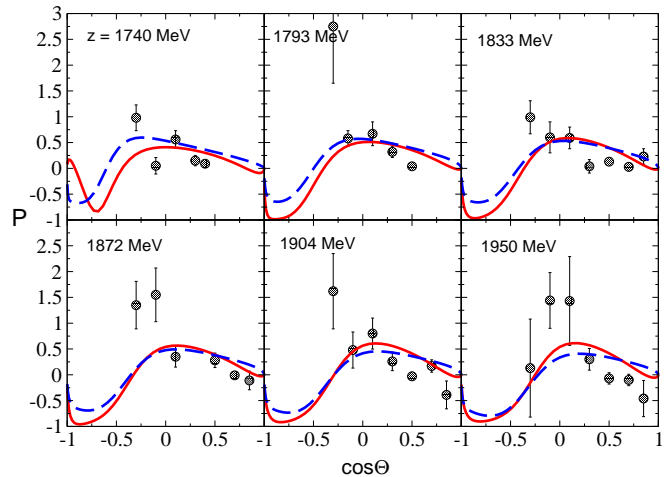}
\end{center}
\caption{Polarization [2/2] of the reaction $\pi^- p\to \eta n$. Solid (red) lines: fit A; dashed (blue) lines: fit B; data from
Ref.~\cite{Baker:1979aw}. Note that the energy for data from Ref.~\cite{Baker:1979aw} has been lowered, cf. Sec.\ref{database}. The data for $P$
were not included in the fit.}
\label{fig:enpol2}     
\end{figure}

Polarization data for the reaction $\pi^- p\to\eta n$ have only been published by Baker {\it et al.}~\cite{Baker:1979aw} in form of 12 polarized
cross sections from $z=1740$~MeV to $z=2235$~MeV. Here, in principle, the same beam calibration problems as in Brown {\it et
al.}~\cite{Brown:1979ii} could occur, since the same experimental apparatus and a similar beam line were used. Moreover, the results from the
latter experiment enter the analysis by Baker in the event selection of the final data sample. In Fig.~\ref{fig:enpol1} we still show results
for the polarized cross section.  The form of the polarization does not depend strongly on the energy, so that a $4\%$ error in the beam
momentum would not be outright noticeable. Because of the problems mentioned, the polarized cross sections entered our analysis with a very low
weight although the fits shown in Figs.~\ref{fig:enpol1} and \ref{fig:enpol2} exhibit an acceptable description of the data.

\begin{figure}
\begin{center}
\includegraphics[width=0.48\textwidth]{pola_x_dsdo_etaN_1.eps}\\
\vspace*{-0.07cm}
\hspace*{0.31cm}\includegraphics[width=0.466\textwidth]{pola_x_dsdo_etaN_2.eps}\\
\end{center}
\caption{Polarization [1/2] of the reaction $\pi^- p\to \eta n$. Solid (red) lines: fit A; dashed (blue) lines: fit B; data from
Ref.~\cite{Baker:1979aw}. Note that the beam energies for data from Ref.~\cite{Baker:1979aw} have been lowered and the data are included in the
fit with low weight (see text).}
\label{fig:enpol1}     
\end{figure}

To summarize, given the problematic data situation for the reaction $\pi^-p\to\eta n$ above $z>1.56$~GeV, we found it difficult to pin down the
partial-wave content at higher energies. 

\begin{figure}
\begin{center}
\includegraphics[width=0.48\textwidth]{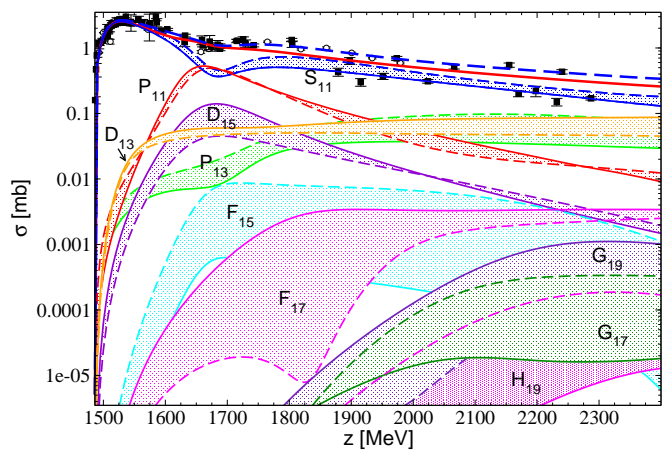}\\
\end{center}
\caption{Partial cross sections of the reaction $\pi^- p\to \eta n$. Solid lines: fit A; dashed lines: fit B; data: as in Fig.~\ref{fig:entot}.}
\label{fig:entotpart}     
\end{figure}

For reasons stated at the end of Sec.~\ref{sec:numerics}, we refrain from showing amplitudes for the inelastic reactions. Instead,  we present
the partial cross sections in Fig.~\ref{fig:entotpart}, where the solid lines represent fit A and the dashed lines denote fit B. 

\begin{figure}[h!]
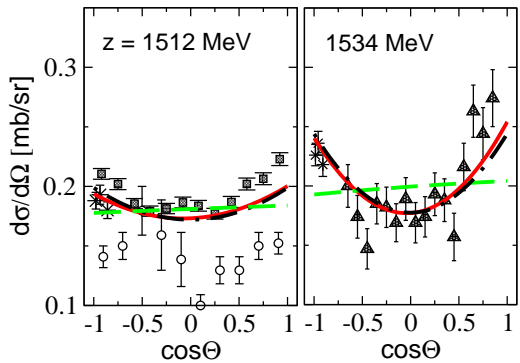

\begin{center}
\includegraphics[width=0.215\textwidth]{dsdo_etaN_1512_u-shape.eps}
\hspace*{-0.2cm}
\includegraphics[width=0.16\textwidth]{dsdo_etaN_1534_u-shape.eps}
\end{center}
\caption{Differential cross section of the reaction $\pi^- p\to \eta n$ at $z=1512$ and $z=1534$~MeV. Only results of fit A are shown. Solid
(red) lines: full result; dashed (green) lines: $S_{11}$, $P_{11}$ and $P_{13}$ contributions only; dash-dotted (black) lines: $S_{11}$ and
$D_{13}$  contributions only; data: as in Fig.~\ref{fig:endiff1}.}
\label{fig:etaN_s+dwave}     
\end{figure}

As can be seen in Fig.~\ref{fig:entotpart}, the strong increase of the total cross section just above the $\eta N$ threshold ($z=1486$~MeV) is
dominated in both fits by the $S_{11}$ wave, which is in line with the known strong branching ratio of the $N(1535)$ 1/2$^-$ resonance to the
$\eta N$ channel (cf. Table~\ref{tab:bra1}). The influence of the $S_{11}$ partial wave remains large over the whole energy range. At $z\sim
1.65$~GeV, the $P_{11}$ partial wave shows noticeable impact but decreases quickly. This resonance plays also a role in $\eta$ photoproduction,
as recently shown in Ref.~\cite{Shklyar:2012js}. In addition, the $D_{15}$ partial wave is strong at $z\sim 1.7$~GeV. At higher energies, the
$P_{13}$ and $D_{13}$ gain some influence besides the $S_{11}$. Higher partial waves like the $F_{17}$, $G_{17}$, $G_{19}$ and $H_{19}$ are very
small overall. The partial-wave content of the total cross section is similar in fit A and B. Note the logarithmic scale of
Fig.~\ref{fig:entotpart}, which causes discrepancies of the two fits for the small partial waves to appear enlarged. The differences in the
$F_{17}$ contribution, in contrast, are already apparent in the elastic $\pi N$ channel, c.f. Fig.~\ref{fig:pin2}. The u-shape of the
differential cross section visible from threshold to around $z\sim 1.6$~GeV is due to the $SD$-wave interference, as can be seen in
Fig.~\ref{fig:etaN_s+dwave}.

%%%%%%%%%%%%%%%%%%%%%%%%%%%%%%%%%%%%%%%%%%%%%%%%%%%%%%%%%%%%%%%%%%%%%%%%%%%%%%%%%%%%%%%%%%%%%%%%%%%%%%%%%%%%%%%%%%%%%%%%%%%%%%%%%%%%%%%%%%%%%%%%

\subsection{Reaction $\boldsymbol{\pi^- p\to K^0\Lambda}$}
\label{sec:kzl}
For the reaction $\pi^-p\to K^0\Lambda$, total and differential cross sections, polarization, and spin rotation parameter measurements from
$z=1626$~MeV up to $z=2405$~MeV are included in the fit. The data base contains differential cross sections at 46 energies and 8 different
experiments from Refs.~\cite{Baker:1978qm, Saxon:1979xu, Binford:1969ts, Dahl:1969ap, Bertanza:1962pt, Knasel:1975rr, Yoder:1963zg,
Goussu:1966ps}. In contrast to the $\eta N$ case, no severe inconsistencies appear in the $K\Lambda$ differential cross sections and systematic
uncertainties were applied as given by the respective authors in the original publications. 

\begin{figure*}[h!]
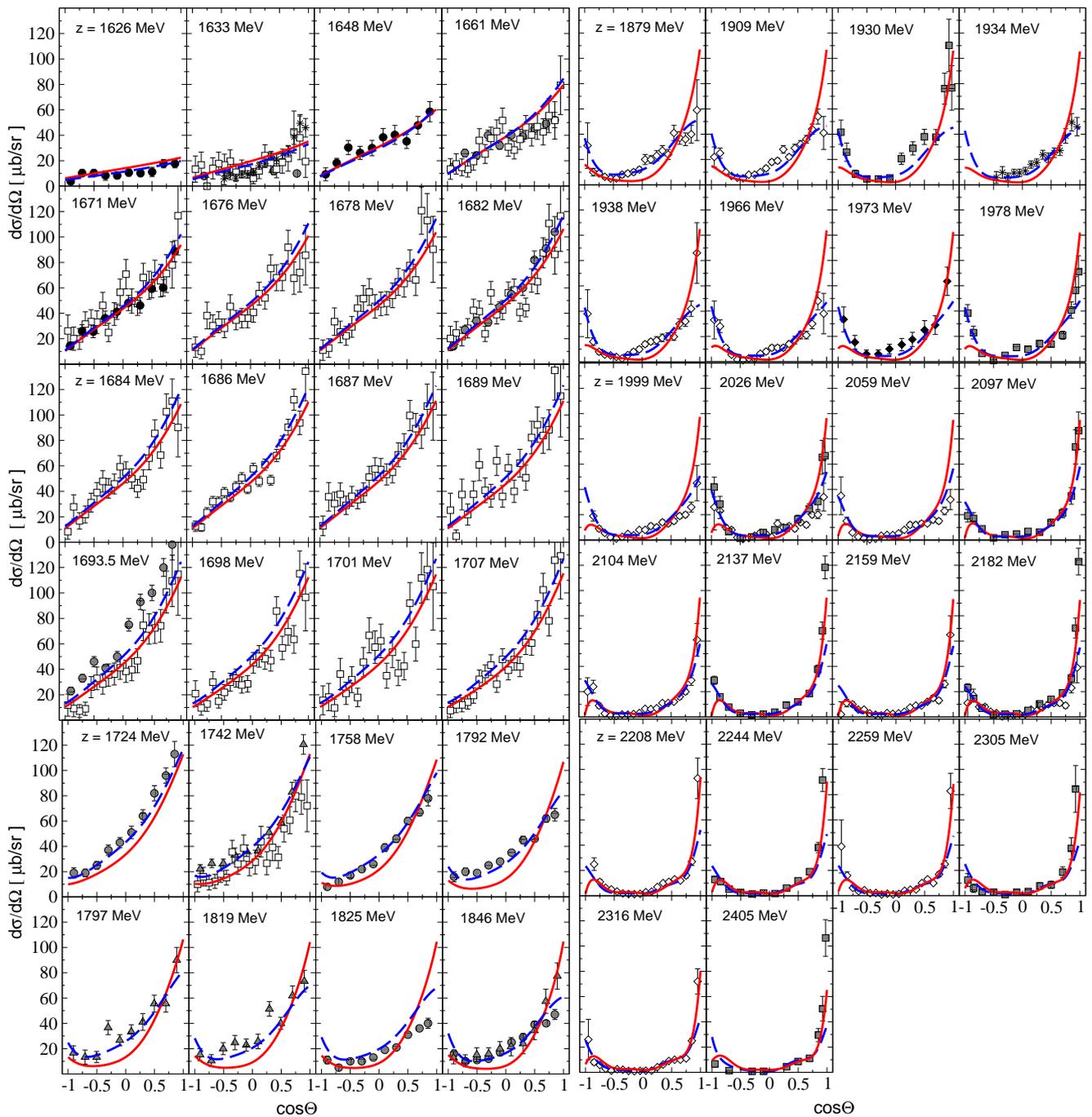

\includegraphics[width=0.506\textwidth]{dsdo_K0Lam_1.eps}
\includegraphics[width=0.458\textwidth]{dsdo_K0Lam_4.eps}\\
\vspace*{-0.1cm}
\includegraphics[width=0.506\textwidth]{dsdo_K0Lam_2.eps}
\includegraphics[width=0.458\textwidth]{dsdo_K0Lam_5.eps}\\
\vspace*{-0.1cm}
\includegraphics[width=0.506\textwidth]{dsdo_K0Lam_3.eps}
\includegraphics[width=0.458\textwidth]{dsdo_K0Lam_6.eps}
\caption{Differential cross section of the reaction $\pi^- p\to K^0\Lambda$. Solid (red) lines: fit A; dashed (blue) lines: fit B; data:
filled circles from Ref.~\cite{Bertanza:1962pt}; empty squares from Ref.~\cite{Knasel:1975rr}; partially filled circles from
Ref.~\cite{Baker:1978qm}; stars from Ref.~\cite{Yoder:1963zg}; triangles from Ref.~\cite{Binford:1969ts}; empty diamonds from
Ref.~\cite{Saxon:1979xu}; partially filled squares from Ref.~\cite{Dahl:1969ap}; filled diamonds from Ref.~\cite{Goussu:1966ps}. }
\label{fig:kldiff1}     
\end{figure*}

\begin{figure}
\begin{center}
\includegraphics[width=0.48\textwidth]{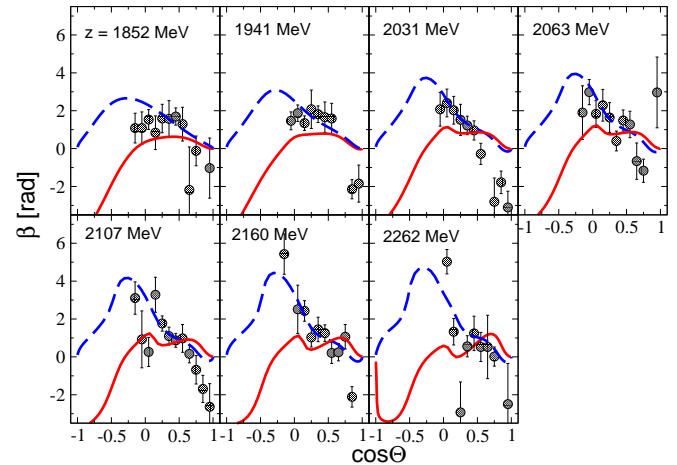}\\
\end{center}
\caption{Spin-rotation parameter $\beta$ of the reaction $\pi^- p\to K^0\Lambda$. Solid (red) lines: fit A; dashed (blue) lines: fit B; data
from Ref.~\cite{Bell:1983dm}. Note that $\beta$ is $2\pi$ cyclic. }
\label{fig:klbeta}     
\end{figure}

For the recoil polarization, we use the data of Refs.~\cite{Baker:1978qm, Saxon:1979xu, Binford:1969ts, Bertanza:1962pt, Bell:1983dm}. 27
energies are covered. At backward angles $\varTheta \geq 90^\circ$, the data exhibit inconsistencies and huge error bars. It should be noted
that several data points take unphysical values larger than 1. A measurement of the spin-rotation parameter $\beta$ was reported in
Ref.~\cite{Bell:1983dm}. All seven data sets at energies from $z=1852$~MeV to $2262$~MeV are included in our analysis; cf.
Fig.~\ref{fig:klbeta}. Note that this observable is $2\pi$ cyclic. Errors were applied as stated by the authors.  The spin-rotation angle can
also be written in terms of the  spin-transfer coefficients $K_{ij}$, 
\be
\tan\beta=\frac{2\,{\rm Im}\,(gh^*)}{|g|^2-|h|^2}=\frac{K_{zx}}{K_{zz}}=-\frac{K_{xz}}{K_{xx}}=\frac{K_{zx}}{K_{xx}}
   =-\frac{K_{xz}}{K_{zz}}\; , \non
\label{betakzz}
\ee
where $g$ is the spin-non-flip and $h$ the spin-flip amplitude (see Ref.~\cite{Doring:2010ap} for a definition of observables and amplitudes).

The data for total cross section of $K\Lambda$ production included in the fit are from Refs.~\cite{Baker:1978qm, Saxon:1979xu, Binford:1969ts,
Dahl:1969ap, Bertanza:1962pt, Knasel:1975rr,  Yoder:1963zg, Goussu:1966ps, Jones:1971zm, VanDyck:1969ay, Keren:1964ra, Eisler:1958,
Miller:1965}. Additionally included data from Ph.D theses are referenced in Ref.~\cite{Landolt}. 

\begin{figure}
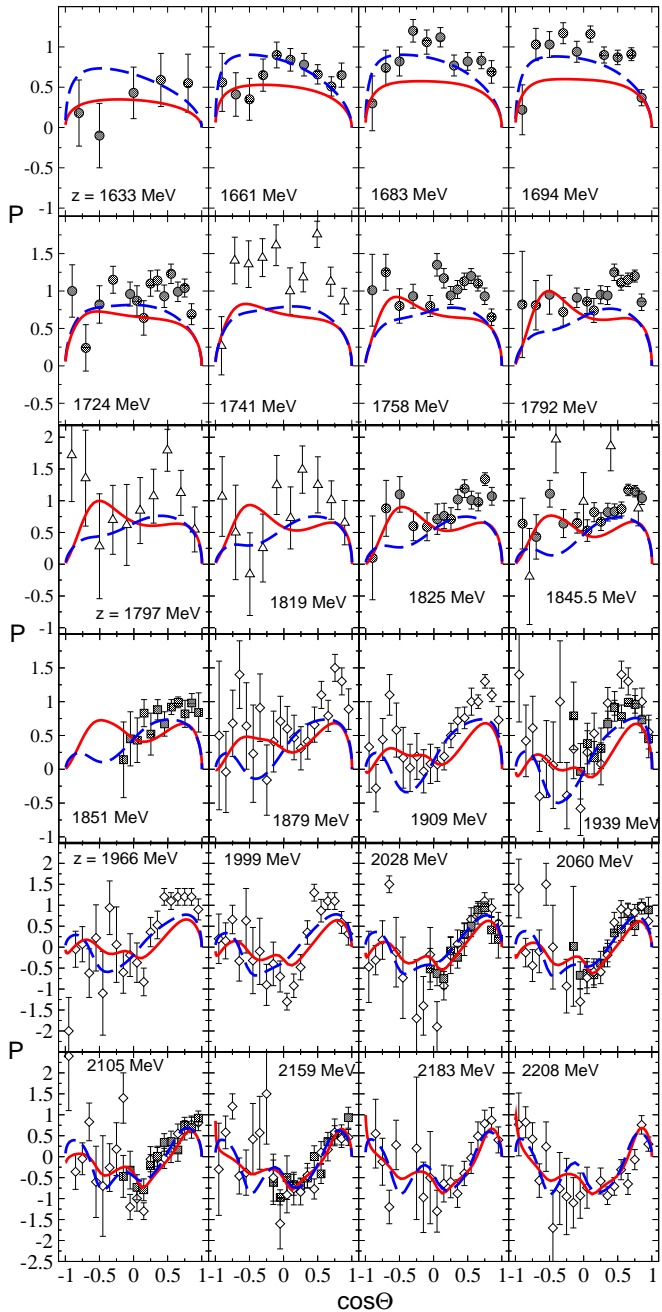

\begin{center}
\includegraphics[width=0.48\textwidth]{pola_K0Lam_1.eps}\\
\vspace*{-0.1cm}
\hspace*{0.03cm}\includegraphics[width=0.48\textwidth]{pola_K0Lam_2.eps}\\
\vspace*{-0.1cm}
\includegraphics[width=0.48\textwidth]{pola_K0Lam_3.eps}
\end{center}
\caption{Polarization of the reaction $\pi^- p\to K^0\Lambda$. Solid (red) lines: fit A; dashed (blue) lines: fit B; data: filled circles from
Ref.~\cite{Baker:1978qm}; empty triangles up from Ref.~\cite{Binford:1969ts}; filled squares from Ref.~\cite{Bell:1983dm}; empty diamonds from
Ref.~\cite{Saxon:1979xu}. }
\label{fig:klpola1}     
\end{figure}

\begin{figure}
\begin{center}
\includegraphics[width=0.32\textwidth]{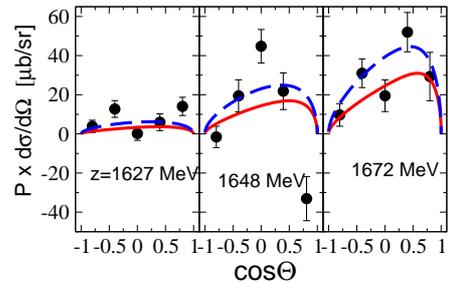}
\end{center}
\caption{Polarized cross section of the reaction $\pi^- p\to K^0\Lambda$. Solid (red) lines: fit A; dashed (blue) lines: fit B; data from
Ref.~\cite{Bertanza:1962pt}.}
\label{fig:klpola2}     
\end{figure}

\begin{figure}
\begin{center}
\includegraphics[width=0.48\textwidth]{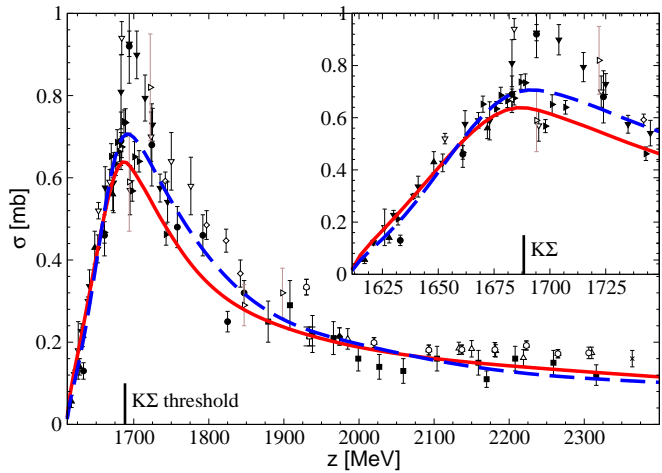}\\
\end{center}
\caption{Total cross section of the reaction $\pi^- p\to K^0 \Lambda$. Solid (red) lines: fit A; dashed (blue) lines: fit B; data: filled
circles from Ref.~\cite{Baker:1978qm}; filled squares from Ref.~\cite{Saxon:1979xu}; empty diamonds from Ref.~\cite{Binford:1969ts}; empty
triangles up from Ref.~\cite{Dahl:1969ap}; filled triangles up from Ref.~\cite{Bertanza:1962pt}; filled triangles down from
Ref.~\cite{Jones:1971zm}; empty triangles down from Ref.~\cite{VanDyck:1969ay}; filled triangles right from Ref.~\cite{Knasel:1975rr}; empty
triangles left from Ref.~\cite{Keren:1964ra}; empty triangles right from Ref.~\cite{Eisler:1958}; empty squares from Ref.~\cite{Yoder:1963zg};
filled diamonds from Ref.~\cite{Goussu:1966ps}; stars from  Ref.~\cite{Miller:1965}; for empty circles and crosses see Ref.~\cite{Landolt}.}
\label{fig:kltot}     
\end{figure}

The current fit result is compared with the data in Figs.~\ref{fig:kldiff1} to \ref{fig:kltot}. The description of the $\pi^-p\to K\Lambda$
differential cross section, polarization, and spin rotation parameter is satisfactory for both fits A and B. At low energies, the description of
the differential cross sections are very similar in fit A and B, while at medium energies fit B seems to be a little better. At higher energies,
the peak in forward direction is more pronounced in fit A, but also clearly visible in fit B. The backward peak, on the other hand, is better
described in fit B. Note that the behavior of fit A at backward angles for higher energies, where it bends downwards at $\text{cos}\,
\varTheta\sim-1$, also appears in some data sets, e.g. at $z=2159$~MeV and $z=2182$~MeV. Here, the data situation seems to be slightly
inconsistent. Differences in fits A and B for the polarization (Fig.~\ref{fig:klpola1}) around $\text{cos}\, \varTheta=-0.5$ show that the data
quality is not good enough to reliably fix the amplitude. The total cross section, shown in Fig.~\ref{fig:kltot}, is also well described,
although there are conflicting measurements around $z\sim 1.7$~GeV. This is reflected in the discrepancies around that energy between fits A and
B. Neither of the two fits, however, reaches the highest data points. It should be noted that the differential cross section in that energy
range, shown in Fig.~\ref{fig:kldiff1}, is well described; this applies especially to fit B. Hence, our result suggests that the total cross
section peaks with $\sigma\sim 0.7$~mb, and that the conflicting data with higher values should be discarded.

In Fig.~\ref{fig:klbeta}, we show the description of the spin-rotation parameter $\beta$. The most apparent discrepancies in fits A and B appear
at backward angles where no data are available to constrain the fit. This demonstrates the need for precise data in all observables and angles
to reliably determine the amplitude.

\begin{figure}
\begin{center}
\includegraphics[width=0.48\textwidth]{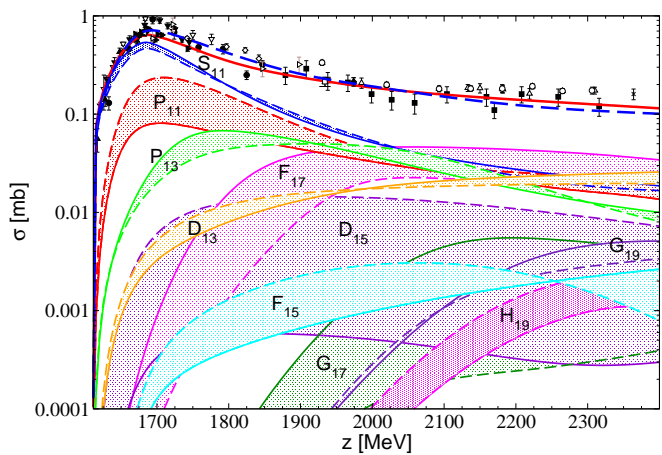}\\
\end{center}
\caption{Partial cross section of the reaction $\pi^- p\to K^0 \Lambda$. Solid lines: fit A; dashed lines: fit B; data: same as in
Fig.~\ref{fig:kltot}.}
\label{fig:kltotpart}     
\end{figure}
The partial cross sections are shown in Fig.~\ref{fig:kltotpart}. The solid lines indicate fit A and the dashed lines represent fit B. As will
be  discussed in Sec.~\ref{sec:p11}, the description of the $K^0\Lambda$ data at low energies requires a $P_{11}$ resonance, which, in turn,
induces a structure in the $\pi N\to\pi N$ amplitude that matches the single-energy solution~\cite{Arndt:2006bf} even though the (unstructured)
energy dependent solution in $\pi N\to \pi N$ was fitted. The strong influence of the $P_{11}$ on the $K^0\Lambda$ channel at low energies is
reflected in the partial cross sections. Of the remaining partial waves only the $P_{13}$ shows some influence at low energies, which is,
however, not comparable to the $P_{11}$ and $S_{11}$. At higher energies above 2~GeV, the $P_{11}$, $S_{11}$ and $P_{13}$ decrease and the
$D_{13}$ and $F_{17}$ become stronger in both fits. 

%%%%%%%%%%%%%%%%%%%%%%%%%%%%%%%%%%%%%%%%%%%%%%%%%%%%%%%%%%%%%%%%%%%%%%%%%%%%%%%%%%%%%%%%%%%%%%%%%%%%%%%%%%%%%%%%%%%%%%%%%%%%%%%%%%%%%%%%%%%%%%%%

\subsection{Reaction $\boldsymbol{\pi^- p\to K^0\Sigma^0}$}
\label{sec:kzsz}
\begin{figure}
\begin{center}
\includegraphics[width=0.48\textwidth]{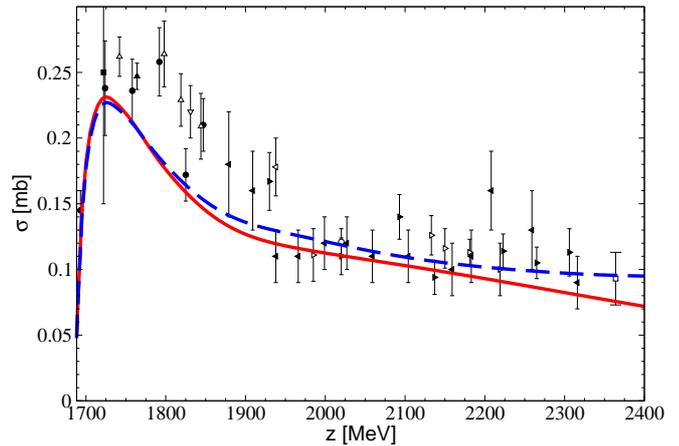}\\
\end{center}
\caption{Total cross section of the reaction $\pi^- p\to K^0 \Sigma^0$. Solid (red) line: fit A; dashed (blue) line: fit B; data: empty
triangles up: Ref.~\cite{Binford:1969ts}; empty triangles right: Ref.~\cite{Dahl:1969ap}; filled squares: Ref.~\cite{Eisler:1958}; empty
triangles left: Ref.~\cite{Goussu:1966ps}; filled triangles up: Ref.~\cite{Anderson:1966} filled circles: Ref.~\cite{Baker:1978bb}; empty
circles: Ref.~\cite{Thomas:1973uh}; filled triangles left: Ref.~\cite{Hart:1979jx}. For empty triangles down, filled triangles right, and empty
squares see \cite{Landolt}.}
\label{fig:kzsztot}     
\end{figure}

\begin{figure}
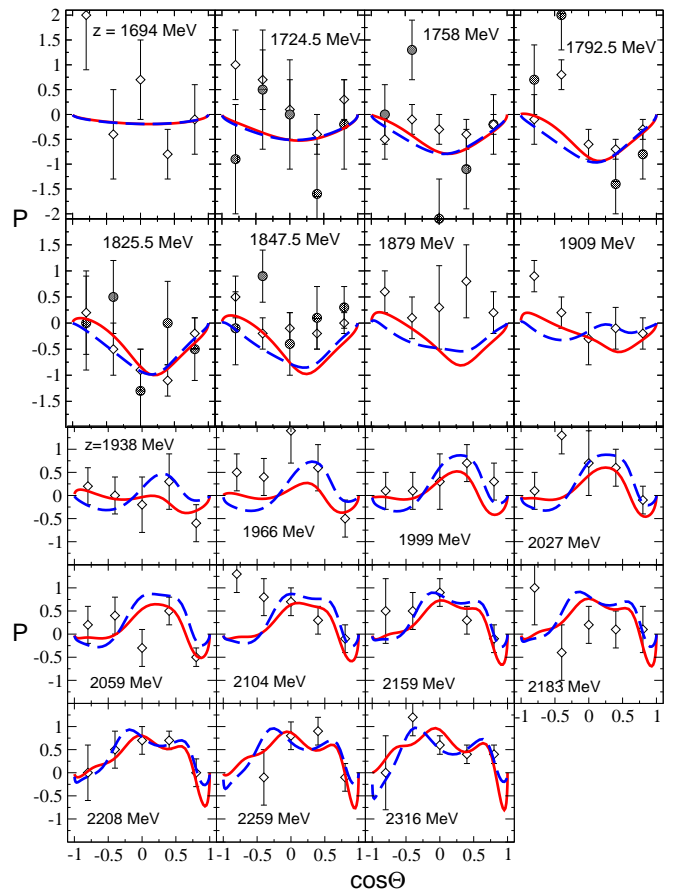

\begin{center}
\includegraphics[width=0.48\textwidth]{pola_K0S0_1.eps}\\
\vspace*{-0.1cm}
\includegraphics[width=0.48\textwidth]{pola_K0S0_2.eps}
\end{center}
\caption{Polarization of the reaction $\pi^- p\to K^0\Sigma^0$. Solid (red) lines: fit A; dashed (blue) lines: fit B; data: circles from
Ref.~\cite{Baker:1978bb}; diamonds from Ref.~\cite{Hart:1979jx}. }
\label{fig:kzszpola}     
\end{figure}

\begin{figure*}
\includegraphics[width=0.506\textwidth]{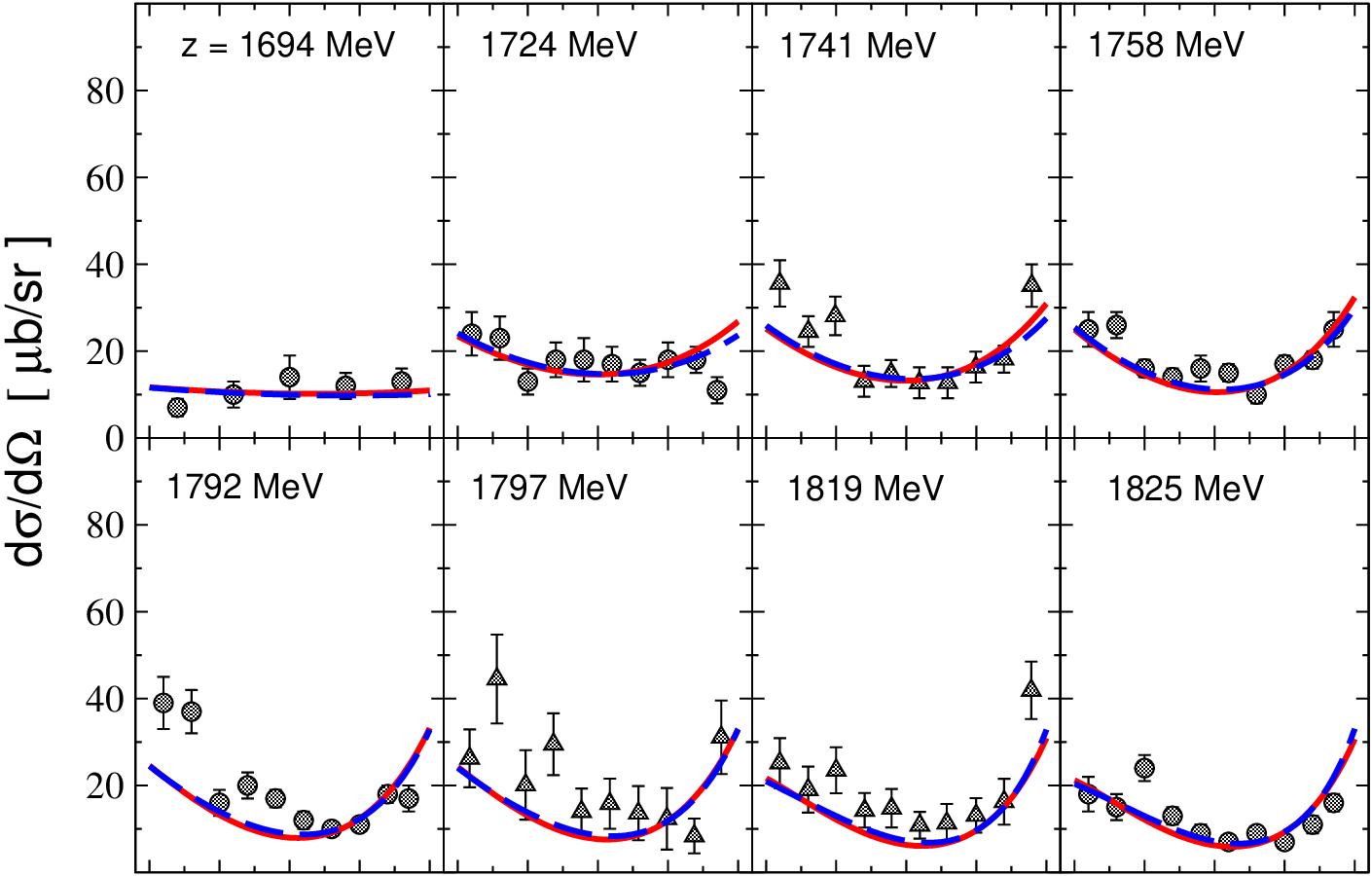}
\includegraphics[width=0.458\textwidth]{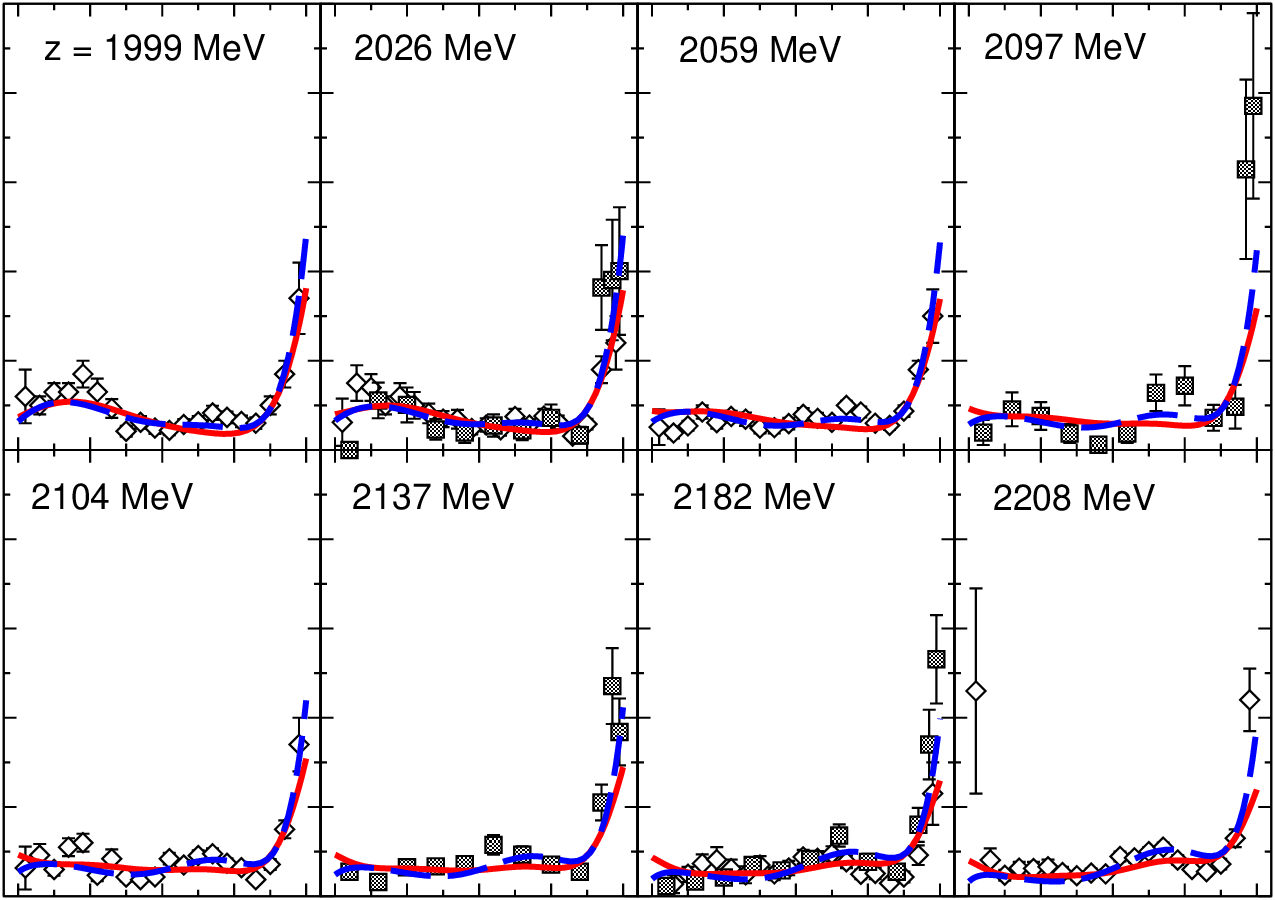}\\
\vspace*{-0.1cm}
\includegraphics[width=0.506\textwidth]{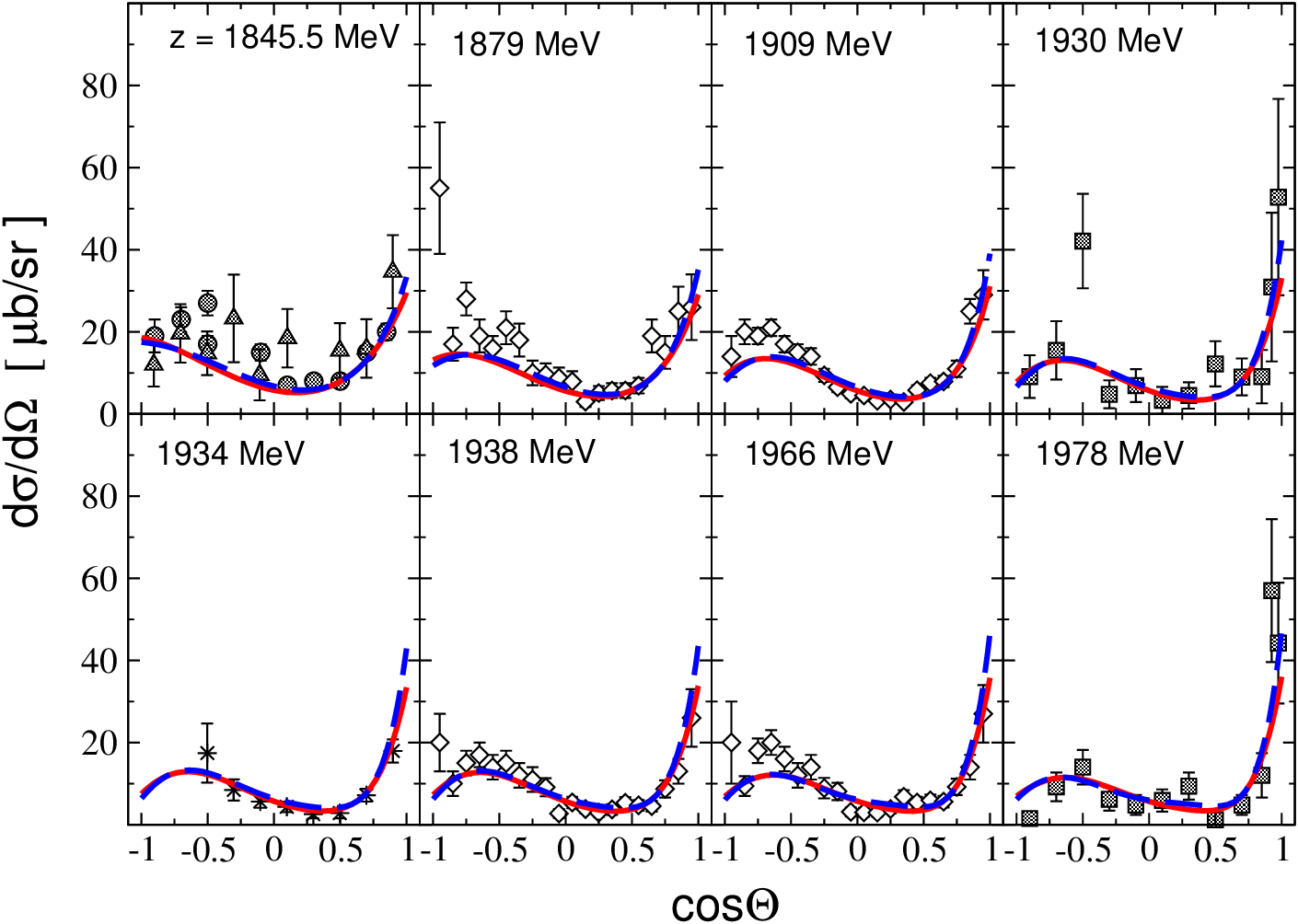}
\includegraphics[width=0.458\textwidth]{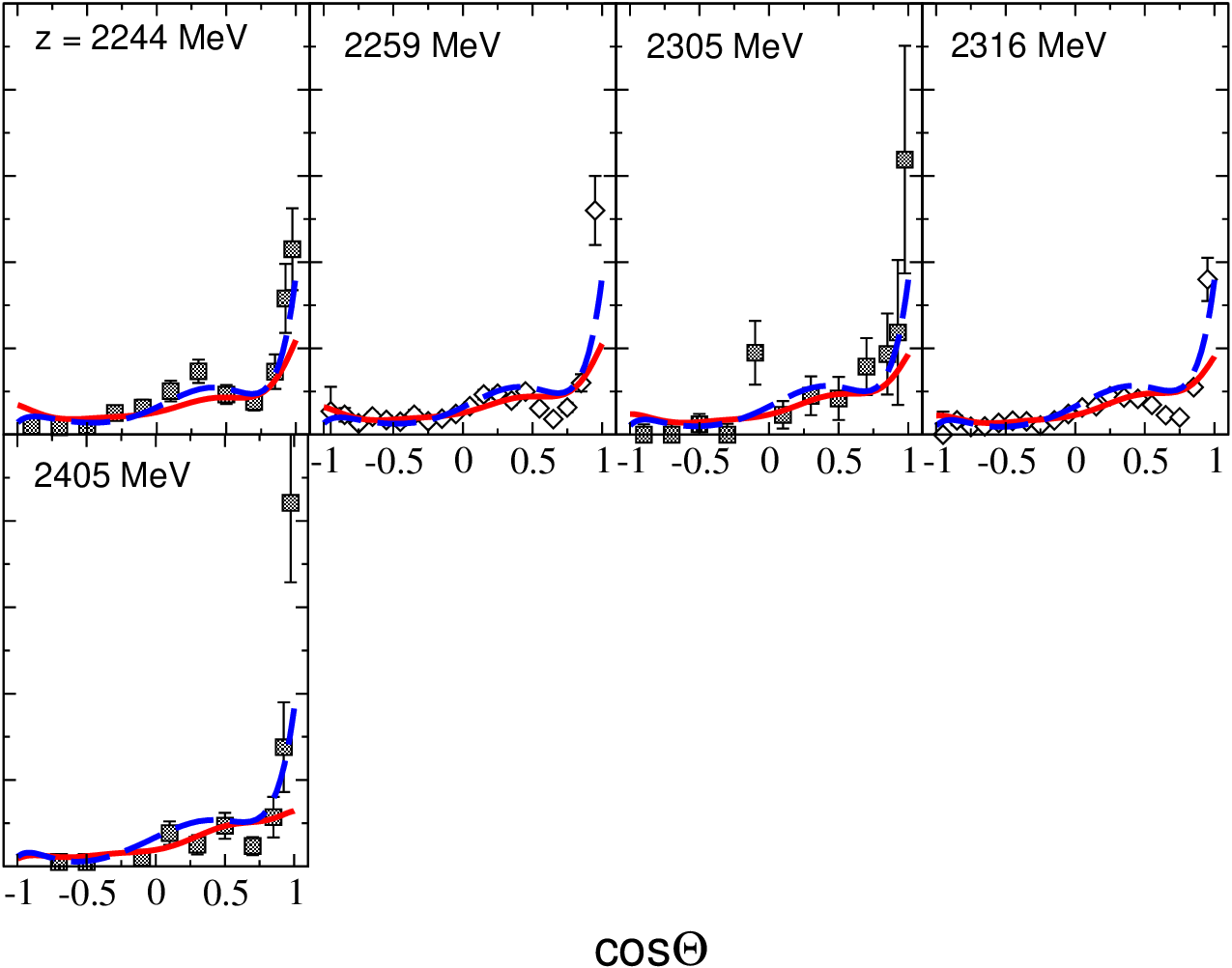}
\caption{Differential cross section of the reaction $\pi^- p\to K^0\Sigma^0$. Solid (red) lines: fit A; dashed (blue) lines: fit B; data:
circles from Ref.~\cite{Baker:1978bb}; up triangles from Ref.~\cite{Binford:1969ts}; squares from Ref.~\cite{Dahl:1969ap}; diamonds from
Ref.~\cite{Hart:1979jx}; stars from Ref.~\cite{Yoder:1963zg}.}
\label{fig:kzszdiff1}     
\end{figure*}

\begin{figure}
\begin{center}
\includegraphics[width=0.48\textwidth]{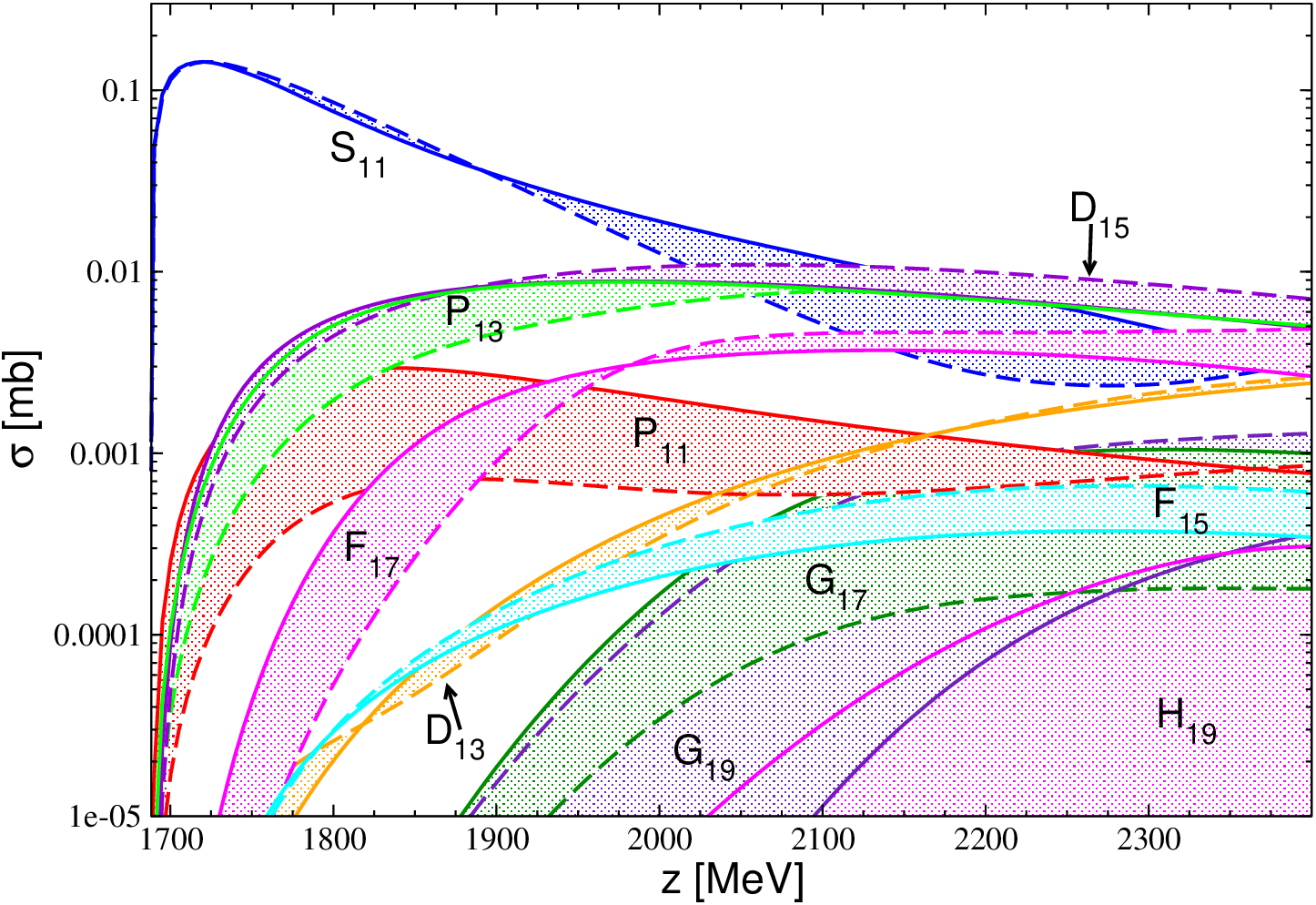}\\
\end{center}
\caption{Partial cross sections of the reaction $\pi^- p\to K^0 \Sigma^0$ in the isospin 1/2 sector. Solid lines: fit A; dashed lines: fit B.
 Note that the partial cross sections for the other $K\Sigma$ final states can be obtained from these
curves, Fig.~\ref{fig:kzsztotpart_3h}, and Eq.~(\ref{csconnect}).}
\label{fig:kzsztotpart_1h}     
\end{figure}

\begin{figure}
\begin{center}
\includegraphics[width=0.48\textwidth]{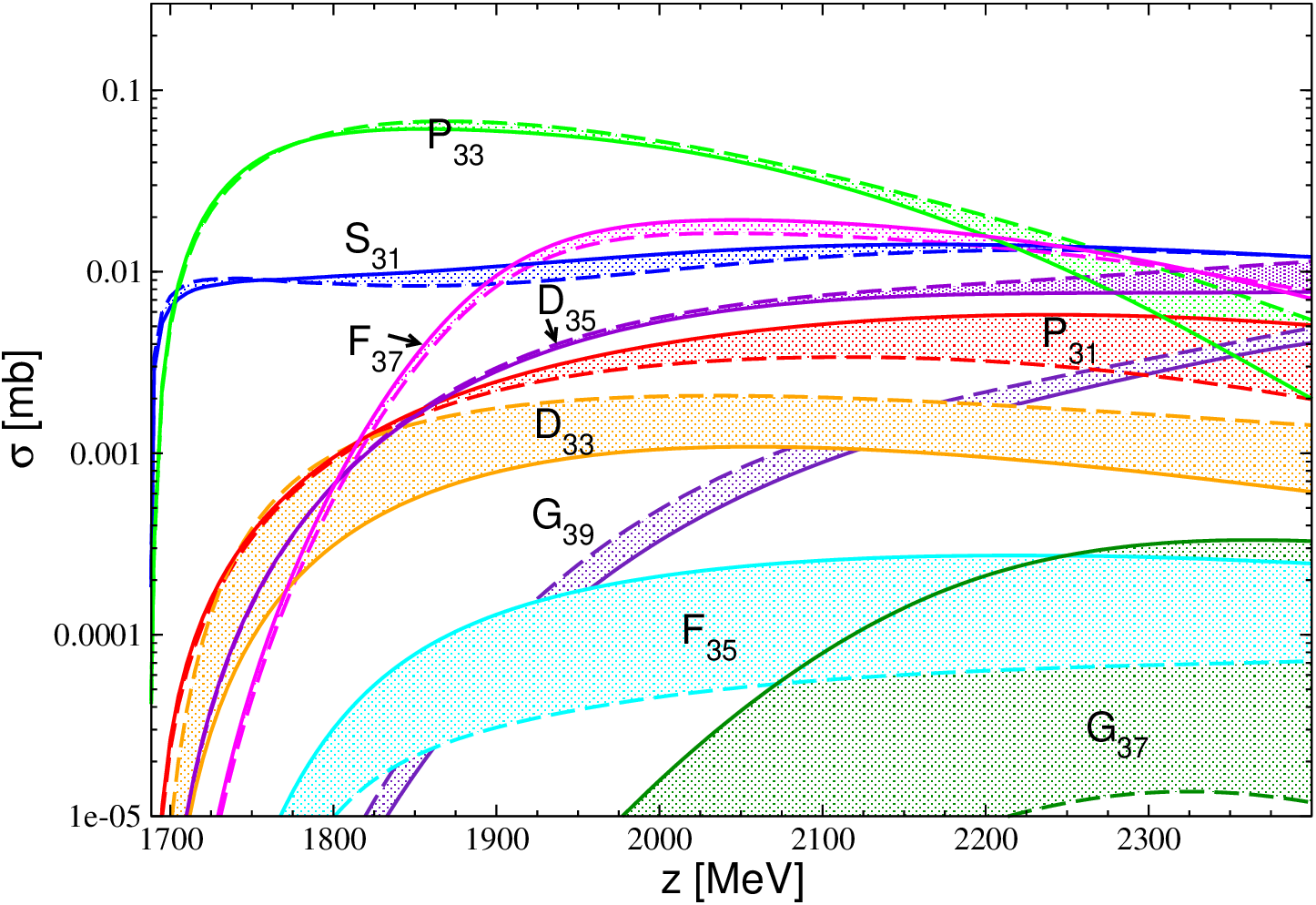}\\
\end{center}
\caption{Partial cross sections of the reaction $\pi^- p\to K^0 \Sigma^0$ in the isospin 3/2 sector. Solid lines: fit A; dashed lines: fit B.
 Note that the partial cross sections for the other $K\Sigma$ final states can be obtained from these
curves, Fig.~\ref{fig:kzsztotpart_1h}, and Eq.~(\ref{csconnect}).}
\label{fig:kzsztotpart_3h}     
\end{figure}

The results for the reaction $\pi^- p\to K^0\Sigma^0$ are shown in Figs.~\ref{fig:kzsztot} to \ref{fig:kzsztotpart_3h}. Here, we take into
account differential cross sections at 29 different energies from $z=1694$~MeV to $z=2405$~MeV from Refs.~\cite{Binford:1969ts, Yoder:1963zg,
Dahl:1969ap, Baker:1978bb, Hart:1979jx}. References~\cite{Baker:1978bb, Hart:1979jx} also provide polarization data. The two data sets, shown in
Fig.~\ref{fig:kzszpola}, do not seem to be consistent at all places and both exhibit large error bars, which reduces the impact on the fit
result. Also, in several places the data take unphysical values larger than 1. For all data in this channel systematic uncertainties were
applied as given in the original publications. The data for the total cross section are taken from Refs.~\cite{Binford:1969ts, Dahl:1969ap,
Eisler:1958, Goussu:1966ps, Anderson:1966, Baker:1978bb, Thomas:1973uh, Hart:1979jx, Landolt}. We do not use the data provided by Livanos {\it
et al.}~\cite{Livanos:1980vj} for this reaction and other $K\Sigma$ final states as they carry large error bars compared to other data.

The current fit results match the differential cross section well. There are only minor differences in fits A and B, predominantly at higher
energies where the forward peak is more pronounced in fit B. Around $z\sim 1.8$~GeV, the total cross section seems to be underestimated a bit by
both fits, but the inspection of the differential cross section in Fig.~\ref{fig:kzszdiff1} shows that there are only a few data points at
backwards angles that are not matched well. In addition, the distribution of the data at backward angles at $z=1792$~MeV and $z=1797$~MeV and
the adjacent energy bins suggests inconsistencies in the data. 

Next, we discuss the partial-wave content of the reaction $\pi^- p\to K^0 \Sigma^0$. The partial cross sections $\sigma^{L_{2I;\,2J}}$ for the
other reactions $\pi^- p\to K^+\Sigma^-$ and $\pi^+ p\to K^+\Sigma^+$, where $L(J)$ is the orbital (total) angular momentum and $I$ the isospin,
are related to the partial cross sections of the $\pi^- p\to K^0 \Sigma^0$ reaction,
\be
\sigma^{L_{1;2J}}_{K^+\Sigma^-}&=& 2\, \sigma^{L_{1;2J}}_{K^0\Sigma^0}, \quad
\sigma^{L_{3;2J}}_{K^+\Sigma^-}= \frac{1}{2} \, \sigma^{L_{3;2J}}_{K^0\Sigma^0},\non
\sigma^{L_{3;2J}}_{K^+\Sigma^+}&=&\frac{9}{2} \, \sigma^{L_{3;2J}}_{K^0\Sigma^0}.
\label{csconnect}
\ee
We therefore only show the partial cross sections for the reaction $\pi^- p\to K^0 \Sigma^0$ in Figs.~\ref{fig:kzsztotpart_1h} and
\ref{fig:kzsztotpart_3h}. The partial cross sections for the final states $K^+\Sigma^-$ and $K^+\Sigma^+$ can be obtained from these curves and
Eq.~(\ref{csconnect}). 

In contrast to $K^0\Lambda$, the isospin $I=1/2$ partial cross sections for the $K\Sigma$ channels (Fig. \ref{fig:kzsztotpart_1h}) are dominated
almost entirely by the $S_{11}$ partial wave in both fits. The $P_{11}$ partial wave plays nearly no role at all in fit B, while in fit A it is
second to the influence of the $S_{11}$ at very low energies. At medium energies, the $D_{15}$ and $P_{13}$ are second to the influence of the
$S_{11}$ in both fits, but not of comparable degree. At higher energies, they exceed the $S_{11}$. Note that Fig.~\ref{fig:kzsztotpart_1h}
suggests that the $D_{15}$ and $P_{13}$ partial waves have the same threshold behavior for fit A, which is, however, not the case as a closer
inspection shows.

For the isospin $I=3/2$ sector (Fig. \ref{fig:kzsztotpart_3h}), the $P_{33}$ is the dominating partial wave over almost the entire energy range
up to $z=2.2$~GeV, except for a narrow region just above threshold where the $S_{31}$ is stronger than the $P_{33}$. This distribution applies
to both fits. In the present analysis, around $z=1950$~MeV the $F_{37}$ partial cross section increases to the same magnitude and becomes even
larger than the $S_{31}$. In the previous analysis of Ref.~\cite{Doring:2010ap}, we also found an increase of the $F_{37}$ at the same energy.
The influence of the remaining partial waves on the total cross section seems to be limited, with the exception of the $D_{35}$ which shows a
similar development as the $F_{37}$, but with a weaker slope. The isospin $I=3/2$ partial-wave content is very similar for fits A and B. The
differences become larger only at very high energies. Note the logarithmic scale of Figs.~\ref{fig:kzsztotpart_1h} and \ref{fig:kzsztotpart_3h},
which causes the discrepancies between the small partial waves to appear enlarged.

%%%%%%%%%%%%%%%%%%%%%%%%%%%%%%%%%%%%%%%%%%%%%%%%%%%%%%%%%%%%%%%%%%%%%%%%%%%%%%%%%%%%%%%%%%%%%%%%%%%%%%%%%%%%%%%%%%%%%%%%%%%%%%%%%%%%%%%%%%%%%%%%

\subsection{Reaction $\boldsymbol{\pi^- p\to K^+\Sigma^-}$}
\label{sec:kpsm}

As mentioned in Sec.~\ref{database}, this reaction is of particular interest due to its $u$-channel sensitivity. Considering the $K^+\Sigma^-$
final state is also crucial to reliably disentangle the isospin content of the $K\Sigma$ final state. See also Sec.~\ref{sec:kpsp} and
Fig.~\ref{fig:nokpsm} where we test the consequences if the data of the reaction $\pi^-p\to K^+\Sigma^-$ are ignored in the fit.

\begin{figure}
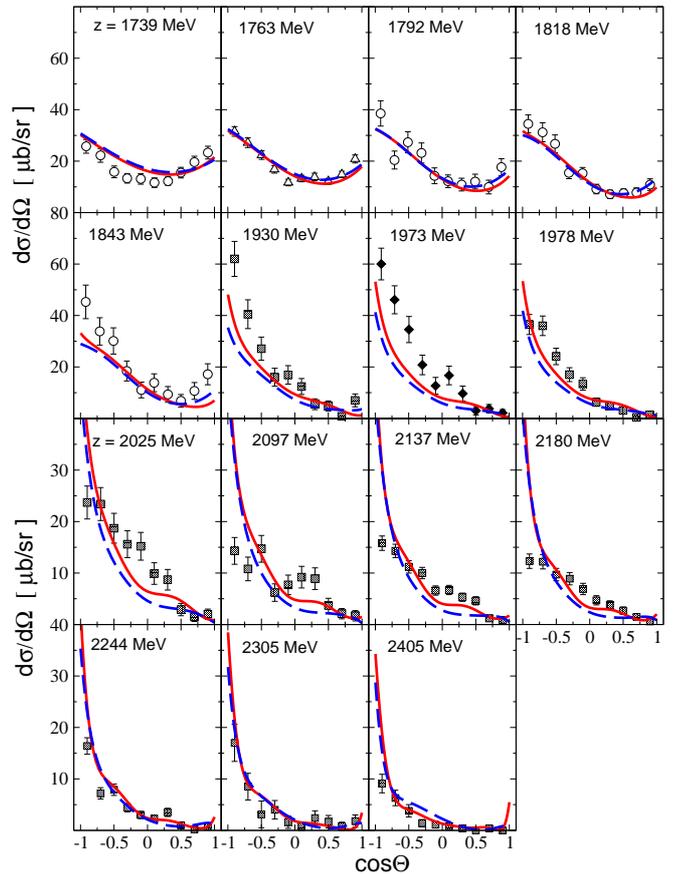

\begin{center}
\includegraphics[width=0.48\textwidth]{dsdo_K+S-_1.eps}\\
\vspace*{-0.13cm}
\hspace*{0.04cm}\includegraphics[width=0.478\textwidth]{dsdo_K+S-_2.eps}
\end{center}
\caption{Differential cross section of the reaction $\pi^- p\to K^+\Sigma^-$. Solid (red) lines: fit A; dashed (blue) lines: fit B; data:
circles from Ref.~\cite{Good:1969rb}; up triangles from Ref.~\cite{Doyle:1968zz}; squares from Ref.~\cite{Dahl:1969ap}; diamonds from
Ref.~\cite{Goussu:1966ps}.}
\label{fig:kpsmdiff1}     
\end{figure}

\begin{figure}
\begin{center}
\includegraphics[width=0.48\textwidth]{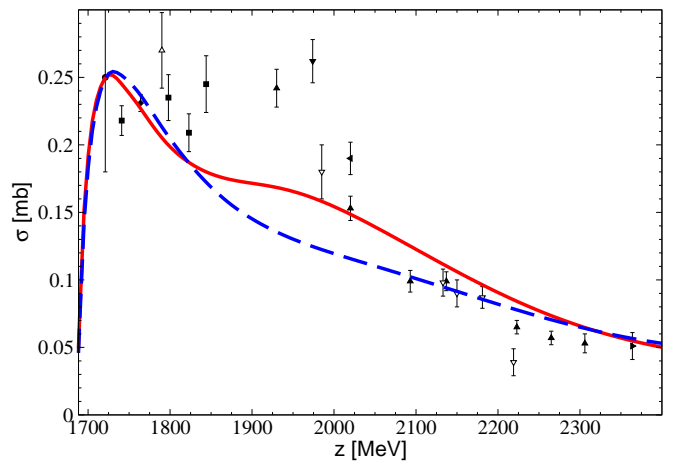}\\
\end{center}
\caption{Total cross section of the reaction $\pi^- p\to K^+ \Sigma^-$. Solid (red) line: fit A; dashed (blue) line: fit B; data: empty
triangles down: Ref.~\cite{Dahl:1969ap}; filled circles: Ref.~\cite{Eisler:1958}; filled triangles down: Ref.~\cite{Goussu:1966ps}; filled
triangles left: Ref.~\cite{Thomas:1973uh}; empty triangles up: Ref.~\cite{Crawford:1959}; filled squares: Ref.~\cite{Good:1969rb}; filled
diamonds: Ref.~\cite{Doyle:1968zz}. For filled triangles up and filled triangles right see Ref.~\cite{Landolt}.} 
\label{fig:kpsmtot}      
\end{figure}

The data base for the reaction $\pi^- p\to K^+\Sigma^-$ comprises only 15 differential cross section sets \cite{Dahl:1969ap, Goussu:1966ps,
Good:1969rb, Doyle:1968zz} from $z=1739$~MeV up to $2405$~MeV. Although no severe inconsistencies are observed, the influence of this reaction
in the fitting procedure is limited due to the small number of data points. For the total cross section, data from Refs.~\cite{Dahl:1969ap,
Eisler:1958, Goussu:1966ps, Thomas:1973uh, Crawford:1959, Good:1969rb, Doyle:1968zz, Landolt} are employed.  The data are shown in
Figs.~\ref{fig:kpsmdiff1} and \ref{fig:kpsmtot}. The data base is quite limited, yet the fact that the total cross section is significantly
smaller than for the $K^+\Sigma^+$ final state makes the description of this channel challenging, setting tight limits on the size of
$u$-channel exchanges, and through rescattering, also of $t$-channel contributions.

The present description is good and very similar in fits A and B at low energies but shows deviations at higher energies, as visible in
Fig.~\ref{fig:kpsmdiff1}. For intermediate energies around $z\sim 1.9$~GeV, the backward peak from $u$-channel exchanges is underestimated, in
fit B even a little more than in fit A, while at higher energies the present results tend to overestimate the data at very backward angles. This
is also visible in the description of the  total cross section in Fig~\ref{fig:kpsmtot}. The data could be conflicting around $z\sim
1.9-2.0$~GeV and the current result clearly represents a compromise in the description of the data at medium and high energies.

As explained in Sec. \ref{sec:kzsz}, the partial-wave content of the reaction $\pi^- p\to K^+ \Sigma^-$ can be obtained from the one of $\pi^-
p\to K^0 \Sigma^0$ (Figs. \ref{fig:kzsztotpart_1h}, \ref{fig:kzsztotpart_3h}), scaled by different isospin factors: as Eq.~(\ref{csconnect})
shows, the $K^+ \Sigma^-$ final state is more sensitive to the isospin $I=1/2$ sector than $K^0\Sigma^0$. This is important for the
determination of the $K\Sigma$ branching ratios of $I=1/2$ resonances.  

As this reaction channel is so important, better data would mean a significant step forward to disentangle the reaction  dynamics in $K\Sigma$
production. This is further motivated by the fact that high precision $K\Sigma$ photoproduction data exist~\cite{McNabb:2003nf, Bradford:2005pt,
Tran:1998qw, Glander:2003jw, Bockhorst:1994jf, Sumihama:2005er, Lleres:2007tx, Zegers:2003ux, Bradford:2006ba, Dey:2010hh, Kohri:2006yx, thesis,
Lawall:2005np, Castelijns:2007qt} and a global analysis of pion- and photon-induced reactions would greatly benefit from better data in the 
$\pi^-p\to K^+\Sigma^-$ reaction. Efforts in this direction are being made at the E19 experiment at J-PARC~\cite{Shirotori:2012ka}.

%%%%%%%%%%%%%%%%%%%%%%%%%%%%%%%%%%%%%%%%%%%%%%%%%%%%%%%%%%%%%%%%%%%%%%%%%%%%%%%%%%%%%%%%%%%%%%%%%%%%%%%%%%%%%%%%%%%%%%%%%%%%%%%%%%%%%%%%%%%%%%%%

\subsection{Reaction $\boldsymbol{\pi^+ p\to K^+\Sigma^+}$}
\label{sec:kpsp}

For the reaction $\pi^+ p\to K^+\Sigma^+$, the same data sample as in our previous analysis \cite{Doring:2010ap} is used, which contains
differential  cross sections and polarization measurements in a range from $z=1729$~MeV to $2318$~MeV for 32 different energies. We employ the
data by Candlin {\it et al.}~\cite{Candlin:1982yv}, supplemented by additional data at lower energies from Refs.~\cite{Winik:1977mm,
Carayannopoulos:1965, Crawford:1962zz, Baltay, Bellamy:1972fa}. The latter are compatible with the data from Ref.~\cite{Candlin:1982yv} in the
overlapping energy region but usually have larger error bars. For higher energies, the polarization was re-measured in Ref.~\cite{Haba:1988rn}
and found to be consistent with that of Ref.~\cite{Candlin:1982yv} up to small deviations. In addition, values for the spin-rotation parameter
$\beta$ by Candlin {\it et al.}~\cite{Candlin:1988pn} are available at two energies and included in the fit. During the measurement of the
spin-rotation parameter, the polarization was re-measured and consistency with the result of Ref.~\cite{Candlin:1982yv} was found once more.
Total cross section data from Refs.~\cite{Candlin:1982yv, Winik:1977mm, Carayannopoulos:1965, Berthelot:1961, Crawford:1962zz, Berthon:1974zd,
Kalmus:1970zx, Daronian:1966, Pan:1970ez, Dagan:1967, Landolt} are also included in the fit. No additional systematic errors beyond the original
ones were assumed for this channel.

\begin{figure}[h]
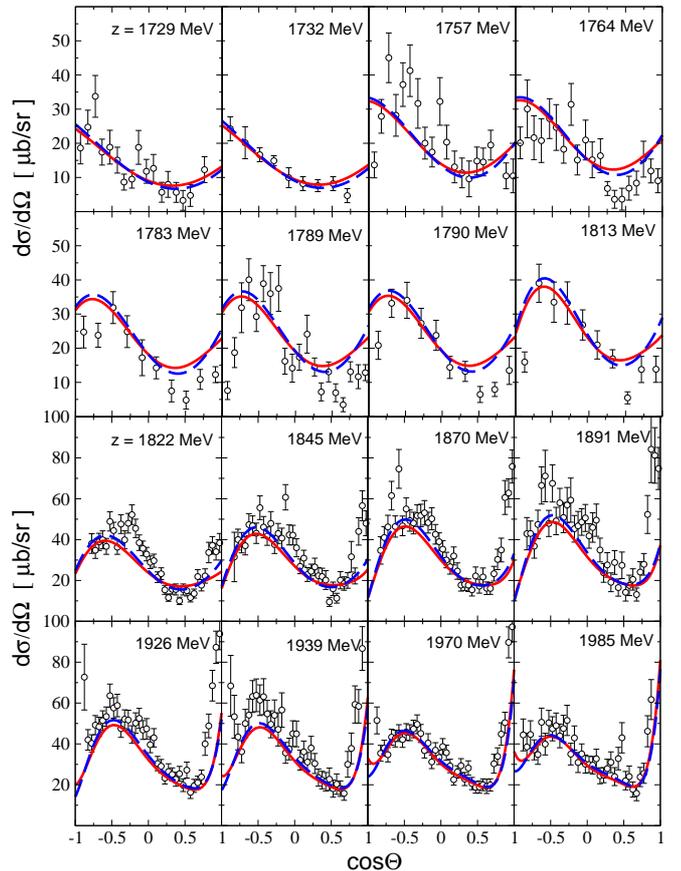

\includegraphics[width=0.48\textwidth]{dsdo_k+s+_low.eps}\\
\vspace*{-0.15cm}
\includegraphics[width=0.48\textwidth]{dsdo_k+s+.eps}
\caption{Differential cross section [1/2] of the reaction $\pi^+ p\to K^+\Sigma^+$. Solid (red) lines: fit A.; dashed (blue) lines: fit B; data:
Ref.~\cite{Candlin:1982yv}, except: $z = 1729,\, 1757,\, 1789$~MeV from Ref.~\cite{Winik:1977mm}; $z = 1732,\, 1783,\, 1813$~MeV from
Ref.~\cite{Carayannopoulos:1965}; $z = 1764$~MeV from Ref.~\cite{Crawford:1962zz}; $z = 1790$~MeV from Ref.~\cite{Baltay}.}
\label{fig:kpspdiff1}     
\end{figure}

\begin{figure}[h]
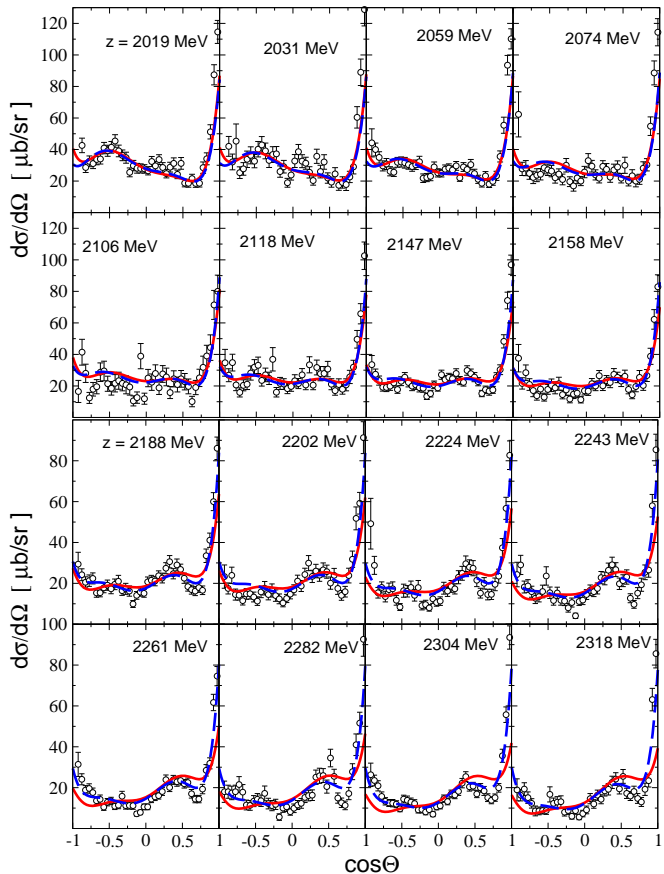

\includegraphics[width=0.48\textwidth]{dsdo_k+s+_up.eps}\\
\vspace*{-0.05cm}
\includegraphics[width=0.48\textwidth]{dsdo_k+s+_high.eps}
\caption{Differential cross section [2/2] of the reaction $\pi^+ p\to K^+\Sigma^+$. Solid (red) lines: fit A; dashed (blue) lines: fit B; data
from Ref.~\cite{Candlin:1982yv}.}
\label{fig:kpspdiff2}     
\end{figure}

\begin{figure}[h]
\begin{center}
\includegraphics[width=0.48\textwidth]{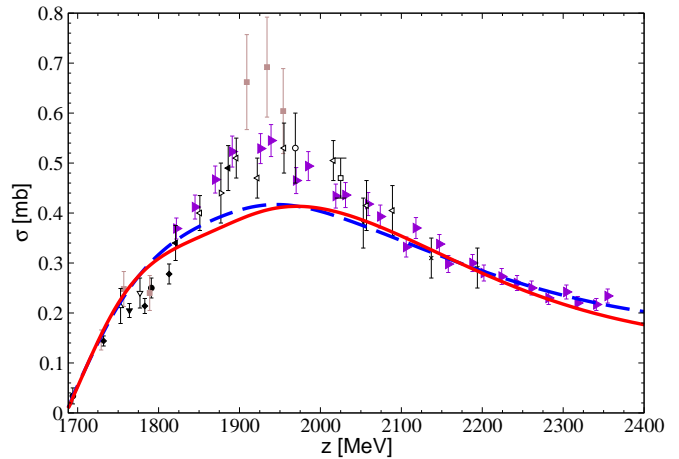}\\
\end{center}
\caption{Total cross section of the reaction $\pi^+ p\to K^+ \Sigma^+$. Solid (red) line: fit A; dashed (blue) line: fit B; data: filled
triangles right: Ref.~\cite{Candlin:1982yv}; filled squares: Ref.~\cite{Winik:1977mm}; filled diamonds: Ref.~\cite{Carayannopoulos:1965}; empty
triangles up: Ref.~\cite{Berthelot:1961}; filled triangles down: Ref.~\cite{Crawford:1962zz}; empty triangles down: Ref.~\cite{Berthon:1974zd};
empty triangles left: Ref.~\cite{Kalmus:1970zx}; empty circles: Ref.~\cite{Daronian:1966}; empty squares: Ref.~\cite{Pan:1970ez}; crosses:
Ref.~\cite{Dagan:1967}. For filled circles, filled triangles left and empty triangles right see Ref.~\cite{Landolt}.}
\label{fig:kpsptot}     
\end{figure}

\begin{figure}[h]
\includegraphics[width=0.48\textwidth]{pola_k+s+_low.eps}\\
\vspace*{-0.02cm}
\includegraphics[width=0.48\textwidth]{pola_k+s+.eps}
\caption{Polarization [1/2] of the reaction $\pi^+ p\to K^+\Sigma^+$. Solid (red) lines: fit A; dashed (blue) lines: fit B; data:
Ref.~\cite{Candlin:1982yv}, except: $z =$ 1729, 1757, 1789 MeV from Ref.~\cite{Winik:1977mm}; $z =$ 1783, 1813 MeV from
Ref.~\cite{Carayannopoulos:1965}; $z =$ 1764 MeV from Ref.~\cite{Crawford:1962zz}; $z =$ 1790 MeV from Ref.~\cite{Baltay}; z = 1732 MeV from
Ref.~\cite{Bellamy:1972fa}.}
\label{fig:kpsppola1}     
\end{figure}

\begin{figure}[h]
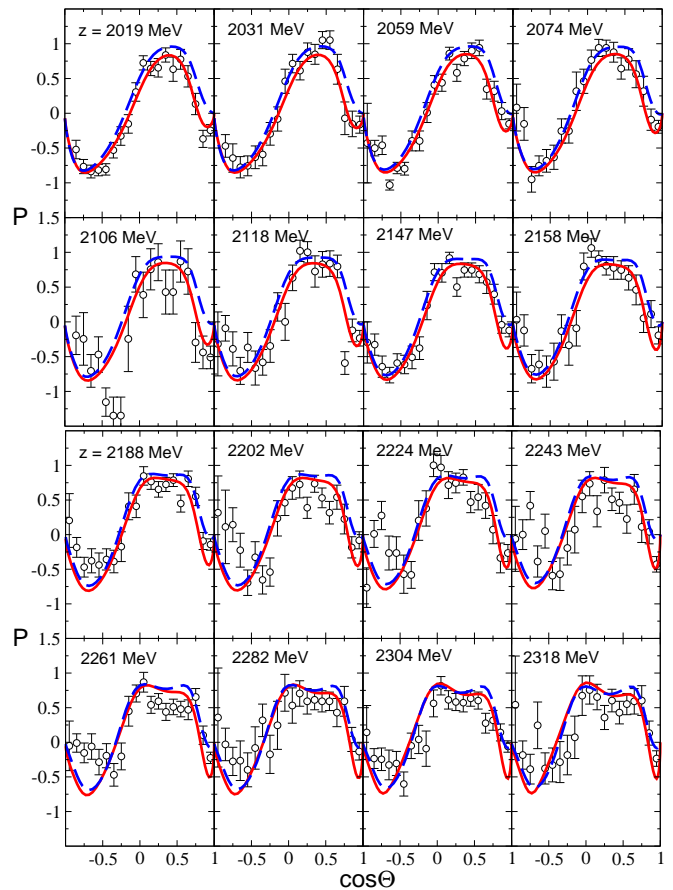

\includegraphics[width=0.48\textwidth]{pola_k+s+_up.eps}\\
\vspace*{-0.02cm}
\includegraphics[width=0.48\textwidth]{pola_k+s+_high.eps}
\caption{Polarization [2/2] of the reaction $\pi^+ p\to K^+\Sigma^+$. Solid (red) lines: fit A; dashed (blue) lines: fit B; data:
Ref.~\cite{Candlin:1982yv}.}
\label{fig:kpsppola2}     
\end{figure}

\begin{figure}[h]
\begin{center}
\includegraphics[width=0.48\textwidth]{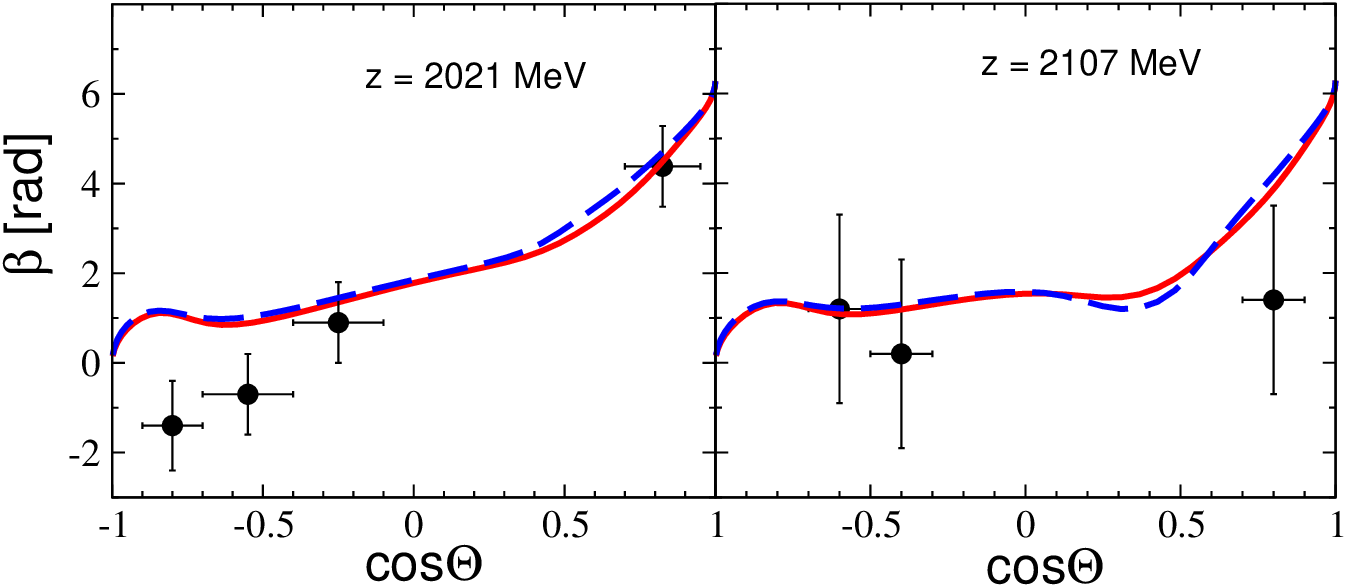}
\end{center}
\caption{Spin-rotation parameter $\beta$ of the reaction $\pi^+ p\to K^+\Sigma^+$. Solid (red) lines: fit A; dashed (blue) lines: fit B; data:
Ref.~\cite{Candlin:1988pn}. Note that $\beta$ is $2\pi$ cyclic.}
\label{fig:kpspbeta}     
\end{figure}

The present results are shown in Figs.~\ref{fig:kpspdiff1} to \ref{fig:kpspbeta} with the solid (red) lines representing fit A and the dashed
(blue) lines representing fit B. Even though we add a large amount of new data and extend the fit to four more channels, we achieve an equally
good result for the differential cross sections, polarization and spin-rotation parameter in the $\pi^+ p\to K^+\Sigma^+$ channel as in our
previous analysis~\cite{Doring:2010ap}. Fits A and B show only very small discrepancies. A slight systematic deviation becomes visible in the
total cross section around $z\sim 1.9$~GeV. The inspection of the differential cross section around that energy, shown in
Fig.~\ref{fig:kpspdiff1}, reveals that the underestimation of the data is mostly due to the pronounced forward peak in the data. In both of the
current fits this forward peak is underestimated at this energy, but well described beyond 2~GeV. 

\begin{figure}[h!]
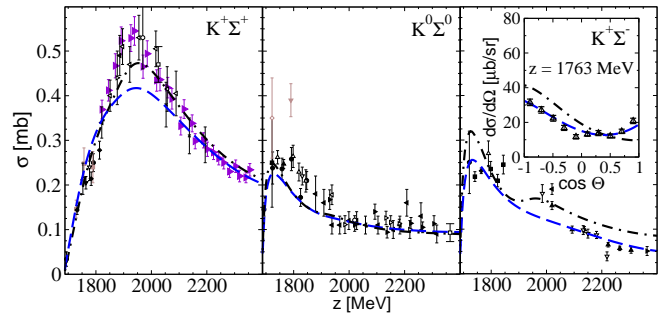

\begin{center}
\includegraphics[width=0.1855\textwidth]{total_cs_KpSp_more_KpSp.eps}  \hspace*{-0.27cm}
\raisebox{-0.268cm}{\includegraphics[width=0.148\textwidth]{total_cs_KzSz_more_KpSp.eps}}   \hspace*{-0.27cm}
\includegraphics[width=0.1473\textwidth]{total_cs_KpSm_more_KpSp.eps}
\end{center}
\caption{Refit (black dash-dotted lines) of a reduced data base after taking out the observables of the reaction $\pi^-p\to K^+\Sigma^-$ (right)
and giving more weight to the $K^+\Sigma^+$ final state (left). The dashed (blue) line shows fit B, for which the $\pi^-p\to K^+\Sigma^-$
observables are included.}
\label{fig:nokpsm}     
\end{figure}

As a test, we have removed the observables of the $K^+\Sigma^-$ reaction from the fitting procedure. This would correspond to the situation in
many analyses where the data of the reaction $\pi^-p\to K^+\Sigma^-$ are not considered. With this change, a refit of $T^\po$ starting with fit
B is performed, and the outcome for the total cross sections of the $K\Sigma$ final states is shown in Fig.~\ref{fig:nokpsm} (dash-dotted lines
vs. the original fit B shown with the dashed lines). As the fit of the reduced data base shows, the $K^+\Sigma^-$ final state is, of course, not
well described any more. In contrast, it becomes now possible to describe the maximum of the $K^+\Sigma^+$ cross section.
This finding emphasizes again the importance to include the data of the reaction $\pi^-p\to K^+\Sigma^-$. 

The partial-wave content of the reaction $\pi^+p\to K^+\Sigma^+$ can be read off Fig. \ref{fig:kzsztotpart_3h}, only scaled by a different
isospin factor [cf. Eq.~(\ref{csconnect})]. These results supersede those of Ref.~\cite{Doring:2010ap}. The $P_{33}$ partial wave dominates
starting almost from threshold up to $z\sim 2.2$~GeV. The $F_{37}$ partial wave with the $\Delta(1950)$ 7/2$^+$ is also well-determined and
important while other partial waves start to become relevant only at higher energies.

%%%%%%%%%%%%%%%%%%%%%%%%%%%%%%%%%%%%%%%%%%%%%%%%%%%%%%%%%%%%%%%%%%%%%%%%%%%%%%%%%%%%%%%%%%%%%%%%%%%%%%%%%%%%%%%%%%%%%%%%%%%%%%%%%%%%%%%%%%%%%%%%

\subsection{Scattering lengths and volumes}
\label{sec:scatlen}

Pion-nucleon scattering lengths and volumes of the current fit result are shown in Table~\ref{tab:scatleng} and compared to values in the
literature.  The notation is $a_{L,\pm}^\pm$ where $L$ is the orbital angular momentum, and an upper plus (minus) sign indicates the isoscalar
(isovector) scattering length/volume. The lower $\pm$ sign indicates the cases $J=L\pm S$ where $J$ (S) is the total angular momentum (nucleon
spin). Of particular interest is the isoscalar scattering length $a^+_{0+}$ that was determined to high accuracy recently in the  calculation of
Ref.~\cite{Baru:2010xn} using chiral perturbation theory. Although in the present approach we cannot match the value from
Ref.~\cite{Baru:2010xn}, we can indeed achieve a very small value for the isoscalar scattering length in both fits. This is not trivial, given
that we have only four parameters for the non-resonant $\pi N\to\pi N$ transitions. Those are the only parameters to directly tune the $\pi N$
background in 20 partial waves from $\pi N$ threshold to beyond $2$~GeV in scattering energy (cf. discussion in Sec.~\ref{sec:pin}).
\begin{table}
\caption{Scattering lengths $a^{\pm}_{0+}$ and volumes $a^{\pm}_{1\pm}$ in units of $10^{-2}/m_{\pi}$ and $10^{-2}/m^3_{\pi}$, compared to ChPT
calculations from Ref.~\cite{Baru:2010xn} (scattering lengths) and Ref.~\cite{Fettes:2000xg} (scattering volumes). Note that for the scattering
volumes in Ref.~\cite{Fettes:2000xg} no uncertainties but three different fit results are given. Here we quote fit 1 together with the maximal
deviation among the three fits. }
\begin{center}
\renewcommand{\arraystretch}{1.50}
\setlength{\tabcolsep}{.2cm}
\begin{tabular}[t]{l|ccc}
\hline \hline 
		& fit A	& fit B		& ChPT  \\ \hline
$a^+_{0+}$	&$-1.64$ &$-0.72 $	&$0.76\,\pm\,0.31$\\
$a^-_{0+}$	&$9.51$ &$10.33 $	&$8.61\,\pm\,0.09$ \\
$a^+_{1-}$	&$-5.62$&$-4.65 $	&$-5.66 \,\pm\,0.53 $\\
$a^-_{1-}$	&$-1.25$&$-0.18 $	&$-1.25\,\pm\,0.23$\\
$a^+_{1+}$	&$11.85$&$12.76 $	&$13.15\,\pm\,0.07$\\
$a^-_{1+}$	&$-7.35$&$-7.88 $	&$-7.99\,\pm\,0.335$\\
\hline \hline
\end{tabular}
\label{tab:scatleng}
\end{center}
\end{table} 
For the isovector scattering length, our values are slightly higher than the ones calculated in Ref.~\cite{Baru:2010xn}. The scattering volumes
are quite well described compared to the fourth-order ChPT calculation of Ref.~\cite{Fettes:2000xg}.

In Table~\ref{tab:scat2}, we show the scattering lengths of other channels and compare the $\pi N$ and $\eta N$ values to those calculated in
Ref.~\cite{Mai:2012wy}  within a  unitary extension of chiral perturbation theory. Fits A and B from our analysis yield similar results. While
the agreement between the results of fit A and of Ref.~\cite{Mai:2012wy} is reasonable in case of the $\pi N$ scattering length, for the real
part of the $\eta N$ scattering length discrepancies can be observed between the current results and those of Ref.~\cite{Mai:2012wy}. 

The present values for the $\eta N$ scattering length can be also compared to the compilation of results given in Refs.~\cite{Arndt:2005dg,
Sibirtsev:2001hz}. The real parts of different analyses usually differ substantially while the imaginary parts are better determined. The
present results for the latter are in line with other determinations. 
\begin{table*}
\caption{Scattering lengths in (in fm) of the present approach (upper rows),  compared with the ChPT calculation from Ref.~\cite{Mai:2012wy} (bottom row).}
\begin{center}
\renewcommand{\arraystretch}{1.20}
\begin{tabular}[t]{l|l|l|ll|ll|ll|ll}
\hline \hline 
&$a_{\pi N}^{(1/2)}$	& $a_{\pi N}^{(3/2)}$		& $a_{\eta N}^{(1/2)}$	&		& $a_{K\Lambda}^{(1/2)}$&		& $a_{K\Sigma}^{(1/2)}$	&		& $a_{K\Sigma}^{(3/2)}$ 	&		\\
fit A&$ 0.25 $	& $-0.16 $ 	& $0.49 $& $+0.24  \,i$		& $0.04 $ &$+0.04  \,i$		& $0.36 $  &$+0.15  \,i$	&  $ -0.30$  &$  +0.04 \,i$	\\
fit B &   $ 0.29$ &$ -0.16$ 	&$0.55 $ &$+0.24\,i$ 			&$0.04 $ &$+0.03\,i$ 			&$0.32 $ &$+0.14 \,i$ 	&$ - 0.30 $ &$+0.05  \,i$              \\
UChPT \cite{Mai:2012wy}&$0.24^{+0.02}_{-0.05}$			&$-0.12^{+0.006}_{-0.006}$				&$0.22^{+0.05}_{-0.02}$	&$+0.24^{+0.15}_{-0.06}\,i$	&	
	&	
& &	&	 &\\
\hline \hline
\end{tabular}
\label{tab:scat2}
\end{center}
\end{table*} 

%%%%%%%%%%%%%%%%%%%%%%%%%%%%%%%%%%%%%%%%%%%%%%%%%%%%%%%%%%%%%%%%%%%%%%%%%%%%%%%%%%%%%%%%%%%%%%%%%%%%%%%%%%%%%%%%%%%%%%%%%%%%%%%%%%%%%%%%%%%%%%%%

\section{Resonance spectrum}
\label{sec:polpos}

%%%%%%%%%%%%%%%%%%%%%%%%%%%%%%%%%%%%%%%%%%%%%%%%%%%%%%%%%%%%%%%%%%%%%%%%%%%%%%%%%%%%%%%%%%%%%%%%%%%%%%%%%%%%%%%%%%%%%%%%%%%%%%%%%%%%%%%%%%%%%%%%

\subsection{Resonance properties}
\label{sec:respro}
In the literature~\cite{Lahiff:1999ur,Matsuyama:2006rp}, the decomposition according to Eq.~(\ref{deco1}) has been used since long ago.
However, in Ref.~\cite{Doring:2009bi} it is shown that this decomposition into pole and non-pole part is problematic in the sense that the pole
part contains finite constant terms $\sim (z-~z_0)^0$ at the pole position $z=z_0$. What is well defined is the expansion of the fully dressed,
physical amplitude in a Laurent expansion around the pole, with the residue term containing all contributions of the pole, and all higher order
terms accounting for the background. 

Thus, a resonance is uniquely characterized by its pole position in the complex energy ($z$) plane, the pole residues for the various channels,
and the Riemann sheet on which it is located. 

Poles can appear on different Riemann sheets (except for the physical sheet of the lowest-lying channel). For example, a resonance situated
above the $\eta N$ threshold, with a sizable coupling to the $\pi N$ channel and vanishing coupling to the $\eta N$ channel would correspond to
poles on both the physical and the unphysical sheet of the $\eta N$ channel, at approximately the same pole position $z_0$ (while both poles
would be on the unphysical $\pi N$ sheet). However, not all poles on different sheets bear physical interest but usually only the one on the
sheet closest to the physical axis. For an example where a pole on a more distant sheet plays a role, one can consider the case of the $N(1535)$
1/2$^-$ resonance and its interplay of poles leading to the cusp at the $\eta N$ threshold~\cite{Doring:2009yv,Doring:2009uc}. Such poles are
sometimes referred to as shadow poles. The selection of the part of the Riemann sheets which is close to the physical axis can be achieved by
rotating the right-hand cuts of all channels towards the negative Im $z$ direction, which is the choice adopted in this study. 

\begin{figure}
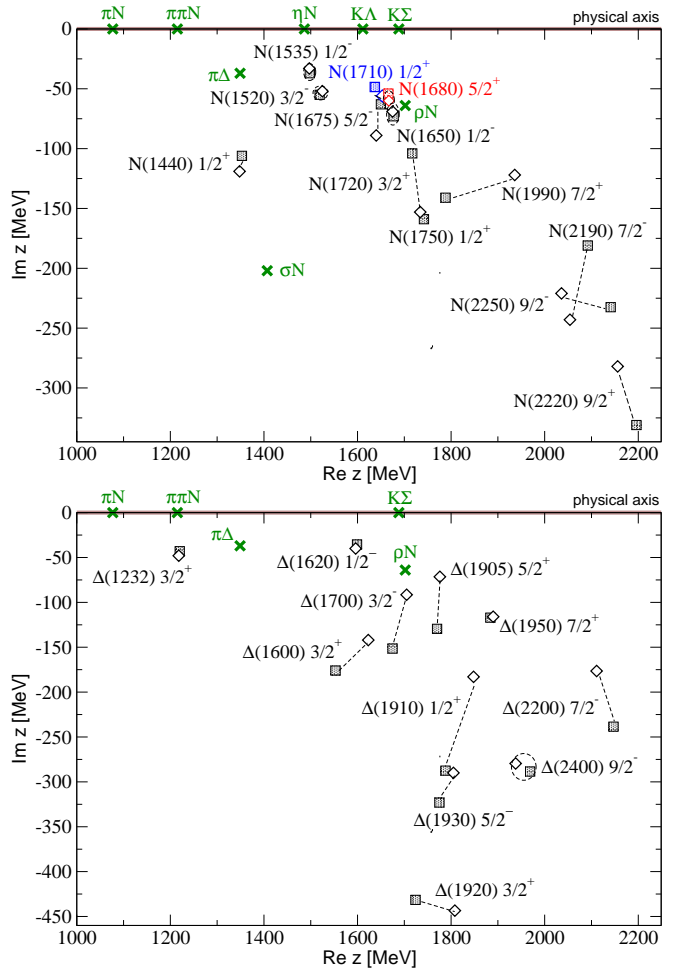

\begin{center}
\includegraphics[width=0.48\textwidth]{cmplx_plane_1h.eps}\\
\includegraphics[width=0.48\textwidth]{cmplx_plane_3h.eps}
\end{center}
\caption{Pole positions for the $I=1/2$ (above) and $I=3/2$ (below) resonances. Squares: fit A; diamonds: fit B. Also, the branch points of the
amplitude are shown (green crosses). Note that all cuts, starting at the branch points, are chosen in the negative Im~$z$ direction, which
uniquely defines the Riemann sheet of the displayed poles.}
\label{fig:new_figure_Deborah}     
\end{figure}

Apart from threshold openings on the physical axis at $z=m_1+m_2$ for the channels with stable particles with masses $m_1$ and $m_2$, i.e. $\pi
N$, $\eta N$, $K\Lambda$, $K\Sigma$, there are threshold openings at $z=2m_\pi+m_N$ for the effective $\pi\pi N$ channels $\pi\Delta,\,\sigma
N,\,\rho N$ that also induce branch points in the complex plane at $z=m_s\pm z_r$, where $m_s$ is the mass of the stable spectator particle and
$z_r$ is the complex resonance pole position of the unstable particle. For the cuts associated with the $\pi\pi N$ channels, we choose the same
convention as for the channels with stable particles, i.e. the cuts are rotated  towards the negative Im $z$ direction. An overview of the
various branch points can be found in Fig.~\ref{fig:new_figure_Deborah}, and Fig.~\ref{fig:anap11} shows the position of the cuts. More details
of the analytic structure can be found in Refs.~\cite{Doring:2009yv,Ceci:2011ae}. The resonance properties have been determined using the
methods of analytic continuation developed in Ref.~\cite{Doring:2009yv} and residues are calculated following the procedure in Appendix C of
Ref.~\cite{Doring:2010ap}.

The choice of cuts defines uniquely that part of the Riemann sheets where all poles extracted in this study are situated (unless stated
otherwise). We refer to this part of the Riemann sheets as {\it second sheet}. The strength of a pole with respect to the different channels is
characterized by the residue $a_{-1,\mu\nu}$ as it appears in the Laurent expansion of the transition amplitude $T^{(2)}_{\mu\nu}$ from channel
$\mu$ to $\nu$ on the second sheet,
\be
T^{(2)}_{\mu\nu}(z)=\frac{a_{-1,\mu\nu}}{z-z_0}+a_{0,\mu\nu}+{\cal O}(z-z_0) \ .
\ee
For $n$ channels, since residues factorize in channel space, one has $n^2$ different residues,
\be
a_{-1,\mu\nu}=g_\mu\,g_\nu
\label{couplings}
\ee
with a unique set of $n$ parameters $g_\mu$ up to one undetermined global sign. To connect to the quantities quoted in the
PDG~\cite{Beringer:1900zz}, the dimensionless meson-baryon amplitude $\tau$ is expressed in terms of the $T$-matrix in the current
normalization,
\be
\tau_{\mu\nu}&=&-\pi\sqrt{\rho_\mu\,\rho_\nu}\,T_{\mu\nu},\quad\rho_{\mu}=\frac{k_{\mu}\,E_{\mu}\,\omega_{\mu}}{z}
\label{taut}
\ee
where $k_{\mu}\,(E_{\mu},\omega_{\mu})$ is the on-shell three-momentum (baryon energy, meson energy) of the initial or final meson-baryon system
$\mu$. The so-called {\it normalized residue}~(NR)~\cite{Beringer:1900zz} is defined for $\tau_{\mu\nu}$, i.e. with Eq.~(\ref{taut}) we obtain
for the transition $\pi N\to \mu$
\be
(NR)_{\pi N\to \mu}=\frac{\pi\sqrt{\rho_{\pi N}\rho_{\mu}}\,g_{\pi N}\,g_{\mu}}{(\Gamma_{\rm tot}/2)}
\ee
with the resonance width $\Gamma_{\rm tot}=-2{\,\rm Im}\,z_0$ [the additional minus sign with respect to Eq.~(\ref{taut}) comes from the
convention of Eq.~(\ref{usualphi}) below]. 

Defining the {\it branching ratio} into channel $\mu$ in the same way as PDG~\cite{Beringer:1900zz}, p. 1268 et sqq., 
\be
\frac{\Gamma_\mu}{\Gamma_{\rm tot}}=\frac{|r_{\pi N\to \mu}|^2}{|r_{\pi N\to\pi N}|(\Gamma_{\rm tot}/2)}\, , 
\label{branching}
\ee
where $r$ is the residue of $\tau(z)$, it is straightforward to show that the {\it transition branching ratio} equals the modulus of the
normalized residue,
\be
\frac{\Gamma^{1/2}_{\pi N}\Gamma^{1/2}_\mu}{\Gamma_{\rm tot}}=|(NR)_{\pi N\to \mu}| \ .
\ee

As for the counting of the residue phase $\theta$~\cite{Beringer:1900zz}, we use H\"ohler's convention~\cite{Hoehler1,Hoehler2} given by
\be
\tau=\tau_B+\frac{|r|\,e^{+i\theta}}{M-z-i\Gamma_{\rm tot}/2}
\label{usualphi}
\ee
for a resonance on top of a background $\tau_B$. For the elastic $\pi N$ residues and their angles as quoted in the PDG, this leads to
\be
|r_{\pi N}|&=& \pi |g_{\pi N}^2\,\rho_{\pi N}|,\non
\theta_{\pi N\to\pi N} &=& \arctan\left[\frac{{\rm Im}\,(g_{\pi N}^2\,\rho_{\pi N})}{{\rm Re}\,(g_{\pi N}^2\,\rho_{\pi N})}\right]\, ,
\label{rerere}
\ee
with the $\arctan$ function sensitive to the quadrants of the complex plane. The phases of the normalized residues into inelastic channels are
defined accordingly. 

\begin{table*} 
\caption{Properties of the $I=1/2$ resonances: Pole positions $z_0$ ($\Gamma_{\rm tot}$ defined as -2Im$z_0$), elastic $\pi N$ residues
$(|r_{\pi N}|,\theta_{\pi N\to\pi N})$, and the normalized residues $(\sqrt{\Gamma_{\pi N}\Gamma_\mu}/\Gamma_{\rm tot},\theta_{\pi N\to \mu})$
of the inelastic reactions $\pi N\to \mu$. For the second Roper pole on the hidden $\pi\Delta$ sheet, shown in brackets, see
Sec.~\ref{sec:Roper}.  (*): not identified with PDG name; (a): dynamically generated, (nf): not found. }
\begin{center}
\renewcommand{\arraystretch}{1.30}
\begin {tabular}{ll|rrrr|rrrr|rrrr|rrrr|rrrr} 
\hline\hline
&&\multicolumn{2}{|l}{Re $z_0$ \hspace*{0.8cm} }
& \multicolumn{2}{l|}{-2Im $z_0$\hspace*{0.2cm} }
& \multicolumn{2}{l}{$|r_{\pi N}|$\hspace*{0.2cm}} 
& \multicolumn{2}{l}{$\theta_{\pi N\to\pi N}$ } 
& \multicolumn{2}{|l}{$\displaystyle{\frac{\Gamma^{1/2}_{\pi N}\Gamma^{1/2}_{\eta N}}{\Gamma_{\rm tot}}}$}
& \multicolumn{2}{l|}{$\theta_{\pi N\to\eta N}$\hspace*{0.1cm}}
& \multicolumn{2}{l}{$\displaystyle{\frac{\Gamma^{1/2}_{\pi N}\Gamma^{1/2}_{K\Lambda}}{\Gamma_{\rm tot}}}$} 
& \multicolumn{2}{l|}{$\theta_{\pi N\to K\Lambda}$\hspace*{0.1cm}}
& \multicolumn{2}{l}{$\displaystyle{\frac{\Gamma^{1/2}_{\pi N}\Gamma^{1/2}_{K\Sigma}}{\Gamma_{\rm tot}}}$} 
& \multicolumn{2}{l}{$\theta_{\pi N\to K\Sigma}$}
\bigstrut[t]\\[0.2cm]
&&\multicolumn{2}{|l}{[MeV]} & \multicolumn{2}{l|}{[MeV]} & \multicolumn{2}{l}{[MeV]} & \multicolumn{2}{l}{[deg]} 
& \multicolumn{2}{|l}{[\%]}  & \multicolumn{2}{l|}{[deg]} & \multicolumn{2}{l}{[\%]}  & \multicolumn{2}{l|}{[deg]} &\multicolumn{2}{l}{[\%]} & \multicolumn{2}{l}{[deg]} \\
		  & fit$\to$		& A	& B	& A	& B	& A	& B	& A	& B	& A	& B	& A	& B	& A	& B	& A	& B	& A	& B	& A	& B	
\bigstrut[t]\\
\hline
 $N (1535)$ 1/2$^-$&   		& 1498	& 1497	& 74	& 66	& 17	& 13	& -37	& -43	& 51	& 48	& 120	& 115	& 7.7	& 8.3	& 68	& 77	& 15	& 34	& -74
& -83	\\
 	 			
 $N (1650)$ 1/2$^-$&   		& 1677	& 1675	& 146 	& 131	& 45 	& 27	& -43	& -38	& 15	& 12	& 57	& 46 	& 25	& 18	& -46	& -43	& 26	& 15	& -63   & -59
\\
 
 $N (1440)$ 1/2$^+_{(a)}$&  		& 1353	& 1348	& 212 	& 238	& 59	& 62	& -103	& -111	& 2	& 5	& -40	& -26	& 2	& 11	& 156	& 152	& 1	& 2	& 67
& 19	\\
 
		  &			&[1357	& 1330	& 228	& 221	& 59	& 55	& -112	& -102	& 2	& 4	& -53	& -20	& 2	& 9	& 133	& 161	& 1	& 2	& 66	& 23]	\\
 
 $N (1710)$ 1/2$^+$&   		& 1637	& 1653	& 97	& 112	& 4	& 8	& -30	& 34	& 24	& 23	& 130	& 164	& 9.4	& 17	& -83	& -41	& 3.9	& 0.1	& -136	& -112
\\
 
 $N (\textit{1750})$  1/2$^+_{(*,a)}$&  		& 1742	& (nf)	& 318	&---	& 8	& ---	& 161	& ---	& 0.5	& --- 	& -140	& ---	& 0.8	& ---	& -170	& ---	& 2.2
& --- 
& 4	& ---	\\
 
 $N (1720)$ 3/2$^+$&  		& 1717	& 1734	& 208	& 306	& 7	& 18	& -76	& -23	& 1.2	& 3.2 	& 98	& 117	& 3.1	& 2.9	& -89	& -63	& 1.7	& 2.2 	& 64
& 90 \\
 
 $N (1520)$ 3/2$^-$&  		& 1519	& 1525	& 110	& 104	& 42	& 36	& -16	& 10	& 3.5	& 3.0 	& 87	& 113 	& 5.8	& 6.3 	& 158	& 177	& 0.8	& 3.6 	& 163
& 164	\\
 
 $N (1675)$ 5/2$^-$&  		& 1650	& 1640	& 126	& 178	& 24	& 34	& -19	& -32	& 6.0	& 3.6	& -40	& -66	& 0.3	& 1.7	& -93 	& -122	& 3.3	& 3.7	& -168
& 179 \\
 
 $N (1680)$ 5/2$^+$&   		& 1666	& 1667	& 108	& 120	& 36	& 41	& -24	& -24	& 0.4	& 1.5	& -47	& -54	& 0.2	& 0.3	& -99	& 72 	& 0.1	& 0.1	& 141
& 141	\\

 $N (1990)$ 7/2$^+$&   		& 1788	& 1936	& 282	& 244	& 4	& 4	& -84	& -87	& 0.4	& 0.5	& -99 	& 90	& 1.7	& 1.5	& -123	& -99	& 0.8	& 1.2	& 28
& 70	\\
 
 $N (2190)$  7/2$^-$&  		& 2092	& 2054	& 363	& 486	& 42	& 44	& -31	& -57	& 0.1	& 0.4	& -28	& 99	& 1.9	& 0.3	& -51	& -75	& 1.3	& 0.5	& -63
& -105 \\
 
 $N (2250)$ 9/2$^-$&  		& 2141	& 2036	& 465	& 442	& 17	& 13	& -67	& -62	& 0.6	& 0.1	& -92	& -96	& 1.1	& 0.7 	& -103	& -106	& 0.3	& 0.7 	& -114
& 62	\\
 
 $N (2220)$ 9/2$^+$&    		& 2196	& 2156	& 662	& 565	& 87	& 46	& -67	& -72	& 0.1	& 0.3	& 63 	& 74	& 0.9	& 0.8	& 53	& 59	& 0.8	& 0.1	& -138
& 52 \\
\hline\hline
\end {tabular}
\end{center}
\label{tab:bra1}
\end{table*}

\begin{table*}
\caption{Residues of the $I=1/2$ resonances for the effective $\pi\pi N$ channels: Normalized residues $(\sqrt{\Gamma_{\pi
N}\Gamma_{\pi\Delta}}/\Gamma_{\rm tot}, \theta_{\pi N\to \pi\Delta})$ for the $\pi\Delta$ channels and coupling constants $g$ for the $\rho N$
and $\sigma N$ channels according to Eq.~(\ref{couplings}).  For the quantum numbers of the $\pi\Delta$ channels (6), (7), the $\rho N$ channels
(2), (3), (4), and the $\sigma N$ channel (8), see Table~\ref{tab:couplscheme}.  (*): not identified with PDG name; (a): dynamically generated, (nf): not found.}
\begin{center}
\renewcommand{\arraystretch}{1.30}
\scalebox{0.9}{
\begin{tabular}{l|rrrr|rrrr|rrrrrr|rr} \hline\hline
 &\multicolumn{4}{|l|}{$\pi\Delta$, channel (6)}
 &\multicolumn{4}{|l|}{$\pi\Delta$, channel (7)}
 &\multicolumn{2}{l}{  $\rho N$,    channel (2)\hspace*{0.6cm}}
 &\multicolumn{2}{l}{  $\rho N$,    channel (3)\hspace*{0.6cm}}
 &\multicolumn{2}{l|}{ $\rho N$,    channel (4)\hspace*{0.6cm}}
 &\multicolumn{2}{l}{  $\sigma N$,  channel (8)}
\bigstrut[t]\\[0.1cm]
&\multicolumn{2}{|l}{\scalebox{0.8}{$\displaystyle{\frac{\Gamma^{1/2}_{\pi N}\Gamma^{1/2}_{\pi\Delta}}{\Gamma_{\rm tot}}}$}}
& \multicolumn{2}{l|}{\scalebox{0.8}{$\theta_{\pi N\to\pi\Delta}$\hspace*{0.1cm}}}
& \multicolumn{2}{|l}{\scalebox{0.8}{$\displaystyle{\frac{\Gamma^{1/2}_{\pi N}\Gamma^{1/2}_{\pi\Delta}}{\Gamma_{\rm tot}}}$}}
& \multicolumn{2}{l|}{\scalebox{0.8}{$\theta_{\pi N\to\pi\Delta}$\hspace*{0.1cm}}}
& \multicolumn{2}{l}{$\displaystyle{g_{\rho N}}$} 
& \multicolumn{2}{l}{$\displaystyle{g_{\rho N}}$} 
& \multicolumn{2}{l|}{$\displaystyle{g_{\rho N}}$} 
& \multicolumn{2}{l}{$\displaystyle{g_{\sigma N}}$} 
\\[0.1cm]
& \multicolumn{2}{|l}{[\%]}  & \multicolumn{2}{l|}{[deg]} 
& \multicolumn{2}{|l}{[\%]}  & \multicolumn{2}{l|}{[deg]} 
& \multicolumn{2}{l}{[10$^{-3}$ MeV$^{-1/2}$]}  
& \multicolumn{2}{l}{[10$^{-3}$ MeV$^{-1/2}$]}  
& \multicolumn{2}{l|}{[10$^{-3}$ MeV$^{-1/2}$]}  
& \multicolumn{2}{l}{[10$^{-3}$ MeV$^{-1/2}$]}  
\\
		  & A	  & B	  & A	  & B	  & A	  & B	  & A	  & B	  & A		  & B		  & A		  & B	  	& A		  & B		  & A		  & B
\bigstrut[t]\\
\hline
 $N (1535)$ 1/2$^-$ & ---   & ---   & ---   & ---   & 24    & 22    & 113   & 114   & -3.7-i5.7	  & -4.1-i2.1	  & --- 	  & ---   	& -0.9+i0.1	  & -1.2+i0.3	  &
-2.3-i2.5  & -1.8-i3.6	\\
 	 	
 $N (1650)$ 1/2$^-$ & ---   & ---   & ---   & ---   & 42	  & 29	  & -37	  & -35	  & -2.5+i2.9	  & -2.3+i10	  & ---		  & ---		& 0.8+i1.9	  & 0.9+i3.4	  &
-2.7+i1.6  & -0.6+i0.7	\\
 
 $N (1440)$ 1/2$^+_{(a)}$ & 19	  & 29	  & 50	  & 66	  & ---   & ---   & ---   & ---   & -0.8+i5.0	  & 2.2+i6.8	  & 2.8-i5.2	  & 2.6-i7.6	& ---		  & ---		  &
-9.0-i29  & 14+i16	\\
 
		  &[21	  & 22	  & 40	  & 81	  & ---	  & ---	  & ---	  & ---	  & -0.5+i5.3	  & 0.8+i4.9	  & 2.5-i5.2	  & 2.6-i7.9	& ---		  & ---		  & -10-i30	  & 14+i18]	\\
 
 $N (1710)$ 1/2$^+$ & 13	  & 30	  & -10	  & 18	  & ---	  & ---	  & ---	  & ---	  & 0.4+i1.8	  & -1.6+i1.7	  & 3.6+i3.3	  & 4.8-i0.9	& ---		  & ---		  &
0-i3.7  & -1.3+i1.6	\\
 
 $N (\textit{1750})$ 1/2$^+_{(*,a)}$  & 20	  & (nf)	  & -150  & ---	  & ---	  & ---	  & ---	  & ---	  & 2.3-i0.5	  & ---	  & -0.5+i1.5	  & ---	& ---		  &
---	
  & 3.3+i3.3	  &---	\\
 
 $N (1720)$ 3/2$^+$ & 23	  & 24	  & -53	  & -50	  & 2	  & 6	  & 112	  & 158	  & -1.3-i0.1	  & -1.0+i0.1	  & 2.7+i4.0	  & 2.6+i5.1	& 0.1-i0.4	  & -0.4-i2.5	  &
-1.6-i1.7  & 1.0+i3.3	\\
 
 $N (1520)$ 3/2$^-$ & 6	  & 5	  & 89	  & 97	  & 56	  & 48	  & -180  & -172  & 1.2+i1.1	  & 2.9+i1.7	  & 2.5-i1.5	  & 4.2-i0.3	& -4.6-i25	  & -3.6-i23	  & 2.4+i3.9
  & -4.1+i9.7	\\
 
 $N (1675)$ 5/2$^-$ & 40	  & 52	  & 149	  & 130	  & $<$1  & 2	  & 	  & -62	  & 0.7-i0.4	  & 0.0-i0.4	  & -4.1-i14	  & -3.2+i1.4	& 0.3-i0.5	  & -0.5+i0.4	  &
-0.4+i0.8  & -0.1+i1.4	\\
 
 $N (1680)$ 5/2$^+$ & 2	  & 3	  & -174  & 166	  & 52	  & 46	  & 151	  & 153	  & 0.0+i0.5	  & 0.1+i0.6	  & 0.5+i0.2	  & 0.4-i0.1	& 2.1-i0.9	  & 0.5-i5.8	  & 1.4-i0.4
  & 2.5-i1.9	\\

 $N (1990)$ 7/2$^+$ & 13	  & 7	  & 91	  & -86	  & 1	  & 1	  & -128  & 82	  & 0.0+i0.0	  & 0.1+i0.2	  & -2.5-i0.4	  & -6.9-i5.9	& 0.2-i0.2	  & -0.7+i0.0	  &
0.3+i0.6  & -0.6-i0.2	\\
 
 $N (2190)$ 7/2$^-$ & 4	  & 2	  & 125	  & 84 	  & 24	  & 16	  & -50	  & -77	  & -0.5-i1.4	  & -0.6-i1.7	  & 0.2-i4.8	  & -0.8-i50	& -1.0-i7.7 	  & -3.6-i10	  & -1.2-i1.3
  & -2.3+i1.9	\\
 
 $N (2250)$ 9/2$^-$ & 26	  & 20	  & 98	  & 101	  & 2	  & 2 	  & -105  & -169  & 1.0-i0.4	  & 0.6-i0.1	  & -0.6+i0.5	  & 1.4+i2.1	& 1.5-i0.3	  & 0.9+i0.0	  &
0.2+i0.3  & 0.4+i0.5	\\
 
 $N (2220)$ 9/2$^+$ & 1  & 2	  & 81	  & 48 	  & 5	  & 26	  & 95	  & -98	  & -1.0+i0.1	  & -0.9+i1.2	  & -3.7-i1.7	  & -2.2-i0.5	& -12-i7.9	  & -5.4-i6.1	  & 0.0-i1.2
  & -1.4-i2.5	\\
\hline\hline
\end {tabular}
}
\end{center}
\label{tab:bra1ppn}
\end{table*}

\begin{table*}
\caption{Properties of the $I=3/2$ resonances: Pole positions $z_0$ ($\Gamma_{\rm tot}$ defined as -2Im$z_0$), elastic $\pi N$ residues
$(|r_{\pi N}|,\theta_{\pi N\to\pi N})$, and the normalized residues  $(\sqrt{\Gamma_{\pi N}\Gamma_\mu}/\Gamma_{\rm tot},\theta_{\pi N\to \mu})$
of the inelastic reactions $\pi N\to K\Sigma$ and $\pi N\to\pi\Delta$. For the quantum numbers of the $\pi\Delta$ channels (6), (7),  see
Table~\ref{tab:couplscheme}; (a): dynamically generated.}
\begin{center}
\renewcommand{\arraystretch}{1.30}
\begin {tabular}{ll|rrrr|rrrr|rrrr|rrrr|rrrr} \hline\hline
&&\multicolumn{4}{|l}{Pole position}
 &\multicolumn{4}{|l}{$\pi N$ Residue}
 &\multicolumn{4}{|l}{$K\Sigma$ channel} 
 &\multicolumn{4}{|l|}{$\pi\Delta$, channel (6)}
 &\multicolumn{4}{|l}{$\pi\Delta$, channel (7)}
\bigstrut[t]\\[0.1cm]
&&\multicolumn{2}{|l}{Re $z_0$ \hspace*{0.8cm} }
& \multicolumn{2}{l|}{-2Im $z_0$\hspace*{0.2cm} }
& \multicolumn{2}{l}{$|r_{\pi N}|$\hspace*{0.2cm}} 
& \multicolumn{2}{l}{$\theta_{\pi N\to\pi N}$ } 
%& \multicolumn{2}{|c}{$\Gamma_{\pi N}/\Gamma_{\rm tot}$}\hspace*{0.3cm}
& \multicolumn{2}{|l}{$\displaystyle{\frac{\Gamma^{1/2}_{\pi N}\Gamma^{1/2}_{K\Sigma}}{\Gamma_{\rm tot}}}$}
& \multicolumn{2}{l|}{$\theta_{\pi N\to K\Sigma}$\hspace*{0.1cm}}
& \multicolumn{2}{l}{$\displaystyle{\frac{\Gamma^{1/2}_{\pi N}\Gamma^{1/2}_{\pi\Delta}}{\Gamma_{\rm tot}}}$} 
& \multicolumn{2}{l|}{$\theta_{\pi N\to \pi\Delta}$\hspace*{0.1cm}}
& \multicolumn{2}{l}{$\displaystyle{\frac{\Gamma^{1/2}_{\pi N}\Gamma^{1/2}_{\pi\Delta}}{\Gamma_{\rm tot}}}$} 
& \multicolumn{2}{l}{$\theta_{\pi N\to \pi\Delta}$}
\\
&&\multicolumn{2}{|l}{[MeV]} & \multicolumn{2}{l|}{[MeV]} & \multicolumn{2}{l}{[MeV]} & \multicolumn{2}{l}{[deg]} 
& \multicolumn{2}{|l}{[\%]}  & \multicolumn{2}{l|}{[deg]} & \multicolumn{2}{l}{[\%]}  & \multicolumn{2}{l|}{[deg]} &\multicolumn{2}{l}{[\%]} & \multicolumn{2}{l}{[deg]} \\
		  & fit$\to$		& A	& B	& A	& B	& A	& B	& A	& B	& A	& B	& A	& B	& A	& B	& A	& B	& A	& B	& A	& B	
\bigstrut[t]\\
\hline
 $\Delta(1620)$	1/2$^-$&	& 1599	& 1596	& 71	& 80	& 17	& 18	& -107	& -107	& 22	& 24	& -107	& -106	& ---	& ---	& ---	& ---	& 57	& 63	& 102	& 101
\\

 $\Delta(1910)$ 1/2$^+$&	& 1788	& 1848	& 575	& 376	& 56	& 20	& -140	& -143	& 4.7	& 1.9	& -144	& -115	& 41	& 22	& 71	& 79	& ---	& ---	& ---
& --- \\

 $\Delta(1232)$ 3/2$^+$	&	& 1220	& 1218	& 86	& 96	& 44	& 50	& -35	& -38	& 	&	&	&	&	&	&	&	&	&	&
&	\\
 
 $\Delta(1600)$ 3/2$^+_{(a)}$&	& 1553	& 1623	& 352	& 284	& 20	& 27	& -158	& -124	& 11	& 13	& -7	& 41	& 28	& 38	& 28	& 74	& 1	& 4
& 15	& -107	\\

[$\Delta(1920)$	3/2$^+$&	& 1724	& 1808	& 863	& 887	& 36	& 19	& 163	& -70	& 16	& 14	& -21	& 50	& 7	& 4 	& 144	& -108	& 1	& $<$1	& -101	&]
\\

 $\Delta(1700)$ 3/2$^-$&	& 1675	& 1705	& 303	& 183	& 24	& 14	& -9	& -4	& 1.5	& 1.6	& -150	& -121	& 5	& 5	& 166	& 173	& 39	& 35	& 149
& 175	\\

 $\Delta(1930)$	5/2$^-$&	& 1775	& 1805	& 646	& 580	& 18	& 14	& -159	& 3	& 3.1	& 1.7	& -3	& 135	& 12	& 11	& 26	& -7	& $<$1	& 1	&
& 155	\\

 $\Delta(1905)$	5/2$^+$&	& 1770	& 1776	& 259	& 143	& 17	& 9	& -59	& -40	& 0.5	& 0.1	& -142	& -99	& 4	& 4	& 130	& -179	& 34	& 29	& 105
& 120	\\

 $\Delta(1950)$	7/2$^+$&	& 1884	& 1890	& 234	& 232	& 58	& 58	& -25	& -19	& 4.0	& 3.8	& -78	& -71	& 55	& 52	& 139	& 149	& 3	& 4	& -84
& -51	\\

 $\Delta (2200)$ 7/2$^-$&	& 2147	& 2111	& 477	& 353	& 17	& 20	& -52	& 7	& 0.6	& 0.1	& -98	& -33	& 2	& 7	& -145	& 107	& 24	& 37	& 111 
& 153	\\

 $\Delta (2400)$ 9/2$^-$&	& 1969	& 1938	& 577	& 559	& 25	& 16	& -80	& -112	& 1.3	& 0.8	& 40	& 6	& 24	& 15	& -98	& -124	& 3	& 1	& 1
& 11	\\
\hline\hline
\end {tabular}
\end{center}
\label{tab:bra2}
\end{table*}

\begin{table*}
\caption{Residues of the $I=3/2$ resonances for the effective $\rho N$ channels according to Eq.~(\ref{couplings}). For the quantum numbers of
the channels (2), (3), (4), see Table~\ref{tab:couplscheme}; (a): dynamically generated. }
\begin{center}
\renewcommand{\arraystretch}{1.30}
\begin {tabular}{ll|rr|rr|rr} \hline\hline
&&\multicolumn{2}{|l}{$\rho N$, channel (2)}
 &\multicolumn{2}{ l}{$\rho N$, channel (3)}
 &\multicolumn{2}{ l}{$\rho N$, channel (4)} 
\bigstrut[t]\\[0.1cm]
&&\multicolumn{2}{l}{$\displaystyle{g_{\rho N}}$[10$^{-3}$ MeV$^{-1/2}$]} 
& \multicolumn{2}{l}{$\displaystyle{g_{\rho N}}$[10$^{-3}$ MeV$^{-1/2}$]} 
& \multicolumn{2}{l}{$\displaystyle{g_{\rho N}}$[10$^{-3}$ MeV$^{-1/2}$]} 
\\
		  	& fit$\to$	& A		& B		& A		& B		& A		& B		
\bigstrut[t]\\
\hline
 $\Delta(1620)$	1/2$^-$&	& 3.8+i3.6	& 3.7+i3.5	& ---		& ---		& 0.9+i0.8	& 1.1+i0.6	\\

 $\Delta(1910)$	1/2$^+$&	& 0.7-i2.0	& 0.3-i0.3	& -1.7+i0.8	& -2.1+i2.8	& ---		& ---		\\

 $\Delta(1232)$	3/2$^+$&	&		&		&		&		&		&		\\
 
 $\Delta(1600)$	3/2$^+_{(a)}$&	& 0.3-i0.4	& 1.6-i2.0	& -2.0-i3.7	& -1.9-i3.4	& 1.1+i1.1	& -0.3-i1.1	\\

[$\Delta(1920)$	3/2$^+$&	& 0.9+i1.6	& 1.8+i1.2	& 4.4+i6.7	& 3.6+i6.5	& 0.7+i1.1& 0.8+i0.5]	\\

 $\Delta(1700)$	3/2$^-$&	& 7.0+i5.7	& 2.5-i0.5	& 6.4+i6.0	& 1.4-i0.5	& -6.3-i5.8	& -3.5-i3.5	\\

 $\Delta(1930)$	5/2$^-$&	& -0.5-i0.2	& -0.4+i0.1	& 1.6+i1.5	& 1.0+i5.0	& 0.8+i0.0	& -0.7+i0.1	\\

 $\Delta(1905)$	5/2$^+$&	& 4.8+i0.6		& 3.0-i0.8		& 4.6+i1.3		& 2.3-i1.2	& -4.1+i 6.5	& -3.5-i0.5	\\

 $\Delta(1950)$	7/2$^+$&	& 0.2-i0.4	& -0.1-i0.5	& 1.8+i0.3	& -2.8-i2.6	& 0.2-i0.3	& 0.3-i0.2	\\

 $\Delta (2200)$ 7/2$^-$&	& 1.3-i1.5	& 1.5-i0.9	& 1.8-i0.2	& 1.3+i0.4	& 1.4+i5.5	& -1.3+i6.1	\\

 $\Delta (2400)$ 9/2$^-$&	& -1.3+i0.6	& -1.1+i0.5	& 1.0+i0.1	& 0.7-i0.3	& -1.3+i0.5	& -1.2+i0.5	\\
\hline\hline
\end {tabular}
\end{center}
\label{tab:bra2ppn}
\end{table*}

\begin{table}
\caption{Branching ratios $\Gamma_{\mu}/\Gamma_{\rm tot}$ of the $I=1/2$ resonances in $\%$ according to Eq.~(\ref{branching}).  (*): not identified with PDG name; (a): dynamically generated, (nf): not found. }
\begin{center}
\renewcommand{\arraystretch}{1.30}
\begin {tabular}{l |rr|rr|rr|rr} \hline\hline

	& \multicolumn{2}{c}{$\frac{\Gamma_{\pi N}}{\Gamma_{\rm tot}} $} &  \multicolumn{2}{c}{$\frac{\Gamma_{\eta N}}{\Gamma_{\rm tot}} $} &  \multicolumn{2}{c}{$\frac{\Gamma_{K\Lambda}}{\Gamma_{\rm tot}} $} &  \multicolumn{2}{c}{$\frac{\Gamma_{K\Sigma}}{\Gamma_{\rm tot}} $} \\

		  		& A		& B		& A		& B		& A		& B  	 	& A		& B	
\bigstrut[t]\\
\hline			

 $N (1535)$ 1/2$^-$&  44	& 40	& 57	& 59	& 1.3 & 1.7	& 4.9 & 29			\\
 	 			
 $N (1650)$ 1/2$^-$&  61 & 41	& 3.7 & 3.6 & 11 & 7.7	& 11 & 5.4	 		\\
 
 $N (1440)$ 1/2$^+_{(a)}$ & 	56 &	52	& 0.1 & 0.5 	& 0.1 & 2.5 	& $<$0.1& 0.1	 \\
  
 $N (1710)$ 1/2$^+$& 8.2 & 14 & 69 &38 & 11 &19 & 1.9 & $<$0.1    \\
 
 $N (\textit{1750})$  1/2$^+_{(*,a)}$& 5.1& (nf) & $<$0.1 & --- & 0.1 &--- & 1.0 &--- 	\\
 
 $N (1720)$ 3/2$^+$&  	6.6 & 11.6 &0.2 & 0.9 & 1.5 & 0.7 & 0.4 & 0.4	\\
 
 $N (1520)$ 3/2$^-$&  	79 & 69 & 0.2 & 0.1 & 0.4 & 0.6 & $<$0.1 & 0.2	\\
 
 $N (1675)$ 5/2$^-$&   39 & 38 &0.9 & 0.4 & $<$0.1 & 0.1 & 0.3 & 0.4		\\
 
 $N (1680)$ 5/2$^+$&  67 & 68 & $<$0.1 & $<$0.1 & $<$0.1 & $<$0.1& $<$0.1&$<$0.1		\\

 $N (1990)$ 7/2$^+$& 	3.2 & 3.6 & 0.1 & 0.1 & 0.9 & 0.6 & 0.2 & 0.4	\\
 
 $N (2190)$  7/2$^-$&  23 & 18 & $<$0.1 & $<$0.1 & 0.2 & $<$0.1 & 0.1 & $<$0.1 		\\
 
 $N (2250)$ 9/2$^-$&  7.5 &5.7 & $<$0.1 & $<$0.1 & 0.2 & 0.1 & $<$0.1 & 0.1		\\
 
 $N (2220)$ 9/2$^+$ & 26 & 16 & $<$0.1 & $<$0.1 & $<$0.1& $<$0.1 & $<$0.1 & $<$0.1  		\\

\hline\hline
\end {tabular}
\end{center}
\label{tab:bra1h}
\end{table}

\begin{table}
\caption{Branching ratios $\Gamma_{\mu}/\Gamma_{\rm tot}$ of the $I=3/2$ resonances in $\%$ according to Eq.~(\ref{branching}).  (a): dynamically generated. }
\begin{center}
\renewcommand{\arraystretch}{1.30}
\begin {tabular}{ll |rr|rr} \hline\hline

	&&  \multicolumn{2}{c}{$\frac{\Gamma_{\pi N}}{\Gamma_{\rm tot}} $} &  \multicolumn{2}{c}{$\frac{\Gamma_{K\Sigma}}{\Gamma_{\rm tot}} $} \\

		  	& fit$\to$	& A		& B		& A		& B		\bigstrut[t]\\
\hline			

 $\Delta(1620)$	1/2$^-$&	& 48 & 45 & 9.9 &13 \\

 $\Delta(1910)$	1/2$^+$&	& 20 & 10 &1.1 & 0.4 \\

 $\Delta(1232)$	3/2$^+$&	& 100 & 100\\
 
 $\Delta(1600)$	3/2$^+_{(a)}$&	& 11& 19 & 11 & 10 \\

[$\Delta(1920)$	3/2$^+$&	&  8.3 & 4.3 & 32 & 44 ]\\

 $\Delta(1700)$	3/2$^-$&	& 16 & 16 &0.1 & 0.2  \\

 $\Delta(1930)$	5/2$^-$&	& 5.6 & 4.8 & 1.7 & 0.6 	\\

 $\Delta(1905)$	5/2$^+$&	& 13 & 12 & $<$0.1 & $<$0.1 	\\

 $\Delta(1950)$	7/2$^+$&	& 50 & 50 & 0.3 & 0.3 	\\

 $\Delta (2200)$ 	7/2$^-$&	&  7.2 & 11 & 0.1 & $<$0.1	\\

 $\Delta (2400)$	 9/2$^-$&	&  8.7 & 5.7 & 0.2 & 0.1 	\\

\hline\hline
\end {tabular}
\end{center}
\label{tab:bra3h}
\end{table}

The extracted pole positions, residues, and branching ratios (normalized residues) are shown in Tables~\ref{tab:bra1} to
\ref{tab:bra3h} and Fig.~\ref{fig:new_figure_Deborah} shows the pole positions in the complex plane. 

The $\rho N$ and $\pi\Delta$ states can couple to a resonance with given $J^P$ in more than one way; these possibilities can be organized in
different channels. The quantum numbers for a value quoted in Tables~\ref{tab:bra1ppn}, \ref{tab:bra2}, and \ref{tab:bra2ppn} can be obtained by
the entry of Table~\ref{tab:couplscheme} corresponding to the channel number and given $J^P$. 

Quoting transition branching ratios for the $\pi\Delta$ channels is physically meaningful because the $\pi\Delta$ state can be approximately
seen as a pair of stable particles, for energies well above the $\pi\Delta$ threshold. However, as this is not the case for the $\sigma N$ and
$\rho N$ channels, we abstain from quoting branching ratios for these latter two channels. Instead, only the plain residues are shown in the
Tables. Strictly speaking, the residue of a resonance decaying into an on-shell $\pi\pi N$ state is a function of the invariant mass of the
unstable particle, $M_I$, (or, equivalently, the three-momentum of the stable spectator particle), and not just one complex number. For the
$\pi\Delta$ state, it is a good approximation to fix $M_I={\rm Re}\,z_0\sim 1232$~MeV (that kinematics is chosen for the values quoted in the
tables). We have checked that, instead of $M_I=1232$~MeV, using the pole mass of our fits ($M_I=1219$~MeV in average), the $\pi N\to
N(1440)1/2^+\to\pi\Delta$ transition branching ratio changes by less than 1\%. The other resonances are heavier than the Roper resonance and the
changes should be even smaller.

Note that the labeling of the resonances in Tables~\ref{tab:bra1} to \ref{tab:bra3h} follows the PDG notation even though the real parts of
the pole position are different; whether the poles determined here can be identified with PDG resonances is discussed in the following sections.

If one defines branching ratios using Breit-Wigner parameterizations~\cite{Beringer:1900zz} their sum equals, by definition, 1, i.e.
$\sum_\mu\Gamma_\mu=\Gamma_{\rm tot}$. The right-hand side of this equation can be determined independently ($\Gamma_{\rm tot}=-2\,i\,{\rm
Im}\,z_0$) and be used as a test of the present results for the branching ratios. Indeed, below the $\pi\pi N$ threshold, the equality holds at
the 1 \% level. However, it should be noted that $\sum_\mu\Gamma_\mu=\Gamma_{\rm tot}$ never holds exactly when defining branching ratios via
residues as done here, even in a manifestly unitary coupled-channel model with only stable intermediate states. This is simply because the
amplitude has non-analytic branch points, required by unitarity, and this information is not contained in the residues. 

%%%%%%%%%%%%%%%%%%%%%%%%%%%%%%%%%%%%%%%%%%%%%%%%%%%%%%%%%%%%%%%%%%%%%%%%%%%%%%%%%%%%%%%%%%%%%%%%%%%%%%%%%%%%%%%%%%%%%%%%%%%%%%%%%%%%%%%%%%%%%%%%

\subsection{Discussion of specific resonances}
\label{casebycase}
The spectrum of resonances determined in the present approach is related to the states found by the GWU/SAID analysis in
Ref.~\cite{Arndt:2006bf} because we fit to the corresponding $\pi N\to\pi N$ partial waves. In that analysis of elastic and charge-exchange
pion-nucleon scattering, relatively few resonances are found and the question is whether with this set of states, in combination with the
background provided by the current approach, the $\pi N\to\eta N$ and $\pi N\to KY$ reactions can be described simultaneously with the $\pi N\to
\pi N$ reaction. It is clear that due to our choice of fitting the $\pi N$ partial waves of the GWU/SAID analysis, the most prominent resonances
of that analysis would be automatically needed here, and this is indeed what we find: those are the excited nucleon states 
\begin{align}
N(1535)\;1/2^- ,&& N(1650)\;1/2^-,&& N(1440)\;1/2^+, \non
N(1720)\;3/2^+,&& N(1520)\;3/2^-,&& N(1675)\;5/2^-, \non 
N(1680)\;5/2^+,&& N(2190)\;7/2^-,&& N(2250)\;9/2^-, \non
N(2220)\;9/2^+.&&               && \nonumber
\end{align}
Note that in Table~\ref{tab:bra1}, in brackets we have also quoted the second pole of the Roper resonance on the hidden $\pi\Delta$ sheet (cf.
discussion in Sec.~\ref{sec:Roper}). 

Furthermore, the following excited $\Delta$ states are found both in the GWU/SAID analysis and in the present study: 
\begin{align}
\Delta(1620)\; 1/2^-,&& \Delta(1910) \;1/2^+,&& \Delta(1232)\; 3/2^+, \non
\Delta(1600)\; 3/2^+,&& \Delta(1700) \;3/2^-,&& \Delta(1930)\; 5/2^-, \non
\Delta(1905)\; 5/2^+,&& \Delta(1950)\; 7/2^+,&& \Delta(2400)\; 9/2^-. \nonumber 
\end{align}
The GWU/SAID analysis lists additional poles in the complex plane that could not be found in the present study,
\begin{align}
N(1860)\; 5/2^+,       && N(2200)\; 11/2^+,&& \Delta(2529)\; 11/2^+ \; . \nonumber
\end{align}
 But the present analysis includes resonances only up to $J=9/2$ and thus does not consider the $J=11/2$
states, and the $\Delta(2529)\; 11/2^+$
lies outside the considered energies anyway. For the second $J^P=5/2\;^+$ state, see Sec.~\ref{sec:fitaprime}.

Apart from the states found in the GWU/SAID analysis, we include a few more genuine $s$-channel states that are needed for a quantitative
description mostly of the inelastic reactions $\pi N\to \eta N$, $K\Lambda$, and $K\Sigma$. Those are
\begin{align}
N(1710)\; 1/2^+,&& N(1990)\; 7/2^+,&& \Delta(1920)\; 3/2^+, \non
\Delta(2200)\; 7/2^-. && && \nonumber
\end{align} 
As Tables~\ref{tab:bra1} and \ref{tab:bra2} show, these states are characterized by a weak coupling to the $\pi N$ channel.

When discussing pole properties in the following, we always refer to the values shown in Tables~\ref{tab:bra1} to \ref{tab:bra3h}. In the discussion of individual resonances, we will highlight particular aspects but we cannot exhaustively compare
every decay channel to the corresponding PDG value~\cite{Beringer:1900zz}.

{\bf $\mathbf{S_{11}}$:} The $N(1535)$ 1/2$^-$ is quite narrow in both the present approach and the GWU/SAID analysis~\cite{Arndt:2006bf}, while
the $N(1650)$ 1/2$^-$ is narrow in the GWU/SAID analysis and much wider in the present fits A and B (cf. also Fig.~\ref{fig:compare_approaches}
in Sec.~\ref{sec:s11_and_bare}); still, those fits describe the SAID solution quite well on the physical axis. In the chiral unitary approach of
Ref.~\cite{Mai:2012wy}, the $N(1535)$ 1/2$^-$ resonance is also narrow [$z_{1535}= (1.547^{+0.004}_{-0.021}-i 0.046^{+0.004}_{-0.017})$~GeV] 
while the width of the $N(1650)$ 1/2$^-$ resonance [$z_{1650}= (1.597^{+0.017}_{-0.020}-i 0.045^{+0.010}_{-0.015})$~GeV] lies in between the
values of this study and the one of the GWU/SAID group. The different values of the widths illustrate the difficulties to extract pole positions
in this partial wave, as will be discussed in Sec.~\ref{sec:s11_and_bare}. The branching ratio of the $N(1535)$ 1/2$^-$ resonance into the $\eta
N$ channel is large (57-59\%), while the one of the $N(1650)$ 1/2$^-$ is much smaller (around 4\%), in line with what is found in other
analyses~\cite{Beringer:1900zz}. The $\pi N\to\eta N$ residues of the two S11 resonance have a non-zero relative phase that indicate a
non-trivial interplay, and the $\pi N\to\eta N$ $S$-wave cross section shown in Fig.~\ref{fig:entotpart} exhibits a dip slightly beyond the
$N(1650)$ 1/2$^-$ resonance position; see also Ref.~\cite{Gasparyan:2003fp} for a discussion.  Both $S_{11}$ resonances show a moderate but
non-vanishing branching into the $\pi\Delta$ channel. 

The $N(1535)$ 1/2$^-$ resonance has a moderate coupling to the $K\Lambda$ channel and a larger one to the $K\Sigma$ channel (cf.
Table~\ref{tab:bra1}). The $N(1650)$ 1/2$^-$ resonance couples strongly to all $KY$ channels. Indeed, the partial cross sections of the
corresponding reactions (Figs.~\ref{fig:entotpart}, \ref{fig:kltotpart}, \ref{fig:kzsztotpart_1h}) show a strong $S$-wave contribution close to
the thresholds and also beyond. It should be stressed however, that strong $S$-wave components in the $KY$ amplitudes cannot be identified with
sub-threshold resonances alone, as sometimes done in the isobar picture. Instead, one has an amplitude with a non-trivial energy dependence in
which sub-threshold resonances play only a moderate role, as demonstrated for pion-induced $\phi$ production~\cite{Doring:2008sv}.

{\bf $\mathbf{P_{11}}$:} For the discussion of the resonance dynamics in the $P_{11}$ partial wave, see Sec.~\ref{sec:p11}. We find in the
current analysis, apart from the (genuine) nucleon and the (dynamically generated) two Roper poles, two more states: a genuine $N(1710)$ 1/2$^+$
state mostly required by the inelastic data, and a dynamically generated, broad $N(\textit{1750})$ 1/2$^+$ state, which is found only in fit A and emerges mostly from the
$\sigma N$ and $\pi\Delta$ dynamics. It should be stressed, however, that the $N(1710)$ 1/2$^+$ is still around $100$~MeV wide and thus cannot
be identified with a potentially exotic narrow state.

{\bf $\mathbf{S_{31}}$:} The width of the $\Delta(1620)$ 1/2$^-$  is relatively small for both fits A and B which can be attributed to slight
deviations of our description from the GWU/SAID partial wave, as can be seen in Fig.~\ref{fig:pin3}. The resonance shows large inelasticities
into the $\pi\Delta$ channel (normalized residue of 57-63\%, compared to the PDG values of 38$\pm$ 9\%).
 
{\bf $\mathbf{P_{31}}$:} The $\Delta(1910)$ 1/2$^+$ resonance is wide, couples only weakly to $\pi N$, and is situated on top of a large,
energy-dependent background. Due to these intrinsic problems, the extracted resonance parameters show large differences between fits A and B.
Within these large boundaries, the properties agree with those of the PDG, in particular with respect to the large inelasticity into
$\pi\Delta$.

{\bf $\mathbf{P_{13}}$:} The $I=1/2$ $P$-wave plays an important role in $KY$ production, as, e.g., the $K\Lambda$ differential cross section
shows. However, there is an interplay between the $N(1720)$ 3/2$^+$ and the $N(1710)$ 1/2$^+$ resonance and individual contributions are more
difficult to pin down. For the transition branching ratios [i.e. moduli of the normalized residues of the $N(1720)$ 3/2$^+$] we obtain the
following numbers (in brackets, the corresponding numbers of Anisovich {\it et al.}~\cite{Anisovich:2012ct,Beringer:1900zz}): $\eta N$:
1.2-3.2\% [$3\pm 2$\%], $K\Lambda$: 2.9-3.1\% [$6\pm 4$\%], $K\Sigma$: 1.7-2.2\%, $\pi\Delta$, $P$-wave: 23-24\% [$29\pm 8$\%], and $\pi\Delta$,
$F$-wave: 2-6\% [$3\pm 3$\%]. 

{\bf $\mathbf{D_{13}}$:} The $\pi N$ residue of the $N(1520)$  3/2$^-$ state is well fixed from $\pi N$ scattering where it appears as a sharp
and prominent resonance. The coupling to $\eta N$ is very small, but well fixed due to the well-known $SD$-interference pattern resulting in the
$u$-shape of the $\pi N\to\eta N$ differential cross section (cf. Fig.~\ref{fig:etaN_s+dwave}). We obtain an $\eta N$ branching ratio of
$0.13-0.15$\% while Arndt {\it et al.} obtain $0.08-0.12$\%~\cite{Arndt:2006bf} and Tiator {\it et al.} $0.08\pm 0.01$\%~\cite{Tiator:1999gr}.
While all these values are comparable, they differ from the result by Penner {\it et al.} of $0.23\pm 0.04$\%~\cite{Penner:2002ma}.

The $N(1520)$  3/2$^-$ resonance is also characterized by a large coupling to the $\rho N$ channel in $S$-wave (obviously to the low-energy tail
of the $\rho$ meson), as reflected by the very large $S$-wave residue of $|g_{\rho N}|>20$ in Table~\ref{tab:bra1ppn}. There is no other state in the
$D_{13}$ partial wave, although our non-resonant part $T^\npo$ exhibits a resonant structure around $z\sim 1.7$~GeV. This is due to a pole
behind the complex $\rho N$ branch point, dynamically generated from the $S$-wave $\rho N$ interaction. The dynamics and meaning of this pole
are similar to the one discussed in Ref.~\cite{Doring:2009yv}. See Refs.~\cite{ Oset:2009vf, Khemchandani:2011et,Khemchandani:2011mf,
Garzon:2012np, Garzon:2013pad} for further discussion of the dynamical nature of this and other states.

{\bf $\mathbf{P_{33}}$:} Apart from the $\Delta(1232)$ 3/2$^+$, we find a wide state that might be identified with the $\Delta(1600)$ 3/2$^+$
from the PDG~\cite{Beringer:1900zz}. As the coupling to the $\pi N$ channel is rather small, compared to the large width, its properties are
difficult to determine, and also the pole properties cited in the PDG~\cite{Beringer:1900zz} show large uncertainties. In our approach, it is
dynamically generated from the $\pi\Delta$ $P$-wave interaction, showing a significantly larger $P$-wave $\pi\Delta$ residue than in the study
by Anisovich {\it et al.}~\cite{Anisovich:2012ct,Beringer:1900zz} (28-38\% vs. $14\pm 10$\%). Note that this state was also found in our
previous analysis~\cite{Doring:2010ap}. The second genuine resonance state included in the $P_{33}$ partial wave, the $\Delta(1920)$ 3/2$^+$, is
characterized by a very high bare mass (cf. Tables~\ref{tab:bare_cou_fit5}, \ref{tab:bare_cou_20Sep_2000}) that undergoes a large
renormalization effect so that the pole moves very far into the complex plane. The pole shows a large branching ratio into $K\Sigma$ and indeed
the $I=3/2$ $K\Sigma$ partial cross section is large over a wide energy range as Fig.~\ref{fig:kzsztotpart_3h} shows. This strength is obviously
required by data, although, due to the large imaginary part of the pole position, one can no longer speak of a resonance signal. 

{\bf $\mathbf{D_{33}}$:} The $\Delta(1700)$ 3/2$^-$ resonance is relatively wide, couples weakly to the $\pi N$ channel, and is situated on a
relatively large, energy-dependent background. Our values for the pole position and $\pi N$ residues agree with the PDG~\cite{Beringer:1900zz}
within the large intrinsic uncertainties.

{\bf $\mathbf{D_{15}}$:} The $N(1675)$ 5/2$^-$ shows a sizeable normalized $\pi N\to\pi\Delta$ residue of 40-52\% in line with the
PDG values~\cite{Beringer:1900zz} (46-50\%) but only a small coupling to the $K\Lambda$ channel. For the reaction $\pi N\to\eta N$, we find that
this resonance plays some role as can be seen in Fig.~\ref{fig:entotpart}, although its mass is in a region where the experimental $\eta N$
data are of particularly poor quality (cf. Sec.~\ref{sec:etan}). 

{\bf $\mathbf{F_{15}}$:} The $N(1680)$ 5/2$^+$ resonance shows only small branchings into the $\eta N$ and $KY$ channels. The role of a second
$F_{15}$ state is discussed in Sec.~\ref{sec:fitaprime}.

{\bf $\mathbf{D_{35}}$:} The real part of the pole position of the $\Delta(1930)$ 5/2$^-$ is in the GWU/SAID partial-wave
analysis~\cite{Arndt:2006bf} around 200~MeV higher than in the present study. The width is larger in the present fits A and B. In any case,
the overall size of the partial wave is small and our fit follows the trend of the GWU/SAID solution (cf. Figs.~\ref{fig:pin3},
\ref{fig:pindel3}); as a consequence, it seems difficult to unambiguously determine the pole position in this partial wave. As can be seen in Fig.~\ref{fig:kzsztotpart_3h}, the $D_{35}$ partial wave plays a minor, but not
insignificant role, in the $K\Sigma$ channel. This is reflected in the branching ratio of the $\Delta(1930)$ 5/2$^-$ into this channel.

{\bf $\mathbf{F_{35}}$:} The $\Delta(1905)F_{35}$ is a relatively inelastic resonance with large branching into $\pi\Delta$ (normalized $\pi
N\to\pi\Delta$ $P$-wave residue of 29-34\% here, compared to $25\pm 6$\% of Anisovich {\it et al.}~\cite{Anisovich:2012ct,Beringer:1900zz}). In
our result the resonance plays no role in $K\Sigma$ production.

{\bf $\mathbf{F_{37}}$:} The pole position of the prominent $\Delta(1950)$ 7/2$^+$ resonance is well determined as
Fig.~\ref{fig:compare_approaches} shows. This resonance is also very important in the $\pi N\to K\Sigma$ reaction: Our value for the normalized
residue and its phase (3.8-4\% and $-78^0$ to $-71^0$) is comparable to the value of Anisovich {\it et
al.}~\cite{Anisovich:2012ct,Beringer:1900zz} ($5\pm 1$\% and $-65^0\pm 25^0$).

{\bf $\mathbf{F_{17}}$, $\mathbf{G_{37}}$:} The resonances included for these partial waves mostly serve to improve the description of the $\pi
N$ elastic scattering. As Figs.~\ref{fig:pin4} and \ref{fig:pindel4} show, the signal of the $G_{37}$ resonance in $\pi N$ scattering is weak in
the sense that the resonance is so wide that the fitted energy region covers only part of the resonance shape. The $G_{37}$ partial wave is
small in $K\Sigma$ production. The $F_{17}$ partial wave is relevant for $K\Lambda$ production at higher energies (cf.
Fig.~\ref{fig:kltotpart}). No resonances are quoted in these partial waves in the GWU/SAID analysis~\cite{Arndt:2006bf} and we cannot claim much
evidence either. 

{\bf $\mathbf{G_{17}}$, $\mathbf{G_{19}}$, $\mathbf{H_{19}}$:} The resonances in these partial waves are wide, but relatively prominent in
elastic $\pi N$ scattering. Our fits to the GWU/SAID analysis~\cite{Arndt:2006bf} are satisfactory and the extracted pole positions and $\pi N$
residues are in the vicinity of the GWU/SAID values. The agreement is better for the $G_{17}$ resonance and slightly worse for the real part of
the $G_{19}$ resonance and the (anyway large) imaginary part of the $H_{19}$ resonance. None of these resonances shows a large branching into
the $\eta N$ and $KY$ channels.

{\bf $\mathbf{G_{39}}$:} The $G_{39}$ resonance is considerably wider in the GWU/SAID partial-wave analysis than here (900~MeV vs. 570~MeV), but
for poles so far in the complex plane, one expects uncertainties to become very large. The normalized residue into $K\Sigma$ is small
(0.8-1.3\%) but seems to be significant, as suggested by the relatively small uncertainty for the partial cross section in
Fig.~\ref{fig:kzsztotpart_3h}.

%%%%%%%%%%%%%%%%%%%%%%%%%%%%%%%%%%%%%%%%%%%%%%%%%%%%%%%%%%%%%%%%%%%%%%%%%%%%%%%%%%%%%%%%%%%%%%%%%%%%%%%%%%%%%%%%%%%%%%%%%%%%%%%%%%%%%%%%%%%%%%%%

\subsection{Fit A$^\prime$: a second $F_{15}$ state}
\label{sec:fitaprime}

\begin{figure}
\begin{center}
\includegraphics[width=0.48\textwidth]{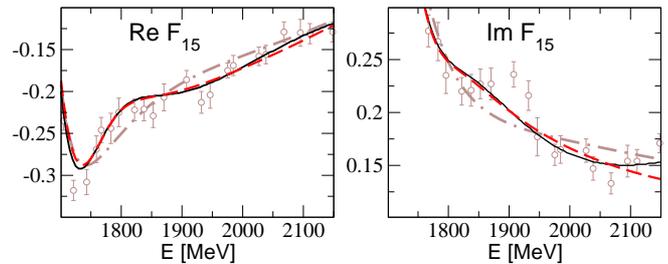}\\
\end{center}
\caption{Detail of the $F_{15}$ partial wave above the prominent $N(1680)$ 5/2$^+$ resonance. Solid (black) lines: energy dependent GWU/SAID
analysis~\cite{Arndt:2006bf}; dashed (red) lines: description obtained by including a second bare $s$-channel state in the $F_{15}$ partial wave
and refitting the entire data base (fit A$^\prime$). The faint dash-dotted lines show the original fit A (same as in Fig.~\ref{fig:pin2}).}
\label{fig:2ndf15}     
\end{figure}

Coming back to the $F_{15}$ partial wave, only one genuine resonance is included in fits A and B, while two resonances are needed in the
GWU/SAID analysis. The second $F_{15}$ resonance state, $N(1860)$ 5/2$^+$ in the PDG, formerly known as $N(2000)$ 5/2$^+$, is not included in
the present analysis; from Figs.~\ref{fig:pin2} and \ref{fig:pindel2} it is obvious that a more quantitative description of the $F_{15}$ may
require this state. As a test, we show in Fig.~\ref{fig:2ndf15} the result when including a second bare $s$-channel state in the $F_{15}$
partial wave and performing a refit (fit A$^\prime$) of the entire data base, starting from fit A and varying all bare resonance masses and
couplings.
The description of the $F_{15}$ partial wave around $z\sim 2$~GeV improves; in fact we achieve an almost perfect match of the GWU/SAID solution.
However, we see no substantial improvement for the other reactions with $\eta N$ and $KY$ final states.  The largest changes are visible in the
$K\Lambda$ polarization data at backward angles, but the data carry large errors (cf. Fig.~\ref{fig:klpola1}). Thus, the inelastic reactions
cannot contribute much to increase the evidence for this state, and more precise polarization data are called for. 

Even though for fit A$^\prime$ we have varied all resonance parameters, the properties of other states change only marginally. Also, the
properties of the strong $N(1680)$ 5/2$^+$ resonance barely change comparing fits A and A$^\prime$ [$z_0(\text{fit A}) =(1666-i 54.1)$~MeV vs.
$z_0(\text{fit A$^\prime$})=(1669-i 57)$~MeV]. In fit A$^\prime$, we obtain $|r_{\pi N}|=39$~MeV compared to $|r_{\pi N}|=42$~MeV in the SAID
solution~\cite{Arndt:2006bf}. The pole position of the second $F_{15}$ resonance $N(1860)$ 5/2$^+$ is at $z_0(\text{fit A$^\prime$})=(1732-i
76.5)$~MeV, while the residue amounts to $|r_{\pi N}|=8$~MeV and $\theta_{\pi N\to\pi N}=-156^\circ$, i.e. is much smaller than the one of the
$N(1680)$ 5/2$^+$.  This is in line with what is naively expected from the shape of the amplitude, i.e the $N(1860)$ 5/2$^+$ resonance appearing
as a small bump to the right of the prominent $N(1680)$ 5/2$^+$. In Ref.~\cite{Arndt:2006bf}, in contrast, the $N(1860)$ 5/2$^+$ has a residue
of $|r_{\pi N}|=60$~MeV. This value, even larger than the one of the $N(1680)$ 5/2$^+$, might have to do with a close-by zero and the special
dynamics of this resonance in the SAID analysis~\cite{workman_private}.

%%%%%%%%%%%%%%%%%%%%%%%%%%%%%%%%%%%%%%%%%%%%%%%%%%%%%%%%%%%%%%%%%%%%%%%%%%%%%%%%%%%%%%%%%%%%%%%%%%%%%%%%%%%%%%%%%%%%%%%%%%%%%%%%%%%%%%%%%%%%%%%%

\subsection{The $\boldsymbol{P_{11}}$ partial wave in $\boldsymbol{\pi N\to\pi N}$}
\label{sec:p11}
\subsubsection{Fits A, B, and the Roper resonance}
\label{sec:Roper}
The $P_{11}$ partial wave and its analytic structure require a detailed analysis. As mentioned in Sec.~\ref{sec:numerics}, we use different
starting conditions for fit A and B. In particular, we demand the nucleon renormalization to be small in fit B.  As a result we observe, in the
course of fitting, differences in the dynamics for the  inelasticities in the $P_{11}$ partial wave, which are closely linked to the properties
of the $\sigma N$ channel.  In the following, we illustrate the interplay between nucleon renormalization and changes in the $\sigma N$ channel.

The $\sigma N$ channel has much influence in the $P_{11}$ partial wave due to the onset of large inelasticities at relatively low energies and
the fact that in $P_{11}$ the two pions and the nucleon of the $\sigma N$ channel are all in relative $S$-waves, i.e. there is no centrifugal
barrier to suppress low-energy contributions.

\begin{figure}
\begin{center}
\includegraphics[width=0.42\textwidth]{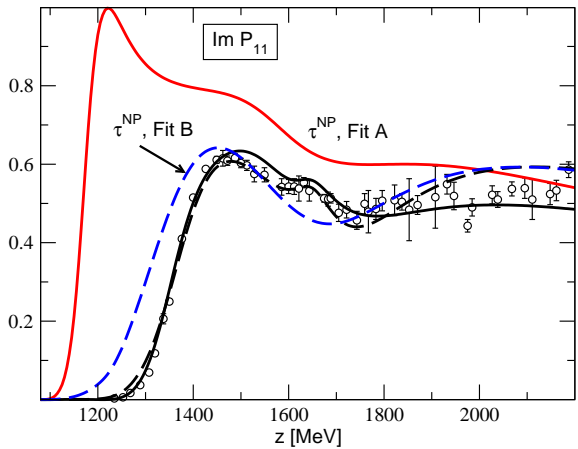}		
\end{center}
\caption{Imaginary part of the $P_{11}$ $\pi N$ amplitude. Besides the full amplitudes $\tau$ for fit A (solid black line) and fit B (dashed
black line), also the corresponding non-pole parts $\tau^\npo$ according to Eqs.~(\ref{deco1}) and (\ref{taut}) are shown. Note that this
decomposition has no physical meaning and serves only for illustration of the renormalization mechanism due to the nucleon pole term.}
\label{fig:p11tnp}     
\end{figure}

By far the largest differences in the parameters found for fits A and B are those  in the $g_{\sigma NN}$ and $g_{\sigma\sigma\sigma}$  coupling
constants, cf. Table~\ref{tab:coupl}. The implications of a small or large $g_{\sigma NN}$, or generally speaking of the two scenarios  A and B,
is illustrated in Fig.~\ref{fig:p11tnp} where we show the non-pole part of the $\tau$ matrix  for both cases [$\tau^\npo$ and $\tau^\po$ are
related to $T^\npo$ and $T^\po$ via Eq.~(\ref{taut})]. 

For fit B (large $g_{\sigma NN}$), the pole in $\tau^\npo$ is obviously already in the vicinity of the full amplitude $\tau$. The pole part,
$\tau^\po$, consists of the nucleon plus a second pole responsible for the $N(1710)$ 1/2$^+$ that we can omit from the discussion for the
moment. When adding the nucleon pole, it repels the pole in $\tau^\npo$ due to the mechanism of resonance repulsion~\cite{Doring:2009bi} and
shifts it to its final position in the full amplitude $\tau=\tau^\po+\tau^\npo$. As this shift is rather small for fit B, so is the effect due
to renormalization of the nucleon pole. Indeed, the bare nucleon mass of $m^b=973$~MeV (cf. Table~\ref{tab:bare_cou_20Sep_2000}) is close to the
physical one. Also, the renormalization of the bare $\pi NN$ coupling ($f_N^{b}=1.012$) is small when compared to the dressed, physical value of
$f_{\pi NN}=0.989$.

For fit A (small $g_{\sigma NN}$), the scenario is different: here, the pole in $\tau^\npo$ is farther away from the Roper pole position as
Fig.~\ref{fig:p11tnp} shows. Due to the discussed mechanism of resonance repulsion, the nucleon renormalization effect is larger and indeed we
obtain a bare nucleon mass of $m^b=1146$~MeV (cf. Table~\ref{tab:bare_cou_fit5}).  The large renormalization of the bare mass in fit A comes
along with a large renormalization of the bare coupling that is $f_N^{b}=0.619$ compared to the physical, dressed value. The size of the
renormalization of fit A is comparable to the one of previous studies~\cite{Krehl:1999km,Gasparyan:2003fp}. For both scenarios A and B, we
obtain a slightly negative phase shift close to the $\pi N$ threshold (cf. Fig.~\ref{fig:pindel1}) which is a remnant of the nucleon bound state
and its interplay with the Roper resonance.

While now the renormalization of the nucleon is understood for fit A and B, the question remains what further differences there are in the
parameterization of the $T^\npo$ amplitude. Tracing back the origin of the pole in $T^\npo$, it turns out that for both fits (and the
previous ones of Refs.~\cite{Schutz:1994ue, Schutz:1994wp, Schutz:1998jx, Krehl:1999km, Gasparyan:2003fp}) a pole in the $P_{11}$ partial wave
is generated from the $t$-channel correlated two-pion exchange with $\rho$ quantum numbers. For fit A (small $g_{\sigma NN}$), adding all other
$t$- and $u$-channel exchanges, one ends up with the pole structure visible in Fig.~\ref{fig:p11tnp}. For fit B (large $g_{\sigma NN}$), the
exchange contributions to and within the $\sigma N$ channel ---see Figs.~\ref{fig:dia1} and \ref{fig:dia2}--- provide so much attraction, that
the pole generated from the $\rho$-exchange is driven far below the nucleon pole position and becomes a spurious state on one of the many
Riemann sheets there (cf. Fig.~\ref{fig:anap11}). The physical amplitude below the nucleon pole and between nucleon pole and $\pi N$
threshold is pole-free. The large $\sigma N$ attraction provides at the same time enough strength for the formation of the pole visible in
$T^\npo$ of Fig.~\ref{fig:p11tnp}.

In summary, while the internal parameterization of the $P_{11}$ partial wave is quite different, the full, physical amplitude is not, and in
both scenarios, it is possible to quantitatively describe the GWU/SAID amplitude. 

\subsubsection{Analytic structure} 
\label{sec:analytic}
\begin{figure}
\begin{center}
\includegraphics[width=0.49\textwidth]{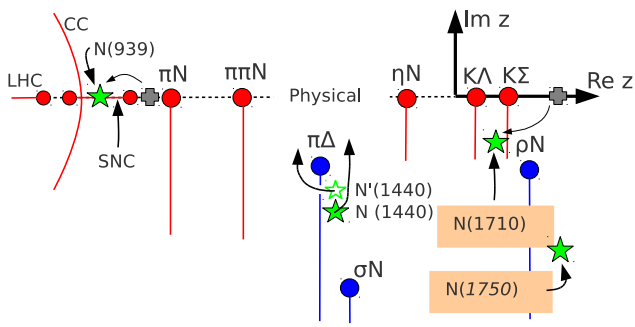}
\end{center}
\caption{Analytic structure of the present fit result (fit B) in the $P_{11}$ partial wave, for complex scattering energies $z$. The circular
cut is labeled ``CC'', the left-hand cut ``LHC''. See text for further explanations. Note that in fit A the bare mass of the nucleon is above
the $\pi N$ threshold.}
\label{fig:anap11}     
\end{figure}

The analytic structure of the current fits A and B is displayed in Fig.~\ref{fig:anap11}. 
Shown are the branch points (thresholds) of the $\pi N$, $\eta N$, $K\Lambda$, and $K\Sigma$ channels on the real, physical $z$-axis, as well as
the $\pi\pi N$ branch points on the real axis and in the complex plane from the $\pi\Delta$, $\sigma N$, and $\rho N$ channels. To search for
poles on the most relevant Riemann sheets, the cuts are chosen to be oriented from the threshold openings into the negative imaginary
$z$-direction as indicated in the figure (see also Sec.~\ref{sec:respro}). In addition, the figure shows the sub-threshold short nucleon cut
(SNC), circular cut (CC) and left-hand cut (LHC).  The structure of cuts and branch points is discussed in detail in Ref.~\cite{Doring:2009yv}. 

The (green) stars indicate the positions of the physical poles. There are two Roper poles on different $\pi\Delta$ sheets, labeled as N(1440)
and N$^\prime$(1440) with the positions given in Table~\ref{tab:bra1}. This two-pole structure is found in several analyses, such as
GWU/SAID~\cite{Arndt:2006bf} and EBAC~\cite{Suzuki:2009nj}, and is due to the fact that poles can repeat on different sheets. The closeness of
the complex $\pi\Delta$ branch point to the resonance poles leads to a non-standard shape that is not of the Breit-Wigner type.

Below the $\pi N$ threshold, the cross indicates the bare nucleon pole and the (green) star the physical position of the nucleon. The same
labeling is used for the second genuine state [the $N(1710)$ 1/2$^+$] with the pole positions given in Table~\ref{tab:bra1}. The properties of
this state are discussed in the following section. In the $\pi N$ amplitude shown in Figs.~\ref{fig:pin1} and \ref{fig:pindel1}, it is visible
as a weak resonance above the Roper state. 

When comparing the pole properties of the Roper resonance for fits A and B according to Tables~\ref{tab:bra1} and \ref{tab:bra1ppn}, it turns
out that for fit B the position is farther in the complex plane and the residue for decay into the $\sigma N$ channel is smaller
than for fit A. In turn, the residue for decay into the $\pi\Delta$ channel is larger for fit B. This demonstrates that the net inelasticity,
that must be approximately the same for fit A and B, is distributed differently among the effective $\pi\pi N$ channels for the two fits.
 
Finally, in fit A we find one additional, dynamically generated pole in the complex plane at around $z=1.75$~GeV and far in the complex plane, called  $N
(\textit{1750})$ 1/2$^+$ in Tables~\ref{tab:bra1} and \ref{tab:bra1ppn}. This state does not appear in fit B and we expect that a dedicated study of the reaction $\pi N\to\pi\pi N$ can shed light on the
question which of the two scenarios A or B provides the more realistic description. 

Here, we just note that the analytic structure of Fig.~\ref{fig:anap11} is different from the one found by the EBAC group in
Ref.~\cite{Suzuki:2009nj}: in that parameterization, there is only one genuine resonance state with a bare mass at $z=1736$~MeV, that, through
the hadronic dressing, is responsible for the two Roper poles plus a pole at $z=(1820-i248)$~MeV.  The nucleon pole is not included in
Ref.~\cite{Suzuki:2009nj}, however, in Ref.~\cite{Kamano:2010ud}, the EBAC group investigated the consequences of a genuine nucleon pole. 

The question arises if there are ways to distinguish the reaction dynamics found here from the one by the EBAC group in
Ref.~\cite{Suzuki:2009nj}. The principal problem is that the only observable quantity is the fully dressed amplitude, while quantities such as
bare masses and couplings depend on the renormalization scheme and the space of included channels, as demonstrated in Sec.~\ref{sec:delta1620}. 
Therefore, while certain scenarios might be ruled out in a dedicated study of $\pi\pi N$ final states, there is, in principle, no way to
distinguish  between internal parameterizations of the amplitude in terms of bare quantities, and, in particular, no meaningful hadronic
``undressing'' of resonances.

%%%%%%%%%%%%%%%%%%%%%%%%%%%%%%%%%%%%%%%%%%%%%%%%%%%%%%%%%%%%%%%%%%%%%%%%%%%%%%%%%%%%%%%%%%%%%%%%%%%%%%%%%%%%%%%%%%%%%%%%%%%%%%%%%%%%%%%%%%%%%%%%

\subsubsection{Properties of the $N(1710)$ 1/2$^+$}
\begin{figure}
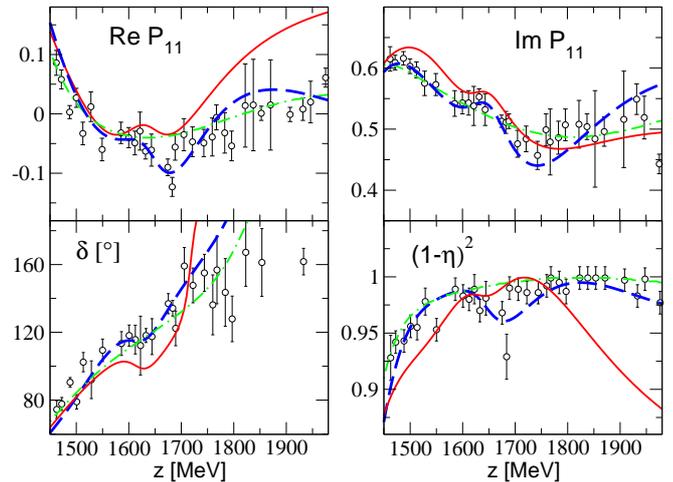

\begin{center}
\includegraphics[width=0.48\textwidth]{p11_1710.eps}\\
\vspace*{-0.1cm}
\includegraphics[width=0.48\textwidth]{del_p11_1710.eps}
\end{center}
\caption{Detail of the partial-wave amplitude (top) and phase shift/inelasticity (bottom) of the elastic $\pi N\to\pi N$ $P_{11}$ wave. The
current results, fit A (solid red lines) and fit B (dashed blue lines), were fitted to the energy-dependent GWU/SAID
solution~\cite{Arndt:2006bf} (dash-dotted green lines). The single-energy solution~\cite{Arndt:2006bf} (points) is {\it not} included in
the fitting procedure.}
\label{p11detail}     
\end{figure}

In Fig.~\ref{p11detail}, the region around $z\sim 1.7$~GeV is displayed in more detail.
The figure shows, apart from fits A and B, the energy-dependent $P_{11}$ partial wave from Ref.~\cite{Arndt:2006bf} that we fit (green
dash-dotted lines). Most intriguingly, our fit B, and to some extent also A, match the single-energy solution from Ref.~\cite{Arndt:2006bf}
 better than the energy-dependent solution (green curve) we fit ---the latter is smooth around $z=1650$~MeV. 

In the present analysis, the inclusion of this second resonance became necessary to improve the $\pi^- p\to K^0\Lambda$ differential cross
section and polarization, and was not required at all for the fit to the energy-dependent $\pi N$ SAID solution.  Note that in the fit
procedure, the resonance position was not forced in any way to its current position, but left to float freely.

Thus, the structure seen in the single-energy solution of the analysis of $\pi N\to\pi N$~\cite{Arndt:2006bf} appears here for an entirely
different reason, namely through coupled-channel effects from the $K\Lambda$ data and, to some weaker extent, the $\eta N$ data. We see that as 
evidence for a resonance at this energy. 

In Ref.~\cite{Ceci:2011ae}, the role of the $\rho N$ branch point, or complex threshold opening, at $z\sim (1.7-i\,0.07)$~GeV, i.e. not so far
away from the discussed resonance, was scrutinized. It is shown that if an approach does not contain such branch points, false resonance signals
may appear. However, the current approach does contain the required $\rho N$, $\sigma N$, and $\pi\Delta$ branch points, and still we observe a
new resonance around $z\sim 1.65$~GeV. We note that the resonance has a considerable branching fraction into $\eta N$ and $K\Lambda$; cf.
Table~\ref{tab:bra1}. See also Figs.~\ref{fig:entotpart} and \ref{fig:kltotpart} for the partial-wave content of the $\pi^-p\to\eta n$ and
$\pi^- p\to K^0\Lambda$ reactions, respectively. For the $\eta n$ final state, the $P_{11}$ resonance is partly responsible for the second
maximum of the total cross section, although it should be stressed that the data situation in that energy range is very poor; see the discussion
in Sec.~\ref{sec:etan}. For the $K^0\Lambda$ final state, Fig.~\ref{fig:kltotpart} shows that in our fit the strong $P$-wave contribution at
$z\sim 1.7$~GeV comes from the $P_{11}$ wave instead of the $P_{13}$ wave that could also play a role here. This is actually demanded mostly by
the polarization data in that energy range. Here, the partial-wave content is fixed better than in the $\eta n$ case because the data are much
better.

The role of a $P_{11}$ resonance at $z\sim 1.7$~GeV in the context of different partial-wave analyses has been discussed extensively in
Ref.~\cite{Ceci:2006ra} (see also references therein). That resonance appears in several quark models~\cite{Capstick:1986bm, Loring:2001kx,
Ronniger:2011td, Melde:2008yr} but also in three-body hadronic calculations~\cite{MartinezTorres:2008kh}. In Ref.~\cite{Ceci:2006ra}, various
fits to the single-energy GWU/SAID solution and existing $\eta N$ partial waves were performed and the need for a $N(1710)$ 1/2$^+$ resonance
was stated. Here, in contrast, we fit directly and simultaneously to the $\eta N$ and $K\Lambda$ data and the energy-{\it dependent} GWU/SAID
solution, i.e. the present fit is unbiased by structures in the single-energy GWU/SAID solution (``data'' points in Fig.~\ref{p11detail}).
Also, instead of being required by the (rather low-quality) $\eta N$ data, the need for a $P_{11}$ state around $z\sim 1.7$~GeV arises from the
$K\Lambda$ data. In any case, it is noteworthy that in two different analyses a $P_{11}$ state around $z\sim 1.7$~GeV is needed.
In a similar context, it should be noted that in the Zagreb analysis actually two poles are found in the complex plane in the region 
Re~$z\sim 1.7$~GeV~\cite{Batinic:1995kr,Batinic:2010zz}. This scenario is similar to the present fit A that also has a second state, though far 
in the complex plane, called $N(\textit{1750})$ 1/2$^+$ (cf. Fig.~\ref{fig:anap11}).

In Refs.~\cite{Azimov:2003bb,Arndt:2003ga}, various $\pi N$ partial waves were scanned in the search for narrow resonances. Two candidates were found; a
stronger signal at $z\sim 1.68$~GeV and a weaker one at $z\sim 1.73$~GeV in the $P_{11}$ partial wave. However, while that first possibility
would be in line with the current findings, the resulting widths were much smaller than the one found here, which is around       110~MeV (cf.
Table~\ref{tab:bra1}). We do not make any attempt to identify the resonance found here with a narrow state of potentially exotic nature; it is
just worth noting that the present approach does contain the $K\Sigma$ channel that has its threshold close to the observed structure; threshold
effects from this channel, are, thus, taken into account. Such threshold effects can become large in $S$-wave and might be even responsible for
the excess of $\eta$ photoproduction on the neutron around these energies~\cite{Doring:2009qr} (for other explanations, see Refs.~\cite{
Anisovich:2008wd,Jaegle:2008ux,Shklyar:2006xw}); we do not observe any structures from $K\Sigma$ in the $P_{11}$ partial wave in the current fit
result. Analyzing $KY$ photoproduction data in the future might shed more light on the issue.

%%%%%%%%%%%%%%%%%%%%%%%%%%%%%%%%%%%%%%%%%%%%%%%%%%%%%%%%%%%%%%%%%%%%%%%%%%%%%%%%%%%%%%%%%%%%%%%%%%%%%%%%%%%%%%%%%%%%%%%%%%%%%%%%%%%%%%%%%%%%%%%%

\subsection{Uncertainties in pole positions and bare parameters}
\label{sec:s11_and_bare}

\begin{figure}
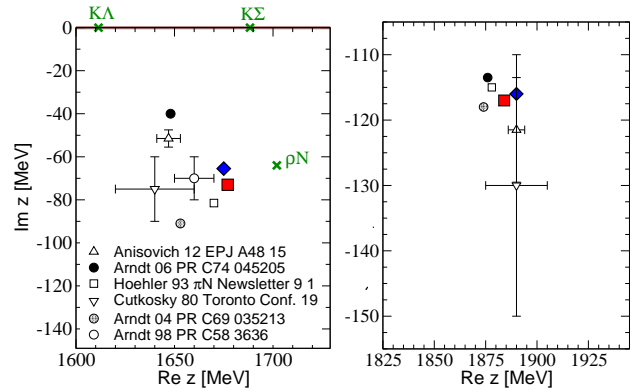

\begin{center}
\includegraphics[width=0.265\textwidth]{cmplx_plane_S11_1650.eps} \hspace*{-0.6cm}
\includegraphics[width=0.24\textwidth]{cmplx_plane_F37_1950.eps}
\end{center}
\caption{Pole positions of the $N(1650)$ 1/2$^-$ (left) and the $\Delta(1950)$ 7/2$^+$ (right) in different analyses~\cite{Anisovich:2011fc,
Arndt:2006bf, Hoehler2, Cutkosky, Arndt:2003if, Arndt:1998nm}. The results of this study are indicated with the (red) square (fit A) and (blue)
diamond (fit B). Also, the $K\Lambda$, $K\Sigma$, and $\rho N$ branch points are shown (green crosses).}
\label{fig:compare_approaches}     
\end{figure}
In Fig.~\ref{fig:compare_approaches}, the present results for the pole positions of the  $N(1650)$ 1/2$^-$ and the $\Delta(1950)$ 7/2$^+$ are
compared to other analyses quoted by the PDG~\cite{Beringer:1900zz}. For the isolated $F$-wave resonance, the results of the different analyses 
agree quite well. By contrast, the results for the $N(1650)$~1/2$^-$ $S$-wave resonance vary much more, because there is another close-by state,
the $N(1535)$ 1/2$^-$. 
\begin{figure}
\begin{center}
\includegraphics[width=0.48\textwidth]{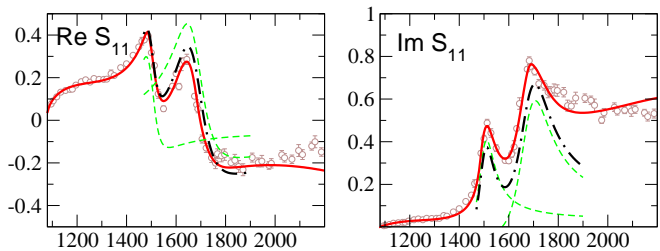}\\
\end{center}
\caption{Present result (fit A, solid lines) together with the Laurent expansions of Eq.~(\ref{pa}) around the different poles in the $S_{11}$
partial wave (dashed lines) and the sum of both terms (dashed-dotted lines).}
\label{fig:pa}     
\end{figure}
To illustrate this effect, in Fig.~\ref{fig:pa} we show the present description of the $S_{11}$ partial wave (fit A, red solid lines) together
with the quantity
\be
\frac{r_{\pi N}}{z-z_0}\ 
\label{pa}
\ee
for both $S_{11}$ resonances (green dashed lines) where $r_{\pi N}$ is the elastic residue and $z_0$ the pole position of the respective
resonance. Obviously, the tail of one resonance provides a strongly energy-dependent background for the other  resonance. Indeed, by summing the
contribution from both resonances, one obtains the dash-dotted curves that resemble the full amplitude quite well.

However, the mutual influence of close-by resonances makes their extraction challenging.  As Table~\ref{tab:bra1} shows, in this study we obtain
a small width for the $N(1535)$ 1/2$^-$ in combination with a rather large width for the $N(1650)$ 1/2$^-$. Note that also the GWU/SAID group
found a relatively narrow width for the $N(1535)$ 1/2$^-$ \cite{Arndt:2006bf} in 2006, although the resonance width determined in
2004~\cite{Arndt:2003if} is larger by 40\%, which again illustrates the difficulties in resonance extraction for the $S_{11}$ partial wave.  

Despite the fact that the extraction of the $S_{11}$ resonances is, for the above reasons, certainly one of the more complicated cases, the real
parts of the pole positions differ only by 1 to 2~MeV in our approach (cf. Tab.~\ref{tab:bra1}) whereas the bare masses are separated by 30~MeV and 127~MeV for the $N(1650)$ 1/2$^-$ and $N(1535)$ 1/2$^-$, respectively
(cf. Tab.~\ref{tab:bare_cou_fit5} and \ref{tab:bare_cou_20Sep_2000}). 

Increased uncertainties for bare masses, compared to pole positions, can be also found for other resonances comparing fits A and B. For example,
the real parts of the pole positions of the $N(1710)$ 1/2$^+$ resonance, that by itself is difficult to pin down, differ only by 16~MeV (cf.
Tab.~\ref{tab:bra1}) while the bare masses are apart by 143~MeV  (cf. Tab.~\ref{tab:bare_cou_fit5} and \ref{tab:bare_cou_20Sep_2000}). Similar
considerations hold for the very prominent $\Delta(1950)$ 7/2$^+$ resonance (differences of 6~MeV in Re~$z_0$ vs. 98~MeV in $m^b$).

This illustrates the high uncertainties tied to bare parameters, cf. also the discussion at the end of Sec.~\ref{sec:renor}. In the following
section, we will further discuss this issue.

%%%%%%%%%%%%%%%%%%%%%%%%%%%%%%%%%%%%%%%%%%%%%%%%%%%%%%%%%%%%%%%%%%%%%%%%%%%%%%%%%%%%%%%%%%%%%%%%%%%%%%%%%%%%%%%%%%%%%%%%%%%%%%%%%%%%%%%%%%%%%%%%

\subsection{Residues vs. bare couplings}
\label{sec:delta1620}
\begin{figure}
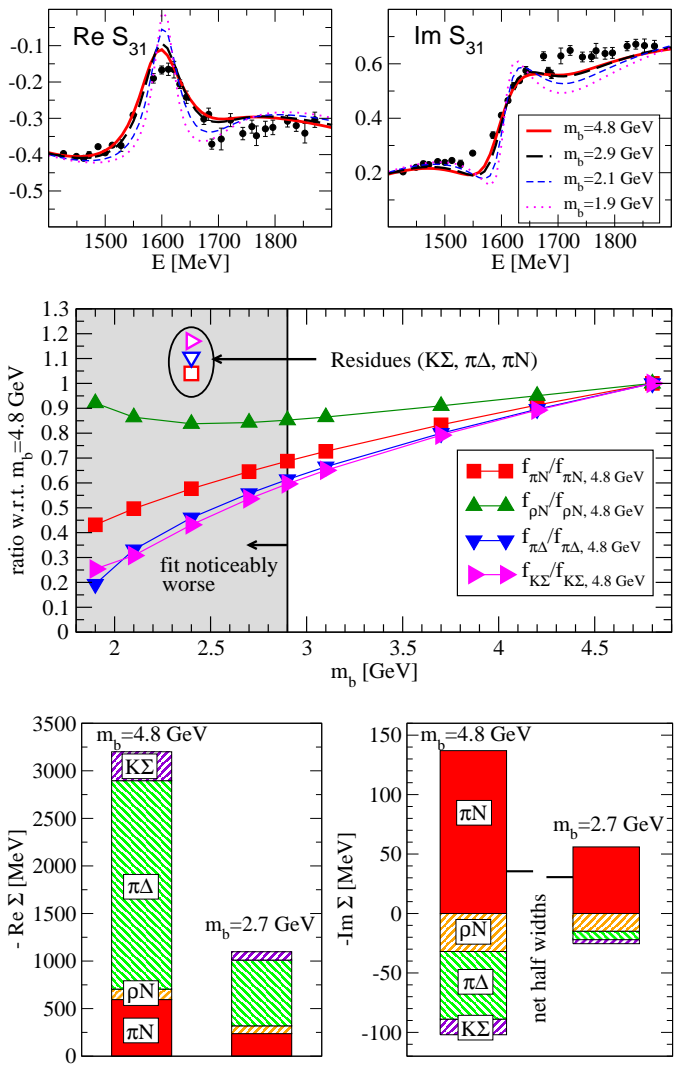

\begin{center}
\includegraphics[width=0.49\textwidth]{s31_vary_mbare.eps}\\[0.3cm]
\includegraphics[width=0.49\textwidth]{chi_square_as_mbare.eps}\\[0.3cm]
\includegraphics[width=0.49\textwidth]{fractions_selfenergy_D1620.eps}
\end{center}
\caption{Upper panel: Dependence of the $S_{31}$ amplitude on the bare mass of the $\Delta(1620)$ 1/2$^-$ (involves refits of the four bare
couplings $f_b$). The case $m^b=4.8$~GeV corresponds to fit A. Middle panel: change of the $f_b$, normalized to the case $m^b=4.8$~GeV (filled
symbols). For $m^b=2.4$~GeV, also the changes of $g$ from the residues [cf. Eq.~(\ref{couplings})] are shown (empty symbols). Lower panel:
contributions to the self energy $\Sigma (z=z_{0})$ from different channels.}
\label{fig:bare_d1620}     
\end{figure}

To point out the large correlations between bare resonance couplings and bare masses, we consider the case of the $\Delta(1620)$ 1/2$^-$
resonance, characterized in the present fits by very high bare masses. For the test, starting from fit A, we reduce the bare mass $m^b$ from its
actual value of 4.8~GeV (cf. Tab.~\ref{tab:bare_cou_fit5}) and then perform a refit of the bare resonance couplings, i.e $f_{\pi N}, \, f_{\rho
N},\, f_{\pi\Delta},\, f_{K\Sigma}$, to the full data base. For different bare masses, we show in the upper panel of Fig.~\ref{fig:bare_d1620}
the results for the amplitude. Reducing the bare mass, the quality of the refits is indistinguishable down to $m^b=2.9$~GeV (changes in the
 $\chi^2$ are less than 1\%). When the bare mass is further reduced, the amplitude shows noticeable deviations. In any case, the real part
of the pole position is shifted by $2$~MeV at $m^b=2.4$~GeV compared to the case $m^b=4.8$~GeV. The imaginary part is shifted by $-5$~MeV. This
means that changes in the bare mass can be almost fully compensated by changes in the bare couplings. 

In the middle panel of Fig.~\ref{fig:bare_d1620}, we show the relative changes of the bare couplings as $m^b$ is changed (filled symbols). Also,
for the case $m^b=2.4$~GeV, the relative changes of the residues are displayed (for this mass, the $\chi^2$ has changed by 2.5\%). While bare
couplings change by up to 60\%, the coupling constant $g$ in the $\pi N$ channel [i.e. the square root of the residue, cf.
Eq.~(\ref{couplings})] changes by 4\% (empty square). The coupling $g$ of the $\pi\Delta$ channel changes by 10\% (empty triangle down), and the
one of the $K\Sigma$ channel by 17\% (empty triangle right). The size of these changes is expected, given that the $\pi N$ amplitude is
precisely known while the $\pi\Delta$ channel has only indirect influence through the resonance width, and the $K\Sigma$ channel is still closed
at these energies (the resonance still contributes to the description of $K\Sigma$ data, and thus to the $\chi^2$, through its finite width). 

In the lower panel of Fig.~\ref{fig:bare_d1620} the contribution of the different channels to the self energy $\Sigma$ is shown according to
Eq.~(\ref{dressed}), i.e. every term in the sum over the channel index $\mu$. The self energy is evaluated at the pole position. As the figure
shows, the real parts of the self energies are very different for different bare masses, fully compensating changes in the bare mass to ensure
an almost unchanged pole position. For the imaginary part, shown to the right, one should note that different channels contribute with different
sign. As indicated with the horizontal black lines, the net half width $\Gamma_{\rm tot}/2=-\text{Im}\,z_0=-\text{Im}\,\Sigma (z=z_0)$ is rather
small and does not change significantly for different bare masses (by $5$~MeV) although the contributions from individual channels change
dramatically.

This numerical example again illustrates the problems tied to bare parameters. On top of this, there are also large correlations between bare
parameters and the cut-off in the form factors of the resonance vertices that we have not investigated for this example. It is, however, clear
that a change in the cut-off can be compensated by one in the bare coupling; thus, there is an additional scheme dependence of the bare coupling
constants (and, of course, also of the bare mass through the self-energy), that further prohibits to attach a physical meaning to these model
parameters.

%%%%%%%%%%%%%%%%%%%%%%%%%%%%%%%%%%%%%%%%%%%%%%%%%%%%%%%%%%%%%%%%%%%%%%%%%%%%%%%%%%%%%%%%%%%%%%%%%%%%%%%%%%%%%%%%%%%%%%%%%%%%%%%%%%%%%%%%%%%%%%%%

\subsection{Outlook: significance of resonance signals and error analysis}
\label{sec:outlook}
The presented fits A and B are the result of extensive searches in parameter space, performed under different starting conditions to ensure that
in the vicinity there are no better minima.  Nonetheless, rather than having several well-distinguished local minima of the $\chi^2$, one
observes many minima that are shallow in the sense that there are strong parameter correlations, for example between bare resonance masses and
coupling constants as discussed in Sec.~\ref{sec:delta1620}. In general, if the contribution to the imaginary part of the resonance self-energy
[cf. Eq.~(\ref{dressed})] is small, as in case for the $\rho N$ channel, a change in the bare mass can be absorbed well by a change of the bare
coupling. Quantifying such correlations is postponed to a later stage of the analysis, but is, in principle, feasible within the current
approach, as has been shown in Ref.~\cite{Doring:2010ap}. 

As for the statistical error (in contrast to the systematic error as estimated here by performing two different fits), in
Ref.~\cite{Doring:2010ap} it has also been demonstrated how the error of the experimental data translates into uncertainties of extracted
resonance pole positions. It should be noted, however, that a true statistical analysis will be only possible once one fits to the actual $\pi
N\to \pi N$ data and not, as was done here and in most other currently active analysis efforts, to the partial-wave solution of the GWU/SAID
group. We have fitted here to their energy-dependent solution as recommended by the GWU/SAID group, but it should be also noted that the error
bars provided in their energy-independent (SES) solution do not have statistical meaning, i.e. they cannot be used for the purpose of
statistical error analysis in a multi-channel approach of the present type. This is also one of the reasons why we refrain from quoting an
overall $\chi^2$. 

In addition to studying the propagation of the statistical error from the data to results such as the partial-wave decomposition and pole
properties, there are systematic uncertainties tied to the selection of experimental data in the fit and the weight of those data in the fitting
procedure. As discussed before, the very different quality of the data makes it unavoidable that some reactions and data sets are weighted
differently from others (apart from different weighting due to additional systematic uncertainties in experiments, as discussed in the context
of particular reactions in previous sections). In other words, the present fits are, unavoidably, the result of a selection process. 

Aside from fitting directly to $\pi N$ data, another possibility to improve the extraction of the resonance content is to include photon-induced
reactions. A big step in this direction has been recently achieved by fitting pion photoproduction data using the gauge-invariant extension of
the present framework~\cite{Huang:2011as}. With the inclusion of $KY$ channels in the present study and the good description found for
pion-induced $\eta N$, $K\Lambda$, and $K\Sigma$ production, it becomes now possible to analyze also the high-precision data on photoproduction
of these final states. Determining the photocouplings to excited baryonic states and their $Q^2$ dependence is also important for other
theoretical approaches that need analyses such as in Refs.~\cite{Chiang:2001as, Lee:1999kd, Drechsel:2007if, Tiator:2011pw} as input, namely
chiral unitary calculations~\cite{Doring:2005bx, Doring:2007rz, Mai:2012wy, Doring:2006ub, Jido:2007sm, Doring:2010rd} that are sometimes
compared to multipoles, in particular for pion photoproduction. A simultaneous fit to both pion- and photon-induced reactions will increase the
reliability of resonance extraction possible in the current approach due to its correct analytic properties.

%%%%%%%%%%%%%%%%%%%%%%%%%%%%%%%%%%%%%%%%%%%%%%%%%%%%%%%%%%%%%%%%%%%%%%%%%%%%%%%%%%%%%%%%%%%%%%%%%%%%%%%%%%%%%%%%%%%%%%%%%%%%%%%%%%%%%%%%%%%%%%%%

\section{Conclusions}
\label{sec:conclusions}

A combined analysis of the reactions $\pi N\to$ $\pi N$, $\eta N$, $K\Lambda$, and the three measured $K\Sigma$ final states $K^+\Sigma^+$,
$K^0\Sigma^0$, and $K^+\Sigma^-$ within a dynamical coupled-channel framework has been performed. For the inelastic reactions, the world data
set from threshold to energies around $2.3$ GeV has been considered.  In the Lagrangian-based calculation, the full off-shell solution of the
Lippmann-Schwinger-type equation provides the correct analytic structure allowing for a reliable continuation into the complex plane to extract
resonance pole positions and residues up to $J^P=9/2^\pm$. The amplitude features also the effective $\pi\pi N$ channels $\pi\Delta$, $\rho N$,
and $\sigma N$ with branch points in the complex plane and a dispersive treatment of $\sigma$ and $\rho$ $t$-channel exchanges.

The Lagrangian-based, field-theoretical framework is particularly suited to perform coupled-channel analyses: the SU(3) fla\-vor symmetry for
the exchange processes allows one to relate different final states, and the $t$- and $u$-channel diagrams, whose explicit inclusion is mandated
by  three-body unitarity arguments, connect also different partial waves and the respective backgrounds. As a result, for all reactions, a
realistic and structured background can be provided that helps map out resonance signals. 

Systematic uncertainties have been roughly estimated by performing two different fits with starting points far from each other in the parameter 
space. Although the parameterization of the amplitudes is quite different, pole positions and residues change much less.  In contrast to these
pole parameters, bare masses and couplings were shown to have no physical meaning.

All of the well-established four-star resonances within the considered energy range were found and their branching into the $\eta N$ and $KY$
channels determined. For resonances less prominently visible in $\pi N$ scattering, a $P_{11}$ state of around 110~MeV width was
found, required mostly by the $K\Lambda$ data and coupled-channel constraints. Indeed, the state might be identified with the $N(1710)$
1/2$^+$. Furthermore, from the $K\Sigma$ data there is need for large
strength in the $P_{33}$ partial wave around $z\sim 1.9$~GeV. A $\Delta(1600)$ 3/2$^+$ state found in a previous analysis could be confirmed.

Apart from determining the resonance spectrum, also scattering lengths and volumes have been calculated which are in fair agreement with
approaches using chiral perturbation theory;  the consequences of adding and removing a second $F_{15}$ state has been tested and the importance
to take the data of all three $K\Sigma$ final states into account  has been demonstrated. 

The present results, in combination with the recent extension to pion photoproduction~\cite{Huang:2011as}, constitute a major step towards the
analysis of high-precision photoproduction data of $\eta N$, $K\Lambda$, and $K\Sigma$ data produced, e.g., at ELSA, JLab, and MAMI, allowing
for a more fundamental understanding of the resonance spectrum.  

%%%%%%%%%%%%%%%%%%%%%%%%%%%%%%%%%%%%%%%%%%%%%%%%%%%%%%%%%%%%%%%%%%%%%%%%%%%%%%%%%%%%%%%%%%%%%%%%%%%%%%%%%%%%%%%%%%%%%%%%%%%%%%%%%%%%%%%%%%%%%%%%

\section*{Acknowledgement}

The numerical calculations were made possible through the grant jikp07 at the JUROPA supercomputer of the Forschungszentrum J\"ulich. This work
is also supported by the EU Integrated Infrastructure Initiative HadronPhysics3 (contract number 283286), by the DFG (Deutsche
Forschungsgemeinschaft, GZ: DO 1302/1-2 and SFB/TR 16, ``Subnuclear Structure of Matter''). F.H. is grateful to the COSY FFE grant No. 41788390
(COSY-058). D.R. thanks the German Academic Exchange Service (DAAD) for financial support within a ``DAAD-Doktorandenstipendium''.  We would
like to thank V. Nikonov, I. Strakovsky and R. Workman for discussions.

%%%%%%%%%%%%%%%%%%%%%%%%%%%%%%%%%%%%%%%%%%%%%%%%%%%%%%%%%%%%%%%%%%%%%%%%%%%%%%%%%%%%%%%%%%%%%%%%%%%%%%%%%%%%%%%%%%%%%%%%%%%%%%%%%%%%%%%%%%%%%%%%
%%%%%%%%%%%%%%%%%%%%%%%%%%%%%%%%%%%%%%%%%%%%%%%%%%%%%%%%%%%%%%%%%%%%%%%%%%%%%%%%%%%%%%%%%%%%%%%%%%%%%%%%%%%%%%%%%%%%%%%%%%%%%%%%%%%%%%%%%%%%%%%%

\appendix

%%%%%%%%%%%%%%%%%%%%%%%%%%%%%%%%%%%%%%%%%%%%%%%%%%%%%%%%%%%%%%%%%%%%%%%%%%%%%%%%%%%%%%%%%%%%%%%%%%%%%%%%%%%%%%%%%%%%%%%%%

\section{Bare resonance parameters}
\label{sec:respar}

In Tables~\ref{tab:bare_cou_fit5} and \ref{tab:bare_cou_20Sep_2000}, the bare resonance parameters are quoted for reproducibility of the fit
result. As explained in Sec.~\ref{sec:renor}, bare masses and coupling constants have no physical meaning, because they depend on the scheme
(i.e., values of the form factors), the number of included channels, and some bare resonance parameters tend to be strongly correlated (cf.
Sec.~\ref{sec:delta1620}). In fact, some bare masses were constrained by hand because large values were obtained in the fit, without noticeable
improvement of the $\chi^2$ compared to lower bare masses. Thus, we are aware that some masses and couplings in Tables~\ref{tab:bare_cou_fit5}
and \ref{tab:bare_cou_20Sep_2000} are rather large, but, considering the above arguments, we regard those high values as unproblematic. We
expect that once $\pi\pi N$ data are included in the fit, the bare masses and couplings tied to the $\pi\pi N$ channel can be fixed more
reliably (their values still being scheme-dependent), but at the current state of the approach it makes no sense to limit these values too
much. 

\begin{table*}
 \caption{Bare resonance parameters fit A: masses $m^b$ and coupling constants $f$. Note that the bare couplings differ sometimes by orders of
magnitude among each other, because the power of the form factors can be different to ensure convergence of the integrals. In particular for the
high partial waves, those powers can be large.}
 \renewcommand{\arraystretch}{1.3}
 \begin{center}
  \begin{tabular}[t]{lc D{.}{.}{5}  D{.}{.}{5}  D{.}{.}{8}  D{.}{.}{5}  D{.}{.}{8}  D{.}{.}{8} }
 \hline\hline
	     & $m_{b} $ [MeV]   &\multicolumn{1}{c}{ $f_{\pi N}$}   &\multicolumn{1}{c}{ $f_{\rho N}$}  &\multicolumn{1}{c}{$f_{\eta N}$}  &\multicolumn{1}{c}{ $f_{\pi\Delta}$}  &\multicolumn{1}{c}{$f_{\Lambda K}$} &\multicolumn{1}{c}{$f_{\Sigma K}$}\\ 
 \hline
 Nucleon	&			$1146  $	&  0.619		& -		& -		& -		& -    		& - 	\\  
 $N (1535)$ 1/2$^-$ &      	$ 2866 $	& 0.324 	& 1.312		& -0.726		& -6.271 		& -0.250  		& -0.081 \\
 $N (1650)$ 1/2$^-$& 	$ 1905 $	& -0.266	& -0.100		& 0.122	   	& -0.835 		& -0.186   	& 0.131\\
 $N (1710)$ 1/2$^+$& 	$2148  $	& 0.082 	& -0.840		& -0.580 	 	& 0.120	 	& 0.542 	 	&-0.122 \\
 $N (1720)$ 3/2$^+$&   	$1936 $	& 0.171 	& -2.190		& -0.084    	& 0.427 		& 0.284  	 	& 0.122 \\
 $N (1520)$ 3/2$^-$&		$2529 $	& 0.085 	& 4.100		& -0.084    	& 0.303		& 0.073  		& -6.74 \cdot 10^{-3} \\
 $N (1675)$ 5/2$^-$&		$3373 $	& 0.634 	& 16.442		& 0.753 		& -2.981 		& 0.103  	 	&-0.876 \\
 $N (1680)$ 5/2$^+$&		$1877 $	& 0.088 	& -0.204		& 4.29 \cdot 10^{-3} & 0.899	& 0.013  		& -8.70 \cdot 10^{-3} \\
 $N (1990)$ 7/2$^+$&		$4000 $	& 0.497 	& 5.722		& 0.202 	 	& -4.839 		& 2.406 		&	 1.169\\ 
 $N (2190)$ 7/2$^-$&   	$2496 $	& 0.0227 	& 1.902 		& 1.29 \cdot 10^{-4}&-0.368	 & 0.013		 & -8.34 \cdot 10^{-3} \\
 $N (2250)$ 9/2$^-$&		$2416 $	& -0.067	& 0.088 		&  -0.025		& 	1.062 	 & -0.120  	& 0.030 \\ 
 $N (2220)$ 9/2$^+$&		$2421 $	& 0.017	& 1.962 		& -1.53 \cdot 10^{-4}& 0.076 	& -4.60 \cdot 10^{-3} &-3.82\cdot 10^{-3}\\
 $\Delta (1620)$ 1/2$^-$&	$4800 $	& 0.690 	& -1.848 	& - 		& 9.566		& -   	& -1.078 \\
 $\Delta (1910)$ 1/2$^+$&	$3700 $	& -0.099 	& 1.000 		& - 		& -0.304 	& - 		& 0.248\\
 $\Delta (1232)$ 3/2$^+$&	$1506 $	& -1.433 	& 2.171 		& - 		& -0.198 	& - 		& 1.058 \\
 $\Delta (1920)$ 3/2$^+$&	$3253 $	& 0.690 	& -0.657 	& - 		& -1.101 	& -    	& -0.479 \\
 $\Delta (1700)$ 3/2$^-$&	$2160 $	& 0.060 	& 0.048		& -  		& -0.512 	& - 		& -0.039 \\
 $\Delta (1930)$ 5/2$^-$&	$2802 $	& -0.318 	& 2.523		& - 		& -1.134		& -  	& -1.246 \\
 $\Delta (1905)$ 5/2$^+$&	$2866 $	& 0.062	& -3.472 	& - 		& -1.665 	& - 		& -0.036 \\
 $\Delta (1950)$ 7/2$^+$&	$2059 $	& 0.577 	& -1.591 	& - 		& 1.508		& - 		& -0.535 \\
 $\Delta (2200)$ 7/2$^-$&	$2521 $	& -0.018 	& 1.384		& -    	& 0.667 	& -   	& 0.010 \\
 $\Delta (2400)$ 9/2$^-$&	$2381 $	& 0.139	& -0.062		& - 		& -0.587 	& -   	& 0.144 \\
 \hline\hline
 \end{tabular}
 \end{center}
 \label{tab:bare_cou_fit5}
 \end{table*}

\begin{table*}
 \caption{Bare resonance parameters fit B: masses $m^b$ and coupling constants $f$. }
 \renewcommand{\arraystretch}{1.3}
 \begin{center}
\begin{tabular}[t]{lc D{.}{.}{5}  D{.}{.}{5} D{.}{.}{8} D{.}{.}{5} D{.}{.}{8} D{.}{.}{8} }
 \hline\hline
	     &$m_{b}$ [MeV]&\multicolumn{1}{c}{ $f_{\pi N}$}&\multicolumn{1}{c}{ $f_{\rho N}$} &\multicolumn{1}{c}{$f_{\eta N}$}& \multicolumn{1}{c}{$f_{\pi\Delta}$}&\multicolumn{1}{c}{$f_{\Lambda K}$}&\multicolumn{1}{c}{$f_{\Sigma K}$} \\ 
 \hline
 Nucleon	&		   	$973 $	& 1.012		& -		& -		& -		& -		& -	\\  
 $N (1535)$ 1/2$^-$&     	$2739 $	& 0.295 	& 1.449 	& -0.691		& -5.708 	& -0.305  	& -0.229 \\
 $N (1650)$ 1/2$^-$& 	$1875 $	& -0.207	& 0.501	& 0.192    		& -0.183 	& -0.149   & 0.171 \\
 $N (1710)$ 1/2$^+$& 	$2005 $	& 0.118 	& 0.0479 	& -0.388  		& 0.095 	& 0.575 	& -0.057 \\
 $N (1720)$ 3/2$^+$&   	$2073 $	& 0.227 	& -1.986	& -0.177  		& 0.525	& 0.259   	& 0.116 \\
 $N (1520)$ 3/2$^-$&		$2248 $	& 0.075 	& 3.581 	& -0.064  		& 0.313	& 0.081  	& -0.015 \\
 $N (1675)$ 5/2$^-$&		$2630 $	& 0.477 	& 4.826 	& 0.865 		& -3.139 	& 0.526  	& -0.613\\
 $N (1680)$ 5/2$^+$&		$1919 $	& 0.083 	& 1.992 	& 0.018		& 0.924	& -0.020 	& -7.94\cdot 10^{-3} \\
 $N (1990)$ 7/2$^+$&		$2571 $	& 0.312 	& 8.065	& -0.100  		& 0.958 	& 0.604 	& 0.494 \\
 $N (2190)$ 7/2$^-$&   	$2592 $	& 0.027 	& 2.264 	& -1.61\cdot 10^{-3} & -0.332 & 1.77\cdot 10^{-3}	&-3.59\cdot 10^{-3} \\
 $N (2250)$ 9/2$^-$&		$2503 $		&  -0.070	& 0.845 	& -7.39\cdot 10^{-3}	& 1.327 & -0.130  	& -0.131 \\
 $N (2220)$ 9/2$^+$&		$2416 $		& 0.017 	& 1.359 	& -8.33\cdot 10^{-4}& -0.659 & -7.84\cdot 10^{-3}	&7.38\cdot 10^{-4}\\
 $\Delta (1620)$ 1/2$^-$&	$4800 $		& 0.803 	& -2.069 	& - 		& 8.574 	& -    	& -1.188 \\
 $\Delta (1910)$ 1/2$^+$&	$3700 $		& -0.419	& 1.000 	& - 		& -0.341 	& - 		& 0.103\\
 $\Delta (1232)$ 3/2$^+$&	$1510 $		& -1.304 	& 2.310 	& - 		& -0.491 	& - 		& 0.999 \\
 $\Delta (1920)$ 3/2$^+$&	$3400 $		& 0.378 	& -0.482 	& - 		& -1.255 	& -    	& -0.211 \\
 $\Delta (1700)$ 3/2$^-$&	$2028 $		& 0.042 	& 1.299 	& -  		& -0.412 	& - 		& -0.065 \\
 $\Delta (1930)$ 5/2$^-$&	$2376 $		& -0.501 	& 4.711	& - 		& 1.118 	& -  		& -0.533 \\
 $\Delta (1905)$ 5/2$^+$&	$2318 $		& 0.056	& 0.911 	& - 		& -1.361 	& - 		& -0.012 \\
 $\Delta (1950)$ 7/2$^+$&	$2157 $		& 0.503 	& 5.866 	& - 		& 1.520 	& - 		& -0.536 \\
 $\Delta (2200)$ 7/2$^-$&	$2545 $		& -0.016 	& 1.595 	& -    	& 0.588 	& -    	& 1.87\cdot 10^{-3} \\
 $\Delta (2400)$ 9/2$^-$&	$2412 $		& 0.157	& 0.036	& - 		& -0.616 	& -    	& 0.165 \\
 \hline\hline
 \end{tabular}
 \end{center}
 \label{tab:bare_cou_20Sep_2000}
 \end{table*}

The coupling constants $f$ of Table~\ref{tab:bare_cou_fit5} and \ref{tab:bare_cou_20Sep_2000} enter the vertices that couple the $N^*$'s and
$\Delta^*$'s to the different channels. The corresponding Lagrangians can be found in Table~8 of Ref.~\cite{Doring:2010ap}. The partial-wave
projected vertex functions, for vertices up to $J=3/2$, are given in Appendix~B of Ref.~\cite{Doring:2010ap}. In this study, we also need the
vertices of $J=9/2$ resonances. For this, we use the prescription analogous to the $J=5/2$ and $J=7/2$ vertices, given in Eq.~(B.3) of
Ref.~\cite{Doring:2010ap}; i.e. the $J=9/2$ vertex functions are obtained from the $J=3/2$ vertex functions according to
\be
(\gamma^{a,c})_{\frac{9}{2}^{-}}	 &=\frac{k^3}{m_b^3}\,(\gamma^{a,c})_{\frac{3}{2}^{+}}	   \non     
(\gamma^{a,c})_{\frac{9}{2}^{+}}	 &=\frac{k^3}{m_b^3}\,(\gamma^{a,c})_{\frac{3}{2}^{-}}		   
\label{higher1}
\ee
with the functions $(\gamma^{a,c})_{\frac{3}{2}^{\pm}}$ from Eq.~(B.2) of Ref.~\cite{Doring:2010ap}. These $\gamma^{a,c}$ contain the
partial-wave projected vertex functions $v$, isospin factors $I_R$ (cf. Table~10 of Ref.~\cite{Doring:2010ap}), and form factors $F$ with a
common  cut-off of $\Lambda=2$~GeV, except for the nucleon cut-off, which is a free parameter in our model and determined to be
$\Lambda_N=1610$~MeV in fit A and $\Lambda_N=958$~MeV in fit B. Equation~(\ref{higher1}) provides the correct dependence on the orbital angular
momentum $L$ for all channels ${\rm MB}=\, \pi N$, $\eta N$, $K\Lambda$, $K\Sigma$, $\rho N$, and $\pi\Delta$; see also
Table~\ref{tab:couplscheme}. In principle, it would also be possible to derive the vertices for $J\geq 5/2$ from Lagrangians, as has been done,
e.g., in Ref.~\cite{Shklyar:2009cx} for the $\pi N$ channel. 

Note that there are different possibilities for the $\pi\Delta$ and $\rho N$ channels to couple to a given $J^P$. For convenience, we show the
complete coupling scheme (applying to $s$-, $t$-, and $u$-channel exchanges) in Table~\ref{tab:couplscheme}. 
\begin{table*}
\caption{Angular momentum structure of the coupled channels in isospin $I=1/2$ up to $J=9/2$. The $I=3/2$ sector is similar up to obvious
isospin selection rules.}
\begin{center}
\renewcommand{\arraystretch}{1.30}
\begin {tabular}{ll|ll|ll|ll|ll|ll} \hline\hline
\bigstrut[t]
$\mu$	&\multicolumn{1}{r}{$J^P=$}
&$\frac{1}{2}^-$&$\frac{1}{2}^+$&$\frac{3}{2}^+$&$\frac{3}{2}^-$&$\frac{5}{2}^-$&$\frac{5}{2}^+$&$\frac{7}{2}^+$&$\frac{7}{2}^-$&$\frac{9}{2}^-$&$\frac{9}{2}^+$\bigstrut[b]\\
\hline
$1$	&$\pi N$ 			&$S_{11}$		&$P_{11}$		&$P_{13}$		&$D_{13}$		&$D_{15}$		&$F_{15}$		&$F_{17}$		&$G_{17}$		&$G_{19}$		&$H_{19}$		\bigstrut[t]\\
$2$	&$\rho N(S=1/2)$		&$S_{11}$		&$P_{11}$		&$P_{13}$		&$D_{13}$		&$D_{15}$		&$F_{15}$		&$F_{17}$		&$G_{17}$		&$G_{19}$		&$H_{19}$		\\
$3$	&$\rho N(S=3/2, |J-L|=1/2)$	&---		&$P_{11}$		&$P_{13}$		&$D_{13}$		&$D_{15}$		&$F_{15}$		&$F_{17}$
	&$G_{17}$		&$G_{19}$		&$H_{19}$		\\
$4$	&$\rho N(S=3/2, |J-L|=3/2)$	&$D_{11}$		&---		&$F_{13}$		&$S_{13}$		&$G_{15}$		&$P_{15}$		&$H_{17}$
	&$D_{17}$		&$I_{19}$		&$F_{19}$		\\
$5$	&$\eta N$ 			&$S_{11}$		&$P_{11}$		&$P_{13}$		&$D_{13}$		&$D_{15}$		&$F_{15}$		&$F_{17}$		&$G_{17}$		&$G_{19}$		&$H_{19}$		\\
$6$	&$\pi\Delta (|J-L|=1/2)$	&---		&$P_{11}$		&$P_{13}$		&$D_{13}$		&$D_{15}$		&$F_{15}$		&$F_{17}$
	&$G_{17}$		&$G_{19}$		&$H_{19}$		\\
$7$	&$\pi\Delta (|J-L|=3/2)$	&$D_{11}$		&---		&$F_{13}$		&$S_{13}$		&$G_{15}$		&$P_{15}$		&$H_{17}$
	&$D_{17}$		&$I_{19}$		&$F_{19}$		\\
$8$	&$\sigma N$			&$P_{11}$		&$S_{11}$		&$D_{13}$		&$P_{13}$		&$F_{15}$		&$D_{15}$		&$G_{17}$		&$F_{17}$		&$H_{19}$		&$G_{19}$		\\
$9$	&$K\Lambda$ 			&$S_{11}$		&$P_{11}$		&$P_{13}$		&$D_{13}$		&$D_{15}$		&$F_{15}$		&$F_{17}$		&$G_{17}$		&$G_{19}$		&$H_{19}$		\\
$10$	&$K\Sigma$ 			&$S_{11}$		&$P_{11}$		&$P_{13}$		&$D_{13}$		&$D_{15}$		&$F_{15}$		&$F_{17}$		&$G_{17}$		&$G_{19}$		&$H_{19}$		\bigstrut[b]\\
\hline\hline
\end {tabular}
\end{center}
\label{tab:couplscheme}
\end{table*}
There, the total spin $S=|\vec S_N+\vec S_\rho|$ is given by the sum of the $\rho$ spin and the nucleon spin, and $L$ is the orbital angular
momentum. The notation of the quantum numbers is the usual one, $L_{2I;2J}$, e.g. $S_{11}$, etc. Different coupling possibilities for the $\rho
N$ and $\pi\Delta$ channels are incorporated here as separate channels $\mu$. 

%%%%%%%%%%%%%%%%%%%%%%%%%%%%%%%%%%%%%%%%%%%%%%%%%%%%%%%%%%%%%%%%%%%%%%%%%%%%%%%%%%%%%%%%%%%%%%%%%%%%%%%%%%%%%%%%%%%%%%%%%%%%%%%%%%%%%%%%%%%%%%%%

\section{Exchange potentials}
\label{sec:ExApp}  

In Appendix~\ref{sec:explicit_exchanges}, we list explicit expressions ${\cal V}$ for all $t$- and $u$-channel exchanges used in this work and
shown in Figs.~\ref{fig:dia1} and \ref{fig:dia2}. The corresponding interaction terms of the Lagrangians can be found in
Table~\ref{tab:lagrangians}. Vertices that are not listed  in Table~\ref{tab:lagrangians} are related to the ones in the table by SU(3) flavor
symmetry. The explicit expressions $g_a,\,g_b$ for those couplings can be found in Sec.~\ref{sec:su3_couplings}. Note that for the $f_0$
exchange the same Lagrangians as for the exchange of a stable $\sigma$ were applied.

\begin{table*}
\caption{Interaction terms of the effective Lagrangians. We use $\vec{\rho}_{\mu\nu}=\delmu\vec{\rho}_{\nu}-\delnu\vec{\rho}_{\mu}$. The vector
notation refers to isospin space.}
\begin{center}
\renewcommand{\arraystretch}{1.30}
\begin{tabular}[t]{c|l|c|l}
\hline \hline
Vertex & $\mathcal{L}_{int}$ &Vertex & $\mathcal{L}_{int}$\\ \hline 
$NN\pi$& $-\frac{g_{NN\pi}}{\mpi}\Psi\gaf\gam\vtau \cdot \delpi\Psi$ &
$NN\omega$& $-g_{NN\omega}\bar{\Psi}[\gam -\frac{\kappa_{\omega}}{2m_{N}}\sigma^{\mu\nu}\partial_{\nu}] \omega_{\mu}\Psi  $ \\
$N\Delta\pi$ & $\frac{g_{N\Delta\pi}}{\mpi}\bar{\Delta}^{\mu}\vec{S}^{\dagger} \cdot \delpi\Psi\;+\; \text{h.c.}$ &
$\omega\pi\rho$ & $\frac{g_{\omega\pi\rho}}{m_{\omega}}\epsilon_{\alpha\beta\mu\nu}\partial^{\alpha}\vec{\rho}^{\beta} 
\cdot \partial^{\mu}\vec{\pi}\omega^{\nu}  $\\
$\rho\pi\pi$ & $-g_{\rho\pi\pi}(\vec{\pi}\times\delpi) \cdot \vec{\rho}^{\mu} $&
$N\Delta\rho$& $ -i \frac{g_{N\Delta\rho}}{m_{\rho}}\bar{\Delta}^{\mu}\gaf\gam\vec{S}^{\dagger} \cdot \vec{\rho}_{\mu\nu}\Psi\;+\; \text{h.c.}$\\
$NN\rho$& $-g_{NN\rho}\Psi[\gam-\frac{\kappa_{\rho}}{2m_{N}}\sigma^{\mu\nu} \partial_{\nu}]\vtau\cdot\vec{\rho}_{\mu}\Psi$&
$\rho\rho\rho$& $g_{NN\rho}(\vec{\rho}_{\mu}\times \vec{\rho}_{\nu}) \cdot \vec{\rho}^{\mu\nu}  $\\
$NN\sigma$ & $ -g_{NN\sigma}\bar{\Psi}\Psi\sigma$&
$NN\rho\rho$& $\frac{\kappa_{\rho}g_{NN\rho}^{2}}{2m_{N}}\bar{\Psi}\sigma^{\mu\nu}\vtau\Psi(\vec{\rho}_{\mu}\times\vec{\rho}_{\nu})  $\\
$\sigma\pi\pi$ & $\frac{g_{\sigma\pi\pi}}{2\mpi}\delpi \cdot \partial^{\mu}\vec{\pi}\sigma$&
$\Delta\Delta\pi$& $\frac{g_{\Delta\Delta\pi}}{\mpi}\bar{\Delta}_{\mu}\gaf\gan\vec{T}\Delta^{\mu}\partial_{\nu}\vec{\pi}  $\\
$\sigma\sigma\sigma$&$-g_{\sigma\sigma\sigma}m_{\sigma}\sigma\sigma\sigma$&
$\Delta\Delta\rho $& $-g_{\Delta\Delta\rho}\bar{\Delta}_{\tau}(\gam-i\frac{\kappa_{\Delta\Delta\rho}}{2m_{\Delta}}\sigma^{\mu\nu}\partial_{\nu})\vec{\rho}_{\mu} \cdot\vec{T}\Delta^{\tau}  $\\
$NN\rho\pi$& $\frac{g_{NN\pi}}{\mpi}2g_{NN\rho}\bar{\Psi}\gaf\gam\vtau\Psi(\vec{\rho}_{\mu}\times\vec{\pi})$&
$NN\eta$& $-\frac{g_{NN\eta}}{\mpi}\bar{\Psi}\gaf\gam\partial_{\mu}\eta\Psi$ \\
$NNa_{1}$&$-\frac{g_{NN\pi}}{\mpi}m_{a_{1}}\bar{\Psi}\gaf\gam\vtau\Psi\vec{a}_{\mu}$&
$NNa_{0} $& $g_{NNa_{0}}\mpi\bar{\Psi}\vtau\Psi \vec{a_{0}}  $\\
$a_{1}\pi\rho$ & $ -\frac{2g_{\pi a_1\rho}}{m_{a_{1}}} [\delpi\times \vec{a}_{\nu}-\partial_{\nu}\vec{\pi}\times \vec{a}_{\mu}] \cdot  [\partial^{\mu}\vec{\rho}^{\nu}-\partial^{\nu}\vec{\rho}^{\mu}]$ &
$\pi\eta a_{0}$& $g_{\pi\eta a_{0}}\mpi\eta\vec{\pi}\cdot\vec{a}_{0}  $ \\
    &$\;\;\;+\frac{2g_{\pi a_1\rho}}{2m_{a_{1}}} [\vec{\pi}\times(\partial_{\mu}\vec{\rho}_{\nu}-\partial_{\nu}\vec{\rho}_{\mu})] \cdot  [\partial^{\mu}\vec{a}^{\nu}-\partial^{\nu}\vec{a}^{\mu}]$&&\\
 \hline \hline
\end{tabular}
\label{tab:lagrangians}
\end{center}
\end{table*}

\begin{figure}
\begin{center}
\includegraphics[height=0.35\textwidth]{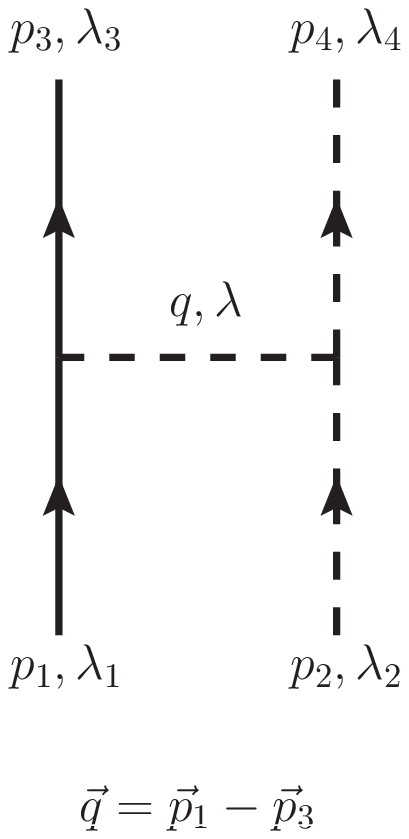} \hspace*{0.5cm}
\includegraphics[height=0.35\textwidth]{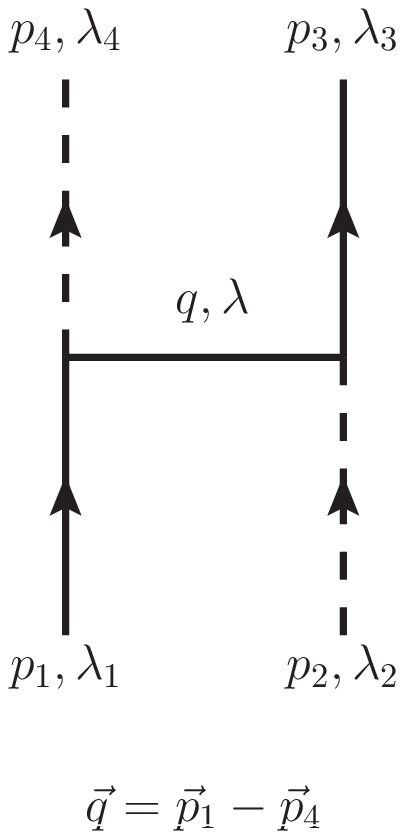}
\end{center}
\caption{$t$- and $u$-channel exchange processes.}
\label{fig:exchanges}
\end{figure}

In the following, we give further details on the notation used in in Appendix~\ref{sec:explicit_exchanges}. The labeling of particles and
helicities are specified in Fig.~\ref{fig:exchanges}. The index $1$ and $3$ ($2$ and $4$) denote the incoming and outgoing baryon (meson).  The
on-shell energies are \begin{eqnarray} E_{i}=\sqrt{\vec{p}_{i}^{\,2}+m_{i}^{2}},\quad \omega_{i}=\sqrt{\vec{p}_{i}^{\,2}+m_{i}^{2}}
\end{eqnarray} for the baryon and the meson, respectively, and $m_1$, $m_3$ ($m_2$, $m_4$) are the masses of the incoming, outgoing baryons
(mesons). In the TOPT framework used in this study, the zeroth component of the initial and final momenta are set to their on-mass-shell values:
$p^{0}_{i}=E_{i}$ or $p^{0}_{i}=\omega_{i}$.

The three-momentum of the intermediate particle is $\vec{q}$. The four-vector $q$ with  $q^{0}=E_{q}$ (baryon exchange) or $q^{0}=\omega_{q}$
(meson exchange) means the exchanged momentum in the first time ordering whereas $\tilde{q}$ indicates the second time ordering with
$\tilde{q}^{0}=-E_{q}$ (baryon exchange) or $\tilde{q}^{0}=-\omega_{q}$ (meson exchange).   For the potentials, we use the notation
$\slashed{p}\equiv \gamma^\mu\,p_\mu$.

Each exchange diagram includes a kinematic normalization factor
\begin{eqnarray}
 N=\frac{1}{(2\pi)^{3}}\frac{1}{2\sqrt{\omega_{2}\omega_{4}}} \ .
\end{eqnarray}
Isospin factors $({\rm IF})$ for the exchange processes (we use the baryon-first convention) can be found in Tables~\ref{tab:isospin2} and
\ref{tab:isospin2con}.  Also, every exchange process ${\cal V}$ quoted in Appendix~\ref{sec:explicit_exchanges} is multiplied with form factors
$F_1,\, F_2$ that depend on the exchanged momentum $\vec q$, quoted in Appendix~\ref{sec:fofatu} and Table~\ref{tab:cutoff_bg}.  The final
expression for the $t$- and $u$-exchanges is then given by
\be
V_{\lambda_1,\lambda_2,\lambda_3,\lambda_4}^{\vec p_1,\vec p_2,\vec p_3,\vec p_4}
=N\,F_1\,F_2 \,({\rm IF})\,{\cal V} \ .
\label{tu}
\ee
To obtain from this the transition potentials $V^\npo$ in Eq.~(\ref{dressed}) in the $JLS$ basis, the projection techniques described in
Refs.~\cite{Krehl:1999km, Gasparyan:2003fp} are used. 

With $m_{ex}$ we denote the mass of the exchange particle. $\epsilon^{\nu}(\vec{p}_i,\lambda_i)$ is the polarization vector of a massive spin 1
particle with momentum $p_i$ and helicity $\lambda_i$.

Furthermore, in the potentials quoted in Appendix~\ref{sec:explicit_exchanges}, the Rarita-Schwinger spinor for spin $3/2$ particles is given by
\begin{equation}
 u^{\mu}(\vec p,\lambda)=\sum_{\lambda_1,\lambda_2}\langle 1\lambda_1,\, \tfrac{1}{2}\lambda_2| \tfrac{3}{2}\lambda \rangle 
  \epsilon^{\mu}(\vec p, \lambda_1) u(\vec p,\lambda_2 )
\end{equation}
and $P^{\mu\nu}$ is the  Rarita-Schwinger tensor of spin $3/2$ particles~\cite{Krehl:1999km},
\begin{multline}
P^{\mu\nu}(q)=(\slashed{q}+m_{ex})\big[-g^{\mu\nu}+\frac{1}{3}\gamma^\mu\gamma^\nu\\
+\frac{2}{3\,m_{ex}^2}\,q^\mu q^\nu -\frac{1}{3m_{ex}}
\left(q^\mu\gamma^\nu-q^\nu\gamma^\mu\right)\big] \ .
\end{multline}
If $P^{\mu\nu}$ appears with the argument $q_{B^*}$ for the exchange of a $J^P=3/2^+$ baryon $B^*=\Delta(1232)$, $\Sigma^*(1385)$, $\Xi^*(1530)$
in Appendix~\ref{sec:explicit_exchanges}, we use the simplified prescription from Ref.~\cite{Schutz:1994wp} in order to avoid problems arising
from the construction of the contact graphs in TOPT~\cite{Schutz:1994wp}. In that case, the zeroth component of the exchanged momentum is
$q_{B^*}^{0}=E_1^{{\rm on}}-\omega_{4}^{{\rm on}}$ with
\begin{eqnarray}
 E_1^{{\rm on}}=\frac{z^{2}+m_{1}^{2}-m_{2}^{2}}{2\,z}\;
  ,\;\;\omega_{4}^{{\rm on}}=\frac{z^{2}-m_{3}^{2}+m_{4}^{2}}{2\,z} \ .
\end{eqnarray}

In case of vector mesons coupling to octet baryons, $f_{BBV}\,=\,g_{BBV}\,\kappa_{V}$ denotes the tensor coupling, $g_{BBV}$ is the vector
coupling (see Appendix~\ref{sec:su3_couplings});  $\epsilon_{\alpha\beta\mu\nu} $ is the totally antisymmetric Levi-Civita tensor with
$\epsilon_{\alpha\beta\mu\nu}=1$ for odd permutations of $\alpha,\,\beta,\,\mu,\,\nu$,  and 
$\sigma^{\mu\nu}=i(\gamma^\mu\gamma^\nu-\gamma^\nu\gamma^\mu)/2$.

Note that for the exchange of $\rho$ and $\sigma$ quantum numbers in the $\pi N\to\pi N$ transitions, we use correlated two-pion exchange. By 
contrast, for the exchange of these quantum numbers involving other channels, no pseudo-data exist to fix these transitions (cf.
Sec.~\ref{sec:formal}) and we resort to the exchange of stable particles with $m_\rho=770$~MeV and $m_\sigma=650$~MeV. For other $t$-channel
processes, the exchanged masses are $m_\omega=783$~MeV, $m_{f_0}=974$~MeV, $m_{a_0}=983$~MeV, $m_{a_1}=1260$~MeV, $m_{K^*}=892$~MeV, and
$m_{\phi}=1020$~MeV. The masses of exchanged baryons take their PDG values~\cite{Beringer:1900zz}; for the unstable $\Xi^*(1530)$, we use
$m_{\Xi^*}=1533$~MeV.

%%%%%%%%%%%%%%%%%%%%%%%%%%%%%%%%%%%%%%%%%%%%%%%%%%%%%%%%%%%%%%%%%%%%%%%%%%%%%%%%%%%%%%%%%%%%%%%%%%%%%%%%%%%%%%%%%%%%%%%%%%%%%%%%%%%%%%%%%%%%%%%%

\begin{widetext}

\subsection{Amplitudes for the exchange diagrams}
\label{sec:explicit_exchanges}
Here, we define the exchange pseudo-potentials ${\cal V}$. Common factors have been isolated according to Eq.~(\ref{tu}). The amplitude for the
correlated $\pi\pi$ exchange in the $\rho$ channel (cf. also Fig.~\ref{fig:dispersion}) can be found in Eq.~(A7) of Ref.~\cite{Krehl:1999km}.
For the correlated $\pi\pi$ exchange in the $\sigma$ channel, see Ref.~\cite{Gasparyan:2003fp}, Eq.~(A3). For this transition, we use the same
subtraction constant of $A_0=25$~MeV$/F_\pi^2$ as in Ref.~\cite{Gasparyan:2003fp}, as we see no need to change its value.

Below, a ``type'' number is introduced for a more compact notation. To obtain the amplitude for a specific diagram (cf. Figs.~\ref{fig:dia1} and
\ref{fig:dia2}), one uses Eq.~(\ref{tu}) with the form factors $F_1$, $F_2$ from Appendix~\ref{sec:fofatu}, $\cal V$ from this section according
to Tabs.~\ref{tab:isospin2} and \ref{tab:isospin2con}, and isospin factors from the same tables.

\begin{description}
 
 \item[Type 1:] $u$-exchange of an $J^P=1/2^+$ octet baryon

\begin{eqnarray}
{\cal V}= \frac{g_{a}g_{b}}{m_{\pi}^{2}}\, \bar{u}(\vec{p}_{3},\lambda_{3})\gaf
\frac{\slashed{p}_{2}}{2E_{q}} 
\left(\frac{\slashed{q}+m_{ex}}{z-E_{q}-\omega_{2}-\omega_{4}}
+\frac{\tilde{\slashed{q}}+m_{ex}}{z-E_{q}-E_{1}-E_{3}}\right)
\gaf \slashed{p}_{4}
u(\vec{p}_{1},\lambda_{1}) 
\end{eqnarray}
 
 \item[Type 2:] $u$-exchange of a decuplet baryon

\begin{eqnarray}
{\cal V}=\frac{g_{a}g_{b}}{m_{\pi}^{2}}\bar{u}(\vec{p}_{3},\lambda_{3})\,\frac{p_{2_{\mu}}
P^{\mu\nu}(q_{B^*})
}{2E_{q}}\left(\frac{1}{z-E_{q}-\omega_{2}-\omega_{4}}+\frac{1}{z-E_{q}-E_{1}-E_{3}}
\right)
 p_{4_{\nu}}\,u(\vec{p}_{1},\lambda_{1}) 
\end{eqnarray}

 \item[Type 3:] $t$-exchange of a scalar meson ($f_0$) 

\begin{eqnarray}
 {\cal V}= \frac{g_a g_b}{2m_{\pi}} (-2 p_{2\mu}p^{\mu}_4)\frac{1}{2\omega_q} 
 \left( \frac{1}{z-\omega_{q}-\omega_{2}-E_3}+\frac{1}{z-\omega_{q}-\omega_{4}-E_{1}} \right) \bar{u}(\vec{p}_{3},\lambda_{3})
  u(\vec{p}_{1},\lambda_{1})
\end{eqnarray}

 \item[Type 4:] $t$-exchange of a vector meson 

\begin{eqnarray}
{\cal V}=  g_{a}\, \bar{u}(\vec{p}_{3},\lambda_{3}) 
\left(\frac{g_{b}\gamma^{\mu}-i\frac{f_{b}}{2m_{N}}\sigma^{\mu\nu}q_{\nu}}{z-\omega_{q}-E_{3}-\omega_{2}} 
+ \frac{g_{b}\gamma^{\mu}-i\frac{f_{b}}{2m_{N}}
\sigma^{\mu\nu}\tilde{q}_{\nu}}{z-\omega_{q}-E_{1}-\omega_{4}} \right) 
\, u(\vec{p}_{1},\lambda_{1})
\frac{(p_{2}+p_{4})_{\mu}}{2\omega_{q}} 
\end{eqnarray}

 \item[Type 5:] Contact interaction in $\pi N\to \rho N$

\begin{equation}
{\cal V}=\;-2\frac{g_{a}}{m_{\pi}} g_{b}
\bar u (\vec{p}_3,\lambda_3)
\gamma^5 \slashed{\epsilon}^*(\vec{p}_4,\lambda_4)
u(\vec{p}_1,\lambda_1) 
\end{equation}

  \item[Type 6:] $t$-exchange of a scalar meson ($a_0$) (scalar coupling)

\begin{eqnarray}
 {\cal V}=\, g_a g_b\, m_{\pi}\, \bar{u}(\vec{p}_{3},\lambda_{3}) u(\vec{p}_{1},\lambda_{1}) 
 \frac{1}{2\omega_q} \left(\frac{1}{z-\omega_{q}-\omega_{2}-E_3}+\frac{1}{z-\omega_{q}-\omega_{4}-E_{1}}\right) 
\end{eqnarray}

  \item[Type 7:] $N$ $u$-exchange in $\pi N\to \rho N$

\begin{eqnarray}
{\cal V}=-i g_{a}\frac{g_{b}}{m_{\pi}} 
\bar u (\vec{p}_3,\lambda_3)
\gamma_5  \slashed{p}_2 &&
\left(\frac{\slashed{q}+m_N}{z-E_q-\omega_2-\omega_4}+
\frac{\tilde{\slashed{q}}+m_N}{z-E_q-E_1-E_3}\right) \\ \nonumber
&\times& \frac{1}{2E_q} \big[\slashed{\epsilon}^*(\vec{p}_4,\lambda_4)
-i\frac{\kappa_{\rho}}{2m_N}\sigma^{\mu\nu}p_{4_\nu}\epsilon^*_{\mu}(\vec{p}_4,\lambda_4)\big]
u(\vec{p}_1,\lambda_1)
\end{eqnarray}

  \item[Type 8:] $\pi$ $t$-exchange in $\pi N\to \rho N$

\begin{eqnarray}
{\cal V}=- \frac{g_a}{m_{\pi}} g_{b}
\bar u (\vec{p}_3,\lambda_3) \gamma^5
\left(\frac{\slashed q(p_2-q)_{\nu}}{2\omega_q(z-\omega_q-E_3-\omega_2)}\right.\left.
+\frac{\tilde{\slashed{q}}(p_2-\tilde{q})_{\nu}}{2\omega_q(z-\omega_q-E_1-\omega_4)}-\gamma^0\delta^0_{\nu}\right)
\epsilon^{*,\nu}(\vec{p}_4,\lambda_4) u(\vec{p}_1,\lambda_1) 
\end{eqnarray}

 \item[Type 9:] $t$-exchange of a pseudovector meson 

\begin{eqnarray}
 {\cal V}= 4  \frac{g_a}{m_{\pi}} g_{b} \bar u(\vec{p}_3,\lambda_3)\gamma^5\gamma_{\mu}&& 
\bigg(  \frac{-g^{\mu\nu} +\frac{q^{\mu}q^{\nu}}{m_{a1}^2 }}{2\omega_q(z-\omega_q-E_3-\omega_2)}  
\Big[(p_2+\frac{q}{2})_{\tau} p^{\tau}_4\epsilon^*_{\nu}(\vec{p}_4,\lambda_4) -
(p_2+\frac{q}{2})^{\tau}\epsilon^*_{\tau}(\vec{p}_4,\lambda_4)p_{4_{\nu}}\ \Big] \nonumber\\ 
&+& \Big[(p_2+\frac{\tilde{q}}{2})_{\tau} p^{\tau}_4\epsilon^*_{\nu}(\vec{p}_4,\lambda_4) -
(p_2+\frac{\tilde{q}}{2})^{\tau}\epsilon^*_{\tau}(\vec{p}_4,\lambda_4)p_{4_{\nu}}\ \Big] 
\frac{-g^{\mu\nu} +\frac{\tilde{q}^{\mu}\tilde{q}^{\nu}}{m_{a1}^2 }}{2\omega_q(z-\omega_q-E_1-\omega_4)}
\nonumber \\ &+& \frac{1}{m^2_{a1}}\delta^0_{\mu} \Big[ p_{2\nu}p_{4}^{\nu}\epsilon^*_{0}(\vec{p}_4,\lambda_4) 
- p_{2}^{\nu}\epsilon^*_{\nu}(\vec{p}_4,\lambda_4)p_{4_0}
\Big]\;\bigg)\; u (\vec{p}_1,\lambda_1)
\end{eqnarray}

 \item[Type 10:] $\omega$ $t$-exchange in $\pi N\to \rho N$ 

\begin{eqnarray}
{\cal V}= \frac{g_{a}}{m_{\omega}}\,
 \bar{u}(\vec{p}_{3},\lambda_{3})\Bigl(\frac{g_{b}\gamma^{\nu}-i\frac{f_{b}}{2m_{N}}\sigma^{\nu\tau}q_{\tau}}{z-\omega_{q}-E_{3}-\omega_{2}} 
 + \frac{g_{b}\gamma^{\nu}-i\frac{f_{b}}{2m_{N}}\sigma^{\nu\tau}\tilde{q}_{\tau}}{z-\omega_{q}-E_{1}-\omega_{4}} \Bigr) 
\, u(\vec{p}_{1},\lambda_{1})\, \epsilon_{\alpha\beta\mu\nu}\, p^{\alpha}_4 \, \epsilon^{*\beta}(\vec{p}_4, \lambda_4)\, p_2^{\mu}
\frac{1}{2\omega_{q}} 
\end{eqnarray}

\item[Type 11:] $N$ $u$-exchange in $\pi N\to \pi\Delta$ 

\begin{eqnarray}
{\cal V}= - \frac{g_{a}g_{b}}{m_{\pi}^{2}}\, \bar{u}^{\mu}(\vec{p}_{3},\lambda_{3}) \frac{p_{2\mu}}{2E_{q}} 
\left(\frac{\slashed{q}+m_{ex}}{z-E_{q}-\omega_{2}-\omega_{4}} +\frac{\tilde{\slashed{q}}+m_{ex}}{z-E_{q}-E_{1}-E_{3}}\right) 
\gaf \slashed{p}_{4} u(\vec{p}_{1},\lambda_{1}) 
\end{eqnarray}

\item[Type 12:] $\rho$ $t$-exchange in $\pi N\to \pi\Delta$

\begin{eqnarray}
 {\cal V}= -i \frac{g_a g_b}{m_{\rho}} \bar{u}^{\mu}(\vec{p}_{3},\lambda_{3}) \gamma^5 (q_{\mu}\gamma_{\nu}-\slashed{q}g_{\mu\nu})
  u(\vec{p}_{1},\lambda_{1}) \frac{1}{2\omega_q} \left(
\frac{1}{z-\omega_q -\omega_2- E_3} + \frac{1}{z-\omega_q- \omega_4-E_1} \right) (p_2+p_4)^{\nu}
\end{eqnarray}

\item[Type 13:] $\Delta$ $u$-exchange in $\pi N\to \pi\Delta$ 

\begin{eqnarray}
{\cal V}=\frac{g_{a}g_{b}}{m_{\pi}^{2}}\bar{u}_{\mu}(\vec{p}_{3},\lambda_{3})\,\gamma^5 \slashed{p}_{2}\frac{
P^{\mu\nu}(q_{B^*})
}{2E_{q}}\left(\frac{1}{z-E_{q}-\omega_{2}-\omega_{4}}+\frac{1}{z-E_{q}-E_{1}-E_{3}}
\right)
 p_{4\nu}\,u(\vec{p}_{1},\lambda_{1}) 
\end{eqnarray}

\item[Type 14:] $N$ $u$-exchange in $\pi N\to \sigma N$ 

\begin{eqnarray}
{\cal V}= -i \frac{g_{a}g_{b}}{m_{\pi}}\, \bar{u}(\vec{p}_{3},\lambda_{3}) \gamma^5 \frac{\slashed{p}_2}{2E_{q}} 
\left(\frac{\slashed{q}+m_{ex}}{z-E_{q}-\omega_{2}-\omega_{4}} +\frac{\tilde{\slashed{q}}+m_{ex}}{z-E_{q}-E_{1}-E_{3}}\right)
 u(\vec{p}_{1},\lambda_{1}) 
\end{eqnarray}

\item[Type 15:] $\pi$ $t$-exchange in $\pi N\to\sigma N$ 

\begin{eqnarray}
 {\cal V}= i \frac{g_a g_b}{m^2_{\pi}} \bar{u}(\vec{p}_{3},\lambda_{3}) \frac{1}{2\omega_q} 
\left( \frac{\gamma^5 \slashed{q}q_{\mu}}{z-\omega_{q}-\omega_{2}-E_3}\,p_2^{\mu} + \frac{\gamma^5
 \tilde{\slashed{q}}\tilde{q}_{\mu}}{z-\omega_{q}-\omega_{4}-E_1}\,p_2^{\mu} + 2\omega_q
\gamma^5\gamma^0p^0_2\right) u(\vec{p}_{1},\lambda_{1}) 
\end{eqnarray}

\item[Type 16:] $\Delta$ $u$-exchange in $\pi N\to \rho N$

\begin{eqnarray}
 {\cal V}&=&i \frac{g_a g_b}{m_{\pi}m_{\rho}} \bar{u}(\vec{p}_{3},\lambda_{3})p_{2\mu} 
 \left( \frac{P^{\mu\nu}(q)}{2E_q(z-E_q-\omega_2-\omega_4)}+
\frac{P^{\mu\nu}(\tilde q)}{2E_q(z-E_q-E_1-E_3)}  \right)\nonumber \\ &&\times  \gamma^5
\big(\slashed\epsilon^*(\vec p_4, \lambda_4)p_{4\nu}-\slashed p_4 \epsilon_{\nu}^*(\vec p_4,
\lambda_4) \big)
u(\vec{p}_{1},\lambda_{1})
\end{eqnarray}

\item[Type 17:] $N$ $u$-exchange in $\rho N\to \rho N$ 

\begin{eqnarray}
 {\cal V}&=& g_a \bar{u}(\vec{p}_{3},\lambda_{3}) [ g_b \gamma^{\mu} + i \frac{f_b}{2m_N} \sigma^{\mu\nu}
  p_{2\nu}]\epsilon_{\mu}(\vec{p}_2,\lambda_2) \frac{1}{2E_q} \left(
\frac{\slashed{q}+m_{ex}}{z-E_{q}-\omega_{2}-\omega_{4}} +\frac{\tilde{\slashed{q}}+m_{ex}}{z-E_{q}-E_{1}-E_{3}} \right)\nonumber \\ 
&&\times \; [\gamma^{\tau} - i \frac{\kappa_{\rho}}{2m_N}
\sigma^{\tau\nu} p_{4\nu}] \epsilon^*_{\tau}(\vec{p}_4,\lambda_4) u(\vec{p}_{1},\lambda_{1})
\end{eqnarray}

 \item[Type 18:] Contact interaction in $\rho N\to \rho N$ 

\begin{equation}
{\cal V}=\,\frac{g_{a}}{m_{N}}f_{b} \bar u (\vec{p}_3,\lambda_3) \sigma^{\mu\nu} \epsilon_{\mu}(\vec{p}_2,\lambda_2)
 \epsilon^*_{\nu}(\vec{p}_4,\lambda_4) u(\vec{p}_1,\lambda_1) 
\end{equation}  

\item[Type 19:] $\rho$ $t$-exchange in $\rho N\to \rho N$ 

\begin{eqnarray}
{\cal V}&=& \; -i g_a   \;\; \Bigg( \bar u (\vec{p}_3,\lambda_3) \;\; \frac{[g_b \gamma^{\mu}- i \frac{f_b}{2m_N}\sigma^{\mu\nu}q_{\nu}] }{2\omega_q(z-\omega_{q}-E_3-\omega_2)} 
\non &\times& \Big[\epsilon^{\tau}(\vec{p}_2,\lambda_2) \epsilon^*_{\tau}(\vec{p}_4,\lambda_4)(-p_4-p_2)_{\mu} + (q+p_4)^{\tau} \epsilon_{\tau}(\vec{p}_2,\lambda_2)
 \epsilon^*_{\mu}(\vec{p}_4,\lambda_4) +  (p_2-q)^{\tau}\epsilon^*_{\tau}(\vec{p}_4,\lambda_4) \epsilon_{\mu}(\vec{p}_2,\lambda_2) \Big] \non
 &+& \Big[ \epsilon^{\tau}(\vec{p}_2,\lambda_2) \epsilon^*_{\tau}(\vec{p}_4,\lambda_4)(-p_4-p_2)_{\mu} + (\tilde q+p_4)^{\tau} \epsilon_{\tau}(\vec{p}_2,\lambda_2)
 \epsilon^*_{\mu}(\vec{p}_4,\lambda_4) +  (p_2-\tilde q)^{\tau}\epsilon^*_{\tau}(\vec{p}_4,\lambda_4) \epsilon_{\mu}(\vec{p}_2,\lambda_2) \Big] \non
 &\times& \bar u (\vec{p}_3,\lambda_3) \frac{[g_b \gamma^{\mu}- i \frac{f_b}{2m_N}\sigma^{\mu\nu}\tilde q_{\nu}] }{2\omega_q(z-\omega_{q}-E_1-\omega_4)} \Bigg) 
 u(\vec{p}_1,\lambda_1)
\end{eqnarray}

\item[Type 20:] $\Delta$ $u$-exchange in $\rho N\to \rho N$ 

\begin{eqnarray}
 {\cal V}&=& \frac{g_a g_b}{m_{\rho}^2} \bar u (\vec{p}_3,\lambda_3) \gamma^5 \big(\gamma_{\kappa}p_{4\mu}-\slashed p_4 g_{\mu\kappa}\big)
\epsilon^{\kappa*}(\vec{p}_4,\lambda_4 ) \left( \frac{P^{\mu\nu}(q)}{2E_q(z-E_q-\omega_2-\omega_4)}+ \frac{P^{\mu\nu}
(\tilde q)}{2E_q(z-E_q-E_1-E_3)}   \right)\nonumber \\
&&\times \gamma^5 \big(\gamma_{\tau}p_{2\nu}-\slashed p_2 g_{\nu\tau}) \epsilon^{\tau}(\vec{p}_2,\lambda_2 \big) u(\vec{p}_1,\lambda_1)
\end{eqnarray}

\item[Type 21:] $N$ $u$-exchange in $\pi\Delta\to \pi\Delta$ 

\begin{eqnarray}
{\cal V}= \frac{g_{a}g_{b}}{m_{\pi}^2}\, \bar{u}^{\mu}(\vec{p}_{3},\lambda_{3}) \frac{p_{2\mu}}{2E_{q}} 
\left(\frac{\slashed{q}+m_{ex}}{z-E_{q}-\omega_{2}-\omega_{4}} +\frac{\tilde{\slashed{q}}+m_{ex}}{z-E_{q}-E_{1}-E_{3}}\right)
 p_{4\nu}u^{\nu}(\vec{p}_{1},\lambda_{1}) 
\end{eqnarray}

\item[Type 22:] $\rho$ $t$-exchange in $\pi\Delta\to \pi\Delta$ 

\begin{eqnarray}
{\cal V}= i\,g_{a}\, \bar{u}^{\tau}(\vec{p}_{3},\lambda_{3}) 
\left(\frac{g_{b}\gamma_{\mu}-i\frac{f_{b}}{2m_{\Delta}}\sigma_{\mu\nu}q^{\nu}}{z-\omega_{q}-E_{3}-\omega_{2}} 
+ \frac{g_{b}\gamma_{\mu}-i\frac{f_{b}}{2m_{\Delta}}
\sigma_{\mu\nu}\tilde{q}^{\nu}}{z-\omega_{q}-E_{1}-\omega_{4}} \right) 
\, u_{\tau}(\vec{p}_{1},\lambda_{1})
\frac{(p_{2}+p_{4})^{\mu}}{2\omega_{q}} 
\end{eqnarray}

\item[Type 23:] $\Delta$ $u$-exchange in $\pi\Delta\to \pi\Delta$

\begin{eqnarray}
 {\cal V}=\frac{g_{a}g_{b}}{m_{\pi}^{2}}\bar{u}_{\mu}(\vec{p}_{3},\lambda_{3})\,\gamma^5 \slashed{p}_{2}\frac{
P^{\mu\nu}(q_{B^*})
}{2E_{q}}\left(\frac{1}{z-E_{q}-\omega_{2}-\omega_{4}}+\frac{1}{z-E_{q}-E_{1}-E_{3}}
\right)
 \gamma^5 \slashed{p}_4\,u_{\nu}(\vec{p}_{1},\lambda_{1}) 
\end{eqnarray}

\item[Type 24:] $\pi$ $t$-exchange in $\rho N\to \pi\Delta$ 

\begin{eqnarray}
 {\cal V}&=& \frac{g_a g_b}{m_{\pi}} \bar{u}_{\mu}(\vec{p}_{3},\lambda_{3}) \frac{1}{2\omega_q} \nonumber \\&&\times 
 \left( \frac{q^{\mu}\,(p_4-q)_{\nu}}{z-\omega_q -E_3-\omega_2}  \epsilon^{\nu}(\vec{p}_{1},\lambda_{1}) 
 + \frac{\tilde q^{\mu}\,(p_4-\tilde q)_{\nu}}{z-\omega_q -E_1-\omega_4}  
 \epsilon^{\nu}(\vec{p}_{2},\lambda_{2}) -\delta^{\mu}_{\;0}\epsilon^0(\vec{p}_{2},\lambda_{2})\,2\omega_q \right) 
u(\vec{p}_{1},\lambda_{1})
\end{eqnarray}

\item[Type 25:] $N$ $u$-exchange in $\rho N\to \pi\Delta$ 

\begin{eqnarray}
 {\cal V}&=& -i\frac{g_a g_b}{m_{\pi}m_{\rho}} \bar{u}^{\mu}(\vec{p}_{3},\lambda_{3})\, 
 \Big[p_{2\mu}\gamma^5 \slashed \epsilon
(-\vec{p}_{2},\lambda_{2})-\epsilon_{\mu}(\vec{p}_{2},\lambda_{2})\gamma^5\slashed p_2\Big]\, \frac{1}{2E_q}
\left(\frac{\slashed q+m_N}{z-E_q-\omega_2-\omega_4}+ \frac{\tilde{\slashed
q}+m_N}{z-E_q-E_1-E_3} \right)\nonumber\\&&\times \,\gamma^5 \slashed p_2 \,u(\vec{p}_{1},\lambda_{1})
\end{eqnarray}

\item[Type 26:]  $\pi$ $t$-exchange in $\sigma N\to \pi\Delta$ 

\begin{eqnarray}
 {\cal V}=i\, \frac{g_a g_b}{m_{\pi}^2}\bar{u}^{\mu}(\vec{p}_{3},\lambda_{3}) \frac{1}{2\omega_q}\left( \frac{q_{\mu}}{z-\omega_q 
 -\omega_2-E_3}q_{\nu}p^{\nu}_4+\frac{\tilde q_{\mu}}{z-\omega_q
-\omega_4-E_1}\tilde q_{\nu}p^{\nu}_4 + \delta^0_{\mu}p_{4_0} 2\omega_q\right) u(\vec{p}_{1},\lambda_{1})
\end{eqnarray}

\item[Type 27:] $N$ $u$-exchange in $\sigma N\to\sigma N$

\begin{eqnarray}
 {\cal V}= g_a g_b\, \bar{u}(\vec{p}_{3},\lambda_{3})\,\frac{1}{2E_{q}} 
\left(\frac{\slashed{q}+m_{ex}}{z-E_{q}-\omega_{2}-\omega_{4}} 
+\frac{\tilde{\slashed{q}}+m_{ex}}{z-E_{q}-E_{1}-E_{3}}\right)u(\vec{p}_{1},\lambda_{1}) 
\end{eqnarray}

\item[Type 28:] $\sigma$ $t$-exchange in $\sigma N\to\sigma N$

\begin{eqnarray}
 {\cal V}= 6\;g_a g_b\, m_{\sigma} \bar{u}(\vec{p}_{3},\lambda_{3})\,u(\vec{p}_{1},\lambda_{1}) \left( \frac{1}{z-\omega_{q}-E_{3}-\omega_{2}}+ 
 \frac{1}{z-\omega_{q}-E_{1}-\omega_{4}} \right)
\end{eqnarray}

\end{description}
\end{widetext}

\begin{table}
\caption{Couplings for the calculation of the $t$-/$u$-channel and contact ({\it ct}) diagrams used in this study. The first two columns specify
the diagram with the exchanged particle ({\it Ex}), cf. Figs.~\ref{fig:dia1} and \ref{fig:dia2}. The third column specifies the diagram type
according to the listing in Sec.~\ref{sec:explicit_exchanges}, the columns 4-6 show which coupling constants to be used in these expressions.
The values of these couplings are quoted in Sec.~\ref{sec:su3_couplings} and Table~\ref{tab:coupl}. The final expressions for the $t$- and
$u$-channel diagrams, given by Eq.~(\ref{tu}), contain also the isospin factors (IF), shown here in columns 7 and 8. }
\begin{center} 
\renewcommand{\arraystretch}{1.44}
\begin{tabular}[t]{llllllrr}
\hline\hline
  Transition 			& Ex 		& Type	& $g_a$			&$g_b$	 		&$f_b$ 			&IF$(\frac{1}{2})$ 	& IF$(\frac{3}{2})$  \bigstrut[b]\\
\hline
$\pi N\rightarrow \pi N$  	& $N$		& 1 	&$g_{\pi NN}$  		&$g_{\pi NN}$		& 			&$-1$ 			&$2$ \bigstrut[t]\\ 
				& $\sigma$	&\multicolumn{4}{l}{correlated $\pi\pi$}				      	&$1$		      &$1$ \\ 
				& $\rho$  	&\multicolumn{4}{l}{correlated $\pi\pi$}				      	&$2i$		      &$-i$ \\ 
				& $\Delta$  	& 2	&$g_{\pi \Delta N}$	& $g_{\pi \Delta N}$	& 			&$\frac{4}{3}$ 		&$\frac{1}{3}$ \\
$\pi N\rightarrow \rho N$ 	& $N$  		& 7	&$g_{\pi NN}$ 		&$g_{NN\rho}$ 		& 			&$-1$ 			&$2$ \\
				& ct		& 5	&$g_{\pi NN}$		&$g_{NN\rho}$ 		& 			&$-2i$ 			&$i$  \\
				& $\pi$		& 8	&$g_{NN\pi}$		&$ g_{\rho\pi\pi}$ 	&  			&$-2i$ 			&$i$ \\
				& $\omega$	&10 	&$g_{\pi\omega\rho}$ 	&$g_{NN\omega}$ 	&$f_{NN\omega}$		&1 			& 1  \\
				& $a_1$		& 9 	&$g_{NNa_1}$ 		&$g_{\pi\rho a_1}$ 	& 			&$-2i$ 			&$i$  \\
				& $\Delta$	&16 	&$g_{N\Delta\pi}$ 	&$g_{N\Delta\rho}$ 	& 			&$\frac{4}{3}$ 		&$\frac{1}{3}$ \\
$\pi N \to \eta N$		& $N$		& 1 	&$g_{\pi NN}$ 		&$g_{NN\eta}$ 		& 			&$\sqrt{3} $ 		& 0 \\
				& $a_0$		& 6 	&$g_{\pi a_0\eta}$ 	&$g_{Na_0N}$ 		& 			&$\sqrt{3}$ 		& 0 \\
$\pi N\to \pi\Delta$		& $N$		&11 	&$g_{\pi NN}$ 		&$g_{N\Delta\pi}$ 	&			&$-\sqrt{\frac{8}{3}}$  &$\sqrt{\frac{5}{3}}$  \\
				& $\rho$	&12 	&$g_{\pi\pi\rho}$ 	&$g_{\Delta N\rho}$ 	& 			& $\sqrt{\frac{2}{3}} $ & $\sqrt{\frac{5}{3}} $  \\
				& $\Delta$	&13 	&$g_{\Delta N\pi}$ 	&$g_{\Delta\Delta\pi}$ 	& 			&$-\frac{5}{3}\sqrt{\frac{2}{3}} $ &$-\frac{10}{3\sqrt{15}} $  \\
$\pi N\to\sigma N$		& $N$		&14 	&$g_{NN\sigma}$  	&$g_{\pi NN}$ 		& 			&$\sqrt{3}$ 		&0  \\
				& $\pi$		&15	&$g_{\pi\pi\sigma}$ 	&$g_{\pi NN}$		& 			&$\sqrt{3}$ 		& 0 \\
$\pi N\rightarrow \Lambda K$	& $\kst$ 	& 4 	&$g_{\pi K^*K}$		&$g_{NK^*\Lambda}$	&$f_{NK^*\Lambda}$	&$\sqrt{3}$		&$0$\\
				& $\Sigma$ 	& 1 	&$g_{\pi\Sigma\Lambda}$	&$g_{N\Sigma K}$	&			&$\sqrt{3}$		&$0$\\
				& $\Sigma^{*}$ 	& 2 	&$g_{\pi\Sigma^*\Lambda}$&$g_{N\Sigma^* K}$	&			&$\sqrt{3}$		&$0$\\ 
$\Lambda K\rightarrow \Lambda K$& $f_0$ 	& 3 	&$g_{KKf_0}$ 		&$g_{\Lambda\Lambda f_0}$& 			&$1$ 			&0\\
				& $\omega$ 	& 4	&$g_{K\omega K}$	&$g_{\Lambda\omega\Lambda}$&$f_{\Lambda\omega\Lambda}$&$1$		&0\\
				& $\phi$ 	& 4 	&$g_{K\phi K}$		&$g_{\Lambda\phi\Lambda}$&$f_{\Lambda\phi\Lambda}$&$1$			&0\\
				& $\Xi$  	& 1 	&$g_{K\Xi\Lambda}$	&$g_{K\Xi\lambda}$	&			&$ 1$			&0\\
				& $\Xi^*$ 	& 2 	&$g_{K\Xi^*\Lambda}$	&$g_{K\Xi^*\Lambda}$	&			&$1 $			&0 \\
$\pi N\rightarrow \Sigma K$	& $\kst$   	& 4 	&$g_{\pi K^*K}$		&$g_{NK^*\Sigma}$	&$f_{NK^*\Sigma}$	&$1$			&$2$\\
				& $\Sigma$ 	& 1 	&$g_{\pi\Sigma\Sigma}$	&$g_{N\Sigma K}$	&			&$2$			&$1$\\
				& $\Lambda$ 	& 1 	&$g_{\pi\Lambda\Sigma}$	&$g_{N\Lambda K}$	&			&$-1$			&$1$\\
				& $\Sigma^{*}$ 	& 2 	&$g_{\pi\Sigma^*\Sigma}$&$g_{N\Sigma^* K}$	&			&$2$			&$1$\\ 
$\Lambda K\rightarrow\Sigma K$	& $\rho$ 	& 4 	&$g_{K\rho K}$		&$g_{\Lambda\rho\Sigma}$&$f_{\Lambda\rho\Sigma}$&$-\sqrt{3}$		&$0$\\
				& $\Xi$ 	& 1 	&$g_{K\Xi\Sigma}$	&$g_{\Lambda\Xi K}$	&			&$\sqrt{3} $		&0\\
				& $\Xi^*$ 	& 2 	&$g_{K\Xi^*\Sigma}$	&$g_{\Lambda\Xi^* K}$	&			&$\sqrt{3} $		&0  \bigstrut[b]\\
\hline
\hline 
\end{tabular}
\label{tab:isospin2}
\end{center}
\end{table}  

\begin{center}
\begin{table}
\caption{Continuation of Table~\ref{tab:isospin2}.}
\renewcommand{\arraystretch}{1.33}
\begin{tabular}[t]{llllllrr}
\hline \hline
  Transition		       & Ex &Type& $g_a$	       &$g_b$	       &$f_b$	       & IF$(\frac{1}{2})$     & IF$(\frac{3}{2})$  \bigstrut[b]\\ \hline
$\Sigma K\rightarrow \Sigma K$	&$f_0$ &3& $g_{KKf_0}$&$g_{\Sigma\Sigma f_0}$ &&$1$&$1$\bigstrut[t]\\
				&$\omega$ & 4&$g_{K\omega K}$&$g_{\Sigma\omega\Sigma}$&$f_{\Sigma\omega\Sigma}$&$1$&$1$\\
				&$\phi$ &4 &$g_{K\phi K}$&$g_{\Sigma\phi\Sigma}$&$f_{\Sigma\phi\Sigma}$&$1$&$1$\\
				&$\rho$ &4 &$g_{K\rho K}$&$g_{\Sigma\rho\Sigma}$&$f_{\Sigma\rho\Sigma}$&$2$&$-1$\\
				&$\Xi$ &1 &$g_{K\Xi\Sigma}$&$g_{K\Xi\Sigma}$&&$ -1$&$2$\\
				&$\Xi^*$ &2 &$g_{K\Xi^*\Sigma}$&$g_{K\Xi^*\Sigma}$&&$ -1$&$2$ \\
$\rho N\to \rho N$		&$N$&17&$g_{NN\rho}$ &$g_{NN\rho}$ &$f_{NN\rho}$ & $-1$&$ 2$  \\
				&ct& 18&$g_{NN\rho}$ & &$f_{NN\rho}$ & $-2i$&$i$  \\
				&$\rho$&19 &$g_{\rho\rho\rho}$ &$g_{NN\rho}$ & $f_{NN\rho}$&$2i$ & $-i$ \\
				&$\Delta$&20 &$g_{N\Delta\rho}$ &$g_{N\Delta\rho}$ & & $\frac{4}{3}$& $\frac{1}{3}$  \\
$\rho N\to \pi\Delta$		&$\pi$&24 &$g_{\pi\pi\rho}$ &$g_{\Delta N\pi}$ & &$-\sqrt{\frac{2}{3}} i$ & $-\sqrt{\frac{5}{3}} i$  \\
				&$N$&25 &$g_{NN\pi}$ &$g_{\Delta N\rho}$ & &$-\sqrt{\frac{8}{3}}$ & $\sqrt{\frac{5}{3}}$ \\
$\eta N\to \eta N$		&$N$&1  &$g_{\eta NN}$ &$g_{\eta NN}$ & &$1$ &$0$  \\
				&$f_0$&6 &$g_{\eta\eta f_0}$ &$g_{NN f_0}$ & &1 &0  \\
$\eta N\to K\Lambda$		&$\kst$&4 &$g_{\eta\kst K}$ &$g_{N\kst\Lambda}$ &$f_{N\kst\Lambda}$ & 1&0  \\
				&$\Lambda$&1 &$g_{\eta\Lambda\Lambda}$ &$g_{N\Lambda K}$ & & 1&0  \\
$\eta N\to K\Sigma$		&$\kst$&4 &$g_{\eta\kst K}$ &$g_{N\kst\Sigma}$ &$f_{K\kst\Sigma}$ &$ -\sqrt{3}$ &0  \\
				&$\Sigma$&1 &$g_{\eta\Sigma\Sigma}$ &$g_{N\Sigma K}$ & & $ -\sqrt{3}$& 0 \\
				&$\Sigma^*$&2 &$g_{\eta\Sigma^*\Sigma}$ &$g_{N\Sigst K}$ & &$ -\sqrt{3}$  & 0 \\
$\pi\Delta\to\pi\Delta$		&$N$& 21&$g_{\pi N\Delta}$ &$g_{\pi N\Delta}$ & &$\frac{1}{3}$ &$-\frac{2}{3}$  \\
				&$\rho$&22 &$g_{\pi\pi\rho}$ &$g_{\Delta\Delta\rho}$ &$f_{\Delta\Delta\rho}$ &$ \frac{5}{3}i$ & $ \frac{2}{3}i$  \\
				&$\Delta$&23 &$g_{\Delta\Delta\pi}$ &$g_{\Delta\Delta\pi}$ & &$-\frac{10}{9}$ &$\frac{11}{9}$  \\
$\sigma N\to\pi\Delta$		&$\pi$&26 &$g_{\pi\pi\sigma}$ &$g_{\Delta\pi N}$ & &$-\sqrt{2}$ &0  \\
$\sigma N\to\sigma N$		&$N$& 27 &$g_{NN\sigma}$ &$g_{NN\sigma}$ & & 1&0  \\
				&$\sigma$&28 &$g_{\sigma\sigma\sigma}$ & $g_{NN\sigma}$& & 1&0  \bigstrut[b]\\
\hline
\hline 
\end{tabular}
\label{tab:isospin2con}
\end{table}
\end{center}

%%%%%%%%%%%%%%%%%%%%%%%%%%%%%%%%%%%%%%%%%%%%%%%%%%%%%%%%%%%%%%%%%%%%%%%%%%%%%%%%%%%%%%%%%%%%%%%%%%%%%%%%%%%%%%%%%%%%%%%%%%%%%%%%%%%%%%%%%%%%%%%%

\subsection{Couplings of $t$- and $u$-channel exchanges}
\label{sec:su3_couplings}

\begin{table}
\caption{Couplings not fixed by SU(3) symmetry. The values marked with $*$ were obtained in this study by fitting to data. For the large size of
$g_{NN\sigma}$ in fit B, see the discussion in Sec.~\ref{sec:p11}. For the sign of $g_{\omega\pi\rho}$ see Ref.~\cite{Nakayama:2006ps}.}
\begin{center}
\renewcommand{\arraystretch}{1.6}
\scalebox{0.96}{
\begin{tabular}[t]{lrrl|lrrl}
\hline \hline 
 Coupling 				& fit A		& fit B			& Ref. 				&Coupl.			&fit A 		& fit B 		& Ref.
		\\ \hline
$g_{NN\sigma}$				&13.85		& 39.31			& * 				&$g_{\sigma\sigma\sigma}$	& 1.75		& 5.58			& *
		\\
$g_{\Sigma\Sigma f_0}$			&\multicolumn{2}{c}{9.41}		& * 				&$g_{\Lambda\Lambda f_0}$	&1.56		&1.59 			& *
		\\
$g_{\pi\rho a_1}$ 			&\multicolumn{2}{c}{$\sim g_{NN\rho}$}	&\cite{Wess:1967jq,Gasparyan:2003fp}	&$g_{NNa_1}$			&\multicolumn{2}{c}{$\sim g_{NN\pi}$}
&\cite{Wess:1967jq,Gasparyan:2003fp}\\
$g_{\pi\eta a_0}^{1/2} g_{NNa_0}^{1/2}$	&\multicolumn{2}{c}{10.03} 		&\cite{Schutz:1998jx} 		&$g_{\omega\pi\rho}$ 		&\multicolumn{2}{c}{-10.0}		& *
		\\
$g_{\eta\eta f_0}^{1/2} g_{NNf_0}^{1/2}$	&\multicolumn{2}{c}{3.305} 		& * 				&$g_{\pi\pi\sigma}$  		&\multicolumn{2}{c}{1.77} 	
&\cite{Krehl:1997kg}	\\
$g_{KKf_0}$				&\multicolumn{2}{c}{1.336}		&\cite{MuellerGroeling:1990cw}	&$g_{\rho\rho\rho}$		&\multicolumn{2}{c}{$\sim g_{NN\rho}$}
&\cite{Gasparyan:2003fp}\\
$g_{\Delta N\rho}$			&\multicolumn{2}{c}{16.03}		&\cite{Janssen:1996kx} 		&$g_{\Delta\Delta\pi}$		&\multicolumn{2}{c}{1.78}	
&\cite{Schutz:1998jx}  	\\
$g_{\Delta\Delta\rho}$ 			&\multicolumn{2}{c}{7.67}		&\cite{Schutz:1998jx}		&$\kappa_{\Delta\Delta\rho}$	&\multicolumn{2}{c}{6.1} 	
&\cite{Schutz:1998jx} 	\\
\hline \hline
\end{tabular}
}
\label{tab:coupl}
\end{center}
\end{table} 

The coupling constants of several $t$- and $u$-channel exchange diagrams can be related to each other by SU(3) flavor symmetry~\cite{de
Swart:1963gc}. The coupling  of a given meson nonet with the octet baryon current $8_B\otimes 8_B$  yields two products, namely $8_B\otimes
8_B\otimes 1_M$ and $8_B\otimes 8_B\otimes 8_M$. The coupling to the meson octet $8_M$ depends on two parameters, $g_1, \,g_2$, corresponding to
the symmetric and antisymmetric representations of $8_B\otimes 8_B$. They can be re-written in terms of a coupling constant $g$ and a mixing
parameter $\alpha$ in the notation of Ref.~\cite{de Swart:1963gc} used here as well, 	
\begin{eqnarray}
 g=\frac{\sqrt{30}}{40}\,g_{1}+\frac{\sqrt{6}}{24}\,g_{2},\quad \alpha = 
\frac{\sqrt{6}}{24}\,\frac{g_{2}}{g}. 
\end{eqnarray}
$g_{1}$ and $g_{2}$ can be also expressed in terms of the standard $D$ and $F$ couplings \cite{Beringer:1900zz}:
\begin{eqnarray}
 D=\frac{\sqrt{30}}{40}g_{1}\,  , \;\;\; F= \frac{\sqrt{6}}{24}\,g_{2}
\end{eqnarray}
with $\alpha=F/(D+F)$ (definition of $\alpha$ of Ref.~\cite{de Swart:1963gc}).  The coupling of the baryon octets to the meson singlet $1_M$
involves  a further independent coupling constant. 

For vertices involving pseudoscalar mesons, we use \cite{Janssen:1996kx}
\begin{eqnarray}
\alpha_{BBP}= 0.4\;, \quad
\alpha_{PPV}= 1 \;, 
\end{eqnarray}
where the index $P$ denotes a pseudoscalar meson and $V$ means a vector meson.  We identify the $\eta$ with the octet $\eta_8$ and assume the
singlet coupling  to be zero. This leads to a difference of only about 10$\%$ in the estimates of $g_{\Sigma\Sigma\eta}$ and
$g_{\Lambda\Lambda\eta}$, assuming an $\eta -\eta'$ mixing angle of $\theta_p \sim -10^\circ$. The corresponding difference in the coupling
constant $g_{K\eta K^*}$ is only about 2$\%$.  For the coupling $g_{NN\eta}$, following Ref.~\cite{Nakayama:2008tg}, we use a 
phenomenologically small value of $g_{NN\eta}=0.147$ instead of the SU(3) value $g_{NN\eta}= g_{BBP} \frac{(4\alpha_{BBP}-1)}{\sqrt{3}}=0.343$.

For the coupling of vector mesons to octet baryons, as e.g. at the $NN\rho$ vertex,  we use the Lagrangian
\begin{eqnarray}
 \mathcal{L}_{int}=-\bar\Psi\left[g_{NN\rho}\gam-\frac{f_{NN\rho}}{2m_{N}}\sigma^{\mu\nu}\partial_{\nu}\right]
\vtau\,\cdot \vec{\rho}_{\mu}\, \Psi 
\end{eqnarray}
which consists of a vector part with $\gam$ and a tensor part with $\sigma^{\mu\nu}$.  The corresponding coupling constants are  connected
through $f_{NN\rho}=g_{NN\rho}\kappa_{\rho}$~\cite{Holzenkamp:1989tq,Reuber:1993ip}  and we use $\kappa_{\rho}=6.1$~\cite{Janssen:1996kx}.

The couplings of the physical $\omega$ and $\phi$ are obtained by assuming ideal mixing of the SU(3) singlet- and octet states, i.e. of
$\omega_1$ and $\omega_8$.  Furthermore, we assume that the $\phi$ meson does not couple to the nucleon  (OZI rule) to fix the singlet coupling
constant. Note, however, that kaon loops provide a small but non-vanishing  effective $\phi NN$ coupling~\cite{Meissner:1997qt,Doring:2008sv}
that is  neglected in this approach.

With regard to the actual coupling strengths employed in the vector-meson  sector, strict SU(3) symmetry is broken in various ways. First and
foremost, as mentioned before, we do not use genuine $\rho$ exchange  in the $\pi N\rightarrow \pi N$ channel but determine its contribution
from  correlated $\pi\pi$ exchange. Furthermore, the value of $g_{NN\omega}$ as  commonly found/needed in studies of the $NN$ interaction
exceeds its SU(3)  prediction; see the discussions in Refs.~\cite{Reuber:1993ip,Janssen:1996kx}.  Thus, in order to accommodate this aspect, we
work with the hypothetical value $\alpha_{BBV}= 1.15$ (instead of the standard value $\alpha_{BBV}= 1$) based on the $\omega$ coupling constant
given in Ref.~\cite{Janssen:1996kx} which introduces a small SU(3) symmetry breaking in all vector-meson couplings.  Furthermore, we set
$f_{NN\omega}=0$ for simplicity.

The employed relations for the various coupling constants are given by the  following expressions.  Their use in the different $t$- and
$u$-channel expressions of Appendix~\ref{sec:explicit_exchanges} is indicated in the fourth and fifth column  of Tables~\ref{tab:isospin2} and
\ref{tab:isospin2con}.

\begin{itemize}
\item Couplings for octet baryon, octet baryon and pseudoscalar meson:
\begin{eqnarray}
 g_{NN\pi}&=& g_{BBP}\;, \nonumber\\
g_{\Sigma NK}&=& g_{BBP}(1-2\alpha_{BBP})\;,\nonumber\\
g_{\Lambda NK}&=& -\tfrac{\sqrt{3}}{3}\, g_{BBP} (1+2\alpha_{BBP})\;,\nonumber\\
g_{\Sigma\Sigma\pi}&=& 2\,g_{BBP}\alpha_{BBP}\;,\nonumber\\
g_{\Lambda\Sigma \pi}&=&\tfrac{2}{\sqrt{3}}\,g_{BBP}\,(1-\alpha_{BBP})\;\nonumber\\
g_{\Xi\Sigma K}&=& -g_{BBP}\;, \nonumber\\
g_{\Xi\Lambda K}&=& \tfrac{1}{\sqrt{3}}\, g_{BBP} (4\alpha_{BBP}-1)\;, \nonumber \\
g_{\Lambda\Lambda\eta}&=&-\tfrac{2}{\sqrt{3}}\, g_{BBP} (1-\alpha_{BBP})\; \non
g_{\Sigma\Sigma\eta}&=&\tfrac{2}{\sqrt{3}}\, g_{BBP} (1-\alpha_{BBP})\;,
\label{2.21}
\end{eqnarray}
with $g_{BBP}=0.989$ \cite{Janssen:1996kx}.

\item Vector coupling for octet baryon, octet baryon and vector meson:
\begin{eqnarray}
g_{NN\rho}&=&g_{BBV}\;,\nonumber\\
g_{NN\omega}&=& g_{BBV}(4\alpha_{BBV}-1)\;,\nonumber\\
g_{\Lambda N\kst}&=& -\tfrac{1}{\sqrt{3}}g_{BBV}(1+2\alpha_{BBV})\;,\nonumber\\
g_{\Sigma N\kst}&=&g_{BBV}(1-2\alpha_{BBV})\;,\nonumber\\ 
g_{\Lambda\Lambda\omega}&=& \tfrac{2}{3}g_{BBV}(5\alpha_{BBV}-2)\;,\nonumber\\
g_{\Sigma\Sigma\omega}&=&2g_{BBV}\alpha_{BBV}\;,\nonumber\\
g_{\Lambda\Lambda\phi}&=& -\tfrac{\sqrt{2}}{3}g_{BBV}(2\alpha_{BBV}+1)\;,\nonumber\\
g_{\Sigma\Sigma\phi}&=& -\sqrt{2}g_{BBV}(2\alpha_{BBV}-1)\;,\nonumber\\
g_{\Sigma\Sigma\rho}&=&2g_{BBV}\alpha_{BBV}\;,\nonumber\\
g_{\Lambda\Sigma\rho}&=&\tfrac{2}{\sqrt{3}}g_{BBV}(1-\alpha_{BBV})\;,
\label{2.22}
\end{eqnarray}
with $g_{BBV}=3.25$ \cite{Janssen:1996kx}
\item Tensor coupling for octet baryon, octet baryon and vector meson: 
\begin{eqnarray}
f_{NN\rho}&=& g_{NN\rho}\kappa_{\rho}\;,\nonumber\\
f_{\Lambda N\kst}&=& -\tfrac{1}{2\sqrt{3}} f_{NN\omega}-\tfrac{\sqrt{3}}{2} f_{NN\rho}\;,\nonumber\\
f_{\Sigma N\kst}&=& -\tfrac{1}{2} f_{NN\omega}+ \tfrac{1}{2} f_{NN\rho}\;,\nonumber\\
f_{\Lambda\Lambda \omega}&=& \tfrac{5}{6} f_{NN\omega}- \tfrac{1}{2} f_{NN\rho}\;,\nonumber\\
f_{\Sigma\Sigma\omega}&=& \tfrac{1}{2} f_{NN\omega}+\tfrac{1}{2} f_{NN\rho}\;,\nonumber\\
f_{\Lambda\Lambda\phi}&=& -\tfrac{1}{3\sqrt{2}} f_{NN\omega} - \tfrac{1}{\sqrt{2}} f_{NN\rho}\;,\nonumber\\
f_{\Sigma\Sigma\phi}&=& - \tfrac{1}{\sqrt{2}} f_{NN\omega}+ \tfrac{1}{\sqrt{2}} f_{NN\rho}\;,\nonumber\\
f_{\Sigma\Sigma\rho}&=& \tfrac{1}{2} f_{NN\omega}+ \tfrac{1}{2} f_{NN\rho}\;,\nonumber\\
f_{\Lambda\Sigma\rho}&=&-\tfrac{1}{2\sqrt{3}} f_{NN\omega}+ \tfrac{\sqrt{3}}{2} f_{NN\rho}\;,  \label{2.23}
\end{eqnarray}
with $\kappa_{\rho}=6.1$~\cite{Janssen:1996kx} and $f_{NN\omega}=0$.

\item Coupling for pseudoscalar meson, pseudoscalar meson and vector meson:
\begin{eqnarray}
g_{\pi\pi\rho}&=& 2\,g_{PPV}\;,\nonumber\\
g_{KK\rho}&=&g_{PPV}\;,\nonumber\\
g_{K\pi\kst}&=& -g_{PPV}\;,\nonumber\\
g_{KK\omega}&=& g_{PPV}\;,\nonumber\\
g_{KK\phi}&=&\sqrt{2}\, g_{PPV}\;, \non
g_{K\eta K^*}&=&-\sqrt{3}\, g_{PPV}\;, \label{2.24}
\end{eqnarray}

with $g_{PPV}=3.02$ \cite{Janssen:1994wn}.

\item Coupling for decuplet baryon, octet baryon and pseudoscalar meson:
\begin{eqnarray}
g_{\Delta N\pi}&=& g_{DBP}\;,\nonumber\\
g_{\Sigma^{*} NK}&=& -\tfrac{1}{\sqrt{6}} g_{DBP}\;,\nonumber\\
g_{\Sigma^{*}\Sigma\pi}&=&\tfrac{1}{\sqrt{6}}g_{DBP}\;,\nonumber\\
g_{\Sigma^{*}\Lambda\pi}&=& \tfrac{1}{\sqrt{2}} g_{DBP}\;, \non
g_{\Xi^*\Lambda K}&=& \tfrac{1}{\sqrt{2}}\;g_{DBP}\;,\non
g_{\Xi^*\Sigma K}&=& \tfrac{1}{\sqrt{6}}\;g_{DBP}\;,\non
g_{\Sigma^*\Sigma\eta} &=& -\tfrac{1}{\sqrt{2}}\; g_{DBP}\;, \label{2.25}
\end{eqnarray}

with $g^2_{\Delta N\pi}/4\pi=0.36$ \cite{Janssen:1996kx}, $g_{DBP}=g_{\Delta N\pi}$ .
\end{itemize}

The coupling constants for scalar mesons ($f_0$, $\sigma$) are not fixed by SU(3) symmetry.  We take the value of $g_{KKf_0}$ from
Refs.~\cite{MuellerGroeling:1990cw, Holzenkamp:1989tq}, obtained from the hyperon-nucleon interaction.  The couplings $g_{NN\sigma}$,
$g_{\sigma\sigma\sigma}$, $g_{\Sigma\Sigma f_0}$  and $g_{\Lambda\Lambda f_0}$ were treated as free parameters and determined  in the fit.  The
values can be found in Table~\ref{tab:coupl}, together with other couplings, partly taken from the literature and partly fitted in the current
study.

%%%%%%%%%%%%%%%%%%%%%%%%%%%%%%%%%%%%%%%%%%%%%%%%%%%%%%%%%%%%%%%%%%%%%%%%%%%%%%%%%%%%%%%%%%%%%%%%%%%%%%%%%%%%%%%%%%%%%%%%%%%%%%%%%%%%%%%%%%%%%%%%

\subsection{Form factors of $t$- and $u$-channel exchanges}
\label{sec:fofatu}
We use 
\begin{eqnarray}
 F(q)=\left(\frac{\Lambda^2-m_{ex}^2}{\Lambda^2+\vec{q}^{\,2}}\right)^n
\end{eqnarray}
for the form factors $F$ of the $t$- and $u$-channel exchanges in Eq.~(\ref{tu}) where $m_{ex}$ is the mass and  $\vec{q}$ the momentum of the
exchanged particle. Powers $n=1,2$ provide monopole or dipole form factors. The dipole type applies to vertices with $\rho$, $\Delta$, $K^*$,
$\Sigma^*$, $\omega$ and $\phi$. Exceptions are $\Delta$ exchange in $\rho N\to\rho N$, where $n=4$, and the $\pi\pi\rho$ vertex in
$\pi\Delta\to\pi\Delta$ with $\rho$ exchange, where $n=1$. Otherwise the monopole form factor is used. Different powers for the form factors are
applied to ensure the convergence of the integral over the off-shell momenta in the scattering equation.

For the nucleon $u$-exchange at the $\pi NN$ vertex in $\pi N\to \pi N,\, \rho N, \, \pi\Delta,\, \sigma N$ the monopole form factor is modified
in order to be normalized to unity at the Cheng-Dashen point:
\begin{eqnarray}
 F(q)=\frac{\Lambda^2-m_N^2}{\Lambda^2-((m_N^2-m_{\pi}^2)/m_N)^2 + \vec{q}^{\,2}} \;.
\end{eqnarray}
The form factor for the contact terms reads:
\begin{eqnarray}
 F(p_4,p_2)=\left( \frac{\Lambda^2-m_4^2}{\Lambda^2+\vec{p}_4^{\,2}}  \right)^2 
 \left( \frac{\Lambda^2 -m_2^2}{\Lambda^2+\vec{p}_2^{\,2}} \right)^2 \, ,
\end{eqnarray}
where $\vec{p}_2$ ($\vec{p}_4$) denotes the momentum  and $m_2$ ($m_4$) the mass of the incoming (outgoing) meson. The form factors for the
correlated $\pi\pi$ exchange potentials can be found in Ref.~\cite{Gasparyan:2003fp}.

Explicit expressions for the form factors of $s$-channel diagrams can be found in Appendix B of Ref.~\cite{Doring:2010ap}.

\begin{table*}
\caption{Cut-offs $\Lambda$ for the form factors of the exchange diagrams. The columns {\it Ex} specify the exchanged particle and {\it ct}
indicates a contact interaction.}
\begin{center}
\renewcommand{\arraystretch}{1.50}
\setlength{\tabcolsep}{.2cm}
\begin{tabular}[t]{cccc|cccc|cccc}
\hline \hline 
Vertex 			& Ex & \multicolumn{2}{c|}{ $\Lambda$ [MeV]}&Vertex & Ex& \multicolumn{2}{c|}{ $\Lambda$ [MeV]} &Vertex& Ex 	& \multicolumn{2}{c}{ $\Lambda$ [MeV]}\\
	  &	& fit A& fit B		&	&	& fit A& fit B		&	&	& fit A& fit B	\\
\hline 

$\pi NN$                & $N$           &$1060 $&$1130$  &$\rho N\rho N$	&ct	&$1020 $&$900$	&$\pi\Sigst\Sigma$& $ \Sigst$  	& $1740$ & $2320$\\
$NN\sigma $		& $N$		&$1800 $&$2830$	&$\pi\omega\rho$	&$\omega$	&$860 $	&$800 $	& $N\Sigma K $ 	& $\Sigma $  	& $1500$&$1700$ \\
$N\pi N $ 		& $\pi$		& $900 $&$1010$	&$\pi a_1\rho$		&$a_1$		&$1260 $&$2200$	&$N\Sigst K $ 	& $\Sigst $   	& $1640$&$1640$\\
$N\sigma N\;$ in $\sigma N\to\sigma N$	& $\sigma$&$1700$&$2660$&$\pi\pi\rho$	&$\pi$		&$800 $ &$880$	&$\pi\Lambda \Sigma$& $\Lambda $& $1400$&$1400$\\
$N\sigma N\;$ in $\pi N\to\pi N$& $\sigma$& $810$ &$900$&$\rho\rho\rho$		&$\rho$		&$3200 $&$2410 $&$N\Lambda K $ 	& $\Lambda $ 	& $1250$&$1250$ \\
$N\rho N$		& $\rho$	& $920 $&$970 $	&$\rho\Delta N$		&$\Delta$	&$1610 $&$2400$	&$K\rho K$ 	& $\rho$	& $3140$&$2580 $ \\
$N\omega N$		&$\omega$	&$1070 $&$ 1320$&$\rho N \Delta$	&$N$		&$800 $	&$800 $	&$Kf_0 K$	& $f_0$		&$3400$&$3480$ \\
$Na_0N$			&$a_0$		&$1740 $&$3003$ &$NN\eta$		&$N$		&$1260 $&$3050$	&$K\omega K $ 	& $\omega $     &$1310$&$1430 $	 \\
$Na_1N$			&$a_1$		&$1260 $&$ 1260$&$\pi a_0\eta$		&$a_0$		&$2260 $&$2260$	& $K\phi K $	& $\phi $   	&$1620$&$1620 $ \\
$\pi N\Delta$ 		& $N$		&$2150$ &$3080$	&$\eta \kst K$		&$\kst$		&$830 $	&$830 $	&$\Lambda f_0\Lambda$ & $f_0 $  &$1320$&$ 1320$ \\
$N\rho\Delta$		& $\rho$	&$1940$& $1700$	&$\eta f_0\eta$		&$f_0$		&$1470 $&$1500 $& $\Lambda\omega\Lambda$ & $\omega$  &$1100$&$1100 $  \\
$N\pi\Delta$		& $\pi$		& $900 $&$1600$	 &$Nf_0N $		&$f_0$		&$1470 $&$1500$ &$\Lambda\phi\Lambda $& $\phi $   &$1300$&$1300 $\\
$\pi\rho\pi\;$ in $\pi N, \pi\Delta\to\pi\Delta$& $\rho$& $1270$&$1270$&$\eta\Lambda\Lambda$& $\Lambda$&$3190 $&$3190$&$\Lambda\rho\Sigma$& $\rho $&$1710 $ &$1750$\\
$\pi\rho\pi\;$ in $\pi N\to\pi N$& $\rho$&$870 $ &$930$ &$\eta\Sigma\Sigma$	& $\Sigma$	&$2060 $&$2060 $& $\Sigma f_0\Sigma$  	& $f_0 $  &$800$ &$2160 $ \\
$N\Delta\pi$		& $\Delta$	& $1400$&$1860$	&$\eta\Sigma^*\Sigma$	& $\Sigma^*$	&$3040 $&$3040 $ &$\Sigma\omega\Sigma $ & $\omega $ &$1940$ &$1940 $\\
$\pi\Delta\Delta$	& $\Delta$	& $1970$&$2010$	&$\pi\kst K $ 		& $\kst $	&$1400 $&$1760$	&$\Sigma\phi\Sigma $    & $\phi $  & $1660 $&$1660 $\\
$\Delta\rho\Delta$	& $\rho$	& $2100$&$2140$	&$N\kst \Lambda $ 	& $\kst $   	&$1000 $&$1000 $ &$\Sigma\rho\Sigma $    & $\rho $  &$1480$ &$1480 $ \\
$\pi\sigma\pi$		& $\sigma$	& $810 $&$900$	&$N\kst\Sigma $ 	& $ \kst$  	& $1010$&$1030$	&$K\Xi\Lambda$		& $\Xi$	&$1670 $ &$1670 $\\
$\pi\pi\sigma $		& $\pi$		& $930 $&$2410$	&$\pi\Sigma\Lambda $ 	& $ \Sigma$  	& $1350$&$1710 	$&$K\Xi^*\Lambda$	& $\Xi^*$&$1850 $ &$1850 $  \\
$\sigma\sigma\sigma$	& $\sigma$	& $1700$&$1810$	&$\pi\Sigst\Lambda $ 	& $\Sigst $   	&$2740$&$1880 $&$K\Xi\Sigma$		& $\Xi$	&$2340 $ &$2340 $\\
$NN\rho$		&$N$		&$1680 $&$1340 $&$\pi\Sigma\Sigma $ 	& $\Sigma $    	& $1500$&$1500$ &$K\Xi^*\Sigma$		& $\Xi^*$&$3600 $ &$3600 $\\	
$\pi N\rho N$		&ct	&$800 $	&$810 $	&			& 		&	&&				& 		& 	&\\	   
\hline \hline
\end{tabular}
\label{tab:cutoff_bg}
\end{center}
\end{table*} 

The cut-off values of the form factors for $t$- and $u$-channel processes, fitted in the present study, are shown in Table~\ref{tab:cutoff_bg}.
Similar to bare resonance masses and couplings, we do not attribute any physical meaning to these parameters and quote the values only for the
sake of reproducibility. It is clear that in the present approach some values of the cut-offs absorb effects from degrees of freedom not
explicitly included in the model. If the fit prefers a high value for a cut-off, we do not consider the corresponding process to be sensitive to
high momentum physics but rather take it as a sign that some missing ingredient of the model is absorbed in that value. 

%%%%%%%%%%%%%%%%%%%%%%%%%%%%%%%%%%%%%%%%%%%%%%%%%%%%%%%%%%%%%%%%%%%%%%%%%%%%%%%%%%%%%%%%%%%%%%%%%%%%%%%%%%%%%%%%%%%%%%%%%%%%%%%%%%%%%%%%%%%%%%%%

\section{$\boldsymbol \pi N\to\pi N$ phase shifts and inelasticities}
\label{sec:piNdel}
\begin{figure}
\begin{center}
\includegraphics[width=0.48\textwidth]{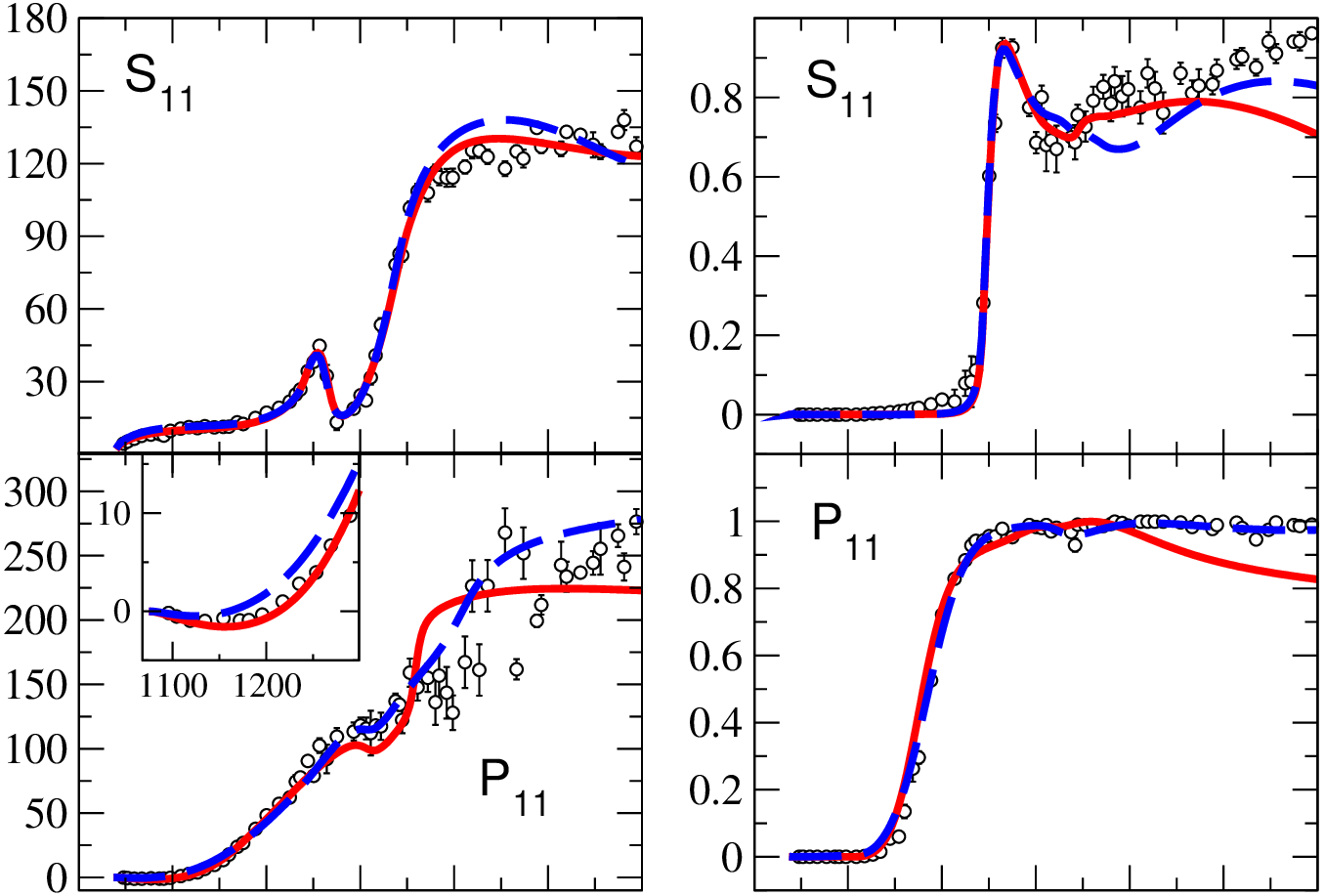}\\
\vspace*{-0.1cm}
\includegraphics[width=0.48\textwidth]{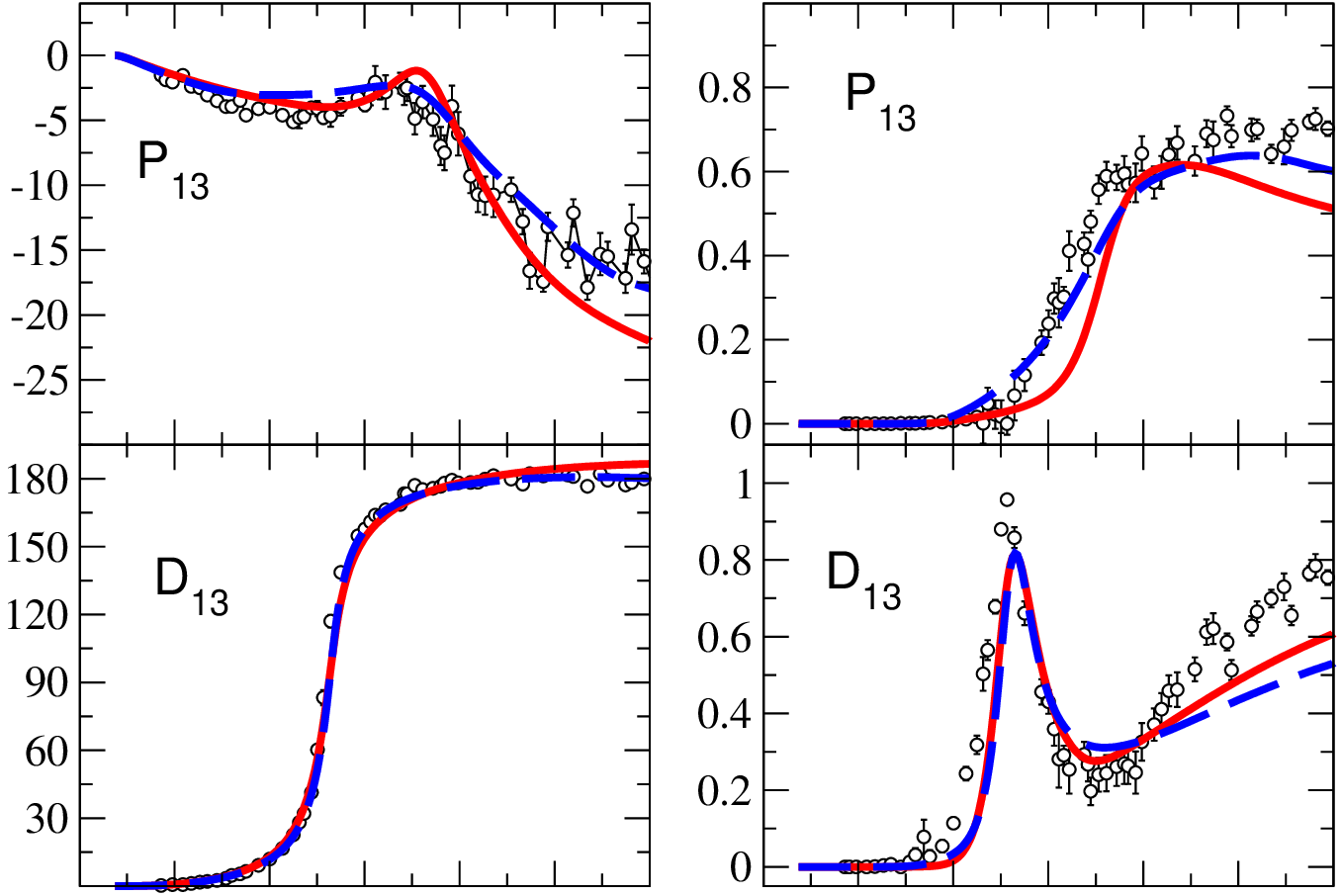}\\
\vspace*{-0.2cm}
\includegraphics[width=0.48\textwidth]{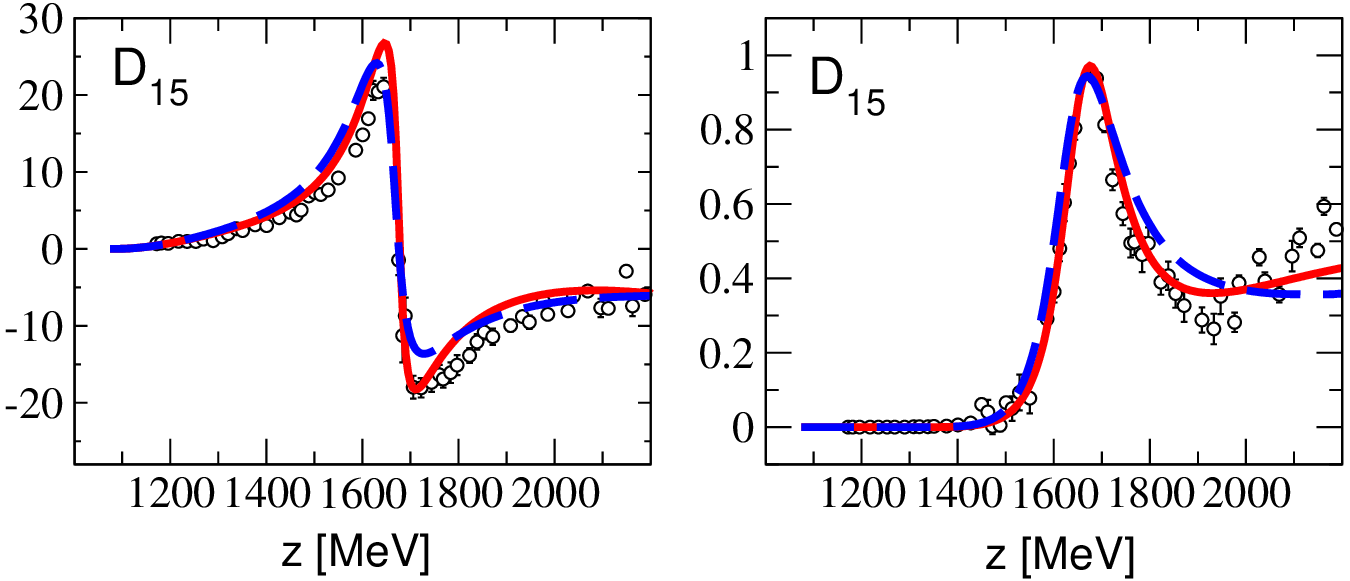}
\end{center}
\caption{Elastic $\pi N\to \pi N$ phase shifts $\delta$ [deg] (left) and inelasticities $(1-\eta)^2$ (right) for isospin $I=1/2$, $S$- to
$D$-waves. Points: GWU/SAID partial-wave analysis (single-energy solution) from Ref.~\cite{Arndt:2006bf}; solid (red) lines: fit A; dashed
(blue) lines: fit B.}
\label{fig:pindel1}     
\end{figure}

\begin{figure}
\begin{center}
\hspace*{-0.05cm}\includegraphics[width=0.482\textwidth]{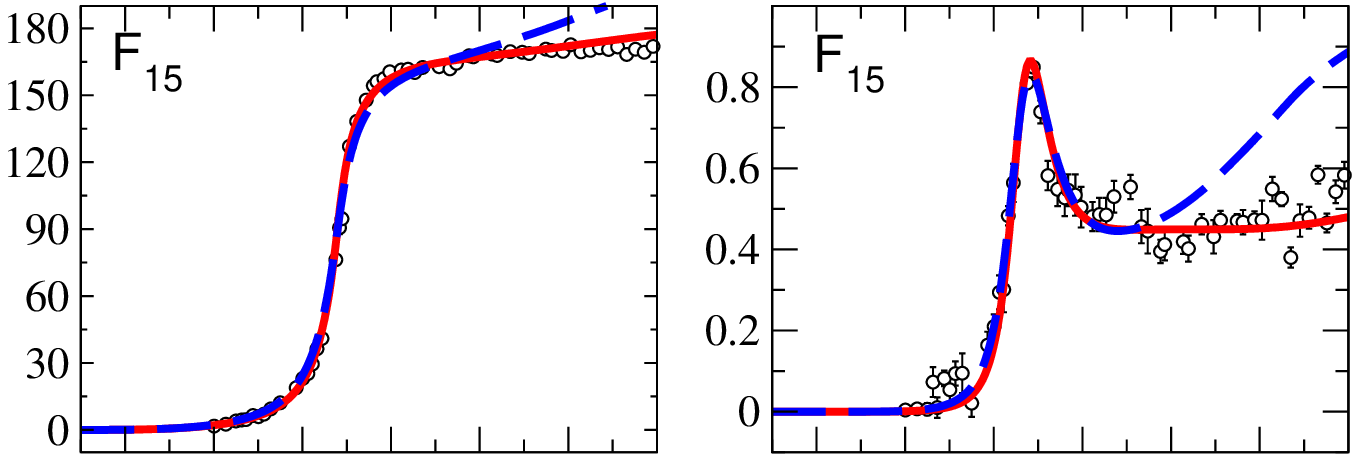}\\
\vspace*{-0.105cm}
\hspace*{0.1cm}\includegraphics[width=0.475\textwidth]{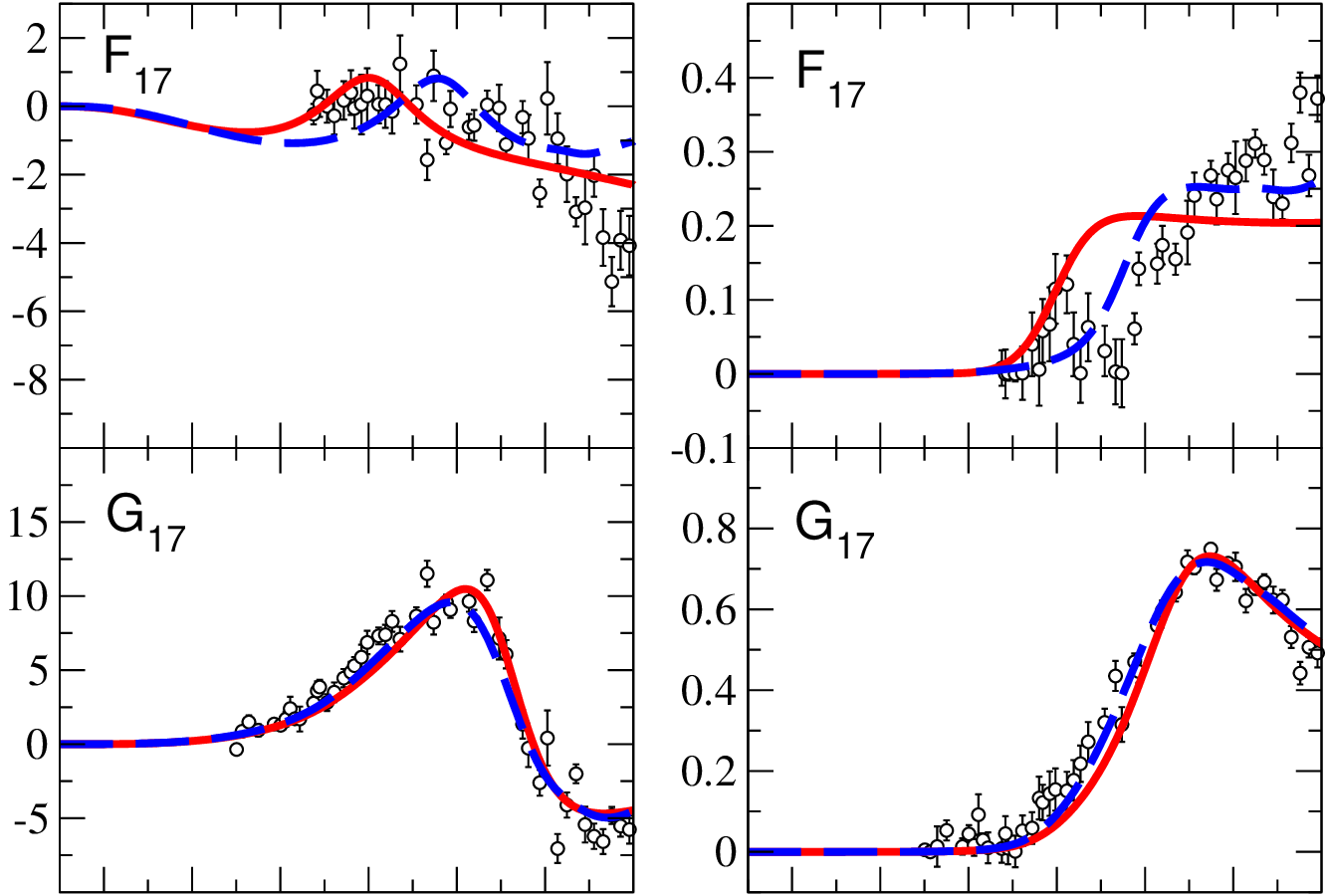}\\
\vspace*{-0.1cm}
\includegraphics[width=0.48\textwidth]{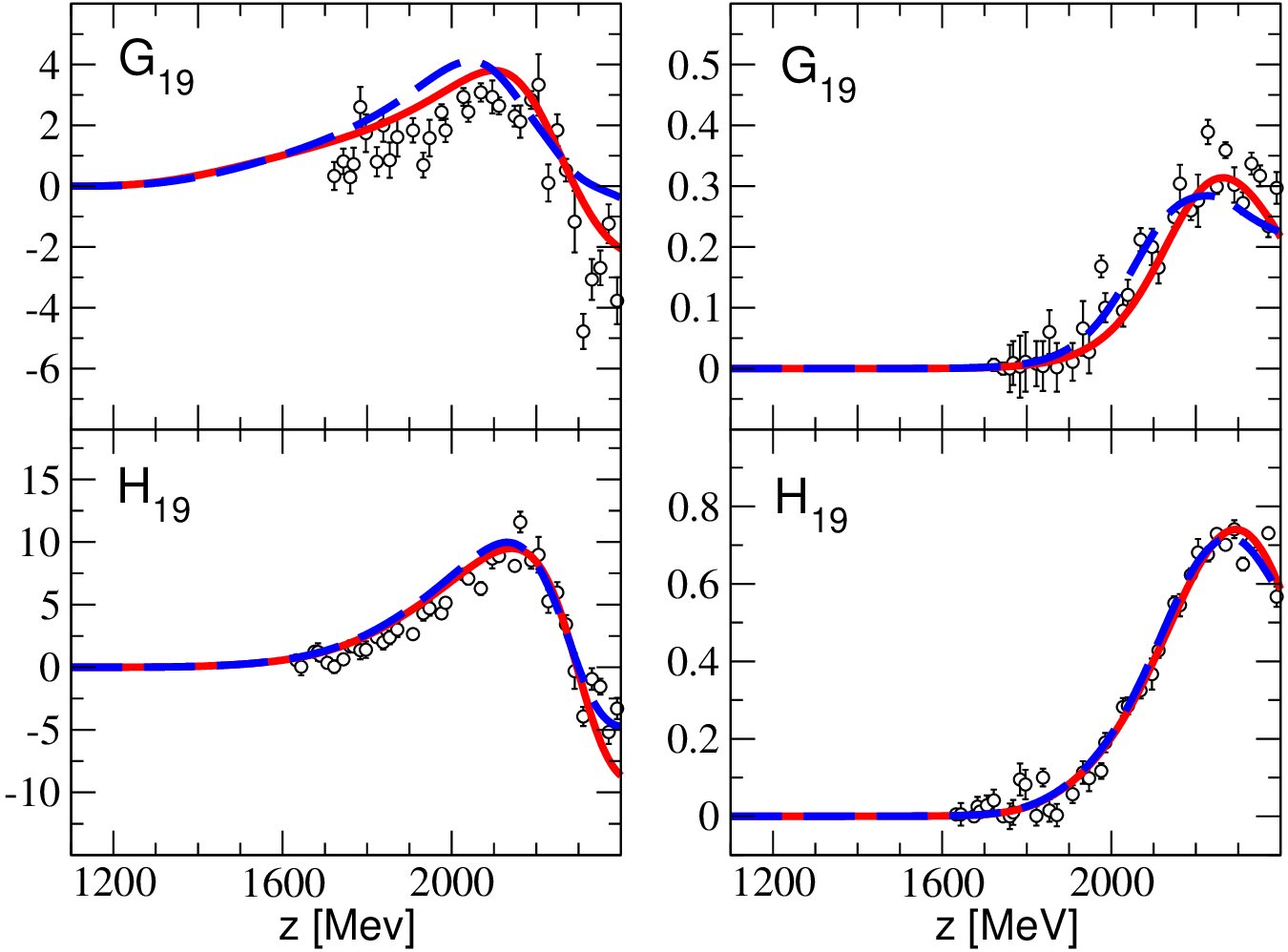}
\end{center}
\caption{Elastic $\pi N\to \pi N$ phase shifts $\delta$ [deg] (left) and inelasticities $(1-\eta)^2$ (right) for isospin $I=1/2$, $F$- to
$H$-waves. Points: GWU/SAID partial-wave analysis (single-energy solution) from Ref.~\cite{Arndt:2006bf}; solid (red) lines: fit A; dashed
(blue) lines: fit B.}
\label{fig:pindel2}   
\end{figure}

\begin{figure}
\begin{center}
\includegraphics[width=0.48\textwidth]{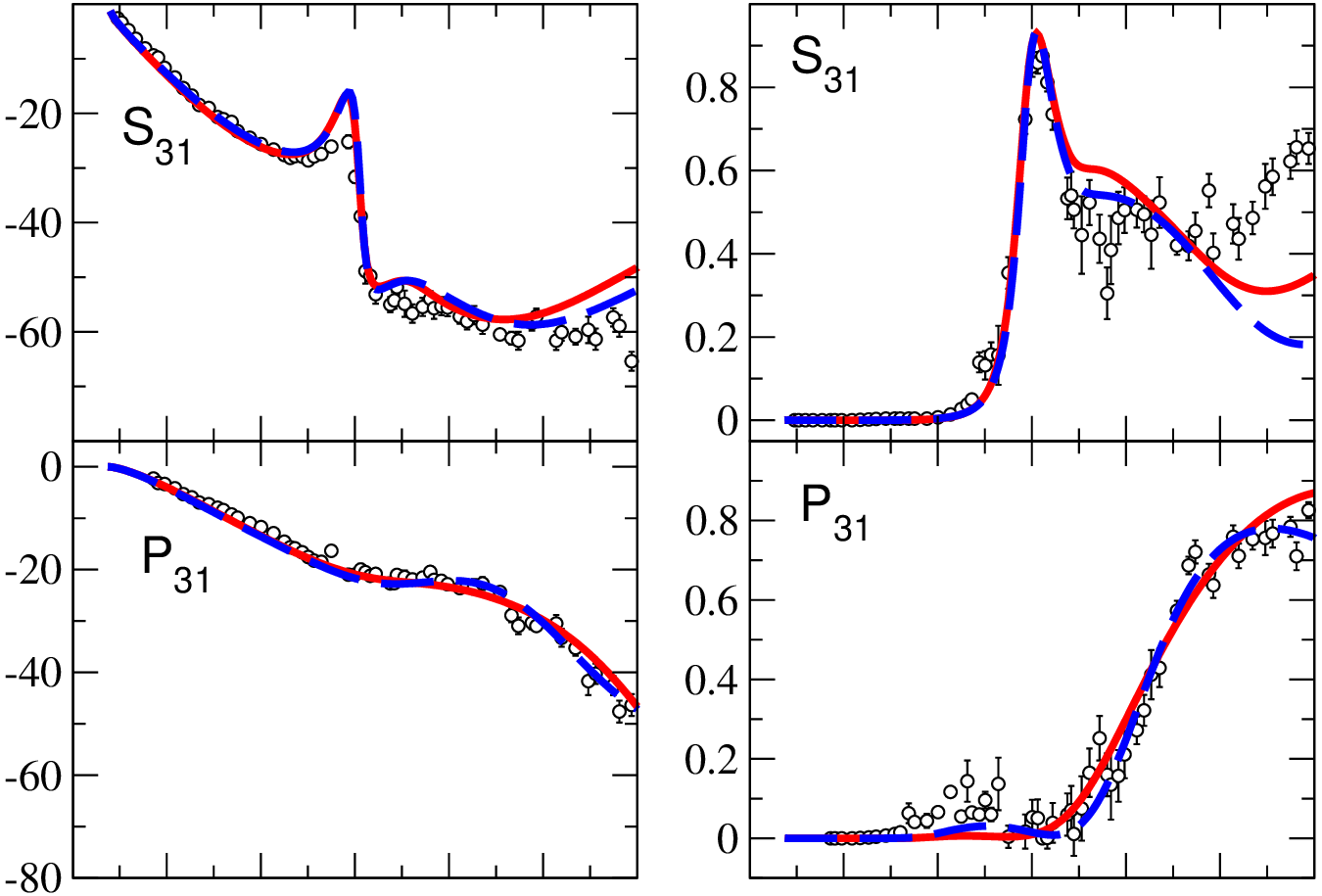}\\
\vspace*{-0.15cm}
\hspace*{-0.05cm}\includegraphics[width=0.482\textwidth]{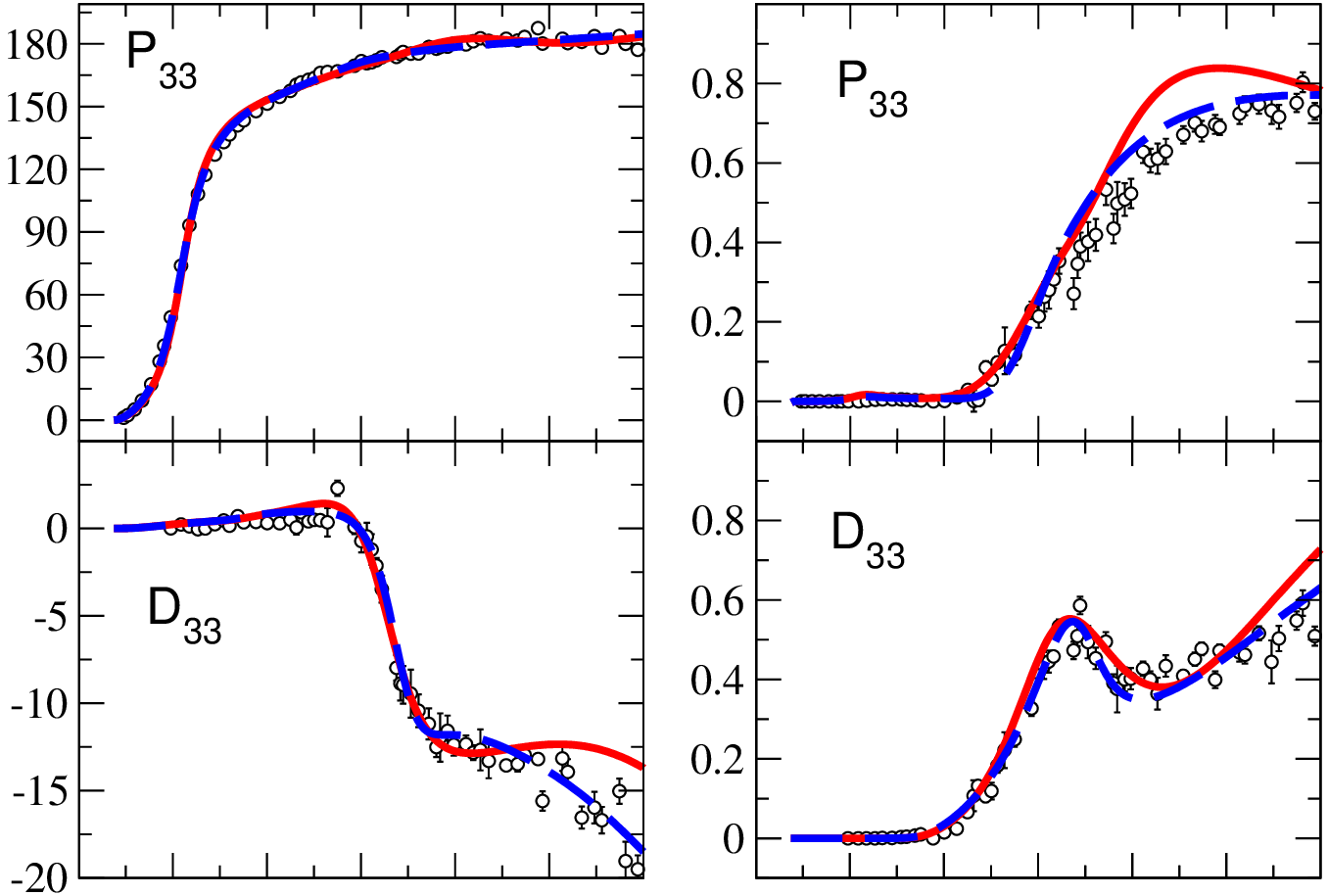}\\
\vspace*{-0.16cm}
\hspace*{0.15cm}\includegraphics[width=0.472\textwidth]{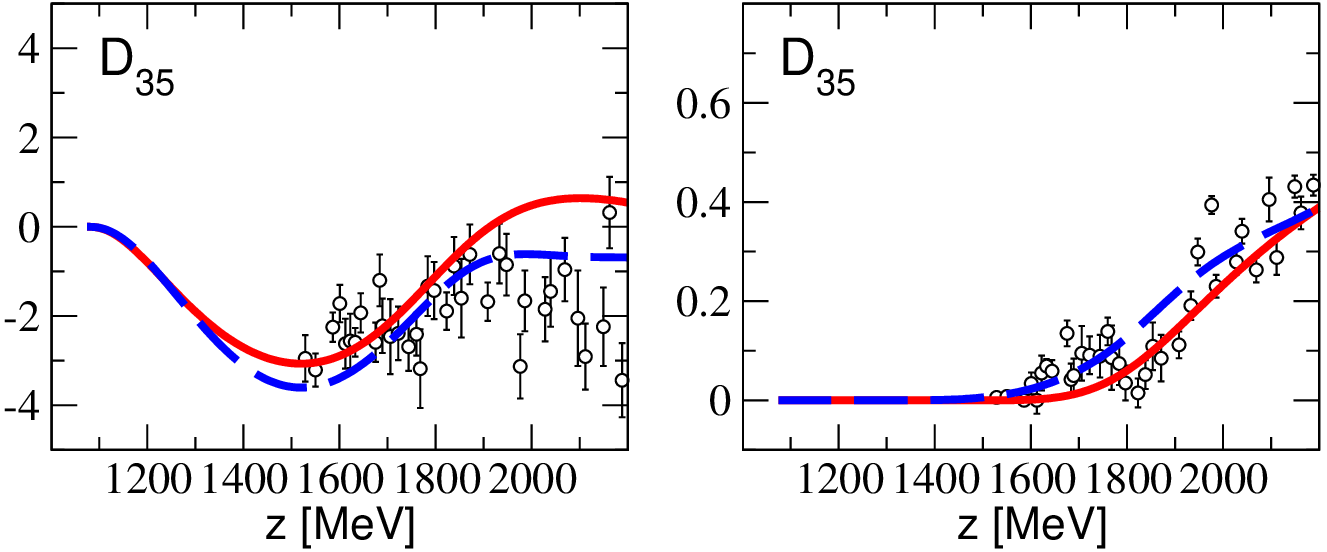}
\end{center}
\caption{Elastic $\pi N\to \pi N$ phase shifts $\delta$ [deg] (left) and inelasticities $(1-\eta)^2$ (right) for isospin $I=3/2$, $S$- to
$D$-waves. Points: GWU/SAID partial-wave analysis (single-energy solution) from Ref.~\cite{Arndt:2006bf}; solid (red) lines: fit A; dashed
(blue) lines: fit B.}
\label{fig:pindel3}     
\end{figure}

\begin{figure}
\begin{center}
\includegraphics[width=0.48\textwidth]{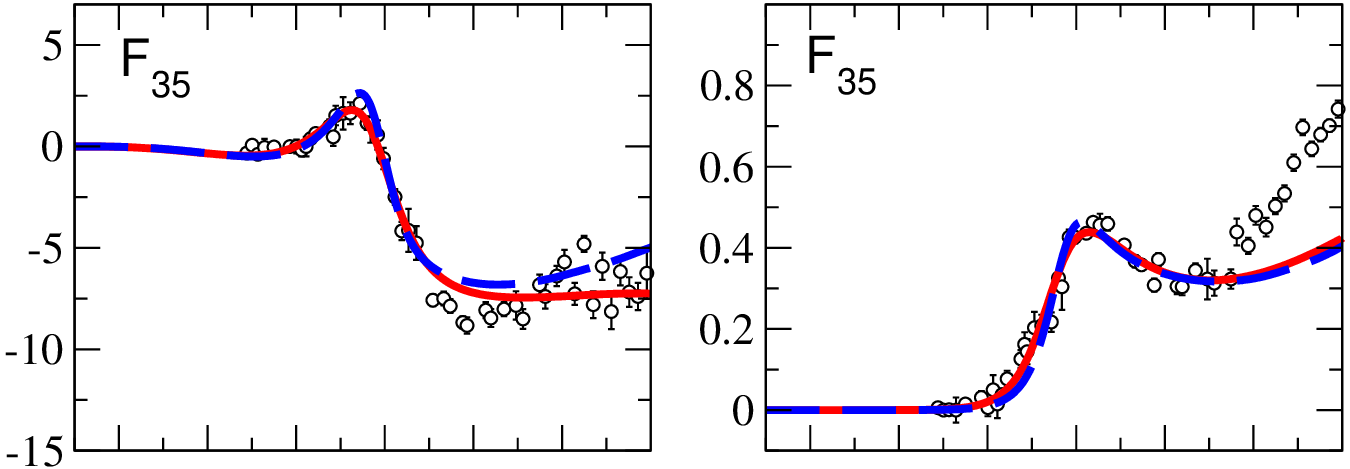}\\
\vspace*{-0.15cm}
\includegraphics[width=0.48\textwidth]{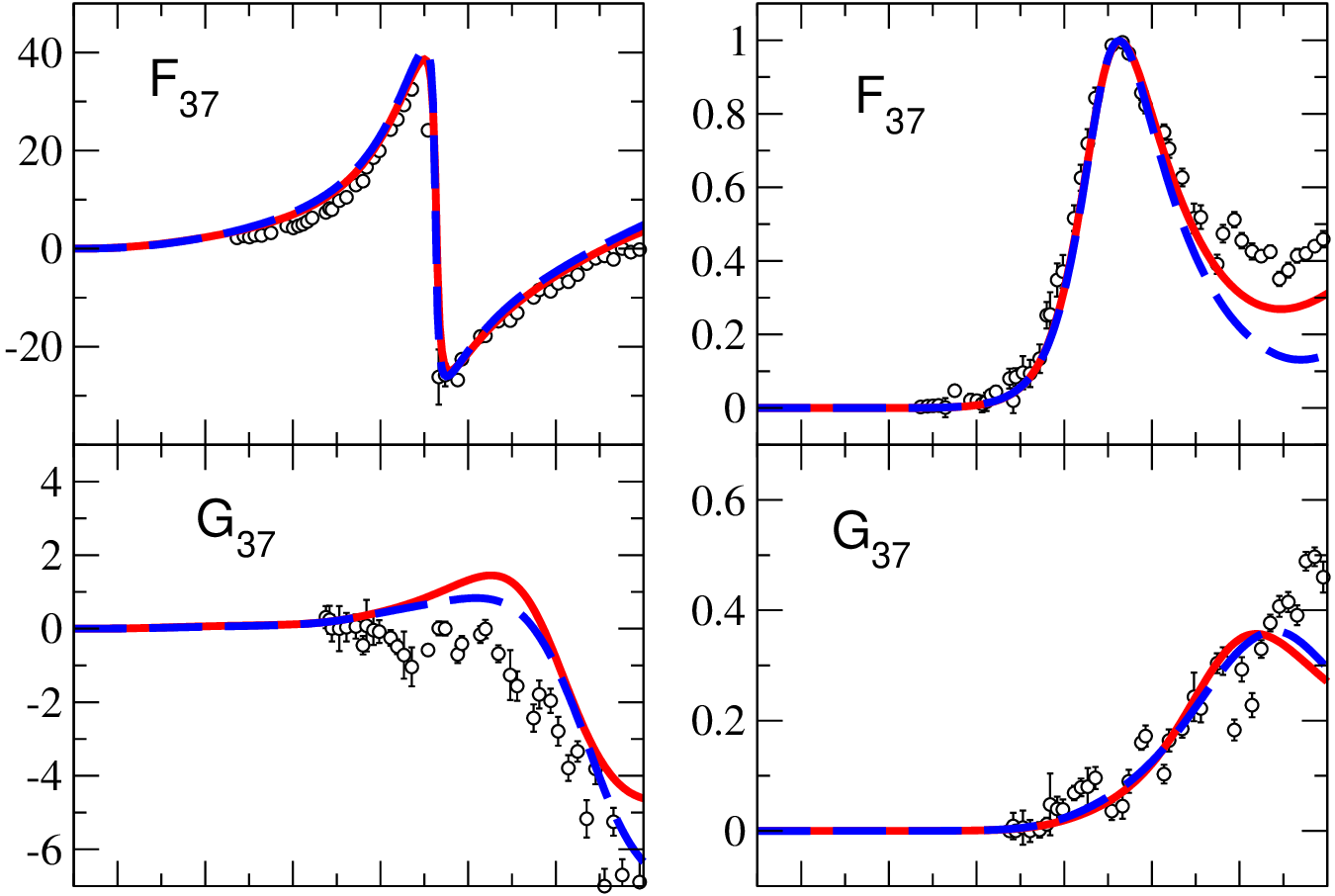}\\
\vspace*{-0.1cm}
\hspace*{0.12cm}\includegraphics[width=0.47\textwidth]{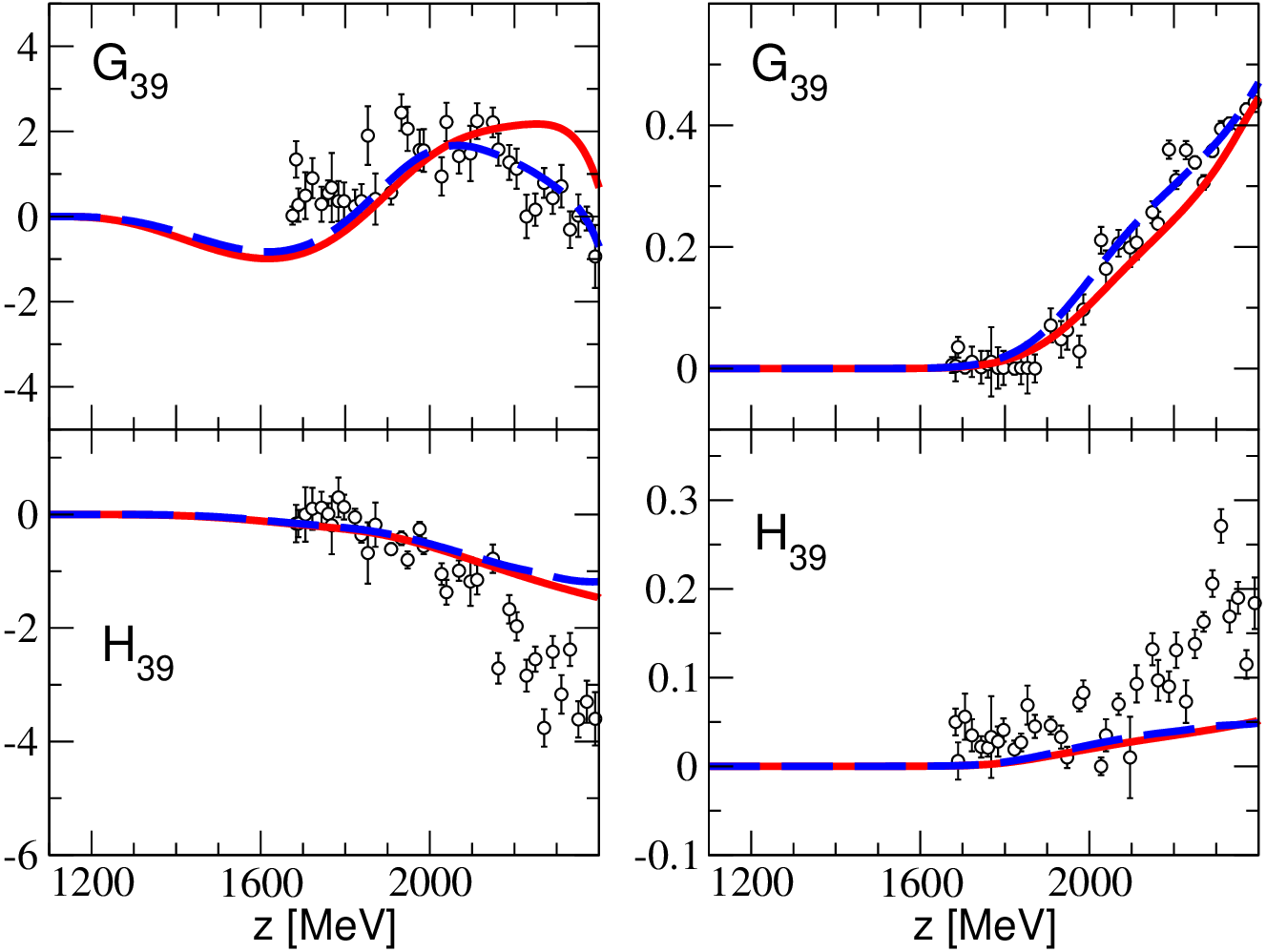}
\end{center}
\caption{Elastic $\pi N\to \pi N$ phase shifts $\delta$ [deg] (left) and inelasticities $(1-\eta)^2$ (right) for isospin $I=3/2$, $F$- to
$H$-waves. Points: GWU/SAID partial-wave analysis (single-energy solution) from Ref.~\cite{Arndt:2006bf}; solid (red) lines: fit A; dashed
(blue) lines: fit B.}
\label{fig:pindel4}     
\end{figure}

In Figs.~\ref{fig:pindel1} to \ref{fig:pindel4} we show the phase shifts and inelasticities for the reaction $\pi N\to \pi N$. The alternative
representation in terms of   partial-wave amplitudes can be found in Sec.~\ref{sec:pin}. There, one can find the discussion on
Figs.~\ref{fig:pindel1} to \ref{fig:pindel4}.
 
%%%%%%%%%%%%%%%%%%%%%%%%%%%%%%%%%%%%%%%%%%%%%%%%%%%%%%%%%%%%%%%%%%%%%%%%%%%%%%%%%%%%%%%%%%%%%%%%%%%%%%%%%%%%%%%%%%%%%%%%%%%%%%%%%%%%%%%%%%%%%%%%

%%%%%%%%%%%%%%%%%%%%%%%%%%%%%%%%%%%%%%%%%%%%%%%%%%%%%%%%%%%%%%%%%%%%%%%%%%%%%%%%%%%%%%%%%%%%%%%%%%%%%%%%%%%%%%%%%%%%%%%%%%%%%%%%%%%%%%%%%%%%%%%%

\end{document}